\documentclass[11pt, subfigure, onehalf]{osudiss-2}

\usepackage{bookmark}
\usepackage{palatino}
\usepackage{fix-cm} 
\usepackage{bm} 
\usepackage[numbers, square, comma, sort&compress]{natbib}

\usepackage[final]{graphicx}
\usepackage{subfigure}

\usepackage{sa-cmds, cancel}
\usepackage{mathtools} 

\hypersetup{plainpages=false,bookmarks=true,breaklinks=true, colorlinks=true, linkcolor=blue}

\newcommand\Regge{{\alpha^{\prime}}}
\newcommand{\ibar}{{\bar{\imath}}}
\newcommand{\jbar}{{\bar{\jmath}}}
\newcommand{\bj}{\bar{\jmath}}
\newcommand{\AdS}{\ensuremath{\text{AdS}}\xspace}
\newcommand{\Oop}{\mathcal{O}}

\newcommand{\YM}{{\scriptscriptstyle\text{YM}}}

\newcommand\lp{\ell_\text{pl}}
\newcommand\Com[2]{\left[#1,\,#2\right]}
\newcommand\bigO{\mathrm{O}}
\newcommand\smallbullet{\ensuremath{\scriptstyle\bullet}}

\newtagform{normalsize}{\normalsize(}{)}
\usetagform{normalsize}

\newenvironment{smalleq}{\begingroup\small\ignorespaces}{%
  \endgroup\ignorespacesafterend}

\author{Steven G. Avery}
\authordegrees{B.S.}
\title{Using the D1D5 CFT to Understand Black Holes}
\unit{Graduate Program in Physics}
\advisorname{Samir D. Mathur} 
\member{Richard Kass}
\member{Yuri Kovchegov}
\member{Stuart Raby}

\begin{document}
\frontmatter

\begin{abstract}
  The bound state of D1-branes and D5-branes in IIB string theory is
  an exceptionally fertile system for the study of black holes. The
  D1D5 system has two dual descriptions: a gravitational and a
  conformal field theory (CFT) description. Here, we focus on using
  the two-dimensional CFT to understand black hole physics.

  After reviewing the D1D5 system, we first show how to perturbatively
  relax the decoupling limit to calculate the emission \emph{out} of
  the AdS/CFT into the asymptotic flat space. We take the effect of
  the neck into account and fix the coupling between the CFT and the
  asymptotic flat space. This calculation is distinguished from other
  AdS--CFT calculations which only work in the strict decoupling limit
  and use the gravitational description to learn about strongly
  coupled field theory.

  We apply the formalism to particular smooth, horizonless
  three-charge nonextremal geometries. In the fuzzball proposal, these
  geometries are interpreted as black hole microstates, but they
  suffer from a classical instability. At first, the instability seems
  problematic in the fuzzball proposal; however, it was argued that if
  one used the D1D5 CFT then the instability could be interpreted as
  precisely the Hawking radiation process for the particular
  microstates. That the instability is classical, and not quantum
  mechanical results from a large Bose enhancement. In this document,
  we perform calculations that confirm this interpretation and
  demonstrate the above emission formalism.

  All of the calculations discussed thus far, and most of the
  calculations in the literature on the D1D5 CFT, are at the
  ``orbifold point'' in moduli space. This point is far from the black
  hole physics of interest, but some calculations agree anyway. To
  understand black holes better it seems likely that moving off of the
  orbifold point will become necessary. We present several
  calculations demonstrating the effect of a single application of the
  marginal deformation operator that moves the D1D5 CFT off its
  orbifold point.  The deformation operator twists two copies of the
  orbifold CFT, which we show produces a ``squeezed state'' with an
  arbitrary number of excitations. Thus, initial high-energy
  excitations can fragment into many low-energy excitations. This
  deformation, should give rise to thermalization and other important
  black hole dynamics.

  Finally, we close with a brief summary and mention some
  opportunities for future work.
\end{abstract}

\dedication{To my grandmother, for telling me that integrals are ``fun.''}
\begin{acknowledgments}

  I am indebted to numerous people who have provided encouragement,
  discussion, and enlightenment throughout my academic career. Perhaps
  most importantly, I am grateful to my advisor, Professor Samir
  Mathur. My understanding has benefited from many conversations about
  physics and mathematics with him. I am grateful to have had the
  opportunity to work with someone with such strong physical intuition
  and insight. On a more personal note, he has helped me navigate
  through the bureaucracies of graduate school and beyond.  I would
  also like to thank Professors Richard Kass, Yuri Kovchegov, and
  Stuart Raby for serving on my candidacy and doctoral committees.
  They and my advisor surely deserve recognition for, if nothing else,
  reading my lengthy candidacy exam paper and this lengthy
  dissertation. Professors Kovchegov and Raby have also had important,
  positive influences on my graduate school experience through their
  classes and informal discussions.

  I am grateful for all of the individuals who have taught me science
  and mathematics over the course of my life. Especially noteworthy
  are my high school math teacher, Mr.~Wells, from whom I learned
  trigonometry, calculus, combinatorics, and some linear algebra; my
  first physics teacher, Mr.~Olivieri, who taught me to think deeply
  about simple things and that I cannot depend on someone else to
  always tell me the answer; my undergraduate advisor, Professor
  Vatche Sahakian, who gave me my first taste of general relativity,
  string theory, and field theory; and my advisor, Professor Samir
  Mathur, whose lectures on particle physics, group theory, general
  relativity, and string theory, I will always cherish. From my time
  as an undergraduate, I am also grateful to Professor John Townsend
  for his encouragement, occasional criticism, and for making me pass
  out of ``Frosh Physics.'' There are numerous others who I have not
  mentioned here whose lectures and teaching have allowed me to
  progress. I would like to thank all of the professors I had at
  Harvey Mudd College; their dedication to undergraduate teaching was
  mostly unappreciated by me at the time---I hope they can forgive my
  occasional nap in class.

  An important part of graduate school is learning to discuss physics.
  I have been privileged with great physics conversations with Archana
  Anandakrishnan, Greg Baker, Iosif Bena, Nikolay Bobev, Konstantin
  Bobkov, Borun Chowdhury, Ben Dundee, Stefano Giusto, Sarang
  Gopalakrishnan, Josh Lapan, Samir Mathur, Jeremy Michelson, Joseph
  Polchinski, Stuart Raby, Yogesh Srivastava, and Anastasios Taliotis.
  Of marked importance in this regard are my collaborators: Borun
  Chowdhury, Samir Mathur, and Jeremy Michelson. While working through
  projects, they have helped my understanding of diverse topics
  immeasurably.

  I should also thank those who read through and offered advice on
  drafts of this dissertation: Borun Chowdhury, Ben Dundee, Sarang
  Gopalakrishnan, Bart Horn, and my wife Elizabeth Avery. Their
  suffering deserves commendation. Others I have consulted about
  various technical points, and I am grateful for their responses: Jan
  de Boer, Borun Chowdhury, Josh Lapan, Emil Martinec, Joseph
  Polchinski, Carlos Tamarit, and Amitabh Virmani. On past projects,
  the results of which are presented here, I or my collaborators have
  also benefited from discussion with Sumit Das, Justin David, Antal
  Jevicki, Yuri Kovchegov, Per Kraus, Oleg Lunin, Emil Martinec, Mohit
  Randeria, Masaki Shigemori, and Yogesh Srivastava. I also extend my
  gratitude to the Kavli Institute for Theoretical Physics (KITP) at
  the University of California, Santa Barbara for supporting me while
  much of this document was written. 
  Relatedly, I am grateful to
  Archana Anandakrishnan and Mike Hinton for being my ``agents'' in
  Columbus, while I was in Santa Barbara. Without their help, I could
  not have fulfilled the graduate school requirements.

  There is more to life than physics, and there are many people that
  have provided support and companionship through my academic journey
  that I have not yet thanked. First and foremost, I am grateful to my
  family. My parents not only foot the bill of my early education, but
  have also provided inestimable encouragement and support throughout
  my life. I am especially grateful to my wife, Elizabeth, whose
  emotional support, encouragement, and patience have sustained me,
  even as she has overcome her own graduate school hurdles.

  I am grateful for the friendship of Jeremy Bridge from high school
  onward. From my time as an undergraduate, I would like to thank the
  rest of ``the triangle'' for their various antics: Eric Malm and
  Jason Murcko. I am also grateful for friendship with Brad Greer, and
  \emph{many} others during my time at Harvey Mudd College. At The
  Ohio State University, I thank all of my officemates; as well as
  Chris Porter, Bill Schneider, Ben Dundee, and Mike Hinton; and the
  rest of the graduate students, for their general camaraderie. I want
  to also thank my KITP officemates and fellow KITP graduate fellows
  for friendship and often asinine conversations: Claudia De Grandi,
  John Biddle, and Sarang Gopalakrishnan. The postdocs at the KITP
  have also been most welcoming.
  
  The work presented here was supported in part by DOE grant DE-FG02-91ER-40690. 
  My stay at the KITP and the writing of this document was supported in part by the National Science Foundation under Grant No.~PHY05-51164.

\end{acknowledgments}

\begin{vita}

\dateitem{January 27, 1983}{Born---Tarzana, CA}
\dateitem{May, 2005}{B.S., Harvey Mudd College, Claremont, CA}
\dateitem{Autumn 2005--Spring 2008}{Graduate Teaching Associate, OSU, Columbus, OH}
\dateitem{Summer 2008--Summer 2010}{Graduate Research Associate, OSU, Columbus, OH}
\dateitem{Autumn 2010}{KITP Graduate Fellow, UCSB, Santa Barbara, CA}

\begin{publist}
\pubitem{S.~G.~Avery and B.~D.~Chowdhury,
   ``Intertwining Relations for the Deformed D1D5 CFT,''
   arXiv:1007.2202 [hep-th].}
\pubitem{S.~G.~Avery, B.~D.~Chowdhury and S.~D.~Mathur,
  ``Excitations in the deformed D1D5 CFT,''
  JHEP {\bf 1006}, 032 (2010)
  [arXiv:1003.2746 [hep-th]].}
\pubitem{S.~G.~Avery, B.~D.~Chowdhury and S.~D.~Mathur,
  ``Deforming the D1D5 CFT away from the orbifold point,''
  JHEP {\bf 1006}, 031 (2010)
  [arXiv:1002.3132 [hep-th]].}
\pubitem{S.~G.~Avery and B.~D.~Chowdhury,
  ``Emission from the D1D5 CFT: Higher Twists,''
  JHEP {\bf 1001}, 087 (2010)
  [arXiv:0907.1663 [hep-th]].}
\pubitem{S.~G.~Avery, B.~D.~Chowdhury and S.~D.~Mathur,
  ``Emission from the D1D5 CFT,''
  JHEP {\bf 0910}, 065 (2009)
  [arXiv:0906.2015 [hep-th]].}
\pubitem{S.~G.~Avery and J.~Michelson,
  ``Mechanics and Quantum Supermechanics of a Monopole Probe Including a Coulomb Potential,''
  Phys.\ Rev.\  D {\bf 77}, 085001 (2008)
  [arXiv:0712.0341 [hep-th]].}
\pubitem{S.~G.~Avery, E.~Malm, and E.~Harley,
  ``The Myth of `The Myth of Fingerprints',''
  The UMAP Journal, Fall 2004, 25(3) 215--230.}
\end{publist}

\begin{fieldsstudy}
\majorfield{Physics}
\end{fieldsstudy}

\end{vita}

\tableofcontents
\clearpage\listoffigures
\clearpage\listoftables

\mainmatter

\chapter{Introduction}\label{ch:intro}

In this chapter we give an overview of the background material, upon
which the rest of the dissertation rests. We begin by reviewing the
basic motivations for and difficulties of quantum gravity, which leads
us to study of black holes in string theory. We give a very basic
introduction to string theory. We then discuss the fuzzball proposal
for the interior structure of black holes. We conclude the
introduction with an outline of the rest of the remaining chapters.

The material presented here is probably insufficient for a complete
understanding of the topics covered, but cites the relevant literature
and serves as review that focuses on the issues pertinent here. It
also serves as an apologia for string theory, which has received some
negative attention in recent years.

\section{Why Quantum Gravity?}\label{sec:whyQG}

The body of this dissertation is framed within the general context of
string theory, which we review in Section~\ref{sec:strings}. Before
jumping into technical details, however, we must first ask: Do we
\emph{really} need a theory of quantum gravity?  One may ask this
question in two ways. First, the pragmatic question of what value such
a theory would have; it almost certainly will not affect the daily
life of a lay person, or even of most physicists---most likely there
will be no experiments in the near future that are able to rigorously
test the details of such a theory.  Second, seeing as even
gravitational waves have not yet been
observed~\cite{gravitational-waves}, one might wonder whether gravity
actually \emph{is} quantized in Nature.  Indeed, Feynman entreats us
to ``keep an open mind''~\cite{feynman-gravity}:
\begin{quote}
  It is still possible that quantum theory does not absolutely
  guarantee that gravity \emph{has} to be quantized. I don't want to
  be misunderstood here---by an open mind I do not mean an empty
  mind---I mean that perhaps if we consider alternative theories which
  do not seem \emph{a priori} justified, and we calculate what things
  would be like if such a theory were true, we might all of a sudden
  discover that's the way it really is!...In this spirit, I would like
  to suggest that it is possible that quantum mechanics fails at large
  distances and for large objects.

\hfill \textit{Feynman, 1962}
\end{quote}
Despite this admonition, there are sundry reasons to believe that
gravity \emph{is} actually quantized, and furthermore that
understanding quantum gravity is of central importance in our ongoing
quest to understand Nature. Classical general relativity leaves us
with many important questions that one would hope quantum gravity
could answer: most prominently, the nature of the initial Big Bang
singularity and the cosmological constant problem. A successful
treatment of these problems with quantum gravity would be of
fundamental importance to the cosmological understanding of the
universe, and might have important astrophysically observable signals.
Moreover, classical general relativity and quantum field theory are
inconsistent without important quantum gravitational corrections.

\subsection{Singularities}

We address both the pragmatic form and the more fundamental form of
the ``why quantum gravity'' question simultaneously. It is
well-established that general relativity is a good description of
classical gravity. In fact, general relativity is the most reasonable
theory of gravitation one could consider that is consistent with
special relativity and the simplest observations. That we get static
gravitational fields implies that gravity is mediated by an
integer-spin bosonic field; that gravity acts over long ranges
suggests that gravity is mediated by a massless field.  Odd-integer
spin fields lead to forces that are both attractive and repulsive.
Since gravity is universally attractive we should consider even-spin
fields. A spin zero field could only couple to the trace of the
stress--energy tensor ${T^\mu}_\mu$ and thus could not couple to the
photon, which has traceless stress--energy tensor.  The fact that
light is deflected by gravity is now well-established.

General relativity is a very successful theory of gravity; however,
there are some issues within the theory that may make one
uncomfortable. Most prominently there are solutions that have
singularities---regions of infinite curvature. The most famous example
is the Schwarzschild black hole with metric (with Newton's constant $G
= 1$)
\begin{equation}\label{eq:schwarzschild}
ds^2 = -\left(1-\frac{2M}{r}\right)dt^2 + \frac{dr^2}{1-\frac{2M}{r}} + r^2d\Omega_2^2,
\end{equation}
where $M$ is the mass of the black hole and $d\Omega_2^2$ is the
differential angular area element. At the Schwarzschild radius $r=2M$,
the coordinates break down, but the underlying spacetime manifold is
perfectly smooth as is clear in other coordinizations. At $r=0$,
however, the geometry is singular, as can be seen by considering a
coordinate invariant measure of curvature,
\begin{equation}\label{eq:Schwarzschild-R}
R^{\mu\nu\lambda \rho}R_{\mu\nu\lambda\rho} = \frac{48 M^2}{r^6}.
\end{equation}
An in-falling observer can reach the geometric singularity at $r=0$ in
finite proper time. This means that an adventurous physicist could
reach the singularity and perform experiments (if he or she could
withstand the tidal forces). The results of such experiments are at
best ill-defined.  The singularity has one saving grace: it is
censored by the event horizon at $r=2M$. The adventurous physicist
could never publish the results of the experiments in a journal
distributed outside the black hole. See
Figure~\ref{fig:penrose-collapse} for a Penrose diagram of spherically
symmetric matter collapsing into a black hole. Note that the
singularity is hidden behind the horizon in dashed red.

One might not be very troubled that singular solutions to Einstein's
equations exist. Perhaps they are never realized in Nature, or perhaps
we should simply discard them as unphysical. Unfortunately, this is a
difficult position to hold. Hawking and Penrose~\cite{hawking-penrose,
  hawking-ellis} demonstrated that singularities arise in general
relativity from perfectly reasonable smooth initial conditions. For
instance, if one starts with a cloud of pressureless dust, then it
will collapse into a
singularity~\cite{oppenheimer-snyder}.\footnote{In a realistic case,
  one expects matter pressure to play a role; however, there is no
  obstacle to making the cloud massive enough to overcome the
  pressure.}  Again, the only consolation would be that some form of
cosmic censorship holds, and all the singularities are hidden behind
event horizons so that we don't have to worry about ``seeing'' the
breakdown of physics.

Even appropriately censored, these singularities are philosophically
troublesome. This is not an esoteric concern, since there is strong
evidence for the existence of astrophysical black holes in
Nature~\cite{blackholes-exist}, and furthermore there is an uncensored
singularity at the original Big Bang event in the
Friedmann--Robertson--Walker--Lema\^{\i}tre metric that best describes
our universe. The metric describing a homogeneous, isotropic and flat
cosmology is~\cite{wald}
\begin{equation}
ds^2 = -d\tau^2 + a^2(\tau)(dx^2 + dy^2 + dz^2),
\end{equation}
with scale factor $a(\tau)$. The exact form of $a(\tau)$ depends on
the matter content of the universe; however, one typically finds that
$a(\tau)$ is proportional to some positive power of $\tau$. From which
it follows that the Ricci scalar diverges at the origin of the
universe, $\tau = 0$:
\begin{equation}
  R = 6\left(\frac{\ddot{a}}{a} + \frac{\dot{a}^2}{a^2}\right)\sim \frac{1}{\tau^2}.
\end{equation}

We find, then, that we cannot simply turn a blind eye to the
singularities that arise in general relativity. Instead it seems that
general relativity is predicting its own downfall, and forcing us to
consider quantum gravity. The unique length scale one can form from
the fundamental constants $G$, $\hbar$, and $c$
\begin{equation}
\lp = \sqrt{\frac{G \hbar}{c^3}},
\end{equation}
the Planck length, is the natural length scale for a theory of quantum
gravity. Thus, when the curvature becomes on the order of $R\sim
1/\lp^2$, we expect large quantum gravity corrections.  Presumably,
any theory of quantum gravity should resolve the singularities that
arise in general relativity (at least those singularities that arise
from smooth initial data~\cite{horowitz-sing}).

There is another cosmological question on which a theory of quantum
gravity might provide insight.  For the Big Bang, we may also desire a
theory of initial conditions.  If a theory of quantum gravity can make
robust physical predictions about the very early universe, then it can
have observable consequences for current experiments measuring the
cosmic microwave background or, alternatively, help explain the
smallness of the cosmological constant~\cite{weinberg-cc}. Finally, we
should note that the most successful theory of cosmology that explains
the homogeneity of the observed universe, namely inflation, is a
quantum gravity effect.

\subsection{Quantum Fields on Classical Geometry}

Above, we argue that the singularities arising in general relativity
encourage the study of quantum gravity. Not only are they
philosophically problematic, but the singularities are also an avenue
for quantum gravity to become an experimental (or at least
observational) field of physics. The above problems manifest
themselves entirely within classical general relativity, and should
presumably be resolved by a theory of quantum gravity.

We now consider what happens when we put quantum fields on curved
background geometrie, which more dramatically illustrates the need for
a theory of quantum gravity. Let us first comment, however, that it is
probably not consistent to think of some aspect of Nature as
fundamentally classical, while the rest is fundamentally quantum.  If,
for example, gravity were classical, it would be difficult to imagine
what gravitational field arises due to an electron that is in a
quantum superposition of different positions. Suppose, for instance,
we claimed that Einstein's equations were
\begin{equation}
G_{\mu\nu} = 8\pi \vev{T_{\mu\nu}},
\end{equation}
then we would get curvature originating from the different superposed
positions. What happens when we make a measurement? The metric must
jump acausally and nonlocally to a new configuration with curvature
originating from only one location~\cite{feynman-gravity, wald}.  If
we insist gravity is classical, then it seems we are forced to discard
quantum mechanics as fundamental. This tension becomes greater in the
sequel.

As we discuss above, black holes are a robust prediction of general
relativity, and there is evidence that they are realized in Nature.
Above, our main concern was the geometric singularity where the
curvature diverges---we now consider the event horizon.  In the
presence of a horizon, Hawking~\cite{hawking-1, hartle-hawking}
demonstrated that black holes are unstable and emit black-body
radiation at infinity.  The energy of the emitted particles is
compensated by negative energy particles that travel into the black
hole, decreasing its mass.  Thus, black holes are quantum mechanically
unstable and will eventually decay, albeit slowly. For instance,
Page~\cite{page-emission} estimates that a solar mass black hole takes
$10^{66}$~years to completely evaporate, which is many many orders of
magnitude longer than the age of the universe (on the order of
$10^{10}$~years).

\subsubsection{Computing Hawking Radiation}

\begin{figure}[!ht]
\begin{center}
\includegraphics[width=8cm]{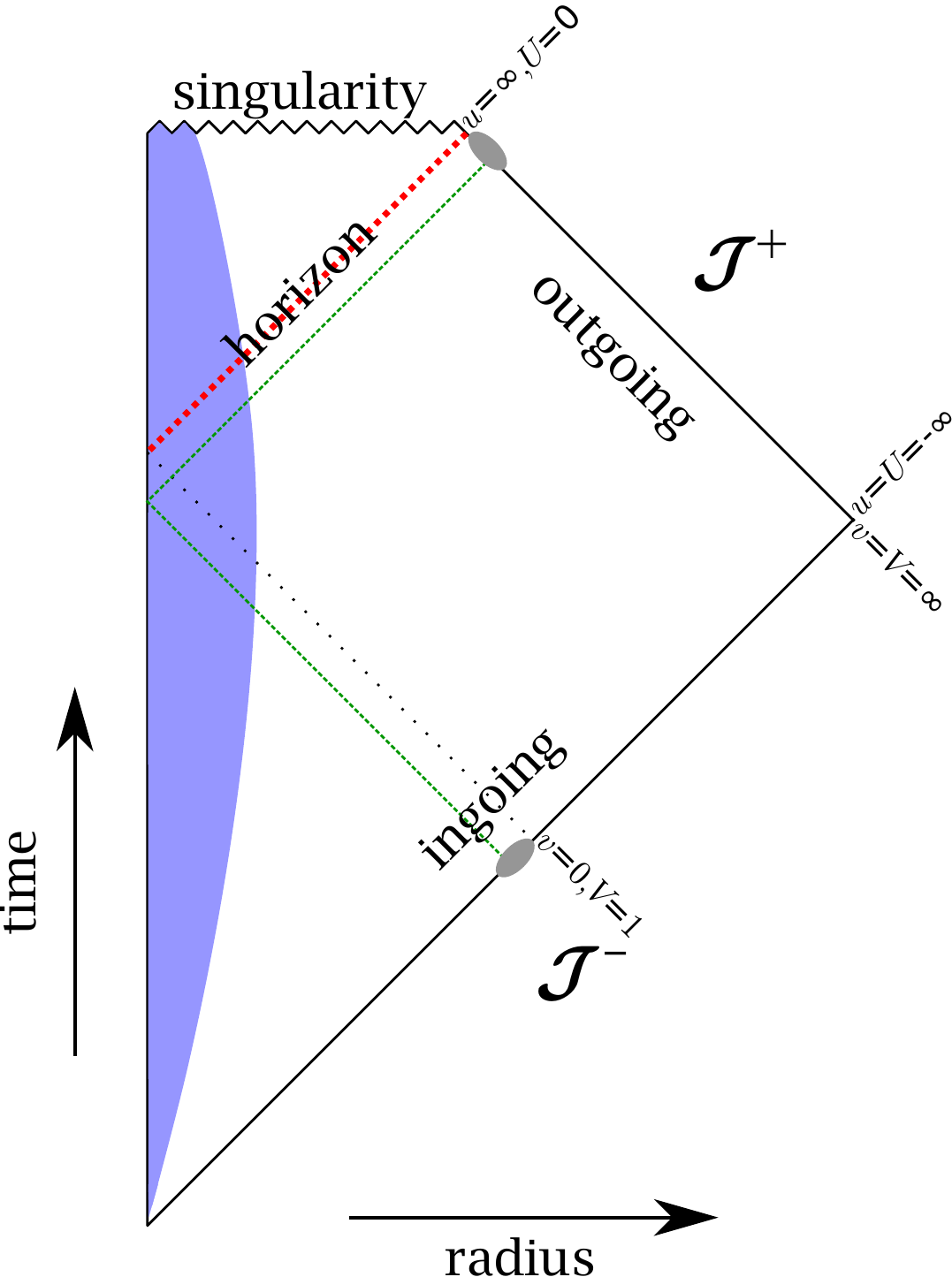}
\caption[Penrose diagram for black hole formation]{The Penrose diagram
  for some matter (light blue) collapsing into a Schwarzschild black
  hole. Outside of the matter, the metric is Schwarzschild. The
  Penrose diagram does \emph{not} correctly indicate \emph{distances}
  between spacetime events; only \emph{angles} are correctly
  indicated. Null geodesics are lines at $\pm 45^\circ$, like the
  green dashed curve illustrating the connection between coordinates
  on future null infinity, $\mathcal{J}^+$, and past null infinity,
  $\mathcal{J}^-$. The gray shading on $\mathcal{J}^+$ and
  $\mathcal{J}^-$ indicates the region where geometric optics is a
  good approximation.}\label{fig:penrose-collapse}
\end{center}
\end{figure}

Let us sketch the essential details needed to understand Hawking
radiation.  Consider a noninteracting, massless, minimally-coupled
scalar field, $\phi$, in the background geometry corresponding to the
formation of a Schwarzschild black hole. The action for $\phi$ is
\begin{equation}
S = \int \drm^4 x\sqrt{-g} g^{\mu\nu}\pd_\mu\phi\pd_\nu\phi,
\end{equation}
where $g_{\mu\nu}$ is the spacetime metric. The Penrose diagram for
the collapse geometry is depicted in
Figure~\ref{fig:penrose-collapse}. Outside of the collapsing matter,
the metric is Schwarzschild, as in Equation~\eqref{eq:schwarzschild}.
For this calculation, we do not need the details of the metric inside
the collapsing matter, which depends on what the matter is.

The equations of motion for the field $\phi$ are given by
\begin{equation}\label{eq:kg-eq}
\square\phi = \frac{1}{\sqrt{-g}}\pd_\mu(\sqrt{-g}g^{\mu\nu}\pd_\nu\phi) = 0.
\end{equation}
For simplicity, we consider only the $s$-wave emission by imposing the ansatz
\begin{equation}
  \phi = \frac{f(r)e^{-i\omega t}}{r}.
\end{equation}
The $s$-wave dominates the radiation spectrum, making this also a
physically reasonably assumption~\cite{hawking-2}. Plugging into the
equations of motion one finds that $f$ must satisfy
\begin{equation}\label{eq:schwarzschild-wave-r}
\frac{r-2m}{r^2}\der{}{r}\left[r(r-2m)\der{}{r}\left(\frac{f(r)}{r}\right)\right] 
           + \omega^2 f(r) = 0.
\end{equation}
We are interested in finding how plane wave solutions near the horizon
evolve into plane wave solutions near asymptotic infinity. Since the
equations of motion are linear, there is a linear relationship between
these two sets of modes. The connection defines a Bogolyubov
transformation between the particle modes near the horizon and the
asymptotic infinity particle modes.

To improve the asymptotic behavior of
Equation~\eqref{eq:schwarzschild-wave-r}, we introduce the
``tortoise'' radial coordinate
\begin{equation}\label{eq:tortoise}
r_* = r + 2 m \log\left|\frac{r-2m}{2m}\right|,
\end{equation}
which gives equations that may be written in the form
\begin{equation}\label{eq:f-tortoise}
f''(r_*) + \left[\omega^2-\frac{2m}{r^3}\left(1-\frac{2m}{r}\right)\right]f(r_*) = 0.
\end{equation}
In the above, $r$ is an implicit function of $r_*$ given in
Equation~\eqref{eq:tortoise}. For both asymptotic infinity $r\approx
\infty$ and near-horizon $r\approx 2m$, Equation~\eqref{eq:f-tortoise}
simplifies and the field behaves as
\begin{equation}
\phi \propto \frac{e^{-i\omega (t \pm r_*)}}{r}.
\end{equation}
Let us define null tortoise coordinates, $u$ and $v$, as
\begin{equation}
u = t - r_*\qquad 
v = t + r_*.
\end{equation}
The null surface $\mathcal{J}^-$ has $u=-\infty$ and is parameterized
by $v$. We define $v=0$ as the point on $\mathcal{J}^-$ connected to
the point where the horizon forms by a null curve. See
Figure~\ref{fig:penrose-collapse}. The null surface $\mathcal{J}^+$
has $v=\infty$ and is parameterized by $u$.

In a spacetime with no global timelike Killing vector, such as that of
a black hole, there is no canonical notion of particle~\cite{wald,
  fulling}.  There are two useful ways to describe the quantum
mechanics of the scalar field. First, we can think about the
``ingoing'' Hilbert space which is defined along the null surface
$\mathcal{J}^-$. This Hilbert space defines our in-states.  We define
the ingoing Hilbert space by expanding $\phi$ in terms of modes on
$\mathcal{J}^-$. Near $v=0$, the expansion takes the form
\begin{equation}
\phi_\omega^{(\text{in})} \sim a_\omega \frac{e^{-i\omega (t+r_*)}}{\sqrt{2\omega}r} 
                             + a_\omega^\dg \frac{e^{i\omega(t+r_*)}}{\sqrt{2\omega}r}.
\end{equation}

The second description of $\phi$ is in terms of the direct product of
an ``outgoing'' Hilbert space and a horizon Hilbert space. The
outgoing Hilbert space is defined in terms of modes along the null
surface $\mathcal{J}^+$, whereas the horizon Hilbert space is defined
in terms of modes along the horizon. We do not need the details of the
horizon here.  Near the horizon, we can expand $\phi$ in terms of
``outgoing modes'' as
\begin{equation}
\phi_\omega^{(\text{out})} \sim b_\omega \frac{e^{-i\omega (t-r_*)}}{\sqrt{2\omega}r} 
                              + b_\omega^\dg \frac{e^{i\omega (t-r_*)}}{\sqrt{2\omega}r}.
\end{equation}

We wish to consider the situation when there are no incoming
particles. This means we want the ingoing vacuum. What we find is that
the ingoing vacuum has nonzero overlap with excited states of the
outgoing Hilbert space. How do we see this? We need to understand how
the ingoing modes at infinity are related to the outgoing modes near
the horizon; we need to relate the $a$ and $a^\dg$'s to the $b$ and
$b^\dg$'s. What we can do is consider an outgoing mode
$\phi_\omega^{(\text{out})}$ and use the equation of
motion~\eqref{eq:kg-eq} to evolve it from the near-horizon outgoing
region into the ingoing region in Figure~\ref{fig:penrose-collapse}.
We find that each outgoing mode evolves into a linear combination of
the ingoing modes, with the coefficients telling us the exact relation
between $a$'s and $b$'s.

The inner product on the ingoing Hilbert space can be stated in terms
of the functions $f$:
\begin{equation}\label{eq:Jm-measure}
\int_{-\infty}^\infty\drm v\, f_\omega(v) f_{\omega'}^*(v) 
 = \int_{-\infty}^\infty\drm v\, \frac{e^{-i(\omega- \omega')v} }{2\sqrt{\omega\omega'}}
 = 2\pi\frac{\delta(\omega-\omega')}{2\omega}.
\end{equation}
We wish to consider an outgoing mode near the horizon on $\mathcal{J}^+$,
\begin{equation}
g_\omega(u) = \frac{e^{-i\omega u}}{\sqrt{2\omega}},
\end{equation}
evolve it backwards in time to the null surface $\mathcal{J}^-$, and
compute the overlap of that function with the in-modes. To evolve the
outgoing mode backwards, we use the geometric optical
approximation---that is, we assume that light rays travel on null
geodesics. As we approach the horizon, the frequency of the outgoing
mode increases making the approximation better and better. In null
Kruskal coordinates,
\begin{equation}
U = -e^{-\frac{u}{4m}}\qquad
V = e^{\frac{v}{4m}},
\end{equation}
and the metric becomes
\begin{equation}
ds^2 = -16m^2 e^{-\frac{r}{2m}}dUdV + r^2d\Omega_2^2.
\end{equation}
From this form, it is clear that null geodesics in the $r$--$t$ plane
are curves of constant $U$ or constant $V$. Let us trace an outgoing
particle at coordinate $u$ on $\mathcal{J}^+$ backward onto a point
$v$ on $\mathcal{J}^-$, following the path shown in
Figure~\ref{fig:penrose-collapse}.  From $\mathcal{J}^-$ we follow a
null geodesic with constant $U$ until we reach $r=0$, then we follow a
null geodesic with constant $V$ from $r=0$ until we reach $v$. This
gives the relation
\begin{equation}
(1-U)V = 1\quad \Longrightarrow \quad v = -4m \log\left(1 + e^{-\frac{u}{4m}}\right)
  \approx -4me^{-\frac{u}{4m}},
\end{equation}
near the horizon. Hence,
\begin{equation}
g_\omega(u)\longrightarrow g_\omega(v) \approx
\begin{cases}
0 & v>0\\
\frac{1}{\sqrt{2\omega}}e^{i4m\omega\log(-\frac{v}{4m})} & v < 0,
\end{cases}
\end{equation}
where it should be clear from Figure~\ref{fig:penrose-collapse} that
propagating an outgoing mode onto $\mathcal{J}^-$ can only give
negative $v$. Let us define functions $\alpha_{\omega\omega'}$ and
$\beta_{\omega\omega'}$ that describe the overlap of $g_\omega$ with
the $f_\omega(v)$ on $\mathcal{J}^-$:
\begin{equation}
g_\omega(v) = \int_0^\infty\frac{\drm \omega'}{2\pi}
  \left(\alpha_{\omega\omega'}f_{\omega'}(v) + \beta_{\omega\omega'}f^*_{\omega'}(v)
    \right).
\end{equation}
We can compute the coefficients $\alpha$ and $\beta$ from the measure
in Equation~\eqref{eq:Jm-measure} to find~\cite{hawking-2}
\begin{equation}\begin{aligned}
\alpha_{\omega\omega'} &\approx \sqrt{\frac{\omega'}{\omega}}
                \int_0^\infty\drm v\, e^{-i\omega' v} \left(\frac{v}{4m}\right)^{i4m\omega}
                 = -\frac{i}{\sqrt{\omega\omega'}}(i4m\omega')^{-i4m\omega}\Gamma(1+i4m\omega)\\
\beta_{\omega\omega'} &\approx \sqrt{\frac{\omega'}{\omega}}
                \int_0^\infty\drm v\, e^{i\omega' v} \left(\frac{v}{4m}\right)^{i4m\omega}
                = \frac{i}{\sqrt{\omega\omega'}}(-i4m\omega')^{-i4m\omega}\Gamma(1+i4m\omega).
\end{aligned}\end{equation}
We see from the above that
\begin{equation}
\beta_{\omega\omega'} = -e^{-4\pi m \omega}\alpha_{\omega\omega'}
\end{equation}
The coefficients tell us that a lowering operator $a$ becomes a linear
combination of raising and lowering operators $b$. The transformation
we have outlined above that relates $a$ and $a^\dg$ operators on the
incoming Hilbert space to $b$ and $b^\dg$ operators on the outgoing
Hilbert space is an example of a Bogolyubov transformation. It follows
from the above~\cite{hawking-1, hawking-2} that if we start with the
incoming vacuum, the expectation value of the number of outgoing
particles is given by
\begin{equation}
N_\omega = {}_a\bra{0}b_\omega^\dg b_\omega\ket{0}_a 
  \sim \frac{1}{\frac{|\alpha|^2}{|\beta|^2} - 1}
  \sim \frac{1}{e^{8\pi m\omega} -1}.
\end{equation}
Thus, we conclude that if we start with the vacuum in the background
of a Schwarzschild black hole, then we end with a Planckian
distribution of outgoing particles corresponding to a black body with
temperature $T = 1/(8\pi m)$. We have skipped over some important
details in this discussion, especially with regard to what happens
with the horizon Hilbert space. A more careful treatment allows one to
see that energy is conserved by negative energy particles falling into
the black hole. See~\cite{hawking-1, hawking-2, hartle-hawking, wald}
for more complete discussions. Let us also comment, as emphasized
in~\cite{visser}, that the Hawking process is quite general and only
requires that there be a future horizon.\footnote{In fact, an
  experimental group has claimed to observe the Hawking process in an
  optical (non-gravitational) analogue~\cite{hawking-in-lab}.}

Prior to Hawking's calculation, Bekenstein~\cite{bekenstein} argued
that black holes have an entropy proportional to the horizon area.
When combined with Hawking's calculation, one fixes the constant of
proportionality and finds the Bekenstein--Hawking entropy,
\begin{equation}\label{eq:SBH}
S_\text{BH} = \frac{A}{4}.
\end{equation}
So even without knowing the microscopic degrees of freedom, we can
indirectly argue that black holes are thermodynamical objects with
temperature and entropy.

\subsubsection{The Information Paradox}\label{sec:info-paradox}

Having derived that black holes emit radiation, let us consider the
following process. We start with a bunch of matter in some quantum
state $\ket{\psi}_\text{in}$ that undergoes gravitational collapse into
a black hole. The black hole then emits radiation thermally until it
entirely evaporates. All that remains of the state
$\ket{\psi}_\text{in}$, then, is thermal radiation---a mixed state
that can only be described by a density matrix!  Thus, we see an
apparent breakdown in quantum mechanics.  Since any black hole formed
by gravitational collapse would not have evaporated in the age of the
universe, one might feel that this is not a problem; however, it may
still be possible that small primordial black holes could have
evaporated, or that one could tunnel into a small black hole
state~\cite{page-emission}. Regardless of whether it is actually
realized, there is nothing to forbid this scenario within our best
physical models of the universe and any inconsistencies still indicate
a failure in our understanding of Nature.
 
This breakdown is frequently described as a loss of unitarity, but
that may be misleading. If the time-evolution operator for a quantum
system is non-unitary, then probability is not conserved.  Typically
this means that degrees of freedom are entering or leaving the system.
For example, one can model alpha-decay with a non-Hermitian
Hamiltonian, where the imaginary part of the energy is the decay rate
corresponding to degrees of freedom leaving the system. With Hawking
radiation, however, the generator of time evolution is not described
by some non-Hermitian Hamiltonian. Instead time evolution could only
be described in terms of a density matrix. This is a much more drastic
situation, since all of our most fundamental laws are formulated in
the language of quantum mechanics.

One may not find the information paradox as stated very compelling.
After all, if one were to burn this dissertation, then the emitted
radiation would presumably be described by a Planckian distribution;
however, that would be a coarse-grained description. In fact, the
radiation would result from some unitary evolution of the quantum
mechanical state of all the atoms in the dissertation. In a similar
spirit, one might think that back-reaction or some small quantum
gravitational effect would restore unitary evolution. These and other
effects certainly do alter the thermal distribution, but one can show
that the information paradox is quite robust and has little to do with
the actual distribution of the emitted particles.

The real problem is that the emitted radiation is entangled with
negative energy particles that fall into the black hole, and once the
black hole evaporates the emitted radiation is entangled with
nothing~\cite{harvey-strominger, preskill-destroy, page-paradox,
  giddings-paradox, mathur-hawking, mathur-hawking-old}.
In~\cite{mathur-hawking, mathur-hawking-old}, it was shown that
quantum mechanics breaks down under the very general assumptions of
locality and that quantum gravitational corrections are confined to
the Planck scale.

Thus, if we assume that Nature is fundamentally quantum mechanical and
local, then we are forced to conclude that quantum gravitational
effects (at least in the presence of a black hole) are \emph{not}
confined to the Planck length. This breakdown of dimensional analysis
indicates that some nontrivial physics is at work. The preceding
discussion of singularities and Hawking radiation suggests
understanding quantum gravity can have observable, and fundamentally
important effects.

Some may question the virtues of pursuing a new theory that is
motivated only by theoretical tensions between existing theories, and
not by experimental data. After all, was it not a failure to consult
with experiments that led Aristotelian physics astray? We suggest a
more apt historical analogue would be to special relativity.
Einstein's main motivation for the introduction of special relativity
was probably not the Michelson--Morley experiment, but rather a
theoretical inconsistency between Galilean relativity and Maxwell's
equations~\cite{sr-history}.

\section{Obstacles to a Theory of Quantum Gravity}

Having shown why quantum gravity is important, let us now explain why
it is difficult. While different people may have different
requirements for a successful theory of quantum gravity, there are two
main obstacles to a quantum description of gravity. First, there is a
conceptual difficulty in formulating a field theory where the field
also defines the underlying spacetime~\cite{wald}.  For instance, for
a quantum bosonic field $\hat{\phi}$ on classical metric $g$,
causality demands that spacelike separated measurements be
independent, meaning
\begin{equation}
\Com{\hat{\phi}(x)}{\hat{\phi}(y)} = 0 \qquad (x-y)^2 > 0,
\end{equation}
where $(x-y)^2$ is the spacetime interval, which depends on
$g_{\mu\nu}(x)$. Clearly, we would like a similar statement to hold
for $\hat{g}_{\mu\nu}$ when we quantize the metric; gravitons should
not violate the causality of the underlying spacetime. We might try to
write something like
\begin{equation}
\Com{\hat{g}_{\mu\nu}(x)}{\hat{g}_{\rho\sigma}(y)} = 0 \qquad (x-y)^2 >0, 
\end{equation}
but this is not a sensible statement. Whether or not $x$ and $y$ are
spacelike depends on the measured value of $g_{\mu\nu}$, but the
commutation relation is supposed to be an operator statement,
independent of the eigenstate measured. More broadly, we see that the
notion of casaulity becomes problematic when spacetime is
quantized~\cite{wald}. Another problem that stems from the dual roles
of $g_{\mu\nu}$ arises in describing topology-changing transitions.
Topology changes should arguably be allowed within a theory of quantum
gravity, but a canonical quantization description seems intractable
since we depend on spacelike foliations of spacetime~\cite{wald}.  One
may try a path integral description, but it is problematic to
associate a transition amplitude with it, since it is not clear what
the gauge-invariant observables are~\cite{wald}.

The second major difficulty is that naive quantizations of general
relativity give nonrenormalizable perturbative expansions. For
instance, if we expand the metric about a flat background
$\eta_{\mu\nu}$\footnote{we put in the factor of $G_N$ to canonically
  normalize the action for the field $\tilde{h}$.}
\begin{equation}
g_{\mu\nu} = \eta_{\mu\nu} + \sqrt{G_N}\tilde{h}_{\mu\nu},
\end{equation}
then the action for $\tilde{h}$ in $4$ dimensions is very
schematically given by
\begin{calc}
S_\text{Einstein--Hilbert} &= \frac{1}{16\pi G_N}\int\drm^4 x\sqrt{-g}R \\
 &\sim\int\drm^4 x (\pd \tilde{h}_{\mu\nu}\,\pd \tilde{h}^{\mu\nu} 
         + \sqrt{G_N}(\pd\tilde{h})^2\tilde{h} + G_N(\pd\tilde{h})^2\tilde{h}^2 + \dots).
\end{calc}
We have only kept a few terms, but these are sufficient to argue that
the action is nonrenormalizable. In $4$ dimensions, the coupling $G_N$
has dimensions
\begin{equation}
[G_N] = (\text{length})^{2}.
\end{equation}
Perturbation theory can be thought of as a power series expansion in
the coupling $G_N$. For instance, consider the cross-section for
graviton--graviton scattering. Cross-sections have dimensions of area
and the only other length scale is the center of mass energy, $E$, so
the perturbative expansion must be of the form
\begin{equation}\label{eq:grav-grav-pert}
\sigma \sim G_N\left(a_0 + a_1 (G_N E^2) + a_2 (G_N E^2)^2 + \dots\right),
\end{equation}
where the $a_i$ are dimensionless constants.  At high energies, then,
$\sigma$ violates the Froissart bound, which states that unitarity
requires the cross-section at high energies grow no faster than
$\log^2 E$~\cite{froissart}. One can bypass this simple dimensional
analysis by introducing another mass scale into the problem: either an
explicit cutoff $\Lambda$ or new massive excitations. From the above,
we see that the new excitations should enter at an energy scale less
than roughly $1/\sqrt{G_N}\sim 1/\lp$.  This is precisely what string
theory does. Alternatively, one might hope that the dimensionless
constants $a_i$ vanish due to some symmetry or happy coincidence---as
it happens they do not, e.g.~\cite{goroff-1, goroff-2, vandeven}.

Historically, some viewed renormalizability as a fundamental
requirement of a well-defined quantum field theory.  A more modern
perspective is that nonrenormalizable field theories (and indeed even
renormalizable field theories) should be viewed as effective
descriptions valid at low energies. As one tries to extrapolate to
higher energies in a nonrenormalizable theory, one gets more and more
new terms with undetermined parameters. The ever-increasing number of
undetermined parameters signals a loss of predicitivity from a
low-energy point of view; however, if we imagine starting from a UV
theory with a finite number of parameters, then we see that all of
those coefficients are fixed by a finite number of UV parameters.
Thus, the distinction between renormalizable and nonrenormalizable
theories simply reflects how sensitive the low-energy effective theory
is to the high-energy (short-distance) physics that we may not know.
For renormalizable theories, all of the short-distance physics can be
encoded in a finite number of variables that can be determined
experimentally, whereas a nonrenormalizable theory, like gravity,
gives an infinite number of low-energy parameters.

The canonical example of this point of view is Fermi weak theory, a
nonrenormalizable but useful description of low-energy weak
interactions. At right about the energy scale predicted by the
Froissart bound argument, one finds new massive degrees of freedom:
the $W$ and $Z$ bosons. Thus, we say that the Glashow--Salam--Weinberg
electroweak theory flows in the infrared, or low-energy, to Fermi weak
theory. Applying these lessons to gravity suggests that rather than
trying to naively quantize general relativity, we should try looking
for a high-energy theory that flows in the infrared to general
relativity.  Of course, finding such a theory with the desired
properties is notoriously difficult.

\subsection{Recent Attempts}

There are at least three active programs attempting to find a
consistent quantization of gravity:\footnote{See~\cite{horava} for
  another attempt, and~\cite{horava-review} for a discussion of its
  current status.}  loop quantum gravity with its many variations,
asymptotically safe gravity, and string theory. Both loop quantum
gravity and asymptotically safe gravity take the attitude that perhaps
the nonrenormalizability suggested by~\eqref{eq:grav-grav-pert}
indicates not a failure of the theory but of the perturbative
expansion. Loop quantum gravity attempts to quantize gravity directly,
without splitting the metric into a background and a small
perturbation~\cite{loops-review}.  Asymptotically safe gravity, on the
other hand, suggests that Equation~\eqref{eq:grav-grav-pert} is an
expansion about the wrong ultraviolet fixed point. We considered the
$\pd \tilde{h}\pd\tilde{h}$ term the free theory about which we
perturbed---this is implicitly an expansion about the Gaussian RG
fixed point.  The asymptotic safety program suggests that there may be
a nontrivial fixed point, for which the interactions are
renormalizable~\cite{weinberg1979, weinbergASG, reuter, niedermaier}.
While the two alternatives to string theory mentioned above are active
areas of research, this dissertation follows the string theory
approach. String theory seems to be the most promising approach, and
has far more tools for answering the questions of interest.

There are a couple of reasons to believe that quantum gravity cannot
be a theory of local quantum fields. For instance, one can think of
general relativity as the gauge theory of diffeomorphisms. Typically,
when quantizing a gauge field theory, the fundamental degrees of
freedom are local, gauge-invariant fields $\phi(x)$; however, in the
case of general relativity $x$ itself transforms under gauge
transformations making such an identification
impossible~\cite{dewitt-2}.  This suggests that a complete theory of
quantum gravity cannot be formulated as a theory of local
fields~\cite{witten-newtonmedal}.

There is a second argument that quantum gravity cannot be a field
theory. Since general relativity predicts that a black hole forms
whenever enough energy is packed into a given region of space, it
follows that the high energy spectrum of gravity should be dominated
by black holes. From the Schwarzschild solution and the
Bekenstein--Hawking formula, Equations~\eqref{eq:schwarzschild}
and~\eqref{eq:SBH}, the entropy grows with energy like
\begin{equation}
S_\text{black hole} \sim E^2.
\end{equation}
On the other hand, the high-energy behavior of any field theory is
expected to be a conformal field theory, whose entropy grows like
\begin{equation}
S_\text{CFT} \sim E^{\frac{3}{4}}
\end{equation}
in four dimensions. Thus, we should not expect a field theory
description of gravity. See~\cite{banks, banks-tasi, shomer} for more
precise and more general arguments that quantum gravity cannot be a
field theory. This leads us to consider string theory.

\section{String Theory}\label{sec:strings}

We now introduce string theory, one of the most promising theories of
quantum gravity. While the full theory remains to be fully defined,
there are certain limits of string theory that are understood very
well. Our best description comes from the worldsheet formalism of a
single string propagating through a fixed background. This is
analogous to the quantum mechanical description of a single particle.
The full description of interacting particles requires quantum field
theory.  Similarly, people think about ``string field theories'' that
have creation and annihilation operators of whole strings, a topic we
discuss no further here. Although the worldsheet description is
restrictive, it contains a lot more physics than the quantum mechanics
of a single particle might suggest.

\begin{figure}[htb]\begin{center}
\includegraphics[height=3cm]{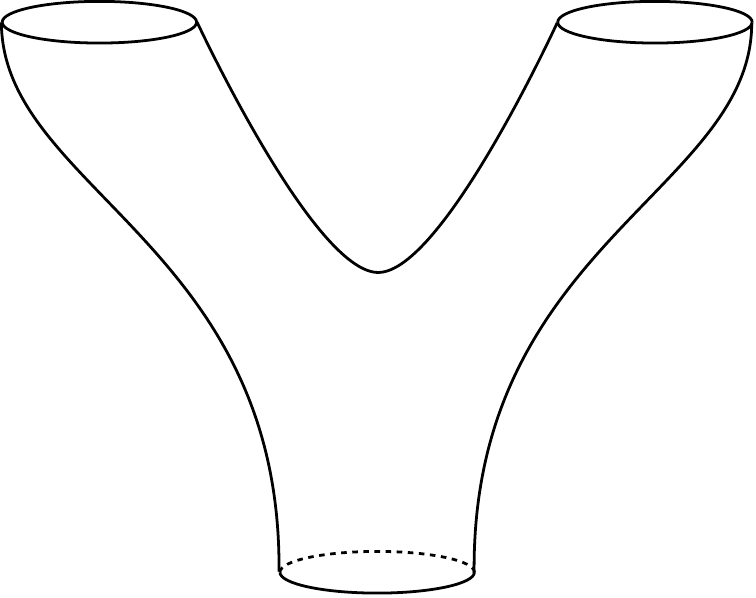}
\caption[The pants diagram]{A closed string splits into two closed
  strings in the ``pants diagram.'' Note that a single worldsheet
  describes this three-point function. Thus the string worldsheet
  description is far richer than the analogous worldline theory of a
  particle.}\label{fig:pants}
\end{center}\end{figure}

Because the string worldsheet can have many different topologies and
excitations, we can actually describe interactions between many
different string states. See Figure~\ref{fig:pants} for an
illustration. Computations of this type have been the most fundamental
tool for exploring the consequences of the theory. What we learn from
analysis of the worldsheet theory is that string theory is a theory of
quantum gravity that is finite, reduces to general relativity in the
low-energy limit, and has no free parameters~\cite{GSW-1, GSW-2,
  polchinski-1, polchinski-2}. One other interesting and unexpected
consequence of the theory is a prediction of the spacetime dimension:
(super)string theory is consistent only in 10 spacetime
dimensions~\cite{GSW-1, GSW-2, polchinski-2}.

That string theory is formulated in 10 dimensions, while we clearly
observe only 4 dimensions might seem like a deal-breaker for the
theory; however, there are at least three reasons why the theory is
still worth pursuing. Firstly, quantum gravity is \emph{hard} as
emphasized above. String theory seems to resolve a lot of the
difficulties in an elegant and consistent manner. This alone merits
the study of string theory. Furthermore, there are enough powerful
tools within string theory that we can use to perform explicit
calculations addressing some of the puzzles about gravity mentioned
above.  Secondly, we can consider string theory on a very small
six-dimensional compact manifold. Such a theory would appear to be
four-dimensional in the length scales that have been explored so far.
The process of compactifying to four-dimensions introduces new degrees
of freedom from the four-dimensional perspective, \`{a} la Kaluza--Klein
theory. By these means, the extra six dimensions give us enough
freedom to potentially recover the Standard Model of particle physics,
giving us the hope that string theory can unify all of known
physics~\cite{GSW-2, polchinski-2}.  Thirdly, string theory allows one
to relate strongly coupled gauge theory correlators in flat space to
weakly coupled gravity calculations in one higher dimension via
``gravitational holography''~\cite{maldacena, gubser-noncrit, witten}.
These tools can give insight into problems in other areas of physics,
for instance, nuclear theory~\cite{ads-qcd} and condensed matter
theory~\cite{hartnoll, mcgreevy}.

\subsection{The Worldsheet}

\begin{figure}[ht]
\begin{center}
\includegraphics[width=7cm]{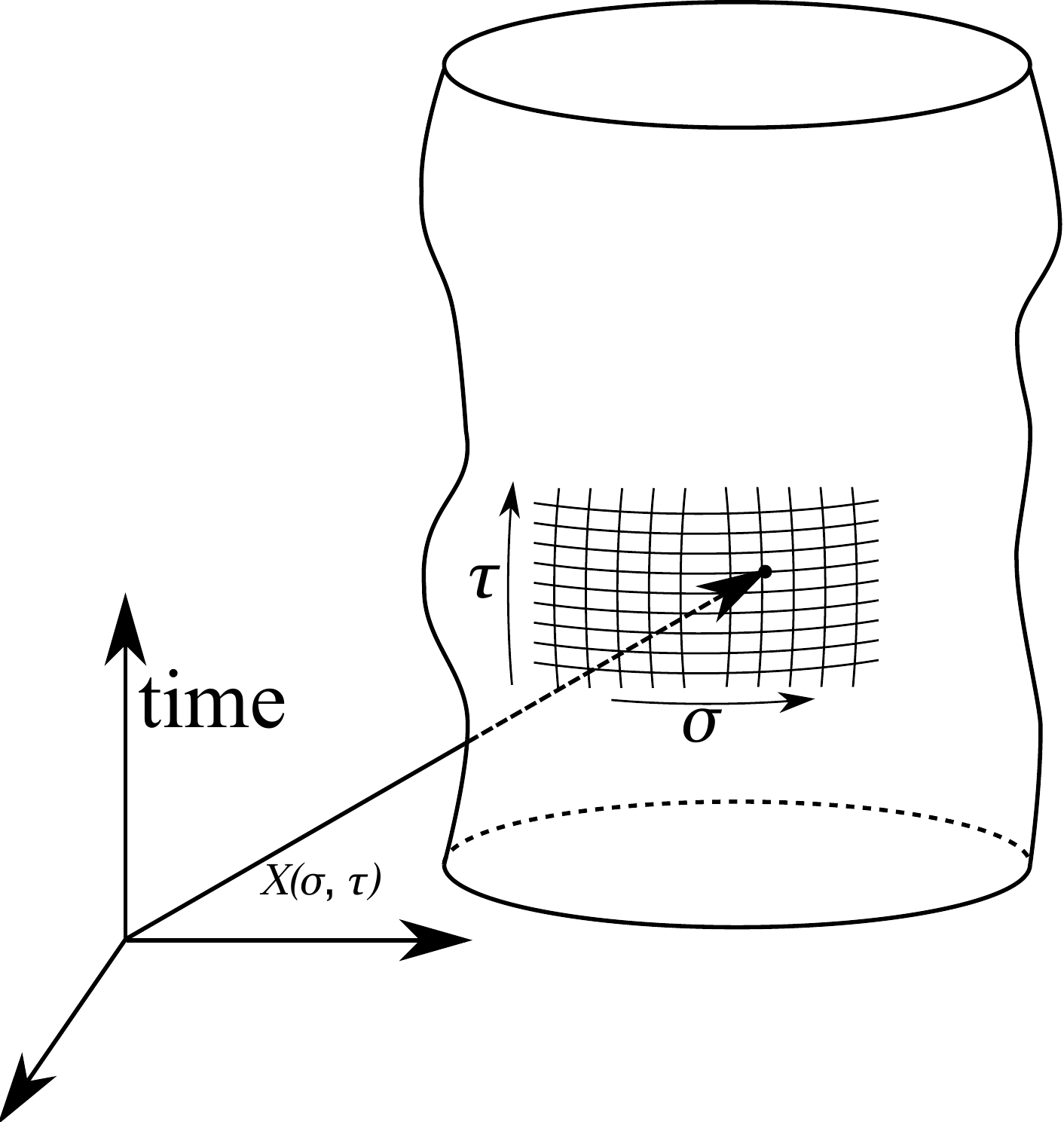}
\caption[Closed string worldsheet]{A closed string propagating through
  time sweeps out a worldsheet. We parameterize the worldsheet with
  functions $X(\sigma,\tau)$. We can think of $X$ as fields living in
  a two-dimensional field theory with coordinates $(\sigma, \tau)$.}
\end{center}
\end{figure}

Let us briefly introduce the worldsheet description of string theory.
For details, see the standard references~\cite{GSW-1, GSW-2,
  polchinski-1, polchinski-2} or, alternatively, some more recent
texts~\cite{johnson, becker, nutshell}.  We can describe the classical
motion of a string through spacetime by giving its path with a Lorentz
vector $X^\mu = X^\mu(\sigma, \tau)$.  The parameters $\sigma$ and
$\tau$ are not coordinates for physical spacetime, and so we are free
to rescale them as we wish.  Most treatments of string theory begin
with the Nambu--Goto action,
\begin{equation}\label{eq:nambu-goto}
S_\text{NG} = -T(\text{area of worldsheet})
  = -T\int\drm\tau\drm\sigma \sqrt{-\det(\pd_a X^\mu\pd_b X_\mu)},
\end{equation}
the action for a relativistic string traveling through flat space.
Here, $a$ and $b$ correspond to either $\sigma$ or $\tau$.  This is
the analogue of the action for a free relativistic point particle:
\begin{equation}
S_{\text{point-particle}} = -m(\text{length of worldline})
  = -m\int\drm\tau\sqrt{-\der{X^\mu}{\tau}\der{X_\mu}{\tau}}.
\end{equation}
The action for a point particle is proportional to the length of the
worldline, the curve it sweeps out through spacetime. The length of
the worldline is also known as the particle's proper time. Similarly,
the Nambu--Goto action is proportional to the area of the string's
worldsheet, the surface it sweeps out through spacetime. The constant
of proportionality is the string tension $T$, which is frequently
exchanged for either the Regge slope parameter $\Regge$ or the string
length $\ell_s$ via
\begin{equation}
T = \frac{1}{2\pi \Regge}\qquad \Regge = \ell_s^2.
\end{equation}
Note that the literature frequently uses different definitions for
$\ell_s$, but we use the above throughout this document. The string
length $\ell_s$ is the natural length scale of string theory.  The
Regge slope parameter gets its name from the linear relation between
the squared-mass $m^2$ and spin $J$ of the quantized string
excitations~\cite{GSW-1}:
\begin{equation}
J = \Regge m^2.
\end{equation}

As one can see from Equation~\eqref{eq:nambu-goto}, the action is
defined in terms of a two-dimensional integral and thus can be thought
of as a two-dimensional field theory. One can make the action look
more like a conventional two-dimensional field theory by writing it in
the form
\begin{equation}\label{eq:polyakov}
S_{\text{Polyakov}} = -\frac{1}{4\pi\Regge}\int\drm\tau\drm\sigma
  \sqrt{-\det\gamma}\gamma^{ab}\pd_a X^\mu\pd_b X_\mu,
\end{equation}
the Polyakov action for the string.\footnote{The action was found by
  Brink, Di Vecchia, and Howe~\cite{brink} and independently by Deser
  and Zumino~\cite{deser-zumino}; however,
  Polyakov~\cite{polyakov-action} ``popularized''
  it~\cite{polchinski-1}.} Note that we have introduced an auxiliary
variable $\gamma_{ab}$, which acts as the metric on the
two-dimensional worldsheet (also called the base space).  One can
integrate out $\gamma_{ab}$ and recover
Equation~\eqref{eq:nambu-goto}. This looks like the action for free
scalars $X^\mu$ in a two-dimensional space with background metric
$\gamma_{ab}$. The action has worldsheet diffeomorphism and Weyl
symmetry. The diffeomorphism invariance gives us the ability to define
new worldsheet coordinates,
\begin{equation}
(\sigma,\tau) \mapsto \big(\sigma'(\sigma, \tau), \tau'(\sigma, \tau)\big)
\end{equation}
with the $X^\mu$ transforming as scalars and $\gamma_{ab}$ as a
two-index tensor.  The Weyl symmetry allows us to make position
dependent rescalings of the worldsheet metric,
\begin{equation}
  \gamma_{ab} \mapsto e^{2\omega(\sigma, \tau)}\gamma_{ab}
\end{equation}
with arbitrary $\omega(\sigma, \tau)$ and $X^\mu$ not transforming.
Rescaling the metric changes the distance between distinct points in
the two-dimensional base space. This symmetry can be thought of as a
reflection of the fact that the string is fundamental, and not
composed of any smaller objects---thus, it looks the same on all
scales.

Let us focus our attention on the closed string, that is, we demand
$X^\mu(\sigma = 0) = X^\mu(\sigma = 2\pi)$. When one quantizes the
Polyakov action with these boundary conditions one finds that the
vacuum state of the two-dimensional theory corresponds to a tachyon in
the ambient spacetime. The tachyon is a problem, which we address
later.  The first excited states form a massless two-index tensor.  We
can break the tensor into a symmetric traceless part $h_{\mu\nu}$, an
antisymmetric part $b_{\mu\nu}$, and the trace $\phi$. The fact that
we have a massless spin-2 particle $h_{\mu\nu}$ suggests that this is
a theory of gravity, since general relativity is essentially the
unique theory resulting from a field theory of massless spin-2
particles~\cite{feynman-gravity, deser-3, deser-2, deser-1,
  kraichnan,gupta, butcher}. By computing the scattering cross-section
of gravitons, for instance, we can fix the 10-dimensional Newton's
constant in terms of the string coupling~\cite{GSW-1, GSW-2,
  polchinski-1, polchinski-2, johnson}:
\begin{equation}
16\pi G_N^{(10)} = (2\pi)^7\Regge^4 g_s^2.
\end{equation}

Since we have massless bosonic fields in spacetime, it is reasonable
to assume that they can condense and acquire a vacuum expectation
value. The essentially unique way to generalize
Equation~\eqref{eq:polyakov} to this case is~\cite{polchinski-1,
  GSW-1}\footnote{The overall minus sign is from an analytic
  continuation to Euclidean signature.}
\begin{equation}\label{eq:string-background}
S = \frac{1}{4\pi\Regge}\int\drm\tau\drm\sigma
  \sqrt{-\det\gamma}\left[
  \big(\gamma^{ab}G_{\mu\nu}(X) + i\epsilon^{ab}B_{\mu\nu}(X)\big)
                     \pd_a X^\mu\pd_b X^\nu + \Regge\Phi(X) R\right],
\end{equation}
where $G_{\mu\nu}$ is the background metric from $h_{\mu\nu}$,
$B_{\mu\nu}$ is the background field from $b_{\mu\nu}$, $\Phi$ is the
background from $\phi$, and $R$ is the Ricci scalar curvature of the
two-dimensional metric $\gamma^{ab}$.  The last term controls how
difficult it is for the string to split apart or join together. For
instance, consider the case that $\Phi$ is a constant $\Phi_0$, then
the last term can be evaluated using the Gauss--Bonnet
theorem:\footnote{We have cheated slightly here. There is an
  additional boundary term that goes into $\chi$. We have implicitly
  assumed here that the worldsheet is a closed manifold, but the Euler
  characteristic dependence we find applies more generally.
  See~\cite{GSW-1, GSW-2, polchinski-1, polchinski-2} for the correct
  treatment.}
\begin{equation}
S_{\Phi_0} = \frac{\Phi_0}{4\pi}
      \int\drm\tau\drm\sigma\sqrt{-\det\gamma}R = \Phi_0\chi,
\end{equation}
where $\chi$ is the Euler characteristic of the worldsheet. 
\begin{figure}[ht]
  \begin{center}
\includegraphics[width=9cm]{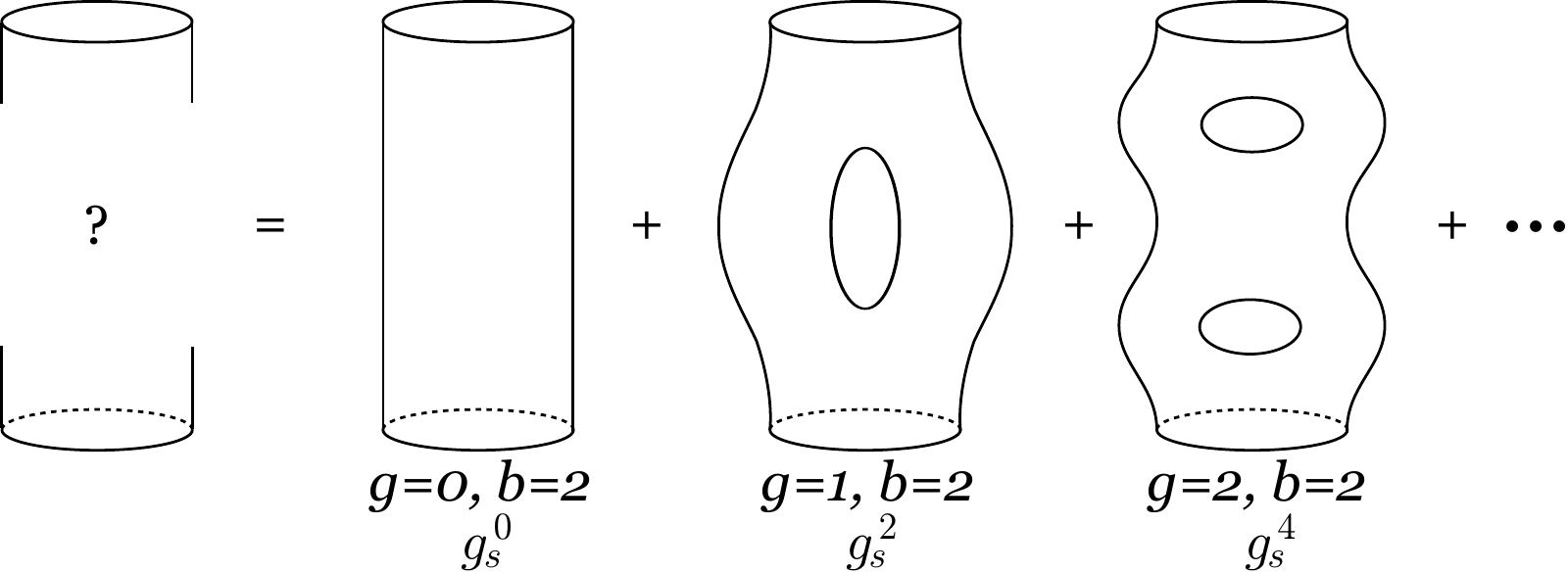}
\caption[The topology expansion of a string correlator]{A cartoon
  illustrating the sum over all possible topologies of the interacting
  string.}\label{fig:topology-expansion}
\end{center}
\end{figure}
The Euler characteristic is determined by the total number of initial
and final states $b$ (boundaries of the worldsheet) and the number of
loops $g$ (the genus of the worldsheet) via
\begin{equation}
\chi = 2 - 2g - b.
\end{equation}
Thus we can expand the partition function in topology as
\begin{equation}\label{eq:string-pert-exp}
Z = \sum_{\chi} \left(e^{\Phi_0}\right)^{-\chi}Z_\chi(\Regge),
\end{equation}
which suggests that we identify the string coupling constant as $g_s =
\exp\Phi_0$. The terms $Z_\chi(\Regge)$ are sums over all intermediate
excitations of the string with fixed topology. In perturbative string
theory, we expand both in $\Regge$, which controls the intermediate
string states, and in $g_s$, which controls the worldsheet topology. The
topological expansion is illustrated in
Figure~\ref{fig:topology-expansion}. For each topology shown, we
perform a sum over all $\Regge$ corrections to the shape.

Now, let us return to the action in
Equation~\eqref{eq:string-background}. As mentioned above the
classical string has a Weyl rescaling symmetry; however, this symmetry
is anomalous in the quantum theory. As it turns out, we must demand
that the quantum theory be Weyl invariant, which means that the
following beta functions must vanish~\cite{polchinski-1}:
\begin{subequations}
\begin{align}
\beta^G_{\mu\nu} &= \Regge R_{\mu\nu} + 2\Regge\nabla_\mu\nabla_\nu\Phi 
              -\frac{\Regge}{4}H_{\mu\lambda\omega}{H_\nu}^{\lambda\omega}
              + \bigO(\Regge^2)\\
\beta^B_{\mu\nu} &= - \frac{\Regge}{2}\nabla^\omega H_{\omega\mu\nu} 
                    + \Regge\nabla^\omega\Phi\,H_{\omega\mu\nu}
                    + \bigO(\Regge^2)\\
\beta^\Phi &= \frac{D-26}{6} - \frac{\Regge}{2}\nabla^2\Phi
                    + \Regge\nabla_\omega\Phi\nabla^\omega\Phi
                    - \frac{\Regge}{24}H_{\mu\nu\lambda}H^{\mu\nu\lambda}
                    + \bigO(\Regge^2),
\end{align}
\end{subequations}
where $H_{\mu\nu\lambda}$ is the field strength of $B_{\mu\nu}$.
Notice that requiring $\beta^G$ to vanish gives Einstein's equations,
$\beta^B$ gives $B_{\mu\nu}$'s equations of motion, and $\beta^\Phi$
fixes the (bosonic string theory) spacetime dimension $D$ to be 26 and
gives $\Phi$'s equations of motion. Consistency demands that the
background field $G_{\mu\nu}$, which the string travels through, must
obey Einstein's equation. This is another way in which we see gravity
emerging from string theory, and a beautiful result of string theory.

The worldsheet theory we have discussed so far describes bosonic
string theory. The theory has two problems: it only has bosonic
degrees of freedom and it has a tachyon in its spectrum. Both of these
problems can be cured by adding fermions to the two-dimensional field
theory in Equation~\eqref{eq:polyakov} in such a way to make the
theory supersymmetric~\cite{GSW-1, GSW-2, polchinski-1, polchinski-2}.

When one adds fermions one must carefully specify the boundary
conditions. Firstly, we can consider closed or open strings: strings
whose ends join to form closed curves in space, or strings with free
endpoints. A closed string implies that $X(\sigma= 0) = X(\sigma =
2\pi)$. On the other hand, for open strings we need to specify
Dirichlet or Neumann boundary conditions for the ends.  Additionally,
one needs to consider the fermions.  For the closed string there are
antiperiodic, called Neveu--Schwarz (NS), boundary conditions; and
periodic, called Ramond (R), boundary conditions around the loop. For
the open string, there are analogous conditions one can impose on the
fermions. The Hilbert space of the worldsheet naturally breaks up into
left-moving modes and right-moving modes, and we can specify the
boundary conditions on the left- and right-moving modes separately.

While the string we are describing is a one-dimensional extended
object, if its size is very small it looks like a point particle with
mass that depends on the internal state of the string. The size of the
string is on the order of the string length.  Since we have not yet
observed stringy behavior in any experiments, we must assume that the
string length is much smaller than the smallest distances probed so
far. Furthermore, it is natural to assume that the string length is
close to the Planck length.\footnote{Although there are many
  exceptions.} Thus, we restrict our attention to the massless modes
and think about what kinds of particles they look like from the
perspective of the ambient spacetime.  The massless NS--NS sector of
the closed string is the same as that of the closed bosonic string
described above. We list the rest of the bosonic massless degrees of
freedom for the two type II strings\footnote{The roman numeral `II'
  refers to two Weyl supersymmetries of the spacetime theory. The `A'
  and `B' label different choices for the chirality of the worldsheet
  fermions. For our purposes here, one can take Table~\ref{tab:IIAIIB}
  as defining the differences.} in Table~\ref{tab:IIAIIB}. The NS--R
and R--NS sectors of the string give fermionic degrees of freedom,
which do not form condensates. If one analyzes the superstring instead
of the bosonic string, one finds that the low-energy limit is
10-dimensional supergravity. In particular, the IIA and IIB string
theories give IIA and IIB supergravity, with the massless spectrum of
the closed string forming the supergravity multiplet.

Looking at Table~\ref{tab:IIAIIB}, let us discuss the physics of the
R--R gauge fields. The $C$ fields are gauge fields that arise from the
R--R sector; the superscript index indicates what kind of form
(antisymmetric tensor) the gauge field is. For instance $C^{(0)}$ is a
Lorentz scalar, while $C^{(2)}$ has two antisymmetrized Lorentz
indices. The corresponding field strength for the gauge field is the
exterior derivative of the form. The plus on $C^{(4)+}$ is a reminder
that the corresponding field strength $\drm C^{(4)+}$ should be
self-(Hodge star)-dual. This is possible since the exterior derivative
of a 4-form is a 5-form, which can be self-dual in 10 dimensions.

\subsection{D-branes}\label{sec:intro:d-branes}

\begin{table}[ht]
\begin{center}
\begin{tabular}{l c c}
 & IIB & IIA\\
\hline
NS--NS & $g_{MN}$, $\phi$, $B_{MN}$ &  $g_{MN}$, $\phi$, $B_{MN}$\\
R--R & $C^{(0)}$, $C^{(2)}$, $C^{(4)+}$ & $C^{(1)}$, $C^{(3)}$
\end{tabular}
\begin{tabular}{l l}
IIB & IIA\\ \hline
D$(-1)\rightarrow C^{(0)}$\hspace{10pt}   & D0$\rightarrow C^{(1)}$\\
D1$\rightarrow C^{(2)}$               & D2$\rightarrow C^{(3)}$\\
D3$\rightarrow C^{(4)+}$              & D4$\rightarrow C^{(3)}$ [M]\\
D5$\rightarrow C^{(2)}$ [M]           & D6$\rightarrow C^{(1)}$ [M]\\
D7$\rightarrow C^{(0)}$ [M]           & D8 \\
D9
\end{tabular}
\end{center}
\caption[Low-energy excitations of IIA and IIB string theory]
{The left table is a summary of the bosonic field content 
  of types IIA and IIB supergravity, which describe the corresponding
  string theories in the low-energy and small string coupling limit. On
  the right, is a summary of the D-brane content of the theories and what
  gauge field the D-brane primarily couples to. The [M] indicates a
  magnetic type coupling and the $+$ superscript is a reminder that the
  corresponding field strength is constrained to be self-dual.}
\label{tab:IIAIIB}
\end{table}

We saw above how the NS--NS sector couples to the string worldsheet.
One may wonder what the R--R $C$ fields couple to when they acquire a
vacuum expectation value (vev). If we have these gauge fields living
in the 10-dimensional spacetime, it is natural to ask what couples to
them and how.  Supergravity has solitonic solutions which are
$(p+1)$-dimensional (including time) extended (but very localized in
the $10-p$ dimensions) objects, which we call D$p$-branes. These
objects couple to the R--R gauge fields as shown in the table on the
right~\cite{polchinski-2, johnson}. A `[M]' indicates that the
D$p$-brane couples to the gauge field magnetically.  If supergravity
has these solutions, then how do they generalize to string theory?

So far, we have restricted our discussion to closed strings---strings
that form closed loops. One could also consider open strings with free
ends, with Neumann or Dirichlet boundary conditions. If one uses
Dirichlet boundary conditions, then one breaks Poincare invariance; an
unhappy state of affairs, \emph{unless} the Dirichlet boundary
condition is on some other object in the theory. In this case,
Poincare invariance is only broken spontaneously. The D$p$-branes are
precisely objects on which open strings may end with Dirichlet boundary
conditions, which is the origin of the letter `D' (for Dirichlet).
Since the D-branes are solitons of some form of string field theory,
one expects their mass to go like one over the string coupling, and
indeed for a D$p$-brane one finds~\cite{johnson} 
\begin{equation}\label{eq:Tp} T_p
= \frac{\text{mass}}{p\text{-volume}} 
   = \frac{1}{(2\pi)^p\Regge^\frac{p+1}{2} g_s}.  
\end{equation}
Thus, when the string's excitations are light and weakly interacting,
the D-branes are heavy and vice-versa.

In string perturbation theory, the defining quality of D$p$-branes is
that they are $p$-spatial dimensional objects that open strings can
have endpoints fixed to; moreover, open strings may \emph{only} end on
D-branes. For example, if an open string has Neumann boundary
conditions in all $9+1$ dimensions, then there must be a
spacetime-filling D9-brane~\cite{johnson, polchinski-1, polchinski-2}.

\begin{figure}
\begin{center}
\includegraphics[height=5cm]{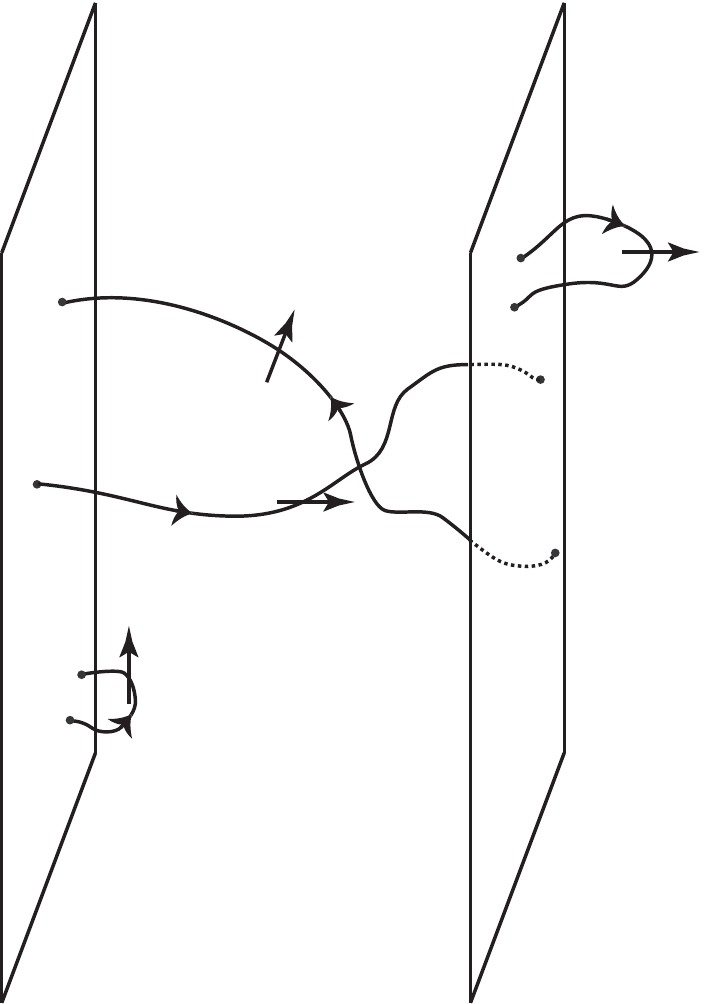}
\caption[Open strings on a D-brane]{An illustration of the different ways that open strings can be 
configured on D-branes. The long arrows indicate the polarization of
the open string, \emph{not} its velocity, which has been suppressed.}
\label{fig:dbrane-open-strings}
\end{center}
\end{figure}

Since D-branes are solitonic objects with mass proportional to the
inverse of the string coupling, they never arise directly in
perturbation theory: one cannot scatter two closed strings and produce
a D-brane; however, if a D-brane is already present in our spacetime,
then we can do a perturbative expansion around this new
background. How does the presence of a D-brane change the perturbation
theory?

Well, for one, the perturbation theory now includes open strings. The
open strings can have both endpoints fixed to the same D-brane or to
two different D-branes. The open strings also have discrete oscillator
modes, and a similar mass spectrum to the closed strings. As with the
closed strings, we focus on the massless ground states of the open
string.  In contradistinction to the closed string, the open string's
ground state polarization only has one Lorentz index~\cite{johnson,
  polchinski-1, polchinski-2}. These different configurations are
shown in Figure~\ref{fig:dbrane-open-strings}.

\begin{figure}
\begin{center}
\subfigure[An open string leaving a D-brane]
	{\includegraphics[height=7.5cm]{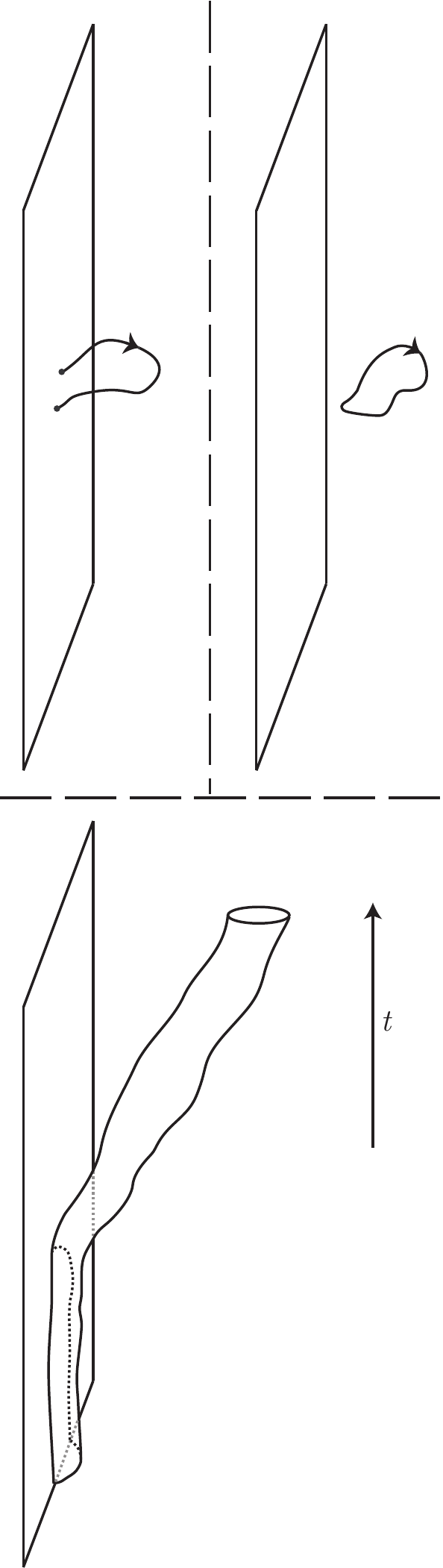}}
\hspace{60pt}
\subfigure[A pair of open strings joining ends]
	{\includegraphics[height=7.5cm]{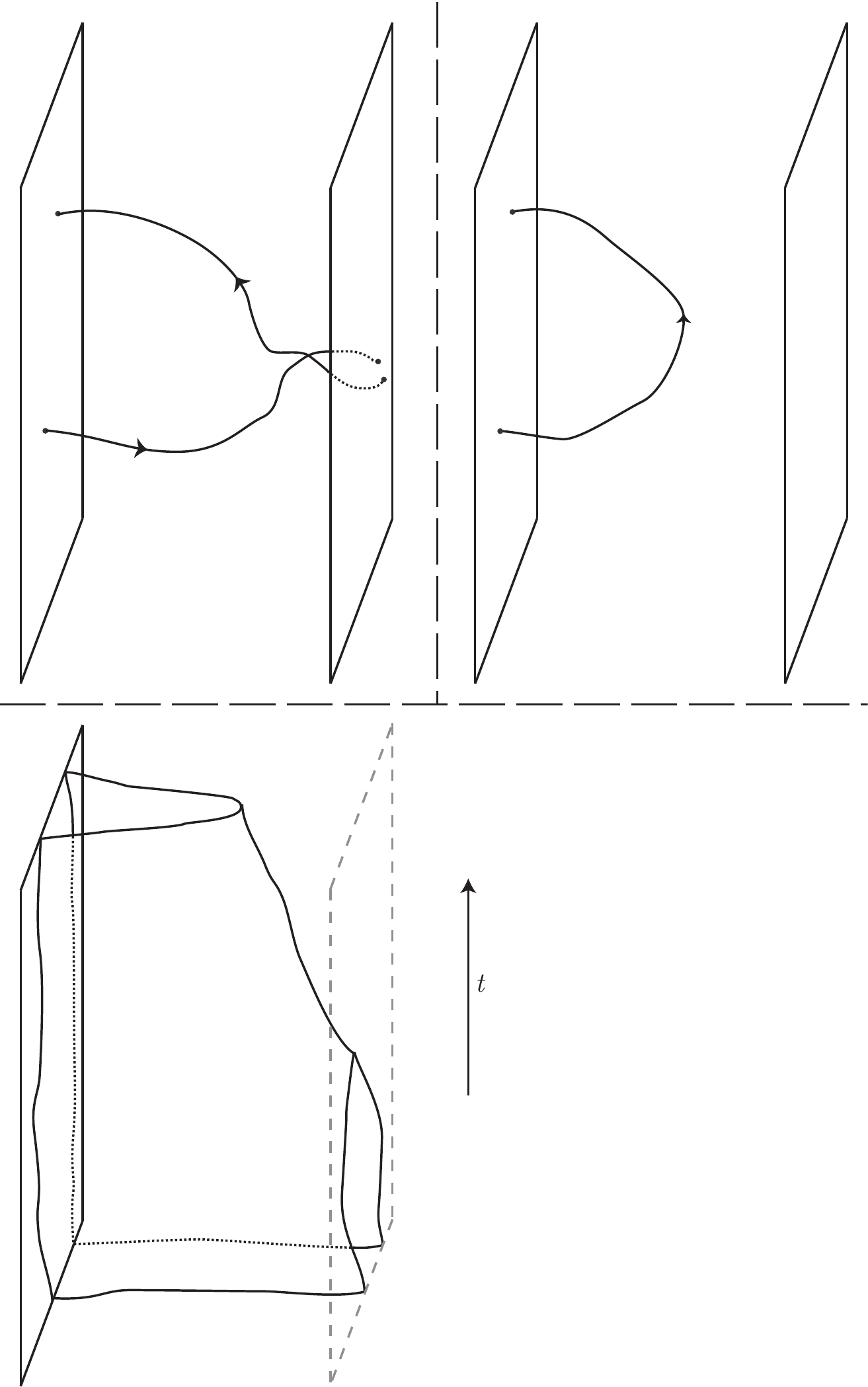}}
\label{fig:open-string-pair}
\vspace{-12pt}
\caption[Open string interactions]{The top figures are before and
  after picture of open strings interacting at order $\sqrt{g_s}$, and
  the bottom figures represent the worldsheets for the same processes.
  The arrowheads indicate the orientation of the string.}
\label{fig:stringPert}
\end{center}
\vspace{-12pt}
\end{figure}

An open string with both ends on the same D-brane can join its ends
together forming a closed string that leaves the D-brane. This process
has a coupling of $\sqrt{g_s}$. Of course, the opposite process can
happen as well---a closed string can bump into a D-brane and become an
open string with end points fixed to the D-brane. Other types of
processes can occur: two correctly oriented open strings connecting to
the same D-brane can meet up and form one open string. In
Figure~\ref{fig:stringPert}, cartoons of these processes are shown.

How are these various processes reflected in the low-energy
supergravity description? Is there some effective field theory
description of the D-brane that captures these dynamics? As it
happens, there is: the Dirac--Born--Infeld (DBI) action with R--R
gauge field couplings~\cite{johnson}
\begin{equation}\label{eq:DBI}\begin{split}
S_{DBI} &= -T_p\int\drm^{p+1}\xi\; e^{-\Phi}
	\sqrt{-\det\big(G^{(\mathrm{ind.})} + B^{(\mathrm{ind.})} + 2\pi\ap F_p\big)}\\
		&\qquad+ i T_p\int C^{(p+1)} + iT_p\int C^{(p-1)}\wedge F_p\cdots,
\end{split}\end{equation}
where $G^{(\mathrm{ind.})}$ and $B^{(\mathrm{ind.})}$ are the induced
metric and R--R gauge field; $F_p$ is a gauge field strength of a
$U(1)$ gauge field living in the world-volume of the D$p$ brane; the
last two and the omitted terms depend slightly on what $p$ is, but
they are the coupling of the D$p$-brane to the R--R gauge fields. This
action is a good approximation when $\ap^{3/2}\pd_a F_{bc}\ll 1$, i.e.
when the D-brane varies slowly in time and space~\cite{nutshell,
  polchinski-1, polchinski-2, johnson}. If one expands the square root
and determinant for small $F$ and $B$ about the metric induced from
flat Minkowski spacetime, then one finds a standard $U(1)$ gauge
theory action. This action only describes the D-brane dynamics, there
is still a full $10$-dimensional bulk supergravity action, as well.

How can we see the string processes described above in
Equation~\eqref{eq:DBI}? The emission of closed strings from the
D-brane is described by the coupling of the various closed string
gauge fields (and gravity) to the D-brane. The open string excitations
of the D-brane can be broken into two classes: those polarized along
the brane and those perpendicular to the brane. The excitations
polarized along the D-brane are described by a $U(1)$ gauge field
living \emph{in} the D$p$-brane's $(p+1)$-dimensional worldvolume,
while the excitations polarized perpendicular to the brane are
described by waves in the position of the brane in the $(10-p)$
transverse dimensions~\cite{johnson, polchinski-1, polchinski-2}.

Open strings connecting two D-branes are not included in the DBI
action, since they will have mass proportional to the separation of
the D-branes and thus are suppressed. This is true unless the
separation between D-branes, in some limit, is zero. If we consider,
for instance, a stack of $N$ parallel, coincident D$p$-branes, then
the open strings that have endpoints on the same D-brane form a $U(1)$
gauge field as before. If that was all we had then the overall gauge
group would just be $U(1)^N$; however, there can now also be open
strings connecting different D-branes, which will connect the $U(1)$s
to each other in nontrivial ways. This leads to a $U(N)$ gauge group
in $(p+1)$ dimensions, with those $U(1)$s embedded as noncommuting
subgroups~\cite{nutshell, johnson, polchinski-1, polchinski-2}.

There is a diagonal $U(1)$ degree of freedom which describes all of
the D$p$-branes moving together in the same way, and transforming all
of the $U(1)$s in the same way. Typically, we are not interested in
this trivial type of motion, and therefore focus on only the $SU(N)$
gauge group left after modding out the diagonal $U(1)$. D-branes, as
it turns out are $^1/_2$-BPS\footnote{They saturate a
  Bogomol'nyi--Prasad--Sommerfield (BPS) bound~\cite{BofBPS, PandS}
  for an object breaking half of the supersymmetry charges.} objects,
spontaneously breaking one-half of the $32$ components of the
$9+1$-dimensional $\Nsc=2$ supersymmetry~\cite{nutshell}.\footnote{The
  `II' in IIA/B refers to the there being two Majorana--Weyl
  supersymmetries.}  Therefore, we expect the effective field theory
describing the D-brane to have $16$ supercharges and an $SU(N)$ gauge
group.

A summary of the total low-energy action would be
\begin{equation}\label{eq:threePartAction}
S = S_\mathrm{bulk} + S_\mathrm{branes} + S_\text{brane-bulk interaction},
\end{equation}
where $S_\mathrm{bulk}$ is the supergravity action,
$S_\mathrm{branes}$ is the $SU(N)$ gauge theory, and the last term is
the coupling of the R--R gauge fields in the bulk supergravity to the
$SU(N)$ field strength of the brane action.

We know roughly what $S_\mathrm{branes}$ and $S_{\mathrm{brane-bulk}}$
look like from Equation~\eqref{eq:DBI}, but we have not discussed what
happens to the supergravity action. In particular, the proposed stack
of D-branes is quite massive in the limits where supergravity is
valid, since $\ap$ and $g_s$ are small, which suggests that the metric
should be distorted by the mass. Since the D-brane's mass is
homogeneously spread over its volume, we make an ansatz for the metric
of the D-branes,
\begin{equation}\label{eq:genericMetric}
ds^2 = f_1(r)dx_\parallel^2 + f_2(r)(dr^2 + r^2d\Omega_{10-p-2}^2),
\end{equation}
where $r$ is a radial coordinate in the perpendicular space, measuring
how far one is from the stack, and the $d\Omega^2$ is the metric for
the constant radius sphere in the perpendicular space. By
$dx_\parallel^2$, we mean a Minkowski dot product of the one-forms
living along the brane directions. We also expect the $C$-field to
which the D-brane primarily couples to be sourced by the D-brane. For
instance, for a stack of $N$ D3 brane~\cite{tasi, maldacena}
\begin{equation}
\frac{1}{f_1(r)} = f_2(r) = \sqrt{1 + \frac{4\pi N g_s\Regge^2}{r^4}}.
\end{equation}
Note that if one takes an appropriate limit in which the $1$ in
$f_2(r)^2$ can be dropped, then the resulting metric is that of
$\AdS_5\times S^5$.

\subsection{The AdS--CFT Correspondence}

From the above considerations, one can arrive at one of the most
remarkable results in string theory.  The AdS--CFT correspondence, or
the statement that string theory in $d+1$-dimen\-sional anti-de Sitter
(AdS) space is \emph{dual} to the physics of a $d$-dimensional
conformal field theory, is one of the most important recent ideas to
arise in theoretical physics. The duality is the most explicit and
powerful instance of gravitational holography. For more comprehensive
treatments, see~\cite{MAGOO, tasi}.

The most heuristic argument for the existence of gravitational
holography was made prior to AdS--CFT~\cite{susskind-holography,
  thooft-holography}. Consider a gravitational theory, and suppose you
want to pack as much information as possible into as small a volume as
possible, i.e. you want to make the perfect hard drive. For typical
physical systems the amount of information one can store grows with
the volume and the energy.  So if one wants to store more information
in a fixed volume, then one ends up making an increasingly massive
hard drive; however, we know that once one packs enough energy into a
given volume one ends up with a black hole. Since the entropy (and
therefore the information one could store) of a black hole scales with
the \emph{area} of the black hole and the black hole has the maximum
entropy a gravitational system can contain in a fixed volume, one
might guess that a $d+1$-dimensional gravitational system should admit
a $d$-dimensional \emph{non}gravitational description.

A second reason to believe the correspondence arises from considering
the large $N$ limit of $SU(N)$ gauge theory.  If one takes
$N\to\infty$ while holding the 't Hooft coupling $\lambda = g_{YM}^2
N$ fixed, then one gets a nontrivial theory~\cite{coleman,
  thooft-largeN}. One can do a double perturbative expansion in $1/N$
and in $\lambda$. When one organizes the perturbative series using 't
Hooft double line notation, one finds that the $1/N$ expansion
corresponds to an expansion in the topology of the diagram. The
expansion, then, takes a remarkably similar form to the string
perturbative expansion in Equation~\eqref{eq:string-pert-exp}. The
extra dimension arises as an energy scale of the gauge
theory.

The above considerations are indirect and quite broad.  We can
precisely state the correspondence by treating the most studied case.
Consider a stack of $N$ D3-branes in IIB string theory such as we
described in Section~\ref{sec:intro:d-branes}. As we discussed, the
open string modes give rise to a $SU(N)$ 3+1-dimensional gauge theory.
On the other hand, the closed string modes give rise to an
$\text{AdS}_5\times S^5$ spacetime with $N$ units of $C^{(4)+}$ flux
on the $S^5$. On the third hand, in string theory there is
open--closed string duality: a string Feynman diagram can be viewed as
either a closed string diagram \emph{or} a higher order open string
diagram. Calculating in both ways gives the same answer.

\begin{figure}
\begin{center}
\includegraphics[width=5cm]{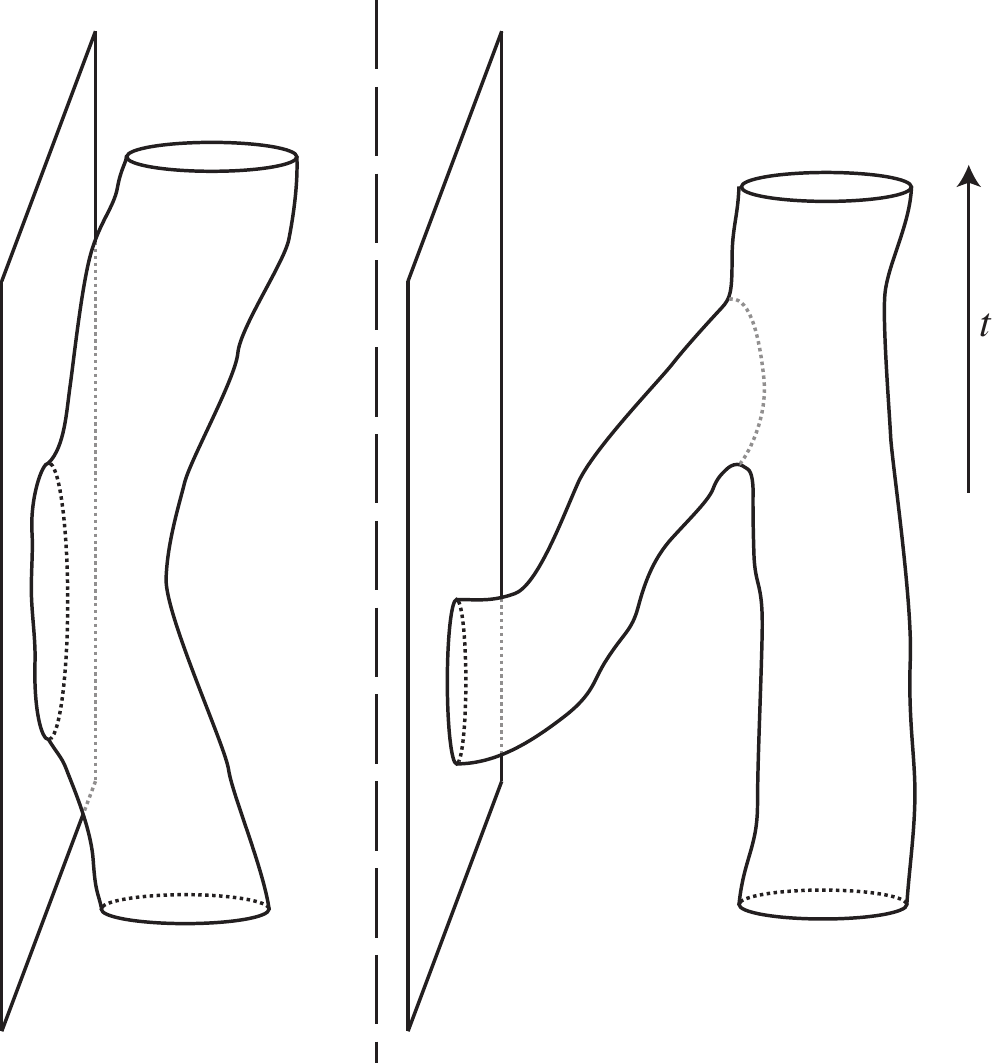}
\caption{Two different ways of picturing a closed string scattering off 
of a D-brane}
\label{fig:dbrane-scattering}
\end{center}
\end{figure}

\begin{figure}
\begin{center}
\includegraphics[height=2.75cm]{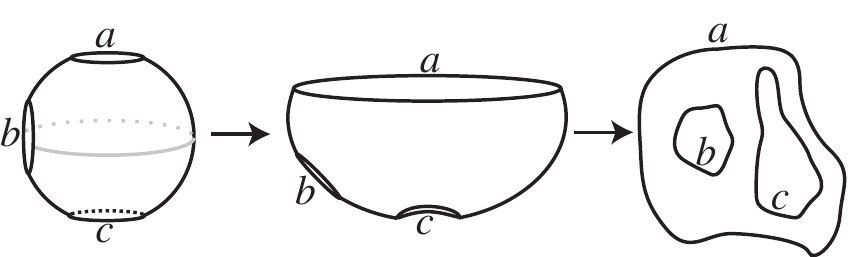}
\caption[An illustration of open--closed string duality]{A cartoon of
  transforming a closed string tree-level Feynman diagram into a
  two-loop open string diagram. We have not been careful to represent
  the conformal nature of the transformation in the
  figure.}\label{fig:closed-open}
\end{center}
\end{figure}

As an example, we consider a closed string scattering off of a
D-brane. To leading order in $g_s$, the relevant diagram is shown in
Figure~\ref{fig:dbrane-scattering}. In the right half of the figure,
we interpret the interaction as the D-brane emitting a closed string
and joining up with a second closed string. This is a tree-level
process, with three external legs. In the left half of the figure if
we run time into the page, we interpret the interaction as a pair of
open strings popping out of the vacuum, one of the legs splitting and
rejoining, and finally the remaining two open strings popping back
into the vacuum. This is a two loop vacuum bubble, requiring a sum
over all possible intermediate open string states.  The distinction is
easier to see if one conformally shrinks the external legs and looks
at the two processes as a sphere with holes in it, and as a disk with
holes in it. Examining Figure~\ref{fig:closed-open}, we can imagine
stretching the top hole in the sphere and smoothing the sphere into
the disk with two holes in it.

We see that tree-level closed string processes are equivalent to open
string loop diagrams, since the same argument can equally well apply
for an interaction with a large number of closed string legs from the
D-brane, as long as the closed string worldsheet does not have any
loops. In string perturbation theory it seems quite natural that the
same diagram can be viewed as a closed string process and as an open
string process; however, if we ask how this duality manifests itself
in the effective field theory description, we become quite confused.
The closed string process should correspond to tree-level supergravity
from $S_\mathrm{bulk}$, while the open string process should
correspond to the $SU(N)$ SYM from $S_\mathrm{branes}$; but they are
both included in the total action. Is the action in
Equation~\eqref{eq:threePartAction} actually \emph{double}-counting in
some sense? Are $S_\mathrm{bulk}$ and $S_\mathrm{branes}$ dual to each
other? Maldacena conjectured that, at least some of the time, the
answer is yes~\cite{maldacena}.

\subsubsection{Defining the Correspondence}

Let us now describe the correspondence for the best studied case,
which arises when considering a stack of $N$ D3-branes. The statement
is that type IIB superstring theory in an $\AdS_5\times S^5$
background with string coupling $g_s$, both manifolds having radius
$R$ and $F^{(5)+}$ satisfying $\int_{S^5}F^{(5)+} = N$, and
four-dimensional $\Nsc=4$ $SU(N)$ SYM with coupling constant $g_\YM$,
are dual to each other~\cite{maldacena, tasi}.  Moreover,
we identify
\begin{equation}
2\pi g_s = g_\YM^2\qquad 
R^4 = 4\pi g_s N \ap^2 \qquad 
\vev{C^{(0)}} = \theta_I,
\end{equation}
where $\theta_I$ is the instanton angle.  The above identifications
are part of what is called the dictionary of the correspondence. 

By duality, we mean that there exists a well-defined one-to-one and
onto mapping of states, operators, and correlation functions in both
theories. The duality is best understood in the large $N$ limit, when
string loops are suppressed. When the 't~Hooft coupling $g_\YM^2 N$ is
large, the string theory is weakly coupled and well-approximated by
supergravity, while the field theory is strongly coupled.  It is
sometimes helpful to write the dictionary in a slightly different
form:
\begin{equation}
\left(\frac{R}{\ell_s}\right)^4 = \lambda\qquad
\left(\frac{R}{\lp^{(10)}}\right)^4 = \frac{\sqrt{2}}{\pi^2}N.
\end{equation}
We see that the 't~Hooft coupling is the AdS radius measured in string
units and $N$ is the AdS radius measured in 10-dimensional Planck
units. This makes the role played by $N$ and $\lambda$ clearer in the
string theory description.  While a complete proof remains to be
found, the AdS--CFT conjecture is extremely useful and is supported by
a large body of evidence~\cite{nutshell, becker, MAGOO}.

In the large $N$ limit, we can elucidate more precisely the nature of
the duality by considering the Laplace equation for fields living in
the $\AdS$. For appropriate boundary conditions, one can bijectively
map the behavior of the field at the boundary of \AdS onto the
corresponding field configuration in the bulk of \AdS. This suggests,
then, that we associate with a field $\phi$ in $\AdS$ which has
boundary value $\phi_0$ a local operator in the dual field theory
$\Oop$. Then the duality becomes the identification~\cite{witten}
\begin{equation}\label{eq:partFunctions}
\left\langle e^{\int_{S^4} \phi_0 \Oop}\right\rangle_\text{CFT} 
 = Z_\mathrm{S}(\phi_0),
\end{equation}
where the classical supergravity partition function is
\begin{equation}
Z_\mathrm{S}(\phi_0) = e^{-S_\mathrm{sugra}(\phi)}.
\end{equation}
We see that the boundary value of $\phi$ acts as a classical source
for $\Oop$ in the CFT partition function. The field theory's Minkowski
space has been written as $S^4$, since we are stating the Euclidean
version of the correspondence.

In this discussion, we focused on the $\AdS_5$--$\text{CFT}_4$
correspondence that comes from considering D3-branes since it is the
most widely studied; however, there are many other incarnations of
\AdS--CFT duality~\cite{MAGOO}. While there are many important
differences between the different versions of the duality, there are
several features that are universal; most importantly, the
identification~\eqref{eq:partFunctions} and the strong--weak nature of
the duality.  The AdS--CFT correspondence is expected to hold more
generally. For instance, \cite{polchinski-holo} conjectures that every
CFT satisfying certain basic requirements should have a dual AdS
description.

In this dissertation, we treat the $\AdS_3$--$\text{CFT}_2$ version of
the correspondence that arises when one considers a bound state of D1
and D5 branes. The two-dimensional CFT is the ``D1D5 CFT'' of the
title.  In Chapter~\ref{ch:d1d5}, we describe both the gravity and the
field theory description of this system. Another important point is
that the majority of interest in AdS--CFT is in using weakly coupled
gravity calculations to understand strongly coupled field theory.
Here, the primary motivation is to use the two-dimensional CFT to
understand strongly coupled gravity.

\section{The Fuzzball Proposal}

Having introduced string theory, which is claimed to be a consistent
theory of quantum gravity, we should return to the issues discussed in
Section~\ref{sec:whyQG} and see what string theory can tell us. String
theory has had some success in addressing both the issues of spacetime
singularities and the information paradox. The AdS--CFT correspondence
has proved a useful tool in a number of these studies. Some classes of
singularities are resolved and others are much better understood
within string theory, but a complete theory remains
elusive~\cite{natsuume, horowitz-sing, johnson-sing, gubser-sing}.

In this dissertation we are more concerned with the information
paradox and the structure of black holes within string theory. The
AdS--CFT proposal and more generally the study of D-branes has proved
invaluable in understanding black holes. For our treatment, we adopt
the fuzzball proposal, which we review below.  For more thorough
treatment of the proposal see~\cite{fuzzballs-elementary,
  fuzzballs-structure, bena-review, skenderis-review, bala-review,
  cv}.

Let us begin our discussion not with the information paradox but with
the entropy puzzle of black holes. We return to the information
paradox later. Recall that in Equation~\eqref{eq:SBH}, we associate an
entropy with a black hole that is proportional to its horizon area.
It is a little unusual to associate an entropy with a classical
solution of a fundamental theory. If we take this entropy seriously
(and Hawking's calculation certainly suggests we should), then we are
immediately forced to ask: where/what are the $e^{S_\text{BH}}$
microstates? This is the entropy puzzle. In a typical thermodynamical
system, say a gas of molecules, the entropy is the logarithm of the
number of microstates that have the same macroscopic properties which
define the system's macrostate. What are the analogous statements for
the black hole?

One of string theory's greatest triumphs addresses this question. One
can consider forming a black hole from a collection of D-branes. When
the gravitational coupling is weak, the D-branes can be described via
a worldvolume gauge theory. On the other hand, when the coupling is
strong, the gauge theory is strongly coupled, but there is a gravity
description as a black hole. One can count the degrees of freedom at
weak coupling and compare to the Bekenstein--Hawking entropy at strong
coupling. In general, one would not expect a weak and strong coupling
answer to agree; however, for certain certain supersymmetric black
holes, these counts are ``protected''\footnote{Typically, it is not
  directly the number of microstates that is protected but an
  ``index.''  See~\cite{sen-review} for more on this point.}  from
changes in coupling. When one performs this calculation, one finds
exact agreement in the limit of large macroscopic charges
(see~\cite{sen-1, sen-2, stromvafa} for early calculations
and~\cite{sen-review} for a recent review).  Within string theory,
then, we see the appearance of $e^{S_\text{BH}}$ microstates; however,
the counting we described leaves us with the question of how the
microstates manifest themselves at strong coupling when there is a
black hole.  For instance, we can take a single microstate of the
weakly coupled gauge theory, and ask what happens to that state as we
turn up the coupling.  Do these microstates of the gauge theory give
rise to a phase space of solutions when the black hole is a good
description?

It appears at first that there is no phase space for the black hole.
We only have the black hole solution and there are various ``no hair''
results~\cite{no-hair-1, no-hair-2} that limit the number of
alternative solutions one could consider---certainly one would not be
able to find $e^{S_\text{BH}}$ states by considering perturbations to
the black hole solution. Even if one could find such perturbations, it
would be difficult to interpret them: each of those ``microstates''
would also have a horizon, and therefore an entropy. Microstates are
pure states and should not have an entropy. 

One might also conclude that all of the microstates must be quantum
mechanical, approaching the naive black hole metric in the classical
limit.  After all, dimensional analysis suggests that the microstates
should be variations of the black hole geometry when the curvature
becomes Planck scale near the singularity; however, as mentioned in
Section~\ref{sec:info-paradox}, this dimensional reasoning does not
seem to allow the information paradox to be resolved. Also, this
scenario seems a bit unphysical since it requires packing a large
amount of phase space into a small physical volume.  In fact, one can
show that if we wish to retain quantum mechanics and locality, the
state at the horizon must depend, in a significant way, on the
microstate that the black hole is in~\cite{mathur-hawking,
  mathur-hawking-old}.

The fuzzball proposal~\cite{lm4, lm5} states that the black hole
solution is an effective coarse-grained description that results from
averaging over $e^{S_\text{BH}}$ horizon-free nonsingular microstates
that have nontrivial structure differing from the ``naive'' black hole
solution up to (at least) the horizon scale. If one starts considering
very stringy states where causality is not well-defined, the notion of
``horizon-free'' becomes ill-defined, and it is currently an open
problem how to precisely formulate this aspect of the fuzzball
proposal for very stringy microstates.  Qualitatively, however, the
statement is simply that \emph{the microstates have nontrivial
  structure up to the horizon} of the corresponding naive black hole
solution. This fact resolves the information paradox: the radiation
leaves from the surface and carries information about the state of the
system. This is just the way any other thermal body radiates. From
this point of view, we can define the fuzzball proposal as the
statement that the state at the horizon is almost orthogonal to the
vacuum~\cite{mathur-hawking}. 

For certain extremal black holes, one can explicitly construct
solutions of supergravity that form a phase space of black hole
microstates, which account for the entropy of the black hole
(see~\cite{lm3, st-1, bena-1, bena-2} and the
reviews~\cite{fuzzballs-elementary, fuzzballs-structure, bena-review,
  skenderis-review, bala-review, cv}).  The solutions are nonsingular,
horizon-free, and differ from each other up to the horizon scale.
Asymptotically they approach the black hole solution. Furthermore,
there is an explicit mapping between solutions and states of the D1D5
CFT. Of course the solutions receive $g_s$ and $\Regge$ corrections,
which are more or less important for different microstates. It remains
an open question whether nonextremal black holes' phase space is
completely captured by a corresponding phase space of supergravity
solutions; see~\cite{recent-bena} and~\cite{recent-deBoer} for recent,
differing perspectives on this issue. Let us emphasize, however, that
the fuzzball proposal does not require the microstates to be
well-described by supergravity. We only need the microstates to differ
from the naive black hole solution and each other up to the would-be
horizon. In fact, we expect generic states of the black hole not to
have a well-defined geometric description; however, there may still be
certain special states that have nice classical descriptions.

Finally, let us comment on the relationship of AdS--CFT duality and
more generally gravitational holography to the fuzzball proposal. One
might wonder why we insist on using a gravitational description if we
expect to have a (strongly coupled) field theory description. Are we
not just using the wrong variables to think about the problem? We know
that the field theory is unitary. Since all of the issues with black
holes and gravity arise when considering the geometrical description,
it is important to resolve those issues in the same language. For
instance, if one is to resolve the Hawking information paradox, one
must show what aspect of the argument fails.  Otherwise, we are faced
with the specter of giving up quantum mechanics, and therefore string
theory and the AdS--CFT correspondence. That is to say, using AdS--CFT
to solve the information paradox with no other explanation of what
breaks in Hawking's calculation \emph{begs the question}.  More
generally, we need an answer to when and why general relativity breaks
down.  Since one can make the curvature arbitrarily weak at the
horizon of the naive black hole by considering very massive black
holes (see Equation~\eqref{eq:Schwarzschild-R}) and locally there does
not seem to be anything interesting happening at the horizon; it is
difficult to introduce new physics at the horizon without
contradicting existing experimental evidence that supports general
relativity.  See~\cite{mathur-hawking, mathur-tunneling} for more on
this point, and its connection to the fuzzball proposal.

\section{Outline}

This dissertation focuses on using the CFT description of the D1D5
system to understand black holes and the fuzzball proposal. Some of
the calculations, however, have more general applications. In
particular, we show how to relax the decoupling limit and allow some
excitations to leak out of the near-horizon AdS/CFT. This is the
content of Chapter~\ref{ch:coupling}. In Chapter~\ref{ch:emission}, we
compute emission out of particular nonextremal microstate geometries.
One of the important results that we learn along the way is that many
different physical processes in the gravity description are related to
the same CFT process. This proves a computational boon for the
immediate goal of finding the emission. The results of these two
chapters demonstrate some interesting aspects of the AdS--CFT
correspondence that have not been extensively studied. In
Chapter~\ref{ch:orbifold} we compute the effect of a marginal
deformation operator in the D1D5 CFT. This operator is believed to be
important for understanding the dynamical physics of black holes:
thermalization, formation, and in-falling observer. While we do not
find concrete answers to these questions, we do see hints. In
particular, we find that states with a few high-energy excitations
have an amplitude to ``fragment'' into many low-energy excitations.

We now outline the remaining chapters of this dissertation in detail.

In Chapter~\ref{ch:d1d5}, we describe the D1D5 system in detail. It
leads to an $\AdS_3$--$\text{CFT}_2$ correspondence, and we describe
both the gravitational theory and the conformal field theory. Of
particular importance is the 20-dimensional moduli space, and the
orbifold point where the CFT is particularly simple. The rest of the
calculations presented in this dissertation are in the orbifold CFT.

In Chapter~\ref{ch:coupling}, we describe how to set up a scattering
problem using AdS--CFT from asymptotic flat space \emph{outside} of
the AdS space. This is different from most AdS--CFT calculations that
do not discuss any asymptotically flat space outside of the AdS. Here,
we are perturbatively relaxing the decoupling limit.

In Chapter~\ref{ch:emission}, we discuss emission from a particular
class of geometries. We introduce the geometries and argue that as per
the fuzzball proposal they should be considered classical
approximations to microstates of the corresponding black hole.  The
geometries are unstable and decay in a calculable way. We reproduce
the gravity spectrum for the emission using the methods in
Chapter~\ref{ch:coupling} and a detailed CFT calculation. We explain
the significance of the calculation for understanding black holes
within the fuzzball proposal.

In Chapter~\ref{ch:orbifold}, we introduce a marginal deformation
operator to the orbifold CFT at first order in perturbation theory. We
introduce a number of tools to calculate the effect of the operator.
In particular we find the effect of the operator on the vacuum is to
produce a squeezed state, and that excitations may be moved across the
operator with Bogolyubov-like coefficients.

Finally, in Chapter~\ref{ch:conclude}, we present some concluding
remarks. We discuss opportunities for future work and our hopes for
greater understanding of black holes through continued study of the
D1D5 system.
\chapter{The D1D5 System}
\label{ch:d1d5}


\begin{table}[ht]
\begin{center}
\begin{tabular}{l||c|c|c|c|c|c|c|c|c|c}
   &   0 & 1 & 2 & 3 & 4 & 5 & 6 & 7 & 8 & 9\\\hline\hline
D5 &  -- & \smallbullet &\smallbullet &\smallbullet &\smallbullet &--&-- &--&--&--\\
D1 &--&\smallbullet&\smallbullet&\smallbullet&\smallbullet&--
      &\smallbullet&\smallbullet&\smallbullet&\smallbullet
\end{tabular}
\caption[The D1D5 brane configuration]{Diagram showing the D1D5 brane
  configuration. The \smallbullet's indicate that the object is
  ``pointlike'' in the corresponding direction and the --'s indicate
  that the object is extended in the corresponding direction. The `0'
  direction is time; the `5' direction is the $S^1$; the `1'--`4'
  directions are the noncompact spatial part of $M^{4,1}$; and the
  `6'--`9' directions are the string scale
  $T^4$.}\label{tab:brane-conf}
\end{center}
\end{table}

In this chapter we discuss the ``D1D5 system'' in generality, quoting
relevant results from the literature and providing the reader with
some necessary background. The basic setup is IIB string theory
compactified on $T^4\times S^1$ with a bound state of D5-branes
wrapping the whole compact space and D1-branes wrapping the circle,
$S^1$.\footnote{More generally, we should consider compactifications
  on compact hyperk\"{a}hler four-dimensional manifolds, $M^4$,
  crossed with a large $S^1$. There are two compact hyperk\"{a}hler
  four-dimensional manifolds: $T^4$ and $K3$. For this dissertation,
  we restrict our attention to $T^4$.} The configuration is summarized
in Table~\ref{tab:brane-conf}.  The $S^1$ direction is distinguished
from the $T^4$ because we want to consider the $S^1$ to be much larger
than the $T^4$.

Just as for the D3-brane system, there are two dual descriptions of
the D1D5 system. Depending on the coupling, one can consider either an
open string (field theory) description of the brane configuration or a
closed string (gravitational) description. In this dissertation, we
focus primarily on the field theory description. One can think about
the field theory in several different ways. Most fundamentally,
perhaps, one can think about the effective theory that arises from the
zero modes of the open strings and its flow into the IR~\cite{dmw02,
  MAGOO}; however, it turns out that the the description one arrives
at by considering open strings has a limited regime of validity.
Alternatively, one can consider starting with the D5-brane worldvolume
gauge theory in which the D1-branes are realized as strings of
instantons~\cite{douglas-branes, vafa-instantons}. In the IR, one need
only describe the behavior of the instantons.  Finally, there is a
specific proposed CFT, the symmetric product CFT that is believed to
describe the IR of a specific point in moduli
space~\cite{vafa-instantons, MAGOO, dijkgraafD1D5, gks,
  seiberg-more-comments, swD1D5, lmD1D5, Jevicki, dvv-1, dvv-2,
  deBoerD1D5}. Much of this chapter's material can be found in the
review~\cite{dmw02}.

\section{IIB on a Torus}

As mentioned, we are interested in compactifying string theory on a
small $T^4$ down to a six-dimensional theory. In the six-dimensional
theory, the D1 and D5 branes appear as strings wrapping the large
$S^1$. Before thinking about the D1D5 system specifically, let us
first turn our attention to IIB string theory on $T^4\times S^1$.

\subsection{IIB on \texorpdfstring{$T^5$}{T5}}

For the moment, let us follow~\cite{lmD1D5} and consider the circle
$S^1$ as part of a five-torus, $T^5$. We then consider how the physics
compactified on the $T^4$ is embedded in the $T^5$ physics. This
approach proves useful in organizing the content of the theories.

The massless, bosonic field content of IIB string theory is summarized
in Table~\ref{tab:IIAIIB}. When we compactify on $T^5$ we get a set of
$U(1)$ gauge fields, $A_\mu$ in the resulting five-dimensional theory.
For this section, we use lower-case Greek indices for the noncompact
five-dimensional Minkowski space, $M^{4,1}$ and capital Latin indices
for the compact $T^5$. We get a $U(1)$ gauge field by considering any
of the fields with only one index in the noncompact space:
\begin{equation}
g_{\mu A}\, (5)\qquad
B_{\mu A}\, (5) \qquad
C^{(2)}_{\mu A}\, (5)\qquad
C^{(4)}_{\mu ABC}\, (10),
\end{equation}
which gives us 25 $U(1)$ gauge fields (the number of degrees of
freedom for each field is shown in parentheses). There are two more
that we missed. If one takes the Hodge star of the exterior derivative
of one of the two-forms, $B^{NS}_{\mu\nu}$ or $C^{(2)}_{\mu\nu}$, then
one gets a two-index field strength. These correspond to two
``magnetic'' U(1) gauge fields.  Thus, we have a total of 27 $U(1)$
gauge fields. They couple to 27 different charges: 5 Kaluza--Klein
(KK) momenta, 5 F1 winding modes, 1 NS5 brane wrapping $T^5$, 5 D1
winding modes, 10 D3 modes, and 1 D5 brane wrapping $T^5$. They
source, respectively, the metric, the NS B-field, the magnetic NS
B-field, the RR 2-form, the RR 4-form, and the magnetic RR 2-form
gauge fields.

How many scalars do we get? We get 42 scalars from
\begin{equation}
g_{AB}\, (15)\qquad
B_{AB}\, (10)\qquad
\phi\,   (1) \qquad
C_{ABCD}\, (5)\qquad
C_{AB}\,   (10)\qquad
C^{(0)}\, (1),
\end{equation}
where we have suppressed the differential form superscript on the
Ramond--Ramond fields. The 42 scalars parameterize the 42-dimensional
moduli space of the 5-dimensional theory compactified on $T^5$. The
moduli space has the local geometry of the 42-dimensional coset
manifold~\cite{polchinski-2, hull-townsend, johnson, lmD1D5}
\begin{equation}
\frac{E_{6(6)}}{Sp(4)},
\end{equation}
where $E_{6(6)}$ is a noncompact version of $E_6$ with $6$ ``sign
flips.''  It has the same \emph{complexified} Lie algebra as $E_6$,
and is 78-dimensional. The symplectic group $Sp(4)$ is 36-dimensional,
so the coset space is 42-dimensional as advertised.  This is not a
global description of moduli space since we must mod out by a discrete
duality group to avoid redundant descriptions.

Type IIB supergravity has a continuous $E_{6(6)}$ symmetry which gets
broken to a discrete $E_{6(6)}(\ints)$ U-duality group in full string
theory~\cite{polchinski-2, hull-townsend, johnson, lmD1D5}. The
U-duality group is generated by $S$ and $T$-duality transformations
that mix the 27 charges.  For instance, $S$-duality exchanges F1 and
D1 winding charges, NS5 and D5 charges. D3-branes are self-dual, and
KK-momenta are invariant under $S$-duality. The 27 charges transform
as a fundamental vector of $E_{6(6)}(\ints)$.

\begin{table}
\begin{center}
\begin{tabular}{c|c|c|c|c}
$SO(5,1)_\text{E}$ rep & $SO(5,5)$ rep & charges & gauge field & mass\\
\hline\hline
tensor & vector (\textbf{10}) & $\{n_5, d_5, f_1, d_1, D^{5ij}\}$ 
  & $\{B^M_{\mu 5}, C^M_{\mu 5}, B_{\mu 5}, C_{\mu 5}, C_{5ij\mu}\}$
  & $1/{\ell_s^2}$\\ \hline
vector & spinor (\textbf{16}) & $\{\vec{w}_{F1}, \vec{w}_{D1}, D_{ijk}, \vec{p}\}$ 
  & $\{B_{\mu i}, C_{\mu i}, C_{\mu ijk}, g_{\mu i}\}$
  & $1/{\ell_s}$\\ \hline
scalar & scalar (\textbf{1}) & $p_5$ & $g_{\mu 5}$ & $\ell_s^0$
\end{tabular}
\caption[The charges and gauge fields of IIB strings compactified on
$T^4$]{Type IIB string theory compactified on $T^5$ has 27 U(1)
  charges in the fundamental representation of the $E_{6(6)}$ duality
  group. This \textbf{27} transforms as a \textbf{10}, \textbf{16},
  and \textbf{1} of the $SO(5,5)$ subgroup that is preserved when we
  take the near-horizon limit. The 27 gauge fields and their
  corresponding sources are shown in the above table along with their
  transformation properties. The $M$ superscript indicates magnetic
  coupling. The mass column shows the scaling as we take $\ell_s$
  small, which corresponds to the near-horizon
  limit.}\label{tab:charges}
\end{center}
\end{table}

\subsection{Relationship with IIB on \texorpdfstring{$T^4$}{T4}}

We have described the structure of the 5-dimensional theory that
results from compactifying IIB string theory on $T^5$. What happens
when we take one of the $S^1$'s to be large and consider the
6-dimensional theory? We keep the $T^4$ at the string scale.  The
$E_{6(6)}$ structure gets broken to $SO(5,5)$, and the \textbf{27} of
$E_{6(6)}$ gets broken into a \textbf{10} (vector), a \textbf{16}
(spinor), and a \textbf{1} (singlet) of $SO(5,5)$. Each representation
has a different mass scaling with $\ell_s$~\cite{lmD1D5}. The way the
different charges break up is summarized in Table~\ref{tab:charges}.
The U-duality group $E_{6(6)}(\ints)$ gets broken to a subgroup
$SO(5,5;\ints) \leq E_{6(6)}(\ints)$.

The moduli space of the 6-dimensional theory is 25-dimensional, and
locally is~\cite{swD1D5, lmD1D5, hull-townsend, polchinski-2, johnson}
\begin{equation}
\frac{SO(5,5)}{SO(5)\times SO(5)}.
\end{equation}
The moduli space is parameterized by (we now use lower-case Latin for
indices in the $T^4$ and `5' for the large $S^1$) the 25 fields:
\begin{equation}\label{eq:25sugra-moduli}
g_{ij}\,(10)\qquad
B_{ij}\,(6)\qquad
\phi\, (1)\qquad
C_{ijkl}\,(1)\qquad
C_{ij}\,(6)\qquad
C^{(0)}\, (1).
\end{equation}

Having given some generalities, let us now more precisely specify the
D1D5 system. Of the heavy $SO(5,5)$ vector charges we want only to
have $d_1$ and $d_5$ charges, so $n_5=f_1=D^{5ij} = 0$. Similarly, we
want the spinor to vanish. Note that we want D1 winding charge only on
the $S^1$, $d_1$, and not winding charge on the $T^4$, $\vec{w}_{D1}$.
At times, we do want to consider some $p_5$ charge. This is sometimes
referred to as the D1D5P system. Thus we restrict our attention to a
three-dimensional subspace of the 27 possible charges. As explained
in~\cite{lmD1D5}, one can use U-duality to rotate $d_1$ and $d_5$ into
each other, preserving the product $d_1d_5$.

\section{Supergravity Description}

We now turn to the closed string description of the D1D5 system.  As
the gravity description, is not our primary interest in this paper, so
our treatment is appropriately brief.  See~\cite{dmw02, johnson} for
more thorough treatments.

\subsection{The Two-charge Extremal Black Hole}

There is a black hole solution that is asymptotically
\begin{equation}
M^{4,1}\times S^1\times T^4,
\end{equation}
and has only D1 and D5 charge. The geometry can be found by using the
``harmonic superposition rule''~\cite{tseytlin-harmonic} and
$T$-duality\footnote{\cite{hassanT} may come in handy.} to get to the
appropriate duality frame (see~\cite{dmw02}, e.g.). The solution can
be written down in the form~\cite{johnson}\footnote{The metric is
  written in the Einstein frame.}
\begin{equation}\begin{aligned}\label{eq:d1d5naive}
\left(\frac{H_1}{H_5}\right)^{\frac{1}{4}}
 ds_E^2 &= \frac{1}{\sqrt{H_1H_5}}\left(-dt^2 + dy^2\right) 
        + \sqrt{H_1H_5}\left(dr^2 + r^2d\Omega_3^2\right)
       + \sqrt{V_{T^4}}\sqrt{\frac{H_1}{H_5}}ds^2_{T^4}\\
e^{\phi} &= g_s^\infty\sqrt{\frac{H_1}{H_5}}\\ 
F^{(3)} &= dC^{(2)}\qquad
F^{(3)}_{rty} = \pd_r H_1^{-1}\qquad 
F^{(3)}_{\theta\phi\chi} = 2Q_5 \sin^2\theta \sin\phi,
\end{aligned}\end{equation}
where
\begin{equation}
H_1 = 1 + \frac{Q_1}{r^2}\qquad
H_5 = 1 + \frac{Q_5}{r^2},
\end{equation}
and $r$ is the radius in the noncompact space; $\theta$, $\phi$ and
$\chi$ are the angular coordinates of $\Omega_3$; and $g_s^\infty$ is the
string coupling at infinity. We coordinatize the $S^1$ with $y$ and
the $ds^2_{T^4}$ is the metric on $T^4$ with unit volume. The solution
is more precisely termed a black ring, since the $S^1$ direction is
large. This is the extremal 2-charge black ring solution. The horizon
coincides with the singularity at $r=0$. We find it convenient to
write the volume of the $T^4$ at infinity as
\begin{equation}\label{eq:v-def}
V_{T^4} = (2\pi\ell_s)^4v_\infty,
\end{equation}
where $\ell_s^2 = \Regge$. The charges $Q_1$ and $Q_5$ are related to
the integer-valued charges $d_1$ and $d_5$ via
\begin{equation}\label{eq:Q-def}
Q_1 = \frac{g_s^\infty\ell_s^2}{v_\infty}d_1\qquad
Q_5 = g_s^\infty \ell_s^2 d_5.
\end{equation}
The solution preserves one-fourth of the supersymmetry of IIB string
theory, as expected for a two-charge solution.

This is the ``naive'' black hole solution, which we discuss in
Chapters~\ref{ch:intro} and~\ref{ch:coupling}. The fuzzball proposal
suggests that this solution results from some form of coarse-graining
over nonsingular, horizonless microstates. For the 2-charge extremal
black ring, all of these solutions have been found~\cite{lm3, st-1,
  bal, mm}.  They are exact solutions of supergravity that are smooth,
horizonless and asymptotically look the same as the naive
solution---they have the same charge and mass. Geometric quantization
of the space of solutions~\cite{rychkov} reproduces the entropy of the
black hole as computed, for instance from the brane
description~\cite{marolf-palmer, bak-2-charge}.\footnote{We would be
  remiss if we did not note that~\cite{sen-revisit} gives a discussion
  of potential problems with the interpretation of these results
  within the fuzzball proposal.}

Let us investigate the asymptotic behavior of the solution. For very
large $r$, $H_1$ and $H_5$ become unity and the solution is flat
5-dimensional Minkowski with an $S^1$ and a $T^4$ as advertised. The
RR 2-form flux is appropriate for the D1 and D5 charges. The string
coupling is $g_s^\infty$.

We can also consider the small $r$ limit when we drop the `1' in $H_1$
and $H_5$. Then, the solution takes the form
\begin{equation}\begin{aligned}\label{eq:near-horizon-metric}
ds_S^2 = \left(\frac{Q_1}{Q_5}\right)^{\frac{1}{4}}ds_E^2
 &= \frac{r^2}{\sqrt{Q_1Q_5}}\left(-dt^2 + dy^2\right)
    + \sqrt{Q_1Q_5}\frac{dr^2}{r^2} + \sqrt{Q_1Q_5}d\Omega_3^2
    + \sqrt{V_{T^4}\frac{Q_1}{Q_5}}ds^2_{T^4}\\
e^{\phi} &= g_s^\infty\sqrt{\frac{Q_1}{Q_5}}.
\end{aligned}\end{equation}
We see that both the $T^4$ volume and the string coupling get scaled
by $Q_1/Q_5$; the $S^3$ gets a fixed size; and the $r$, $t$ and $y$
coordinates form an $\AdS_3$ space. Both the $\AdS_3$ and the $S^3$
have radius $(Q_1Q_5)^\frac{1}{4}$. These facts are for the string
frame metric. One can study the Einstein metric and see that the event
horizon at $r=0$ has vanishing surface gravity and area. The vanishing
surface gravity means that the black hole has zero temperature and
therefore does not Hawking radiate---this is expected for an extremal
hole. The geometry receives $\Regge$ higher derivative corrections,
which give the geometry (in other duality frames, at least) a nonzero
area and therefore entropy~\cite{stromvafa, sen-2, dabholkar-1,
  dabholkar-2, sen-review}. The $(r,t,y)$ part of the metric
in~\eqref{eq:near-horizon-metric} is an extremal BTZ black
hole~\cite{BTZ-1, BTZ-2}, so we can think of the near-horizon geometry
as $\text{BTZ}\times S^3\times T^4$.

In order for classical supergravity to be a good description we need
the curvature to be small (radius of the $\AdS_3$ to be large) with
respect to the string scale. We also need the string coupling to be
small so that we do not need to consider loops. These requirements
translate into a large-charge limit~\cite{MAGOO, dmw02}
\begin{equation}\label{eq:large-charge}
d_1,\, d_5 \gg \frac{1}{g_s} \gg 1.
\end{equation}
In the large-charge limit the entropy can be computed using a CFT
description and agreement is found~\cite{stromvafa, sen-1, sen-2}.

\subsection{Generalizations}

One can similarly construct a 3-charge D1D5P extremal black hole
solution and look for corresponding microstate geometries~\cite{gms1,
  gms2, bena-1, bena-2, bena-4}. The extremal black holes are where
the fuzzball proposal is best tested; however, extremal black holes
are peculiar. They have an infinitely long AdS throat that can
nevertheless be traversed in a finite proper time by massless
particles and they saturate a BPS bound. They have the minimum amount
of mass for the amount of charge they have, so they do not Hawking
radiate. It is interesting, therefore, to consider the fuzzball
proposal for nonextremal black holes. This is the subject of
Chapter~\ref{ch:emission}.

The solution in Equation~\eqref{eq:d1d5naive} and its 3-charge
extremal generalization have none of the 25 moduli listed in
Equation~\eqref{eq:25sugra-moduli} turned on. This point in moduli
space is ``marginally bound''~\cite{swD1D5, dmw02, lmD1D5, johnson,
  witten-higgs}, which means that the D1 and D5 branes can separate
with no additional energy. One can also construct solutions that are
not marginally bound by adding, for example, NS B-field on the
torus~\cite{maldacena-russo, dmwy}. In fact, one can show that the
branes may split apart with no additional energy if and only if both
the appropriate linear combination of the RR scalar $C^{(0)}$ and the
RR 4-form $C^{(4)}_{6789}$, as well as the NS B-field
vanish~\cite{lmD1D5, dijkgraafD1D5}.

For reviews of progress in the fuzzball proposal, we refer the reader
to~\cite{bala-review, fuzzballs-elementary, fuzzballs-structure,
  bena-review, skenderis-review, cv}. For a review of accounting for
the entropy of black holes by performing weak coupling microscopic
counting, see~\cite{sen-review}.

\subsection{The Near-Horizon Limit}

In order for us to make use of the AdS--CFT correspondence, we must
take a near-horizon limit that decouples the asymptotic flat physics
from the AdS physics~\cite{maldacena, MAGOO}. This decoupling limit
corresponds to going to the IR fixed point of the D-brane description,
and is in effect a low-energy limit. As demonstrated above, in this
limit, the supergravity description becomes $\AdS_3\times S^3\times
T^4$. One might expect that by appropriately tuning the 25 moduli in
Equation~\eqref{eq:25sugra-moduli} at asymptotic infinity one could
attain any value of the moduli in the AdS-space. If this were the
case, the near-horizon geometry would have a 25-dimensional moduli
space. In fact, this is \emph{not} the case: The 25-dimensional moduli
space at infinity gets ``attracted'' to a certain submanifold in the
near-horizon limit that depends only on the charges~\cite{attractor-1,
  attractor-2, attractor-3, attractor-4}. The attractor mechanism
results from the extremality of the system and does not require
supersymmetry~\cite{attractor-2, nonsusy-attractor-1,
  nonsusy-attractor-2, nonsusy-attractor-3}.

For the D1D5 system, the 25-dimensional moduli space (at infinity) is
attracted to a 20-dimensional subspace in the near-horizon
limit---five of the scalars get ``fixed,'' and excitations around the
preferred value acquire a mass. One can find the near-horizon moduli
space by minimizing the mass of the bound state\footnote{More
  generally the critical points of the central charge.} with respect
to the moduli~\cite{attractor-4}, in which case one finds the
constraints~\cite{lmD1D5}:
\begin{subequations}\label{eq:attractor-constraints}
\begin{align}
v B_{ij}g^{ik}g^{jl} &= \frac{1}{2}B_{ij}\epsilon^{ijkl}\label{eq:self-dual}\\
v C^{(0)} &= C^{(4)}_{6789} - \frac{1}{8}\epsilon^{ijkl}B_{ij}C^{(2)}_{kl}
   \label{eq:C0C4}\\
v + \frac{1}{8}\epsilon^{ijkl}B_{ij}B_{kl} &= \frac{d_1}{d_5},\label{eq:fixed-volume}
\end{align}
\end{subequations}
where $v$ is the near-horizon analog of $v_\infty$ in
Equation~\eqref{eq:v-def}. It is worth emphasizing that the moduli in
Equation~\eqref{eq:attractor-constraints} are the \emph{near-horizon}
moduli. If we set all of the moduli to zero, then we find
\begin{equation}
v = \frac{d_1}{d_5} = v_\infty\frac{Q_1}{Q_5}
\end{equation}
after using Equation~\eqref{eq:Q-def}. This is consistent with the
solution in Equation~\eqref{eq:d1d5naive}. 

Equation~\eqref{eq:self-dual} gives a self-duality constraint on the
NS B-field; Equation~\eqref{eq:C0C4} fixes a linear combination of the
RR scalar and the RR 4-form; Equation~\eqref{eq:fixed-volume} fixes
the volume of the $T^4$, $v$, at the horizon. Note that the metric in
Equations~\eqref{eq:attractor-constraints} is the string metric, and
that we treat the volume of the $T^4$ as a separate modulus, $v$.
Then, we see that Equation~\eqref{eq:self-dual} forces the NS B-field
to be self-dual (under the Hodge star). The 3 component
anti--self-dual part gets set to zero, and perturbations about that
point are massive.

We would like to find a basis for the 20-dimensional near-horizon
moduli space that solves the
constraints~\eqref{eq:attractor-constraints}.  The basis that we find
convenient to use is listed in Table~\ref{tab:sugra-near-moduli}. The
scalar $\Xi$ is the linear combination of $C^{(0)}$ and
$C^{(4)}_{6789}$ that remains massless (ie. is tangent to the
constraint surface defined by~\eqref{eq:C0C4}). The rightmost column
counts the degrees of freedom, which one sees sum to 20 as claimed. As
mentioned, the near-horizon geometry is $AdS_3\times S^3\times T^4$.
It is helpful to identify the fields in terms of their representation
under the $SO(4)_E$ isometry of the $S^3$ and also the $SO(4)_I$
symmetry of the $T^4$ tangent space. The $SO(4)_I$ symmetry is broken
by the compactification, but it is still useful as an organizing
principle~\cite{Jevicki, dmw02}. The two $SO(4)$ algebras are more
naturally thought of in terms of $SU(2)\times SU(2)$ for later
connection with the CFT description.

\begin{table}[ht]
\begin{center}
\begin{tabular}{c|c|c|c}
Field & $SO(4)_E \simeq SU(2)_L\times SU(2)_R$ & $SO(4)_I \simeq SU(2)_1\times SU(2)_2$ & DOF\\\hline
$g_{ij} - \frac{1}{4}\delta_{ij}g_{kk}$ & (\textbf{1}, \textbf{1})
    & (\textbf{3}, \textbf{3}) & 9\\
$B^+_{ij}$ & (\textbf{1}, \textbf{1}) & (\textbf{3}, \textbf{1})  & 3\\
$C^{(2)}_{ij}$ & (\textbf{1}, \textbf{1})
    & (\textbf{3}, \textbf{1}) $\oplus$ (\textbf{1}, \textbf{3}) & 6\\
$\Xi$ & (\textbf{1}, \textbf{1}) & (\textbf{1}, \textbf{1}) & 1\\
$\phi$ & (\textbf{1}, \textbf{1}) & (\textbf{1}, \textbf{1}) & 1\\\hline
 & & & 20
\end{tabular}
\caption[The near-horizon supergravity moduli]{Table of the
  gravitational near-horizon moduli. These fields do not have any
  preferred value at the horizon. 5 of the 25 moduli at infinity get
  ``fixed'' in the near-horizon limit---these are the 20 remaining
  near-horizon moduli. Note that they are all singlets of the
  $SO(4)_E$ isometry of $S^3$.}\label{tab:sugra-near-moduli}
\end{center}
\end{table}

Having lost 5 of the 25 fields that parameterize the moduli space at
infinity, it is natural to ask what happens to $SO(5,5)/SO(5)\times
SO(5)$ local geometry of moduli space. The 20-dimensional near-horizon
moduli space is locally the coset space~\cite{swD1D5, lmD1D5}
\begin{equation}\label{eq:local-mod-near}
\mathcal{K_\text{sugra}^*} = \frac{SO(5,4)}{SO(5)\times SO(4)} \subset
\frac{SO(5,5)}{SO(5)\times SO(5)},
\end{equation}
where again we must mod out by a discrete duality group that preserve
the background charges, $\Gamma_{\vec{q}}\leq SO(5,5;\ints)$. One can
easily check that $\mathcal{K}_\text{sugra}^*$ is 20-dimensional, as
claimed. We call the full moduli space
\begin{equation}
\mathcal{M}^*_{\text{sugra}, q} 
   = \Gamma_{\vec{q}}\backslash \mathcal{K}^*_\text{sugra}
\end{equation}
for later use.

\section{The Brane Description}

Classical supergravity, as discussed above, is a good description of
the D1D5 system in the large-charge limit (with large $g_sd_1$ and
$g_s d_5$) in Equation~\eqref{eq:large-charge}. From the D-brane
tension~\eqref{eq:Tp} and the DBI action~\eqref{eq:DBI}, we see that
$\sqrt{g_s}$ plays the role of field theory coupling for the brane
description.  This is consistent with the interpretation of
$\sqrt{g_s}$ as the open string coupling constant. In the large-charge
limit, then we want to keep the 't Hooft couplings small. Thus the
brane description should be weakly coupled when
\begin{equation}
1 \ll d_1, d_5\ll \frac{1}{g_s}.
\end{equation}
We wish to find a description in this limit, which should be dual to
the above supergravity description.

As before, we wish to emphasize that AdS--CFT arises from open--closed
string duality. The supergravity description arises from the low-energy
behavior of the closed string modes. We now wish to find the CFT
description that arises from the low-energy behavior of the open
string modes. In fact, the behavior of the open string modes should
first give rise to a nonconformal field theory. We then RG flow to the
IR fixed point CFT. This should (loosely) correspond to taking the
near-horizon limit of the gravity description. Just as there is a
20-dimensional near-horizon moduli space for the gravity description,
there is a 20-dimensional moduli space of CFTs. At the orbifold point,
for instance, one can explicitly identify 20 exactly marginal
perturbations that move one in 20 different directions in CFT space,
and further relate them to the 20 supergravity moduli in
Table~\ref{tab:sugra-near-moduli}. In fact, one can do more---one can
even find 5 irrelevant perturbations that correspond to the 5
directions in moduli space that got fixed in the near-horizon limit.
This helps make the connection between RG flow and the near-horizon
limit more precise. For more on these points
see~\cite{deBoer-attractor}.

Following~\cite{dmw02}, we give two descriptions of the open string
modes. First, we explicitly consider the gauge theory that results
from considering open strings that begin and end on either the D1 or
the D5 branes. This model makes a lot of comments about the D1D5
system and its moduli space more explicit (see e.g.~\cite{gks, dmw02,
  hassan-wadia, maldacena-thesis}). We then recognize that we can use
a ``branes within branes'' description~\cite{douglas-branes,
  vafa-instantons}, where the D1s are realized as instanton
configurations within the D5-brane worldvolume gauge theory. This
description has a larger regime of validity, and more directly
connects with the orbifold model of the D1D5 CFT.

\subsection{Gauge  Theory Description}\label{sec:gauge-desc}

The effective field theory description of D-branes results from the
open string zero modes. Working within perturbative string theory, the
D-branes define boundary conditions for the open strings. There are
three types of strings we may consider: 5--5 strings with both
endpoints on D5 branes; 1--1 strings with both endpoints on D1
branes; and 1--5 and 5--1 strings with one endpoint on a D1 and
one endpoint on a D5.

\subsubsection{The 5--5 Strings}

Let us first consider the 5--5 strings. If we just had a stack of $d_5
= N_5$ D5-branes, then the open string modes give rise to a
5+1-dimensional $U(N_5)$ gauge theory with 16 supercharges. This is
precisely what the 5--5 open strings give. Let us recall the heuristic
picture of how the bosonic degrees of freedom arise. The open strings
with polarization parallel to the branes have Neumann boundary
conditions and give rise to a $U(N_5)$ gauge field.  The open strings
with polarization perpendicular to the branes have Dirichlet boundary
conditions and give adjoint scalars of the 5+1-dimensional theory. The
adjoint scalars describe the transverse oscillations of the D-branes.
When the scalars acquire a vev it corresponds to some of the D-branes
separating in the corresponding direction. If the branes separate in
some direction then the $U(N_5)$ gauge group is broken.  When all of
the D-branes are coincident, the gauge theory is said to be in the
Higgs phase, and when some of the D-branes have separated the gauge
theory is said to be in the Coulomb phase.

The gauge theory should have 16 supercharges, since a stack of
D-branes breaks half of the 32 supercharges of IIB string theory.  In
5+1-dimensions, the 16 supercharges break into $\Nsc = 2$ Weyl
spinors.  One can find the Lagrangian for the 5--5 open string modes
by dimensionally reducing $\Nsc = 1$ 9+1-dimensional $U(N_5)$
super--Yang-Mills theory to 5+1 dimensions. One can see this from
T-duality of a space-filling D9-brane, for instance. The coupling
constant for the worldvolume gauge theory of a D$p$-brane can be
identified as~\cite{johnson}
\begin{equation}
  (g_{\YM,p})^2 = g_s(2\pi)^{p-2}\Regge^{\frac{p-3}{2}}\longrightarrow
g_{\YM,5} = g_s (2\pi)^{3}\ell_s.
\end{equation}

Recall that we are taking the size of $T^4$ to be on the string scale.
Therefore, we should dimensionally reduce the 5+1 dimensional theory
down to 1+1 dimensions parameterized by time and the $S^1$
coordinates, $t$ and $y$. The KK momentum modes should be dropped,
since they are very massive, and we are interested in the low-energy
theory. The coupling in the 1+1-dimensional theory gets a factor of
$v$ from the dimensional reduction.

The literature (e.g.~\cite{dmw02, MAGOO, maldacena-thesis,
  hassan-wadia, johnson}) frequently organizes the field content of
the theory according to a four-dimensional $\Nsc=2$ classification. In
four-dimensions, $\Nsc=2$ corresponds to 8 supercharges (either two
four-component Majorana spinors or two (complex) two-component Weyl
spinors). There are two massless supersymmetry representations that
are relevant to us: a vector multiplet and a hypermultiplet. The
vector multiplet (on-shell) consists of a vector field, two spin-half
Weyl fermions, and a complex scalar. The off-shell multiplet has three
additional auxiliary real scalars (usually called $D$).  The
hypermultiplet consists of two complex scalars, two Weyl spinors, and
two complex auxiliary scalars (usually called $F$). The $\Nsc=2$
four-dimensional supersymmetry has an $SU(2)_R$ symmetry.  In the
vector multiplet, the vector field and complex scalars are singlets,
the Weyl spinors transform as two doublets, and the three auxiliary
$D$-fields transform as a triplet. In the hypermultiplet, the fermions
transform as singlets, the two complex scalars transform as a doublet,
as do the two complex $F$-fields. See e.g.~\cite{sohnius, argyres}.

We can organize the bosonic field content that comes from the 5--5
strings as \linebreak follows~\cite{maldacena-thesis, dmw02}:
\begin{equation}\begin{aligned}\label{eq:5-5fields}
&\text{vector:}\quad & &A^{(5)}_t, A^{(5)}_y, \vec{\Phi}^{(5)}\\
&\text{hyper:}      & &\Phi^{(5)}_i
\end{aligned},\end{equation} where the vector symbol is used to denote
a vector in the four-dimensional spatial part of $M^{4,1}$. The
$\Phi$'s are the scalars resulting from open strings polarized
perpendicular to the $t$--$y$ space, and the $A$'s are the
two-components of the gauge field. All of the above are adjoints of
the $U(N_5)$ gauge group. Note that we must take two degrees of
freedom from $\vec{\Phi}$ along with the two gauge field components to
form a \emph{four}-dimensional vector of the vector multiplet---the
remaining two components of $\vec{\Phi}$ are the two scalars.  

\subsubsection{The 1--1 Strings}

The 1--1 strings in actuality differ very little from the 5--5
strings. If we have $d_1=N_1$ D1 strings, then the opens strings must
give rise to a $U(N_1)$ gauge theory in 1+1 dimensions that has 16
supercharges.  The interpretation of the open string modes parallel
and perpendicular to the branes is the same as above, of course. One
can derive the Lagrangian by dimensionally reducing $\Nsc = 1$
9+1-dimensional $U(N_1)$ super--Yang-Mills down to 1+1 dimensions.
Notice that is precisely what we ended up doing for the 5--5 open
string modes, replacing $N_1$ with $N_5$. The $U(N_1)$ Yang--Mills
coupling is
\begin{equation}
g_{\YM, 1} = \frac{g_s}{2\pi\ell_s^2}.
\end{equation}
We organize the 1--1 field content analogously to
Equation~\eqref{eq:5-5fields},
\begin{equation}\begin{aligned}
&\text{vector:}\quad & &A^{(1)}_t, A^{(1)}_y, \vec{\Phi}^{(1)}\\
&\text{hyper:}      & &\Phi^{(1)}_i
\end{aligned}.\end{equation} 
The fields are labeled in the same fashion as for the 5--5 strings,
and they all take values in the adjoint of $U(N_1)$.

\subsubsection{The 5--1 and 1--5 Strings}

The 5--1 and 1--5 strings are the most important for many of our
purposes.  Note that both the 1--1 and the 5--5 theories have 16
supercharges ($\Nsc =4$ in 3+1), but we expect that the bound state
should have 8 supercharges. The 5--1 and 1--5 strings break the
supersymmetry down to 8 supercharges.

Let us first consider 5--1 strings. These strings are bifundamental,
fundamental under $U(N_5)$ and antifundamental under $U(N_1)$. The
behavior of the string depends on the polarization. In particular, the
NN and DD directions behave differently from the ND or DN
directions.\footnote{N and D refer to Neumann and Dirichlet boundary
  conditions on the end points. Since the string has an orientation ND
  is distinct from DN.  Note that for 1--1 or 5--5 strings we can only
  have NN or DD boundary conditions.} For 5--1 strings, the NN
directions are time and the $S^1$ while the DD directions are the
noncompact space. The ND directions are the $T^4$, and there are no DN
directions. There are two sectors for the open string, NS and R. In
the NS sector, the ground states consist of four zero modes $\psi^i_0$
in the DN direction (the $T^4$)---GSO projection cuts that down to two
bosonic degrees of freedom. The excitations polarized in the NN and DD
directions are massive in the NS sector. In the R sector, only the
excitations in the NN and DD are massless, and give rise to two
on-shell fermionic degrees of freedom after GSO
projection~\cite{johnson, dmw02, maldacena-thesis}.

The 1--5 strings work out very similarly: we get two bosonic degrees
of freedom from the NS sector and two fermionic degrees of freedom
from the R sector. All of the bosonic degrees of freedom are polarized
in the $T^4$ directions, while the fermionic degrees of freedom are
polarized in the time, $S^1$, and noncompact directions.  Combining
the 5--1 string massless excitations with the 1--5 massless
excitations to from a bifundamental hypermultiplet. As above, we focus
on the bosonic content~\cite{dmw02, johnson, maldacena-thesis}:
\begin{equation}
\text{hyper:}\quad\chi^{A},
\end{equation}
a spinor of $SU(2)_1$, where recall $SO(4)_I = SU(2)_1\times SU(2)_2$.
We also have its complex conjugate, $\chi^\dg$.

\subsubsection{The Potential and Moduli Space}

For this section, let us call the bosonic part of the vector
multiplets $Y^{(1)}_a$ and $Y^{(5)}_a$, with $a$ running over the
noncompact $\re^4$ as well as the $t$ and $y$ coordinates. The
Lagrangian for the theory we have outlined can be deduced by
dimensionally reducing $d=6$ $\Nsc=1$ field theory with $U(N_1)$ and
$U(N_5)$ gauge groups. Of primary importance for our discussion here,
is the potential on the bosonic field content.

We quote the result from~\cite{dmw02}; the potential may be written as
the sum of four terms:
\begin{subequations}\label{eq:the-potential}
\begin{align}
V_1 &= -\frac{1}{4g_1^2}\sum_{a,b}\Tr_1\com{Y^{(1)}_a}{Y^{(1)}_b}^2
       - \frac{1}{4g_5^2}\sum_{a,b}\Tr_5\com{Y^{(5)}_a}{Y^{(5)}_b}^2\\
V_2 &= -\frac{1}{4g_1^2}\sum_{i,a}\Tr_1\com{\Phi^{(1)}_i}{Y^{(1)}_a}^2 
       - \frac{1}{g_5^2}\sum_{i,a}\Tr_5\com{\Phi^{(5)}_i}{Y^{(5)}_a}^2\\
V_3 &= \frac{1}{4}\sum_{a}\Tr_1(\chi Y^{(5)}_a - Y^{(1)}_a \chi)
                          (Y^{(5)}_a \chi^\dg - \chi^\dg Y^{(1)}_a)\\
\begin{split}
V_4 &= \frac{1}{4}\Tr_1\left(\chi\, i\Gamma^T_{ij} \chi^\dg 
   + i\com{\Phi^{(1)}_i}{\Phi^{(1)}_j}^+ - \frac{\zeta_{ij}^+}{N_1}\right)^2\\
 &\quad+ \frac{1}{4}\Tr_5\left(\chi^\dg\, i\Gamma_{ij}\chi 
   + i\com{\Phi^{(5)}_i}{\Phi^{(5)}_j} - \frac{\zeta^+_{ij}}{N_5}\right)^2,
\end{split}\end{align}
\end{subequations}
where the `$+$' superscript indicates the self-dual part of an
antisymmetric matrix. We use $i,j$ as indices in the compact $T^4$ and
$a, b$ as indices in the noncompact space parameterized by $\vec{x}$.
The $\zeta_{ij}$ are Fayet--Iliopoulos (FI) parameters, and the
$\Gamma_{ij} = \frac{1}{2}\com{\Gamma_i}{\Gamma_j}$ are the spinor
rotation matrices of $SO(4)_I$. Note that the above expressions may
use slightly different conventions than the above and following
discussion, but this does not affect any of our conclusions.

Since we want to discuss the IR limit, we are interested in
supersymmetric minima satisfying $V= V_1 + V_2 + V_3 + V_4 = 0$. The
space of solutions to $V=0$ are the moduli space of the
theory---different possible vev's that the fields can have. Note that
we are discussing a two-dimensional theory, and one usually expects IR
fluctuations to restore any continuous symmetry breaking (see
e.g.~\cite{coleman-no-GB}), and therefore we should not be speaking of
fields acquiring vevs; however, we use this language with the
understanding that we end up with the correct description with the
space of solutions to $V=0$ becomes the \emph{target space} of a
two-dimensional sigma model~\cite{witten-dynamics}.

\subsubsection{The Higgs and Coulomb Phase}

There are two classes of minima to the potential in
Equation~\eqref{eq:the-potential}: those where hypermultiplets are
zero and those where the vector multiplets are zero. The first is
called the Coulomb branch and the second is called the Higgs branch.
When the $Y_a$ acquire a nonzero vev that corresponds to some of the
branes separating in the noncompact space and breaking the gauge group
down. In the most extreme case we only have $U(1)$ gauge theories when
all of the branes are widely separated, hence the name Coulomb branch.
When the $Y_a$ are zero, in the Higgs branch, the branes all sit at
the origin of the noncompact space, but the hypermultiplets may have a
nonzero vev.  We are interested in describing the physics of the Higgs
branch when we truly have a bound state. When all of fields vanish, we
are at the ``intersection'' of the Higgs and Coulomb branches.

On the Higgs branch, then, $V_1=V_2=V_3=0$ by definition. Thus, we are
only left with the condition that $V_4$ vanish (which can be
interpreted as two $D$-flatness conditions). The Higgs branch is
parameterized by $\Phi^{(1)}_i$, $\Phi^{(5)}_i$, and the hypermultiplet
$\chi$. What role is played by the FI parameters, $\zeta$? One can
show that they are proportional to the amount of NS B-field turned on
in the supergravity description~\cite{dmw02, hassan-wadia}. Thus one
can see that when the NS B-field is non-vanishing we must lie on the
Higgs branch in order for $V=0$, and thus the branes cannot separate.
For a more careful treatment of these issues, see~\cite{dijkgraafD1D5,
  swD1D5, lmD1D5}.

Since $V_4$ is the sum of two positive definite terms, we must demand
that both terms vanish individually. Let us write out the components
of the doublet as
\begin{equation}
\chi= \begin{pmatrix}A\\ B^\dg\end{pmatrix}.
\end{equation}
If one carefully analyzes the solutions to $V_4=0$ modulo gauge
invariance, then one finds that $\chi$ must satisfy
\begin{subequations}\label{eq:hyper-constraint}\begin{align}
\Tr_5(A^\dg A) - \Tr_1(B^\dg B) &= \zeta^+_{69}\\
\Tr_5(AB^T) &= \zeta^+_{67}+i\zeta_{68}^+,
\end{align}\end{subequations}
the traceless parts of the $\Phi_i$'s are completely fixed, and the
traces are free. The 8 degrees of freedom associated with
$\Tr_1\Phi^{(1)}_i$ and $\Tr_5\Phi^{(5)}_i$ are the center of mass
degrees of freedom of the D1 branes and the D5 branes, respectively,
in the $T^4$. This entire analysis was performed
in~\cite{hassan-wadia} and reviewed in~\cite{dmw02}.

\subsubsection{The CFT Limit}

We should now take the extreme IR limit, which should correspond to
taking the near-horizon limit in the closed string description. Before
thinking about what we get from the above gauge theory, let us ask:
What are our basic expectations for the IR theory from the dual
supergravity description? First, we expect the theory to be a CFT,
since it is an IR fixed point and since it should be dual to anti-de
Sitter space whose isometries generate a Virasoro
algebra~\cite{MAGOO}. Second, since it should live on the boundary of
the near-horizon $AdS_3\times S^3\times T^4$, it should be a
two-dimensional conformal field theory. Furthermore it should have 8
supercharges. Finally, one can work out the central charges of the
$\text{CFT}_2$ by calculating the classical Poisson brackets of the
generators of diffeomorphisms that preserve the asymptotic AdS space;
under an infinitesimal transformations $\xi_\mu$ the metric transforms
as $g_{\mu\nu}\mapsto g_{\mu\nu} + \xi_{\mu;\nu} + \xi_{\nu;\mu}$. In
fact this analysis for $\AdS_3$ was performed in
1986~\cite{brown-henneaux}, over a decade before AdS--CFT! In any
case, one finds that the central charge is
\begin{equation}
c= \frac{3R_{\text{AdS}_3}}{2 G^{(3)}_N},
\end{equation}
where $G_N^{(3)}$ is the three dimensional Newton's constant and
$R_{\text{AdS}_3}$ is the AdS radius. If one plugs in with the
solution in Equation~\eqref{eq:near-horizon-metric}, then one finds 
\begin{equation}
c = 6N_1N_5 + \text{(subleading)},
\end{equation}
where the subleading corrections come from an expansion in
$N_1N_5$---we only expect the calculation to be correct in the large
charge limit when supergravity is a good description.

Now, let us follow~\cite{hassan-wadia, dmw02} and think about what the
IR theory for the $U(N_5)\times U(N_1)$ gauge theory outlined above
should be. Note that we already discussed how to dimensionally reduce
to a two-dimensional theory, and the theory has the right number of
supersymmetries. We should expect that the lowest energy excitations
should correspond to moving in ``flat'' directions of the potential.
Since we are interested in a description of the Higgs branch, these
excitations correspond to moving around in the moduli space defined by
$V_4=0$ with the vector multiplets having zero vev. Thus, we expect to
describe collective a collective mode for $A$ and $B$ subject to the
constraint~\eqref{eq:hyper-constraint}, as well as the traces
$\Tr_1\Phi^{(1)}_i$ and $\Tr_5\Phi_i^{(5)}$. The bosonic part of the
conformally invariant point is 
\begin{equation}
\int\drm t\drm y \Tr\left[\pd_\alpha A^\dg \pd^\alpha A 
   + \pd_\alpha B^\dg \pd^\alpha B\right],
\end{equation}
where the index $\alpha$ runs over $t$ and $y$; we also have the
center of mass degrees of freedom. The $A$ and $B$ of $\chi$ are
constrained to live on the target space defined
by~\eqref{eq:hyper-constraint}. The manifold defined
by~\eqref{eq:hyper-constraint} is hyper-K\"{a}hler, which is required
for $\Nsc = (4,4)$ supersymmetry of a two-dimensional sigma
model~\cite{dijkgraafD1D5}.  From the above field content (along with
the fermions that we have neglected slightly) the central charge is
\begin{equation}
c = \bar{c} = 6(N_1N_5 + 1),
\end{equation}
where the extra 6, comes from the center of mass degrees of freedom in
the torus: $\Tr_1\Phi^{(1)}_i$ and $\Tr_5\Phi_i^{(5)}$, and the
superpartners. In this description, the gauge group $U(N_5)\times
U(N_1)$ gets broken down to the discrete subgroup $S_{N_5}\times
S_{N_1}$~\cite{hassan-wadia, dmw02}. 

\subsubsection{Regime of Validity}

The gauge theory derivation we outlined above has many elements of the
final orbifold CFT that we settle on, as we see in what follows;
however, its regime of validity is limited and is believed to be less
accurate than the orbifold CFT. As mentioned, the theory has the same
supersymmetry as $\Nsc = 2$ in four dimensions (or $\Nsc =1$ in six
dimensions) so we used the same multiplet language. Recall that the
bosonic content of the hypermultiplets should transform as a doublet
under the $SU(2)$ R-symmetry of four-dimensional $\Nsc=2$ theory. Thus
we identify $SU(2)_1$ of the torus as the $R$-symmetry of the
four-dimensional theory; however, David, Mandal, and
Wadia~\cite{dmw02} argue that it is not consistent to have compact
hypermultiplets.  Therefore, we must be working in a decompactifying
limit in which the hypermultiplets effectively see a noncompact space
and not the small $T^4$. This means, they argue, that the expectation
values of the hypermultiplets must be much less than the string
scale---ie., we are near the origin of the Higgs branch that connects
with the Coulomb branch~\cite{dmw02}. If one decompactifies the torus
by taking $v\to\infty$, then one finds the theory discussed
in~\cite{witten-higgs}.

Let us emphasize the key points that we learn from the above
description before we abandon it. The first point is that D1D5 system
has a rich moduli space that we can explore quite explicitly via
Equation~\eqref{eq:the-potential}. In particular it has two branches:
a Coulomb branch where the branes can separate and a Higgs branch
where the the branes are bound. Recall that the supergravity
description with no RR fields or NS B-field turned on is marginally
bound and can fragment with no additional energy cost. The correct
description of the intersection of the Coulomb and Higgs branches was
investigated in~\cite{swD1D5, dsD1D5}. In particular, one finds that
although classically the two branches meet, in the CFT description the
two branches are infinitely far away and one cannot get from one to
the other~\cite{swD1D5}.

The second aspect of the gauge theory description that is worth
emphasizing is the qualitative nature of the effective IR CFT we find.
It is a two-dimensional $\Nsc=(4,4)$ sigma model with hyper-K\"{a}hler
target space and central charge $c=6(N_1N_5 + 1)$. All of these facts
remain essentially unchanged in our ultimate description. Let us make
one further comment. Conformal field theories have a moduli space
whose tangent space is spanned by a set of exactly marginal
operators---perturbing the CFT with one of these operators gives a new
CFT with different properties. Moreover, in the case of
two-dimensional $\Nsc=(4,4)$ supersymmetric CFTs (SCFTs) the local
structure of the moduli space is completely determined by
supersymmetry to be of the form~\cite{seiberg-observations, cecotti}
\begin{equation}\label{eq:gen-SCFT-mod}
\frac{SO(4,n)}{SO(4)\times SO(n)},
\end{equation}
where $n$ is fixed by the number of marginal operators. See
also~\cite{deBoer-attractor} for a review of the moduli space of these
SCFTs. Let us clarify: so far we have discussed three moduli spaces.
First we discussed the moduli space of supergravity compactified on
$T^4\times S^1$ and its near-horizon subset. Then, we showed that the
open strings give rise to a $U(N_5)\times U(N_1)$ gauge theory with a
moduli space defined by the vanishing of
Equation~\eqref{eq:the-potential}. When we went to the effective IR
CFT, the gauge theory moduli space became the \emph{target} space of a
two-dimensional sigma model. Now, we just introduced the moduli space
of that two-dimensional sigma model. Since the near-horizon
supergravity should be dual to the IR sigma model, we should expect
that their respective moduli spaces agree.  Indeed, we see that
Equation~\eqref{eq:local-mod-near} is of the form in
Equation~\eqref{eq:gen-SCFT-mod} with $n=5$.

\subsection{Instanton Description}

We now switch to an alternative, more accurate treatment of the D1D5
bound state. Let us begin our discussion with only the $N_5$ D5-branes
wrapping $T^4\times S^1$. The ground state excitations of the open
strings living on the D5-branes give rise, as discussed above, to a
5+1-dimensional $U(N_5)$ with 16 supercharges. In six dimensions, this
corresponds to $\Nsc = 2$ Weyl self-conjugate super-spinors. One can
derive this theory by dimensionally reducing $\Nsc=1$ $U(N_5)$ gauge
theory in ten dimensions on a four-torus. At this point in our
previous discussion we dimensionally reduced on the string scale
$T^4$, arguing that the KK modes are too heavy to excite in the
low-energy description of interest. There are other, non-perturbative,
excitations; however, that one might consider on the $T^4$.

Recall that Yang--Mills theory in four dimensions has instanton
solutions: classical solutions of the Euclidean equations of motion
with finite action. The field strength of these configurations satisfy
a self-duality condition
\begin{equation}
F^{(2)} = \pm *F^{(2)} \quad\longrightarrow\quad
F_{ij} = \pm \frac{1}{2}\epsilon_{ijkl}F^{kl},
\end{equation}
and have nontrivial winding number or Pontryagin index $\nu\in\ints$
defined by
\begin{equation}\label{eq:pontryagin}
32\pi^2\nu = \Tr\left[\int F^{(2)}\wedge F^{(2)}\right] 
       = \frac{1}{2}\Tr\left[\int\drm^4 x\, \epsilon_{ijkl}F^{ij}F^{kl}\right].
\end{equation}
The superscripted `2' indicates that $F$ is a 2-form. The solutions
are localized in time and but otherwise are similar to solitons, hence
the name.  Instantons are treated in many different places in the
literature (see e.g.~\cite{ADHM, coleman, rajaraman}).

There are analogous solutions to the D5-brane worldvolume theory with
field strength self-dual with respect to the $T^4$ directions.  From
the perspective of the full six-dimensional theory these solutions are
dynamical strings wrapping $S^1$ that are localized in $T^4$.
Following the literature, we term these solutions instantons even
though they are not localized in time. These solutions break half of
the 5-brane worldvolume theory's supersymmetries.  The 5-brane field
theory contains a term
\begin{equation}
\int C^{(2)}\wedge \Tr[F\wedge F],
\end{equation}
and therefore the instantons source the RR 2-form in quantized units
from~\eqref{eq:pontryagin}. All of these facts suggest that we
interpret these strings of instantons as D1-branes wrapping
$S^1$~\cite{douglas-branes, vafa-instantons}.

From this discussion, we conclude that we are interested in $N_1$
strings of instantons in the D5-brane worldvolume theory. In
particular, we want an effective IR CFT description of this theory.
There are a set of fermionic and bosonic zero modes associated with
the instanton solutions~\cite{weinberg-yi, coleman, jeremy}. The IR
limit of the theory should be just these zero modes. Put another way,
the $N_1$ strings of instantons of the $U(N_5)$ gauge theory have a
moduli space of solutions that have the same D1-brane charge. Let us
call this space $\mathcal{W}_\text{inst.}$.\footnote{We choose this
  notation so that the target moduli space $\mathcal{W}$ is clearly
  distinct from the CFT moduli space.} From our above discussion, we
see that the IR dynamics of the D1D5 system should be a
two-dimensional $\Nsc = (4,4)$ sigma model with target space
$\mathcal{W}_\text{inst.}$~\cite{vafa-instantons, douglas-branes,
  dijkgraafD1D5}. One can show that this moduli space is
hyper-K\"{a}hler, and that the SCFT has the same local geometry as the
near-horizon geometry in Equation~\eqref{eq:local-mod-near} and
central charge $c=6N_1N_5$~\cite{dijkgraafD1D5, swD1D5, MAGOO, dmw02}.
The center of mass zero modes are separate degrees of freedom as in
Section~\ref{sec:gauge-desc}. They do not play a role in any of the
physics we describe here, so we do not mention them further. This SCFT
has a moduli space $\mathcal{M}_\text{SCFT}$ that one can show exactly
coincides with $\mathcal{M}_\text{sugra}^*$~\cite{dijkgraafD1D5}. This
description, as before, is on the Higgs branch. The intersection with
the Coulomb branch manifests itself as points in
$\mathcal{M}_\text{sugra}^*$ where there is a ``small instanton
singularity,'' which correspond to a D1 brane being emitted, for
example~\cite{johnson, swD1D5, lmD1D5}.

In~\cite{vafa-instantons, witten-higgs}, it is argued that the
instanton moduli space $\mathcal{W}_\text{inst.}$ is a smooth
deformation of a symmetric product of a four-torus $\tilde{T}^4$
\begin{equation}\label{eq:sym-product}
  \text{Sym}_{N_1N_5}(\tilde{T}^4) = \frac{(\tilde{T}^4)^{N_1N_5}}{S_{N_1N_5}},
\end{equation}
where $S_n$ is the symmetric group of degree $n$; and furthermore,
that there is a point in the moduli space, $\mathcal{M}_\text{SCFT}$
or equivalently $\mathcal{M}_\text{sugra}^*$, where
$\mathcal{W}_\text{inst.}$ takes this form. Heuristically, we can
motivate this description. Consider the case of $N_5=1$, in which
case, we have $N_1$ strings of instantons wrapping $S^1$.  The
low-energy theory describes the instanton dynamics inside the D5-brane
worldvolume. Thus we must specify their position within the $T^4$;
however, there is no physical distinction between permutations of the
instanton labels $1$ to $N_1$ and so we should mod out by all possible
permutations, $S_{N_1}$. Of course, this argument is too simplistic to
be rigorous since among other things the instantons have finite size.
Moreover, as emphasized in~\cite{dmw02}, the $\tilde{T}^4$ can be
different from the compactification torus, $T^4$. Since we do not use
the detailed structure of either of the tori, we drop the
distinguishing tilde for the rest of the dissertation. The connection
between the two tori is discussed in~\cite{gks}.

The point in moduli space where $\mathcal{W}_\text{inst.}$ takes the
form~\eqref{eq:sym-product} is referred to as the ``symmetric product
point'' or the ``orbifold point,'' interchangeably. Arguments for its
existence and location within moduli space are advanced
in~\cite{lmD1D5, dijkgraafD1D5, swD1D5, gks, dmw02}.  In particular,
Larsen and Martinec~\cite{lmD1D5} identify the orbifold point with
canonical charges where $d_5 = N_5 = 1$ is $g_s = 0$ and $C^{(0)} =
1/2$. One can see that this far away from the regime of validity for
supergravity~\eqref{eq:large-charge}.

The SCFT with target space $(T^4)^{N_1N_5}/S_{N_1N_5}$, we call the
``orbifold model'' or ``orbifold CFT.'' All of the calculations in the
chapters that follow are in this CFT. We describe the details of the
orbifold CFT in Section~\ref{sec:orbifold}.

\section{The Correspondence}\label{sec:correspondence}

\begin{figure}[ht]
\begin{center}
\subfigure[~The NS vacuum state]{\label{fig:nsvacuumCFT}
	\raisebox{40pt}{\includegraphics[width=6cm]{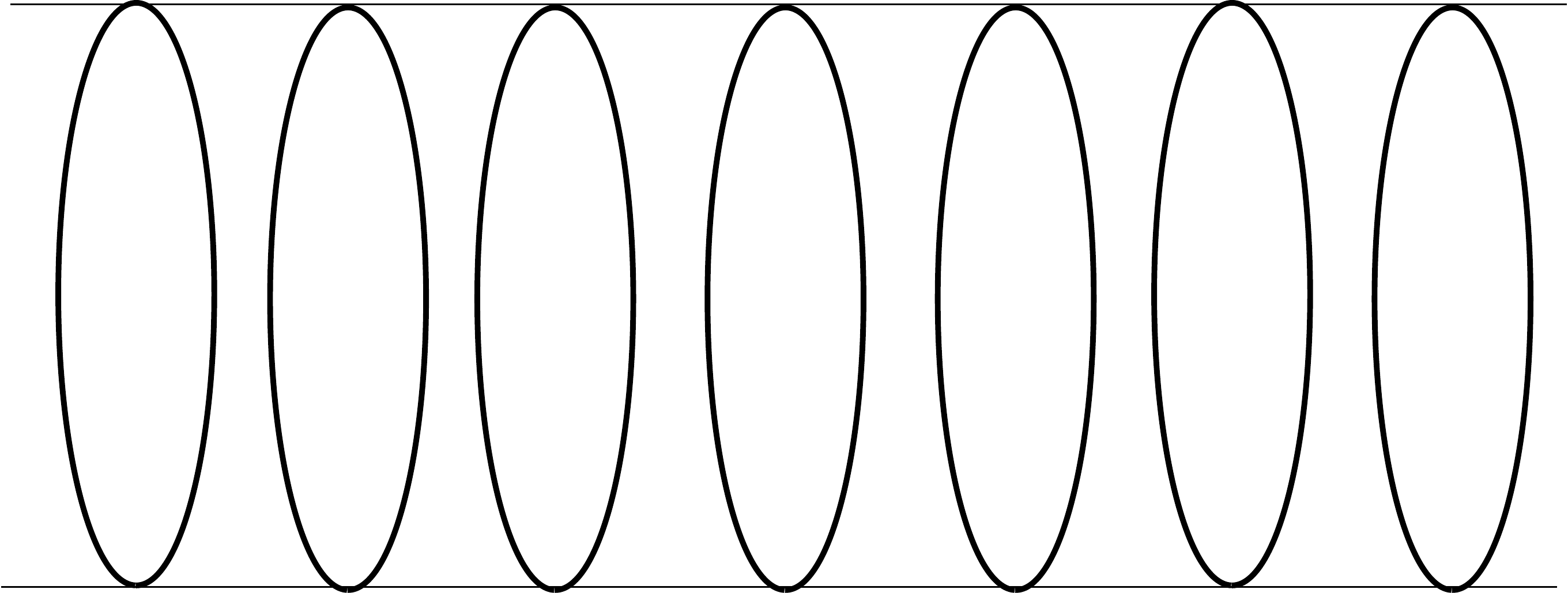}}}
\hspace{15pt}
\subfigure[~The gravity dual of the NS vacuum state]{\label{fig:nsvacuumGrav}
	\includegraphics[width=5cm]{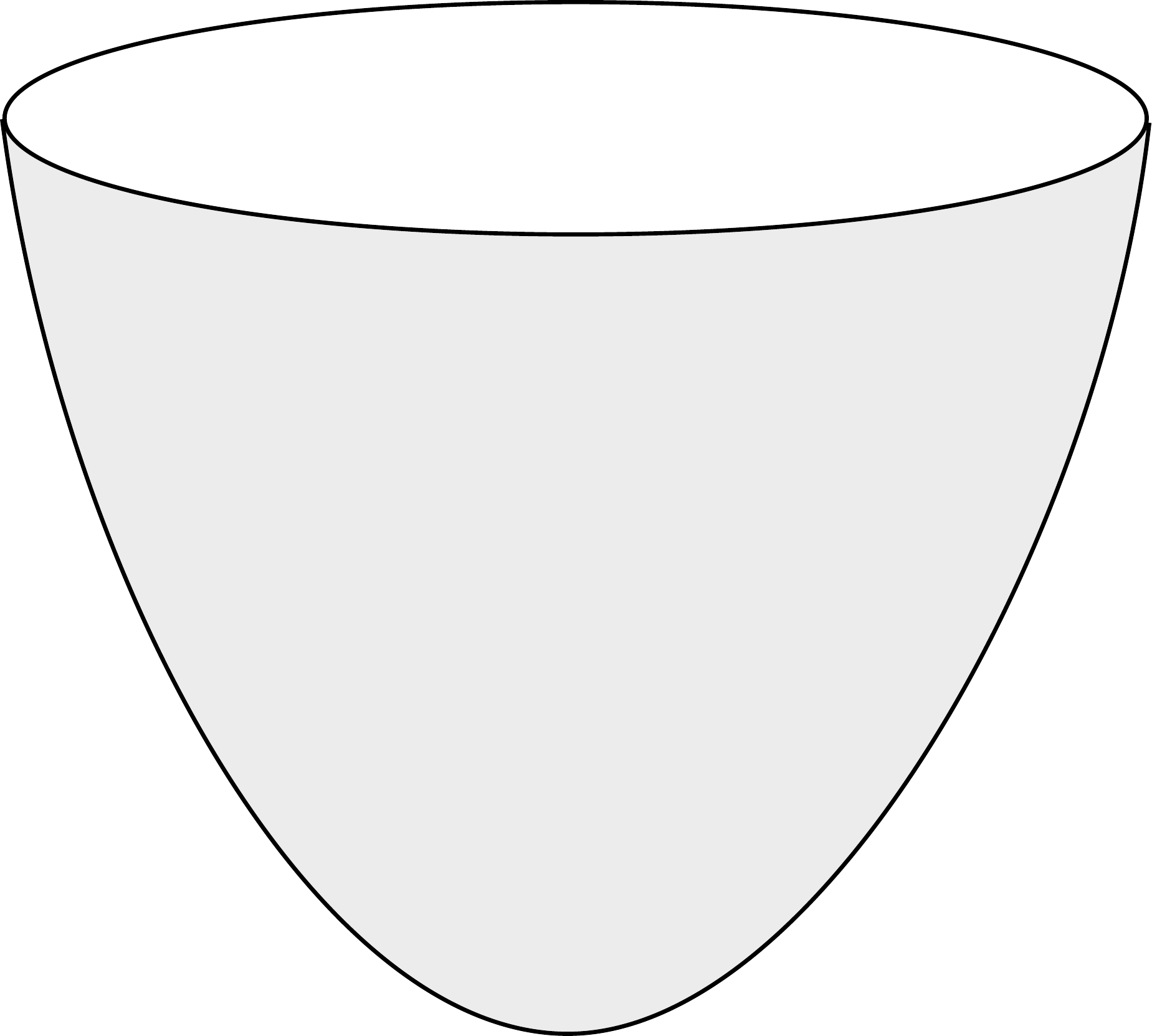}}
      \caption[The NS vacuum and global AdS]{(a) The NS vacuum state in the CFT and (b) its
        gravity dual, which is global $AdS$.  The NS vacuum is the
        simplest possible state having no twists, no excitations, and
        no base spin.\label{fig:nsvacuum}}
\end{center}
\end{figure}

Before describing the orbifold CFT in detail, let us pause to
explicitly state the proposed $\AdS_3$--$\text{CFT}_2$ correspondence
that is being proposed, filling in some parts of the dictionary and
addressing some subtleties. The claim is that type IIB string theory
compactified on $T^4$ in $\AdS_3\times S^3$ is dual to two-dimensional
$\Nsc=(4,4)$ SCFT with central charge $c= \bar{c} = 6N_1N_5$ and
target space $\mathcal{W}_\text{inst.}$.\footnote{As discussed, there
  is also the center of mass degrees of freedom, which we suppress.}
The CFT lives at the two-dimensional boundary of $\AdS_3$. The
fermions of the CFT derive their periodicity from the AdS space. If we
have global AdS space, then the CFT is in the NS sector (anti-periodic
boundary conditions) since global AdS has a contractible cycle and
going around the $S^1$ at the boundary looks like a $2\pi$ rotation at
a point in AdS space. See Figure~\ref{fig:nsvacuumGrav} for a
depiction of global $\AdS_3$. The CFT also has a R sector (periodic
boundary conditions) that can be related to the NS sector via
\emph{spectral flow}~\cite{spectral} (see
Section~\ref{sec:spectral-flow}). The NS vacuum is dual to global
$\AdS_3$, whereas the R vacuum is dual to the two-charge extremal
black hole in Equation~\eqref{eq:near-horizon-metric}. This is
consistent since the geometry does not have a contractible
cycle~\cite{MAGOO}.

In our discussion so far we have focused on the moduli spaces of the
SCFT and the supergravity, emphasizing the role of the
near-horizon/low-energy limit. We have not been careful to discuss and
show the agreement of the symmetries of the two theories. Recall that
the D1D5 system is 1/4-BPS, and therefore supergravity solutions have
8 Killing spinors, ie. eight supersymmetries. When one takes the
near-horizon limit, however, one goes to a more symmetric space
$\AdS_3\times S^3$ which preserves 16 supersymmetries~\cite{MAGOO}.
This doubling of the supersymmetry in the near-horizon limit also
happens for the D3-brane case, for instance. The $\Nsc=(4,4)$
superalgebra; however, only has 8 (real) supersymmetries.
Fortunately, when we take the IR limit, we end up with a $\Nsc =
(4,4)$ \emph{superconformal} symmetry that has twice the number of
fermionic symmetries. This illustrates again the connection between
the near-horizon limit of the gravity description and the IR limit of
the field theory description. The commutator of the special conformal
operator with the supercharges gives new fermionic symmetries, usually
denoted $S$. The $S$'s are not supercharges since they anticommute to
give a special conformal transformation and not a translation. Thus,
the $\Nsc=(4,4)$ superconformal algebra has 16 fermionic symmetries,
of which 8 are supersymmetries.

The $\AdS_3\times S^3\times T^4$ has an $SO(2,2)\simeq SL(2,\re)\times
SL(2, \re)$ isometry of the $\AdS_3$, an $SO(4)_E$ isometry of $S^3$,
and $SO(4)_I$ of the torus which is broken by the compactification.
The conformal group in two dimensions is infinite-dimensional with
Virasoro generators $L_n$ and $\bar{L}_n$ for all $n\in\ints$;
however, only the subalgebra spanned by $n=-1, 0, 1$ are well-behaved
symmetries globally~\cite{difrancesco}. This subalgebra generates an
$SL(2,\re)\times SL(2, \re)$ group of transformations that can be
identified with the isometries of $\AdS_3$. These are the
transformations under which the vacuum is invariant in the
SCFT~\cite{MAGOO}. The $\Nsc=(4,4)$ superconformal algebra has an
$SO(4)$ $R$-symmetry which is identified with the $SO(4)_E$ isometry
of $S^3$. The $\Nsc=(4,4)$ superconformal algebra also has an $SO(4)$
outer automorphism (see e.g.~\cite{spectral, deBoer-attractor}) that
we identify with $SO(4)_I$ of the torus. As discussed above, both the
geometry and the SCFT have the same number of fermionic symmetries;
putting the bosonic and fermionic symmetries together, one can show
that both the geometry and the SCFT have the same supergroup,
$SU(1,1|2)\times SU(1,1|2)$~\cite{MAGOO, dmw02}.

\section{The Orbifold Model of the D1D5 CFT}\label{sec:orbifold}

In this section we explicitly describe the orbifold model of the D1D5
CFT. For a quick summary of our conventions, see
Appendix~\ref{ap:CFT-notation}. As already mentioned, this is a
two-dimen\-sional $\Nsc = (4,4)$ superconformal sigma model with target
space $(T^4)^{N_1N_5}/S_{N_1N_5}$. The orbifold model is the analog of
free super Yang--Mills in the more familiar $\AdS_5$--$\text{CFT}_4$
duality. The base space of the sigma model is the cylinder
parameterized by $(t, y)$ (the $S^1$ and time of the supergravity). We
take $y$ to have periodicity $2\pi R$; $R$ is the radius of the $S^1$.

\subsection{The Symmetries and Action}

The symmetries of our theory are $SO(4)_E \simeq SU(2)_L\times
SU(2)_R$ and the $SO(4)_I\simeq SU(2)_1\times SU(2)_2$ rotations of
the torus. Indices correspond to the following representations
\begin{gather*}
\alpha, \beta \qquad \text{doublet of $SU(2)_L$} \hspace{60pt}
\dot{\alpha}, \dot{\beta} \qquad \text{doublet of $SU(2)_R$}\\
A, B \qquad \text{doublet of $SU(2)_1$}\hspace{60pt}
\dot{A}, \dot{B} \qquad \text{doublet of $SU(2)_2$}\\
i, j \qquad \text{vector of $SO(4)_I$}\hspace{60pt}
a, b \qquad \text{vector of $SO(4)_E$}.
\end{gather*}
Note that these conventions differ slightly from what we use above.
The two-dimensional theory can be written in terms of bosons
$X^i_{(r)}(t,y)$ and two-component Majorana-Weyl fermions
$\Psi^{aA}_{(r)}(t,y)$.\footnote{Each two-component $\Psi$ (with fixed
  $i$, $A$, and $(r)$) has one real degree of freedom once we demand
  that $\Psi$ satisfy a reality condition (Majorana) and have positive
  chirality (Weyl).} The subscripted parenthetical index is a ``copy''
index. Since the target space is the symmetric product of $N_1N_5$
$T^4$s, we have $N_1 N_5$ copies of a sigma model with target space
$T^4$ which we symmetrize. The symmetrization $S_{N_1N_5}$ we can
think of as a discrete gauge symmetry. Each copy has four real bosons
and eight real fermions. Once these are broken into left and right
sectors of a CFT, we get $c=\bar{c}=6$ central charge from each copy
which is consistent with our previous discussion. We frequently
suppress the copy index in our discussion and calculations.

The action can be written roughly in the form (up to numerical
factors)
\begin{equation}
S \sim \sum_{r=1}^{N_1N_5}\int\drm t\int_0^{2\pi R}\drm y
    \left[\pd_\rho X_{(r)}^i\pd^\rho X_{(r)}^i 
          + i\epsilon_{\dot{A}\dot{B}}\bar{\Psi}^{a\dot{A}}_{(r)} 
             \gamma^\rho\pd_\rho\Psi^{a\dot{B}}_{(r)}\right],
\end{equation}
where $\rho$ runs over $t$ and $y$, and $\gamma^\rho$ is the Dirac
gamma matrix. We mention this only to make contact with our previous
discussion. We find it more convenient to Wick rotate to Euclidean
time and map the cylinder to a dimensionless complex plane, breaking
the theory into left and right-movers. We define dimensionless
Euclidean coordinates via
\begin{equation}
\tau = \frac{it}{R}\qquad \sigma = \frac{y}{R},
\end{equation}
and map to the complex plane with coordinates $z$ and $\bar{z}$ via
\begin{equation}
z = e^{\tau + i\sigma}\qquad \bar{z} = e^{\tau - i\sigma}.
\end{equation}
Holomorphic functions of $z$ are the ``left-movers'' and
anti-holomorphic functions are the ``right-movers.''

We also break the two-component fermions into single-component
left-moving and right-moving fermions, $\psi^{\alpha
  \dot{A}}_{(r)}(z)$ and $\bar{\psi}^{\dot{\alpha}
  \dot{A}}_{(r)}(\bar{z})$, satisfying a reality constraint
\begin{equation}
\psi^\dg_{\alpha \dot{A}} = 
   -\epsilon_{\alpha \beta}\epsilon_{\dot{A}\dot{B}}\psi^{\beta B}\qquad
\bar{\psi}^\dg_{\dot{\alpha} \dot{A}} 
   = -\epsilon_{\dot{\alpha}\dot{\beta}}\epsilon_{\dot{A}\dot{B}}
           \bar{\psi}^{\dot{\beta} \dot{B}}.
\end{equation}
We can also break the vector of $SO(4)_I$ into doublets of $SU(2)_1$
and $SU(2)_2$, writing the bosons as
\begin{equation}
X_{A\dot{A}(r)}(z,\bar{z}).
\end{equation}

The action for a single copy in this notation can be written as
\begin{equation}
S = \frac{1}{4\pi}\int\drm^2z\left[\pd X^i(z)\pdb X^i(\bar{z}) 
  - \psi^{\alpha \dot{A}}\pd \psi_{\alpha \dot{A}} 
  - \bar{\psi}^{\dot{\alpha} \dot{A}}\pdb \bar{\psi}_{\dot{\alpha} \dot{A}}
        \right],
\end{equation}
where the conventions for raising and lowering indices is in
Appendix~\ref{ap:CFT-notation}. The action is not particularly
interesting since it is a free theory in two dimensions. The theory
has an OPE current algebra. On the left sector (the right sector is
analogous) we have the stress--energy current, the supercurrents, and
the $SU(2)_L$ currents which are give by $T(z)$,
$G^{\alpha A}(z)$, and $J^a(z)$ respectively. Note that $a=1, 2,
3$ or $a=+, -, 3$ is an $SU(2)_L$ triplet index. The currents have
modes labeled by $L_n$, $G^{\alpha A}_{m}$, and $J^a_n$
respectively. All of these currents and modes have an implicit copy
index; however, since we mod out by $S_{N_1N_5}$ it is only the
diagonal sum over all copies that survives as a symmetry of the full
theory.

As we discuss in Section~\ref{sec:correspondence}, while the $L_n$,
$G^{\alpha A}_{n}$, and $J^a_n$ span an infinite-dimensional
algebra; only a finite subalgebra generate globally defined
symmetries. These are the generators that annihilate the NS vacuum:
$\{L_{\pm 1}, L_0, G^{\alpha A}_{\pm \frac{1}{2}}, J^a_0\}$. These
generators also form a basis for the anomaly-free subalgebra. Note
that the NS sector corresponds to anti-periodic boundary conditions
for the fermions on the \emph{cylinder}; when one maps to the complex
plane, there is a Jacobian factor that switches the sign so that
periodic fermions in the $z$-plane correspond to anti-periodic
fermions on the cylinder and vice-versa. In the NS sector, then, the
fermions and supersymmetry generators have half-integer modes and
integer modes in the R sector. The Cartan subalgebra is spanned by
$\{L_0, J^3_0\}$. A state's eigenvalue of $L_0$ is its conformal
weight or scaling dimension usually denoted $h$ or $\Delta$. Since
$J^a_0$ are the generators of $SU(2)_L$, we have the usual $SU(2)$
representation theory. In particular, we can label states by their
Casimir $(J^a_0)^2$ in addition to their $J^3_0$ eigenvalues, which we
generally denote by $j$ and $m$, respectively. Of course, there are
the analogous eigenvalues on the right sector, which have bars on top,
as usual.

\subsection{Twist Operators}

\begin{figure}[ht]
\begin{center}
\subfigure[~Untwisted component strings]{\label{fig:twist1}
	\includegraphics[width=6cm]{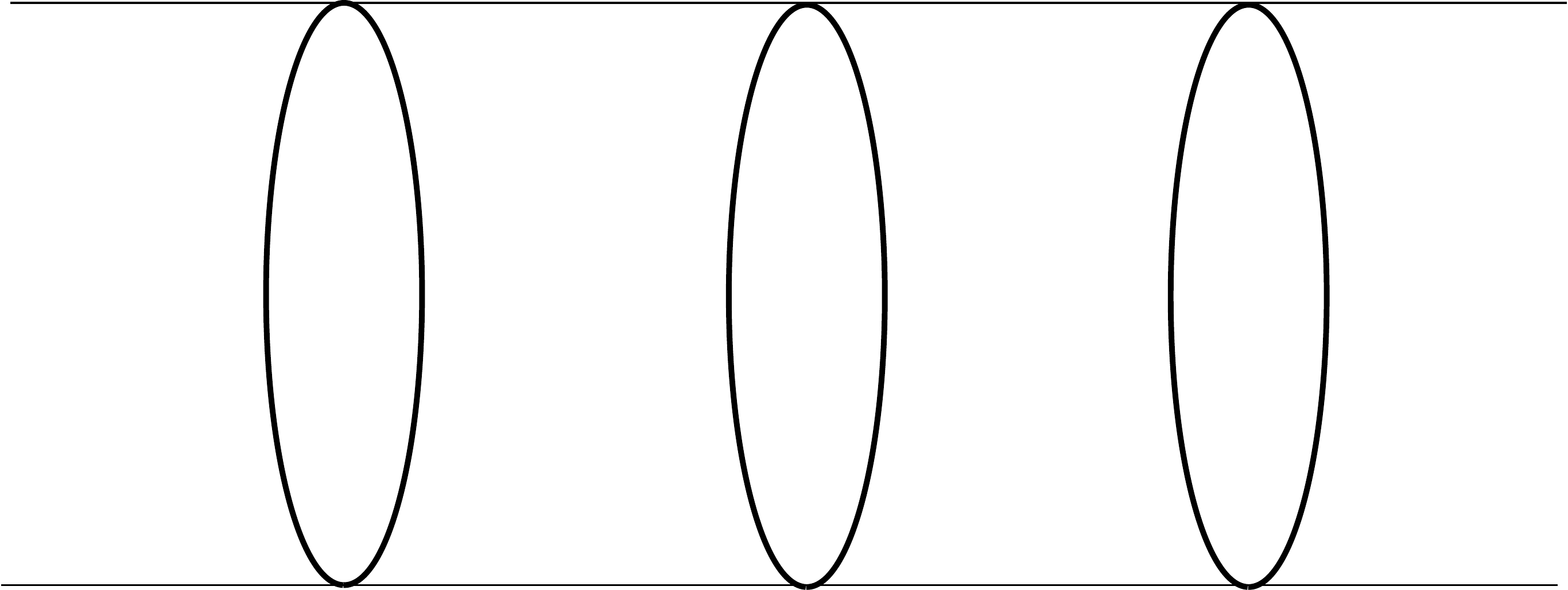}}
\raisebox{34pt}{$\xrightarrow{\hspace{5pt}{\displaystyle \sigma_3}\hspace{3pt}}$}
\subfigure[~The twisted component string]{\label{fig:twist2}
	\includegraphics[width=6cm]{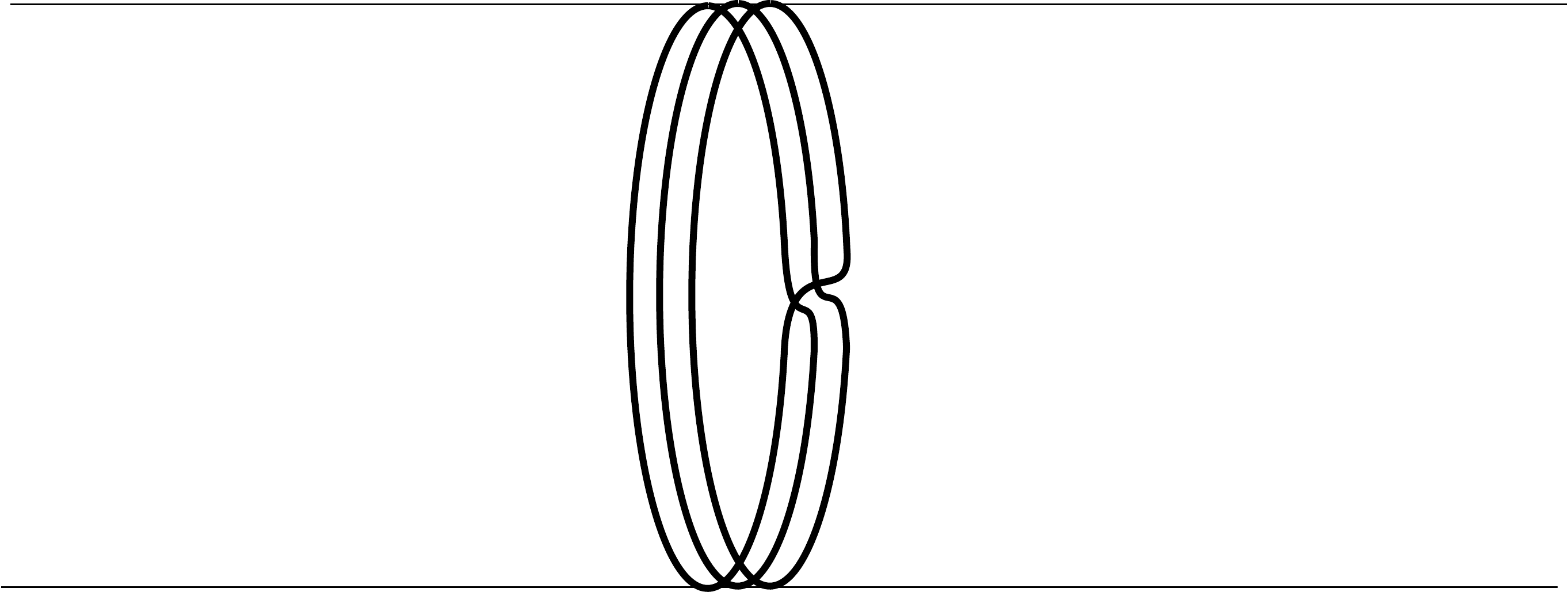}}
      \caption[The twist operator $\sigma_3$.]{The twist operator $\sigma_3$.  Each loop represents a
        copy of the CFT wrapping the $S^1$. The twist operator
        joins these copies into one single copy of the CFT living on a
        circle of three times the length of the original circle.
        \label{fig:twist}}
\end{center}
\end{figure}

Since we orbifold by the symmetric group $S_{N_1N_5}$, we generate
``twist sectors,'' which can be obtained by acting with ``twist
operators'' $\sigma_n$ on an untwisted state. Suppose we insert a
twist operator at a point $z$ in the base space. As we circle the
point $z$, different copies of $T^4$ get mapped into each other. Let
us denote the copy number by a subscript $a=1, 2, \dots n$. The twist
operator is labeled by the permutation it generates. For instance,
every time one circles the twist operator
\begin{equation}
\sigma_{(123\dots n)},
\label{qone}
\end{equation} 
the fields $X^i_{(r)}$ get mapped as
\begin{equation}
X^i_{(1)} \rightarrow
X^2_{(2)} \rightarrow
\cdots
\rightarrow
X^i_{(n)} \rightarrow X^i_{(1)},
\end{equation}
and the other copies of $X^i_{(a)}$ are unchanged. More explicitly, if
we insert $\sigma_{(123\dots n)}(z_0)$, then the bosons have boundary
conditions
\begin{equation}\begin{split}
X^i_{(r)}\big(z_0 + z e^{2\pi i}\big) &= X^i_{(r+1)}\big(z_0 + z\big)\qquad r=1,\dots, n-1\\
X^i_{(n)}\big(z_0 + z e^{2\pi i}\big) &= X^i_{(1)}\big(z_0 + z\big),
\end{split}\end{equation}
with copies $r=n+1$ through $N_1N_5$ unaffected. We have a similar
action on the fermionic fields. We depict this twisting in
Fig.~\ref{fig:twist}. Each set of linked copies of the CFT is called
one ``component string.'' Because of the altered boundary conditions,
in $n$-twisted sector there are new $1/n$-moded excitations. These
fractional modes are important for black hole physics and are in the
background of much of the work presented in this dissertation.  

We often abbreviate a twist operator like the one in
Equation~\eqref{qone} with $\sigma_n$ for simplicity (we have to give
the indices involved in the permutation explicitly when we use
$\sigma_n$ in a correlator). We call these operators, $\sigma_n$,
``bare twists'' to distinguish them from operators which have
additional $SU(2)_L$ charge added to the bare twist forming operators
that are chiral primaries for the supersymmetric CFT. See
Appendix~\ref{ap:twist-corr} (or for the original
references~\cite{lm1, lm2}) for how to compute correlators of twist
operators. By calculating the difference in Casimir energy, one can
show that the bare twist operators, $\sigma_n$, have weight~\cite{lm1}
\begin{equation}
\Delta_n = \bar{\Delta}_n = \frac{c}{24}\left(n - \frac{1}{n}\right),
\end{equation}
where for us $c=6$, the central charge of a \emph{single} copy,
\emph{not} the total central charge, $c_{\text{tot.}} = 6N_1N_5$.

\subsection{Chiral Primaries and Short Multiplets}\label{sec:short}

Let us discuss some representation theory of the NS sector of the
$\Nsc=(4,4)$ algebra.  The theory is very similar to the $\Nsc=(2,2)$
case~\cite{vafa-warner, deBoer-attractor}. We call a state a
\emph{Virasoro primary} if it is killed by all the positive Virasoro
generators. Virasoro primaries are a useful concept since they
correspond to operators that transform nicely under general conformal
transformations.  If we just use the word primary, this is the sense
in which we mean it. Finally, we call a state $\ket{\phi}$ a
\emph{global primary} if it is killed by all of the positive modes of
the anomaly-free subalgebra:
\begin{equation}
L_{+1}\ket{\phi} = G^{\alpha A}_{+\frac{1}{2}}\ket{\phi} = 0.
\end{equation}
The global primaries are useful because they are killed by all of the
positive modes of the subalgebra that corresponds to the symmetries of
the supergravity. We are interested in exploring the representation
theory of the anomaly-free subalgebra. Note that the literature does
not typically distinguish between these two concepts.\footnote{The
  $\Nsc = (2,2)$ literature (e.g.~\cite{deBoer-attractor}) sometimes
  define a superconformal primary as a state that is annihilated by all
  of the positive modes of the chiral algebra. We do not find such a
  concept useful here.}

Let us now consider the anti-commutator of the fermionic symmetry
generators $G^{\alpha A}_{\pm 1/2}$ from
Appendix~\ref{ap:CFT-notation}:
\begin{subequations}\begin{align}
\ac{G^{-A}_{+\frac{1}{2}}}{G^{+B}_{-\frac{1}{2}}} &= \epsilon^{AB}(J_0^3 - L_0)\\
\ac{G^{+A}_{+\frac{1}{2}}}{G^{-B}_{-\frac{1}{2}}} &= \epsilon^{AB}(J_0^3 + L_0).
\end{align}\end{subequations}
From these one can show that for any state $\ket{\psi}$ with weight
$h$ and $J_0^3$ charge $m$
\begin{subequations}\begin{align}
\sum_B \big| G^{+B}_{-\frac{1}{2}}\ket{\psi}\big|^2 
+ \sum_B \big|G^{-B}_{\frac{1}{2}}\ket{\psi}\big|^2 &= 2(h-m)\label{eq:chiral-sum}\\
\sum_B \big| G^{-B}_{-\frac{1}{2}}\ket{\psi}\big|^2 
+ \sum_B \big|G^{+B}_{\frac{1}{2}}\ket{\psi}\big|^2 &= 2(h+m).\label{eq:antichiral-sum}
\end{align}\end{subequations}
Since in any unitary theory the left-hand sides of
Equations~\eqref{eq:chiral-sum} and~\eqref{eq:antichiral-sum} are the
sum of positive definite quantities, we immediately derive a bound on
all physical states of the theory,
\begin{equation}\label{eq:BPS-bound}
h \geq |m| \qquad \longrightarrow\qquad h \geq j.
\end{equation}

We call a state $\ket{\chi}$ \emph{chiral} if it satisfies
\begin{equation}
G^{+A}_{-\frac{1}{2}}\ket{\chi} = 0\qquad A=1,2.
\end{equation}
The chiral (Virasoro) primary states are of especial importance. They
are the analog of the highest weight state for
$SU(2)$~\cite{vafa-warner}, their energies as well as their two and
three-point functions are protected as one moves in moduli
space~\cite{MAGOO, deBoer-attractor}, and they are used to identify
the duals of supergravity particles in the bulk.  Furthermore, under
spectral flow chiral primary operators map to Ramond vacua.

Chiral primary states are precisely the states that saturate the
$h\geq m$ bound.  That is, chiral primary implies $h=m$, and $h=m$
implies chiral primary. In fact, $h=m$ implies both (Virasoro) primary
and global primary.  That $h=m$ implies chiral, can be seen from
Equation~\eqref{eq:chiral-sum}.  Thus, chiral global primary, chiral
(Virasoro) primary, and $h=m$ are all equivalent. From the bound we
immediately see that chiral primaries are also the highest weight
states of the $SU(2)_L$ multiplet, $h=j=m$. The $G^{\alpha A}_{\pm
  \frac{3}{2}}$ anti-commutator can be used to derive a bound on the
weight $h$ of chiral primaries:
\begin{equation}\label{eq:BPS-max-bound}
h_\text{c.p.} \leq \frac{c_\text{tot.}}{6} = N_1N_5.
\end{equation}
In fact, we can make a much stronger statement for the orbifold model.
By looking at general $G$ anticommutators in the $n$-twisted sector,
where there are $1/n$-fractional modes, one can show that chiral
primaries in the $n$-twisted sector must have weight bounded by
\begin{equation}\label{eq:cp-n-bound}
\frac{n-1}{2} \leq h_\text{c.p.}\leq \frac{n+1}{2}.
\end{equation}
Indeed, this is precisely what is demonstrated quite explicitly below.
Note that the maximum twist is $n=N_1N_5$, which gives a much tighter
bound than Equation~\eqref{eq:BPS-max-bound}.

Of importance later in the dissertation, is that supergravity
particles can be identified as the anomaly-free subalgebra descendants
of chiral primaries; that is, chiral primary operators which are acted
upon only by $L_{-1}$, $J_0^-$, or $G^{-A}_{-\frac{1}{2}}$. The other
generators of the anomaly-free subalgebra annihilate the state since
$h=m$ implies global primary.  

\begin{figure}[ht]
\begin{center}
  \subfigure[~A short multiplet in the $h$--$m$ plane]{\label{fig:h-m}
\includegraphics[width=9cm]{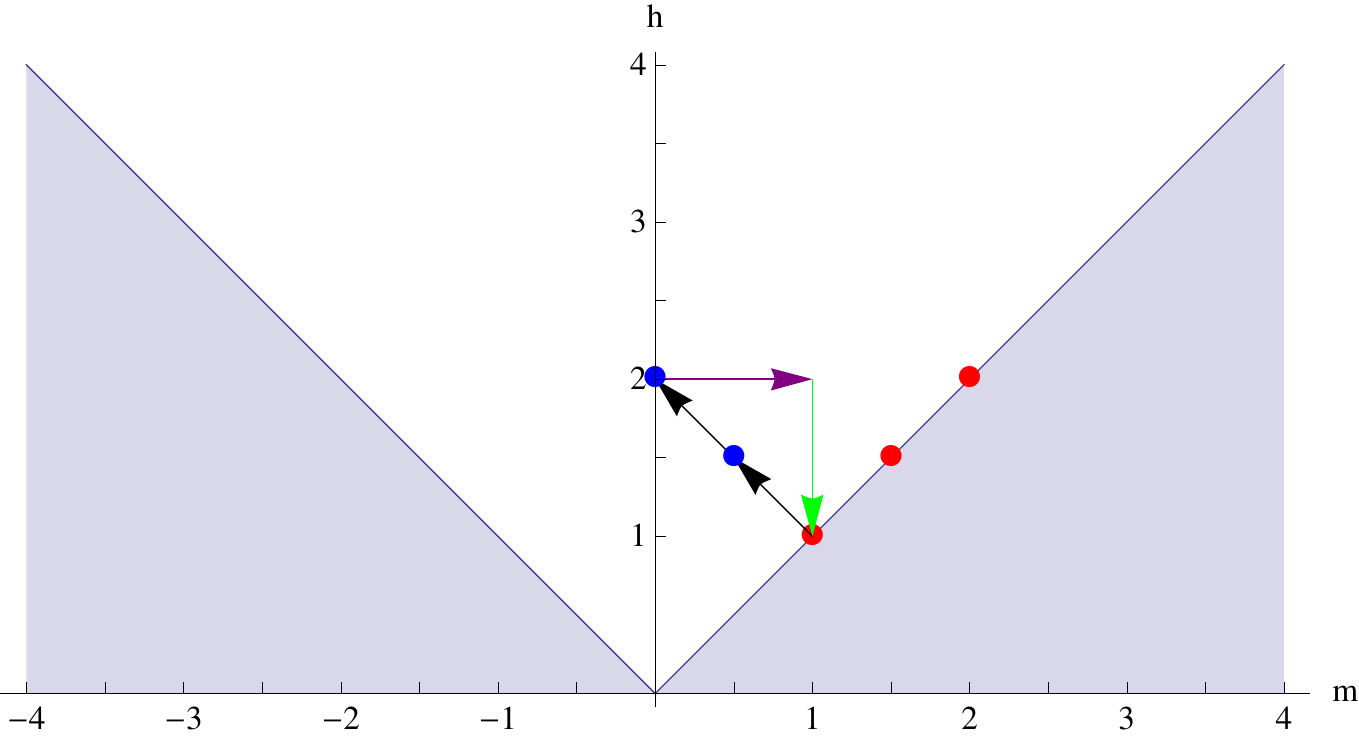}}
\subfigure[~The short multiplets in the $h$--$j$ plane]{\label{fig:h-j}
\includegraphics[width=5cm]{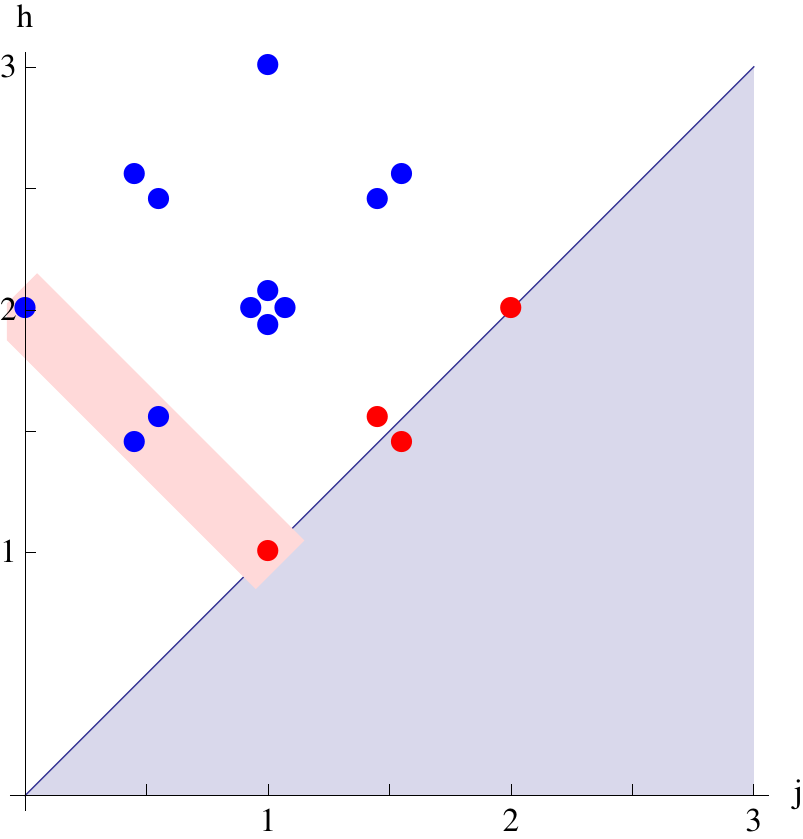}}
\caption[The short multiplets]{On the left, we show the $h$--$m$ plane
  for the NS sector. Unitarity excludes the shaded region. The red
  dots are the chiral primary states of the $3$-twisted sector.  The
  middle red dot corresponds to two distinct chiral primaries. The
  blue dots represent states one attains by acting with
  $G^{-A}_{-1/2}$ on the $h=j=1$ chiral primary. Their are two at
  $h=m=1/2$. The purple arrow represents the action of $J_0^+$ and the
  green arrow represents the action of $L_1$. As discussed in the
  text, this gives the same state with $h=1$ and $m=2$, which must be
  projected out. The $h$--$j$ plane on the right, shows the states of
  the short multiplets (up to application of $J_0^-$ and $L_{-1}$).
  The dots are displaced slightly off the vertices to show the
  degeneracy. One of the short multiplets is highlighted in light
  red.}\label{fig:cps}
\end{center}
\end{figure}

In the singly twisted sector, there are four chiral primaries,
\begin{equation}
\vac_{NS}\qquad 
\psi^{+\dot{A}}_{-\frac{1}{2}}\vac_{NS}\qquad  
J^+_{-1}\vac_{NS} \propto 
  \psi^{+\dot{1}}_{-\frac{1}{2}}\psi^{+\dot{2}}_{-\frac{1}{2}}\vac_{NS},
\end{equation}
one with $h=j=0$, two with $h=j=\frac{1}{2}$, and one with $h=j=1$.
The last one saturates the bound for the singly twisted sector. To
make more chiral primaries we must look at twist operators. This
structure is repeated for the twisted chiral primaries; there is a
minimal charge chiral primary to which we can apply two fermion
operators to get a total of four chiral primaries in each twist
sector. See Figure~\ref{fig:cps} to see the chiral primaries (red
dots) of the $3$-twisted sector in the weight--charge plane.

\begin{table}\begin{center}
\begin{tabular}{c|c|c}
 state           & $h=j=m$ & degeneracy\\\hline
 $\ket{c_0}$     & $\frac{n-1}{2}$ & 1\\
 $\psi^{+\dot{A}}_{-\frac{1}{2}}\ket{c_0}$ & $\frac{n}{2}$ & 2\\
 $J^+_{-1}\ket{c_0}$ & $\frac{n+1}{2}$ & 1
\end{tabular}
\caption[The chiral primaries in the $n$-twisted sector]{In the
  $n$-twisted sector there are four chiral primaries. Their structure
  is summarized above. The lowest and highest weight chiral primaries
  saturate the bound in Equation~\eqref{eq:cp-n-bound}. Each of the
  four chiral primaries gives a short multiplet and can be mapped to a
  distinct R ground state by spectral flow.}\label{tab:cps}
\end{center}\end{table}

Let us also introduce the supermultiplets of the theory, the
representations of\linebreak  the anomaly-free subalgebra. Consider a global
primary state---a state annihilated by $G^{\alpha A}_{+\frac{1}{2}}$
and $L_1$. We can generate new states by acting with combinations of
the four $G^{\alpha A}_{-\frac{1}{2}}$.  This gives $1+ 4+ 6+ 4+ 1=
16$ states (each term in the sum corresponds to a different number of
$G_{-\frac{1}{2}}$s being applied). On each of these states we can act
with $L_{-1}$ an arbitrary number of times to increase the weight.
These are the supermultiplets.

There are also \emph{short multiplets}, for which some of the
$G^{\alpha A}_{-\frac{1}{2}}$'s annihilate the state.  These are
precisely the chiral primaries, and their descendents under the
anomaly-free subalgebra, which we identify as the duals of
supergravity particles.  We can only act with half of the
$G_{-\frac{1}{2}}$'s, so we can generate four states by acting with
$G^{-A}_{-\frac{1}{2}}$. The chiral primary must be the top member of
its $SU(2)_L$ multiplet from the bound~\eqref{eq:BPS-bound}. One can
show that the two Virasoro primaries created by a single application
of $G^{-A}_{-\frac{1}{2}}$ on a chiral primary $\ket{c}$ must be
annihilated by $J_0^+$. This can easily be seen from the commutator of
$J_0^+$ and $G^{-A}_{-\frac{1}{2}}$:
\begin{equation}
J_0^+ G^{-A}_{-\frac{1}{2}}\ket{c} 
  = G^{+A}_{-\frac{1}{2}}\ket{c} + G^{-A}_{-\frac{1}{2}}J_0^+\ket{c}
  = 0,
\end{equation}
where the two terms separately vanish from the chirality of $\ket{c}$
and it being the highest weight state of $SU(2)_L$. One can also
demonstrate that both states are annihilated by all the positive
Virasoro generators. Thus, both descendants of a chiral primary by a
single susy charge are the top members of the $SU(2)_L$ multiplet and
are Virasoro primaries. These states are not chiral and they are not
global primaries.

Let us now consider the state with both applications of
$G^{-A}_{-\frac{1}{2}}$. One can show that
\begin{calc}
L_{1}J_0^+ \big(G^{-1}_{-\frac{1}{2}}G^{-2}_{-\frac{1}{2}}\ket{c}\big)
  &= L_1 \left(G^{+1}_{-\frac{1}{2}}G^{-2}_{-\frac{1}{2}}\ket{c} + 
     G^{-1}_{-\frac{1}{2}}J_0^+G^{-2}_{-\frac{1}{2}}\ket{c}\right)\\
  &= -L_1 L_{-1}\ket{c}\\
  &= -2 h \ket{c}.
\end{calc}
Thus we see that \emph{part of} two applications
$G^{-A}_{-\frac{1}{2}}$ on $\ket{c}$ is equivalent to
$J_0^-L_{-1}\ket{c}$. For a chiral primary with weight $h=j=m$, we
take the combination
\begin{equation}\label{eq:double-G}
G_{-\frac{1}{2}}^{-1}G_{-\frac{1}{2}}^{-2}\ket{c} 
  + \frac{1}{2h} J_0^{-}L_{-1}\ket{c},
\end{equation}
which one can show is the top member of a spin-$(c-1)$ $SU(2)_L$
multiplet and is killed by $L_{+1}$. This state is only $SL(2,\re)$
primary.

See Figure~\ref{fig:cps} for a depiction of the 3-twisted short
multiplets in the $h$--$m$ and $h$--$j$ plane. Each point in the
$h$--$j$ plane in Figure~\ref{fig:h-j} is the top member of a
$SU(2)_L$ multiplet that can be filled out by repeated application of
$J_0^-$. We can also act with $L_{-1}$ an arbitrary number of times to
generate a new state which will also be the top member of its
$SU(2)_L$ multiplet.

We can summarize the discussion as follows. In each twist sector their
is a minimal weight chiral primary. We can apply two different fermion
modes or both to generate 3 more chiral primaries.  Each of those four
chiral primaries gives a distinct short multiplet.  Each short
multiplet has four states that are annihilated by $L_{1}$ and are the
top member of their $SU(2)_L$ multiplet: a chiral primary $\ket{c}$,
$G^{-1}_{-\frac{1}{2}}\ket{c}$, $G^{-2}_{-\frac{1}{2}}\ket{c}$, and
the state in Equation~\eqref{eq:double-G}. To each of those four
states we can apply $L_{-1}$ an arbitrary number of times and fill out
the rest of the $SU(2)_L$ multiplets by repeated application of
$J_0^-$. See Tables~\ref{tab:short} and~\ref{tab:cps}. There are also
special ultra-short multiplets with chiral primaries of weight less
than or equal to $h=1/2$. For these multiplets, the state in
Equation~\eqref{eq:double-G} is missing.

\begin{table}\begin{center}
\begin{tabular}{l|c|c|c|c}
   & state     & $h$ & $j$ & $m$ \\\hline
CP & $\ket{c}$ & $h$ & $h$ & $h$ \\
P  & $G^{-A}_{-\frac{1}{2}}\ket{c}$ 
               & $h+\frac{1}{2}$ & $h-\frac{1}{2}$ & $h-\frac{1}{2}$\\
S  & $G^{-1}_{-\frac{1}{2}}G^{-2}_{-\frac{1}{2}}\ket{c} 
        + \frac{1}{h}J_0^-L_{-1}\ket{c}$
               & $h+1$ & $h-1$ & $h-1$     
\end{tabular}
\caption[The structure of a short multiplet]{We illustrate the basic
  structure of a short multiplet. One can start with a chiral primary
  (CP) $\ket{c}$, apply two different supercharges to get two Virasoro
  primaries (P), and finally apply both supercharges to a $SL(2,\re)$
  primary (S) (after projecting out). Each of those four states is the
  top member of an $SU(2)_L$ multiplet and has lowest conformal weight
  with respect to $SL(2,\re)$.  One can apply $L_{-1}$ an arbitrary
  number of times and apply $J_0^-$ to fill out the $SU(2)_L$
  multiplet. This completes the short multiplet.}\label{tab:short}
\end{center}\end{table}

See~\cite{MAGOO, vafa-warner, gunaydin-ads3-rep, deBoer-attractor} for
more details. Let us emphasize that we are only discussing the left
sector here, but there are parallel statements for the right sector.

\subsubsection{The Basic Chiral Primary Operators \texorpdfstring{$\sigma^0_{l+1}$}{sigma-nought}}

Let us recall the construction of chiral primary operators introduced
in~\cite{lm2}. Start with the NS vacuum state $\vac_{NS}$, which is in
the completely untwisted sector, where all the component strings are
``singly wound.''  The gravity dual is global AdS.

Now suppose in this gravity dual we want to add one supergravity
quantum carrying angular momentum $\frac{l}{2}$ in each of the two
factors of $SU(2)_L\times SU(2)_R$ (scalars must have $j=\jbar$).  We
take a set of $l+1$ copies of the CFT and join them by using a twist
operator $\sigma_{l+1}$ into one ``multiply wound'' component string.
Let us take the lowest energy state in this twist sector. This state
has dimensions~\cite{lm2}
\begin{equation}
h=\bar h= \frac{c}{24}\left[(l+1) - \frac{1}{(l+1)}\right],
\label{start}
\end{equation}
but it does not yet have any charge, so it is not a chiral primary; it
has more dimension than charge. The current operator $J^+$ carries
positive charge, and we can apply contour integrals of $J^+$ to our
state to raise its charge. Since all operators in the theory have
$h>|m|$, one might think that we cannot reach a chiral primary with
$h=j=m$ if we start with the state (\ref{start}); however, on the
twisted component string we can apply \emph{fractional} modes
$J^a_{\frac{k}{l+1}}$ of the current operators, because any contour
integral around the twist operator insertion has to close only after
going around the insertion $l+1$ times.

Before we apply these fractionally-moded current operators, there is
one more point to note. For the case of $l+1$ even one finds that the
twist operator $\sigma_n$ yields anti-periodic boundary conditions for
the fermion field when we traverse around the twist insertion $l+1$
times. Since we wanted the fermion field to return to itself after
going $l+1$ times around the insertion, we must insert a spin field to
change the periodicity of these fermions. The construction of these
spin fields was explained in detail in~\cite{lm2}, but for now we just
denote the twist with spin field insertions (for both left and right
fermions) as $(S_{l+1}^+\bar{S}_{l+1}^+ \sigma_{l+1})$. With this
notation we find that the chiral primaries are given by
\begin{equation}
\sigma^{0}_{l+1}=
\begin{cases}
J^+_{-\frac{l-1}{l+1}}J^+_{-\frac{l-3}{l+1}}\cdots J^+_{-\frac{1}{l+1}}
\bar{J}^+_{-\frac{l-1}{l+1}}\bar{J}^+_{-\frac{l-3}{l+1}}\cdots
\bar{J}^+_{-\frac{1}{l+1}} \sigma_{l+1} & \text{$(l+1)$ odd}\\
J^+_{-\frac{l-1}{l+1}}J^+_{-\frac{l-3}{l+1}}\cdots J^+_{-\frac{2}{l+1}}
\bar{J}^+_{-\frac{l-1}{l+1}}\bar{J}^+_{-\frac{l-3}{l+1}}\cdots
\bar{J}^+_{-\frac{2}{l+1}}(S_{l+1}^+\bar{S}_{l+1}^+ \sigma_{l+1})
             & \text{$(l+1)$ even}.
\end{cases}
\label{chiral}
\end{equation}
This construction generates chiral primaries with dimensions and charges
\begin{equation}
\sigma^{0}_{l+1}:~~~h=m=\frac{l}{2}, ~~~\bar h=\bar{m}=\frac{l}{2}.
\end{equation}
Note that $J^+\sim \psi^{+\dot{1}}\psi^{+\dot{2}}$, and the current
operators in~\eqref{chiral} fill up the left and right moving ``Fermi
seas'' up to a Fermi level (here we write only the left sector)
\begin{equation}
\sigma^0_{l+1}\sim 
\begin{cases}
\psi^{+\dot{1}}_{-\frac{l-1}{2(l+1)}}\psi^{+\dot{2}}_{-\frac{l-1}{2(l+1)}}
\psi^{+\dot{1}}_{-\frac{l-3}{2(l+1)}}\psi^{+\dot{2}}_{-\frac{l-3}{2(l+1)}}
\cdots \psi^{+\dot{1}}_{-\frac{1}{2(l+1)}}\psi^{+\dot{2}}_{-\frac{1}{2(l+1)}}
     \sigma_{l+1}
                 & \text{$(l+1)$ odd}\\
   \psi^{+\dot{1}}_{-\frac{l-1}{2(l+1)}}\psi^{+\dot{2}}_{-\frac{l-1}{2(l+1)}} 
\psi^{+\dot{1}}_{-\frac{l-3}{2(l+1)}}\psi^{+\dot{2}}_{-\frac{l-3}{2(l+1)}}
\cdots \psi^{+\dot{1}}_{-\frac{1}{l+1}}\psi^{+\dot{2}}_{-\frac{1}{l+1}}
   (S_{l+1}^+  \sigma_{l+1})
                 & \text{$(l+1)$ even}.
              \end{cases}
\label{fermion}
\end{equation}

\subsubsection{Additional Chiral Primaries}

Above we described the construction of the simplest chiral primary
$\sigma^0_{l+1}$. We can make additional chiral primaries as follows:
\begin{enumerate}
\item The next available fermion level for the fermion $\psi^{+\dot
    1}$ in~\eqref{fermion} is $\psi^{+\dot 1}_{-\frac{1}{2}}$. If we
  fill this level, we raise both dimension and charge by $\frac{1}{2}$,
  so we get another chiral primary.
\item We can do the same with the fermion  $\psi^{+\dot 2}$.
\item We can add both fermions $\psi^{+\dot 1},\psi^{+\dot 2}$, which is equivalent to an application of $J^+_{-1}$.
\end{enumerate}
Of course we can make analogous excitations to the right sector as
well. This gives a total of $4\times 4=16$ chiral primaries (more
precisely chiral primary--chiral primary) in a given twist sector.
This exhausts all the possible chiral primaries for this system.

\subsubsection{Anti-chiral Primaries}

We define anti-chiral primaries as states with
\begin{equation}
h=-m, ~~~\bar h=-\bar m.
\end{equation}
To construct these states, we again start with a twist operator
$\sigma_{l+1}$ and the apply modes of $J^- $ instead of
$J^+$. Proceeding in the same way as for chiral primaries, we get the
anti-chiral primary (denoted with a tilde over the $\sigma$) as
\begin{equation}
\tilde\sigma^0_{l+1}=
\begin{cases}
J^-_{-\frac{l-1}{l+1}}J^-_{-\frac{l-3}{l+1}}\cdots J^-_{-\frac{1}{l+1}}
\bar{J}^-_{-\frac{l-1}{l+1}}\bar{J}^-_{-\frac{l-3}{l+1}}\cdots
\bar{J}^-_{-\frac{1}{l+1}} \sigma_{l+1} 
       & \text{$(l+1)$ odd}\\
       J^-_{-\frac{l-1}{l+1}}J^-_{-\frac{l-3}{l+1}}\cdots J^-_{-\frac{2}{l+1}}
\bar{J}^-_{-\frac{l-1}{l+1}}\bar{J}^-_{-\frac{l-3}{l+1}}\cdots
\bar{J}^-_{-\frac{2}{l+1}}(S_{l+1}^-\bar{S}_{l+1}^- \sigma_{l+1}) 
       & \text{$(l+1)$ even}.
\end{cases}
\end{equation} 
We can construct additional anti-chiral primaries just as in the case
of chiral primaries. A chiral primary has a nonvanishing 2-point
function with its corresponding anti-chiral primary. The chiral and
anti-chiral twist operators are normalized such that the 2-point
function is unity at unit separation. 

\subsection{Marginal Deformations}\label{sec:marg-def}

The orbifold point in moduli space does not have a good supergravity
description~\cite{MAGOO, dmw02, lmD1D5, dijkgraafD1D5}. Since we are
interested in using the CFT description to study black hole physics,
we need gravity to be a reasonable description so that the concept of
black hole is well-defined. We have two options to deal with this
problem. We can compute quantities in the orbifold CFT, and hope that
the quantities are protected by a non-renormalization theorem so that
they are not affected by changes in the moduli; or, alternatively, we
can add marginal deformations to the orbifold CFT that move one toward
points in moduli space that do have a good supergravity description.
The calculations of Chapter~\ref{ch:emission} are of the first kind.
In Chapter~\ref{ch:orbifold}, we consider the effect of a single
application of a marginal deformation. This corresponds to first order
in perturbation theory. Since the orbifold point is far from a good
supergravity description, we do not expect perturbation theory to be
sufficient for getting to a supergravity point. We hope, however, that
such analysis may give some insight. Much of that analysis remains to
be completed, at the time of this writing.

We want to find the exactly marginal deformation to the orbifold CFT
that preserve the $\Nsc=(4,4)$ supersymmetry. To be marginal, the
operator should have weight $h=\bar{h} = 1$. If we take any state with
weight $h=\bar{h}=1$; however, there is no guarantee that its weight
won't get renormalized as we change the moduli. Therefore it should be
in a short multiplet. If it is to preserve the $\Nsc=(4,4)$
supersymmetry then it must be a singlet of the $R$-symmetry, $SO(4)_E
= SU(2)_L\times SU(2)_R$. These requirements imply that we should look
for chiral primary--chiral primaries (CP--CP) with weight
$\bar{h}=\bj=\bar{m}=h=j=m=1/2$, and apply
$G^{-A}_{-1/2}\bar{G}^{\dot{-}B}_{-1/2}$ to form a $SO(4)_E$
singlet~\cite{gava-narain, deBoer-attractor, dmw02}.  

There are exactly five CP--CP operators with weight $h=\bar{h}=1/2$:
\begin{equation}
\psi^{+\dot{A}}\bar{\psi}^{\dot{+}\dot{B}}\qquad \sigma_2^{+\dot{+}}.
\end{equation}
Each of the five CP--CP operators gives four deformation operators
following our discussion of short multiplets. Note that applying
$G^{-A}_{-1/2}\bar{G}^{\dot{-}B}_{-1/2}$ to the fermion CP--CP
operators just gives
\begin{equation}
\pd X^{A\dot{A}}\pdb X^{B\dot{B}} \sim \pd X^i\pdb X^j.
\end{equation}
These may be broken into irreducible representations of $SO(4)_I$.
There is the trace, the antisymmetric part, and the symmetric
traceless part. There are four twist deformations~\cite{dmw02,
  gava-narain}
\begin{equation}
\mathcal{T}^{AB} = G^{-A}_{-1/2}\bar{G}^{\dot{-}B}_{-1/2}\sigma_2^{+\dot{+}}
\end{equation}
which can be broken into a singlet and a triplet of $SU(2)_1$ of
$SO(4)_I$ that we call $\mathcal{T}^0$ and $\mathcal{T}^1$,
respectively. In total, then, there are 20 exactly marginal
deformations that preserve the $\Nsc=(4,4)$ superconformal symmetry,
which gives a 20 dimensional moduli space, as expected.  From their
basic properties, the 20 deformations can be identified with the 20
supergravity moduli in Table~\ref{tab:sugra-near-moduli}~\cite{lmD1D5,
  dmw02}.  The deformations are listed in Table~\ref{tab:marg-def}
with their corresponding supergravity field. We are particularly
interested in the 4 twist deformations that move the theory away from
the orbifold point---individual twist sectors are no longer
eigenstates of the Hamiltonian. These are the object of study in
Chapter~\ref{ch:orbifold}.

\begin{table}
\begin{center}
\begin{tabular}{c|c|c|c}
CFT  & SUGRA & $SO(4)_I \simeq SU(2)_1\times SU(2)_2$ & DOF\\\hline
$\pd X^{(i}\pdb X^{j)} - \frac{1}{4}\delta^{ij}\pd X^i\pdb X_i$ 
   & $g_{ij} - \frac{1}{4}\delta_{ij} g_{kk}$ & (\textbf{3}, \textbf{3}) & 9\\
$\mathcal{T}^{1}$ & $B_{ij}^+$ &  (\textbf{3}, \textbf{1}) & 3\\
$\pd X^{[i}\pdb X^{j]}$ & $C_{ij}^{(2)}$
                        & (\textbf{3}, \textbf{1}) $\oplus$ (\textbf{1}, \textbf{3}) & 6\\
$\mathcal{T}^0$ & $\Xi$ & (\textbf{1}, \textbf{1}) & 1\\
$\pd X^i\pdb X_i$ & $\phi$ & (\textbf{1}, \textbf{1}) & 1\\\hline
 & & & 20
\end{tabular}
\caption[Moduli of the orbifold CFT]{Table of the marginal deformation
  operators of the orbifold CFT along with the corresponding
  near-horizon supergravity moduli.}\label{tab:marg-def}
\end{center}
\end{table}

One can also identify the fixed supergravity moduli in the
CFT~\cite{deBoer-attractor, ms-stringy}. These are operators that
preserve the $\Nsc = (4,4)$ supersymmetry and the $SO(4)_E$
$R$-symmetry but break conformal invariance. To make the irrelevant
operators that satisfy these conditions, one starts with $h=\bar{h}=1$
CP--CP states and applies both supercharges on the left and the right
(and projects out to the irreducible part). These states have weight
$h=\bar{h}=2$, and so break the conformal symmetry. There are 6 CP--CP
states with $h=\bar{h}=1$:
\begin{equation}
J_{-1}^+\bar{J}_{-1}^+\vac_{NS}\qquad 
\psi^{+\dot{A}}_{-\frac{1}{2}}\bar{\psi}^{\dot{+}\dot{B}}_{-\frac{1}{2}}\ket{\sigma_2^0}\qquad
\ket{\sigma_3^0}.
\end{equation}
Each of the corresponding 6 CP--CP operators gives rise to one
$h=\bar{h}=2$ irrelevant deformation. These correspond to the 5 fixed
moduli and the fixed size of the $S^3$~\cite{deBoer-attractor}.

\subsection{Spectral Flow}

Spectral flow~\cite{spectral} maps amplitudes in the CFT to amplitudes
in another CFT; under this map dimensions and charges of the left
sector change as (we write only the left sector)
\begin{equation}\label{eq:spectral}
h'=h+\alpha j +\frac{c\alpha^2}{24}\qquad
m'=m+\frac{c\alpha}{12},
\end{equation}
for spectral flow parameter $\alpha\in\re$. We can also separately
spectral flow the right sector with spectral flow parameter
$\bar{\alpha}$ (where the bar is not indicating complex conjugation).
Spectral flow with odd $\alpha$ (in the singly twisted sector)
exchanges NS and R boundary conditions of the left fermions, and
spectral flow with even $\alpha$ preserves the boundary conditions. In
the singly twisted sector, spectral flow by non-integer units gives
more general fermion boundary conditions that are not useful to
consider here.

For a single copy, the center Al charge is $c=6$. Setting $\alpha=-1$
gives 
\begin{equation}
\ket{0^-_R}:\quad h=\frac{1}{4}\qquad  m=-\frac{1}{2} 
\end{equation} 
which is one of the Ramond ground states of the singly wound component
string. For more on spectral flow, see Section~\ref{sec:spectral-flow}
in the Appendix.

\subsection{The Ramond Sector}

Most of our discussion of the orbifold CFT so far is in the NS sector.
The NS sector is a bit simpler to treat in some ways since the
fermions are periodic \emph{in the complex plane}, but the black hole
physics of interest is in the R sector. Fortunately, we can use
spectral flow to map problems in the NS sector to problems in the R
sector. We make use of this technique in Chapter~\ref{ch:emission}.
Thus, we do not need to use the details of the R sector most of the
time. Let us comment, however, that each CP--CP in the NS sector
spectral flows to a distinct R ground state. Thus each component
string in the R sector has a $16=4\cdot 4$ degeneracy from four left
and four right fermion zero modes. Moreover, one sees that twisting
does not cost any additional energy.  We see, then, that the R ground
state has a large degeneracy.  This corresponds to the microstates of
the corresponding extremal black hole. See
Appendix~\ref{ap:CFT-notation} for more details on the Ramond sector.

\section{Overview}

In this chapter, we have described the D1D5 CFT on $T^4$. Throughout
our discussion, we have emphasized the role of the moduli space. We
introduced the supergravity description and its asymptotic and
near-horizon moduli spaces. We then went through several descriptions
of the open string degrees of freedom. We looked at the ground states
of open strings connecting the different branes, which with the
symmetries of the theory suffice to give the low-energy effective
action. This action has a potential with two branches. The Coulomb
branch corresponds to some of the branes separating and it is not of
interest in this dissertation. This description has a limited regime
of validity. We then transitioned to a description in which the D1s
are realized as instantons in the D5 worldvolume theory. Finally we
argued that at some point in moduli space, the low-energy effective
theory should be the orbifold CFT.

The whole discussion serves to motivate the orbifold CFT, which we
then summarized. Most of the calculations presented in this
dissertation are in the orbifold CFT. The fact that the orbifold CFT
has twist sectors, results in most of the nontrivial physics. Of
importance, is the identification of the 20 marginal deformation
operators that correspond to the 20 directions in the tangent space of
the moduli space.

\chapter{Coupling the CFT to Flat Space}
\label{ch:coupling}

In this short chapter, we show how to use AdS--CFT to relate CFT
amplitudes to the emission or absorption by a collection of branes
into the asymptotic flat space. In essence, we are perturbatively
relaxing the decoupling limit by adding an irrelevant operator to the
CFT that couples to flat space degrees of freedom. A similar
calculation was performed in~\cite{dmw99}; however, we demonstrate
a much more general construction---taking into account the effect of
the ``neck'' and considering general $\AdS_{d+1}$--$\text{CFT}_d$.
This material was developed in~\cite{acm1}.

Constructing the desired formalism requires two main steps. For the
first, note that the CFT describes only the physics in the
``near-horizon region'' of the branes; vertex operators in the CFT
create excitations that must travel through the ``neck'' of the
D-brane geometry and then escape to infinity as traveling waves.  Thus
we set up a general formalism that relates CFT amplitudes to
absorption/emission rates observed from infinity.

Traditionally, one uses AdS--CFT to compute correlation functions in
the CFT and compare them to quantities computed in the AdS geometry,
but we are interested in finding the interactions of the brane system
with quanta coming in from or leaving to \emph{flat} infinity.  Thus
we must consider the full metric of the branes, where at large $r$ the
$AdS$ region changes to a ``neck'' and finally to flat space.  

In Section~\ref{sec:geometry}, we review the structure of the
geometric description of branes in string theory. Then, in
Section~\ref{sec:AdS-CFT-norm} we fix the coupling between gravity
fields and operators in the CFT; first fixing the normalization of the
fields and operators, then demanding the two-point functions agree. In
Sections~\ref{sec:outer} and~\ref{sec:intermediate}, we discuss the
behavior of minimal scalars in the asymptotic flat space and the
intermediate transition to AdS, respectively. Finally, in
Section~\ref{sec:interaction} we write down the interacting action
that couples modes in the asymptotic flat space to the CFT dual to the
inner AdS region. We apply the interaction to case of particle
emission in Section~\ref{sec:emission-formula}, deriving a formula
which we use in Chapter~\ref{ch:emission}.

\section{The Geometry}\label{sec:geometry}

\begin{figure}[ht]
\begin{center}
\subfigure[]
{\label{fig:throats-a}
	\includegraphics[width=7cm]{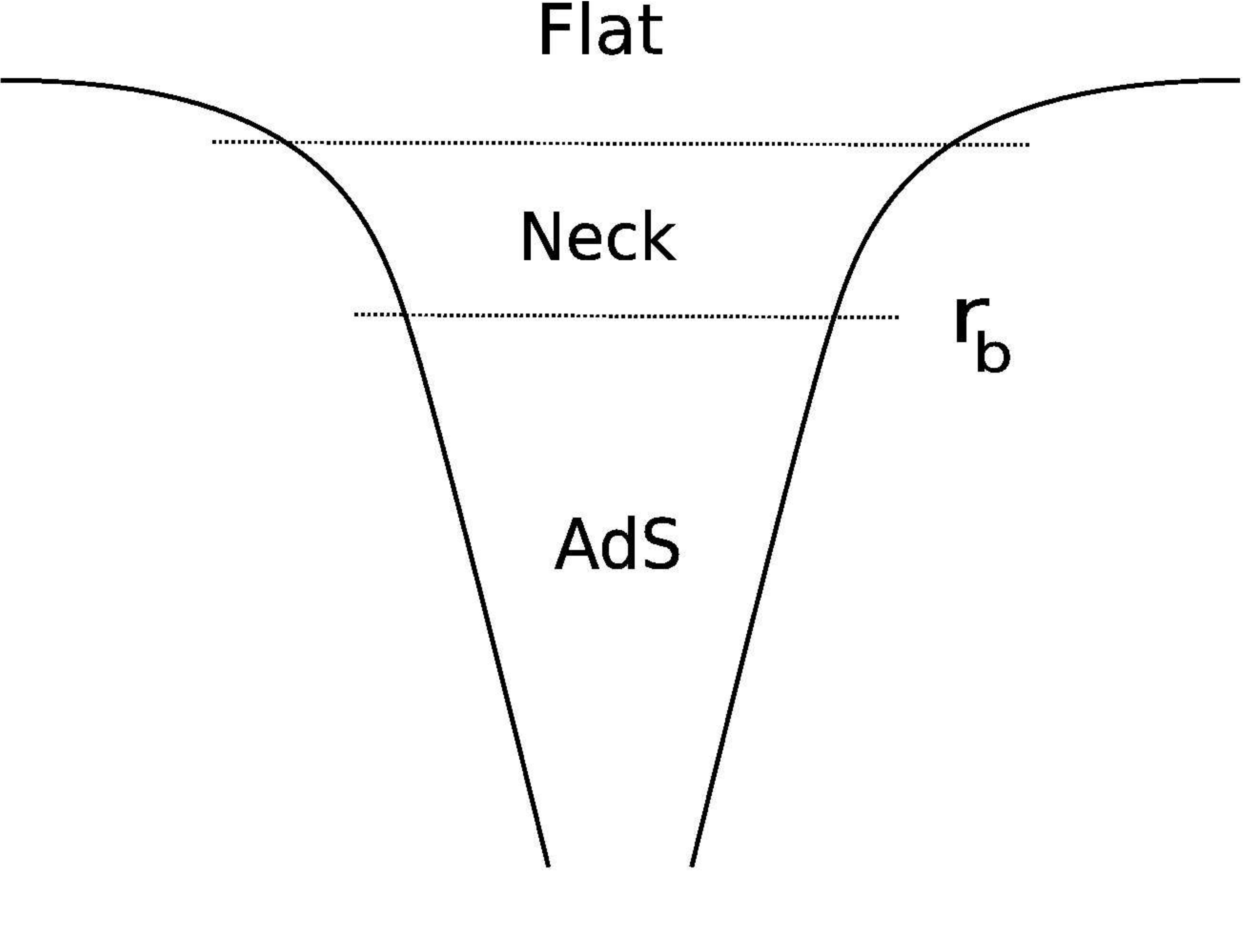}}
\hspace{15pt}
\subfigure[]{\label{fig:throats-b}
	\includegraphics[width=7cm]{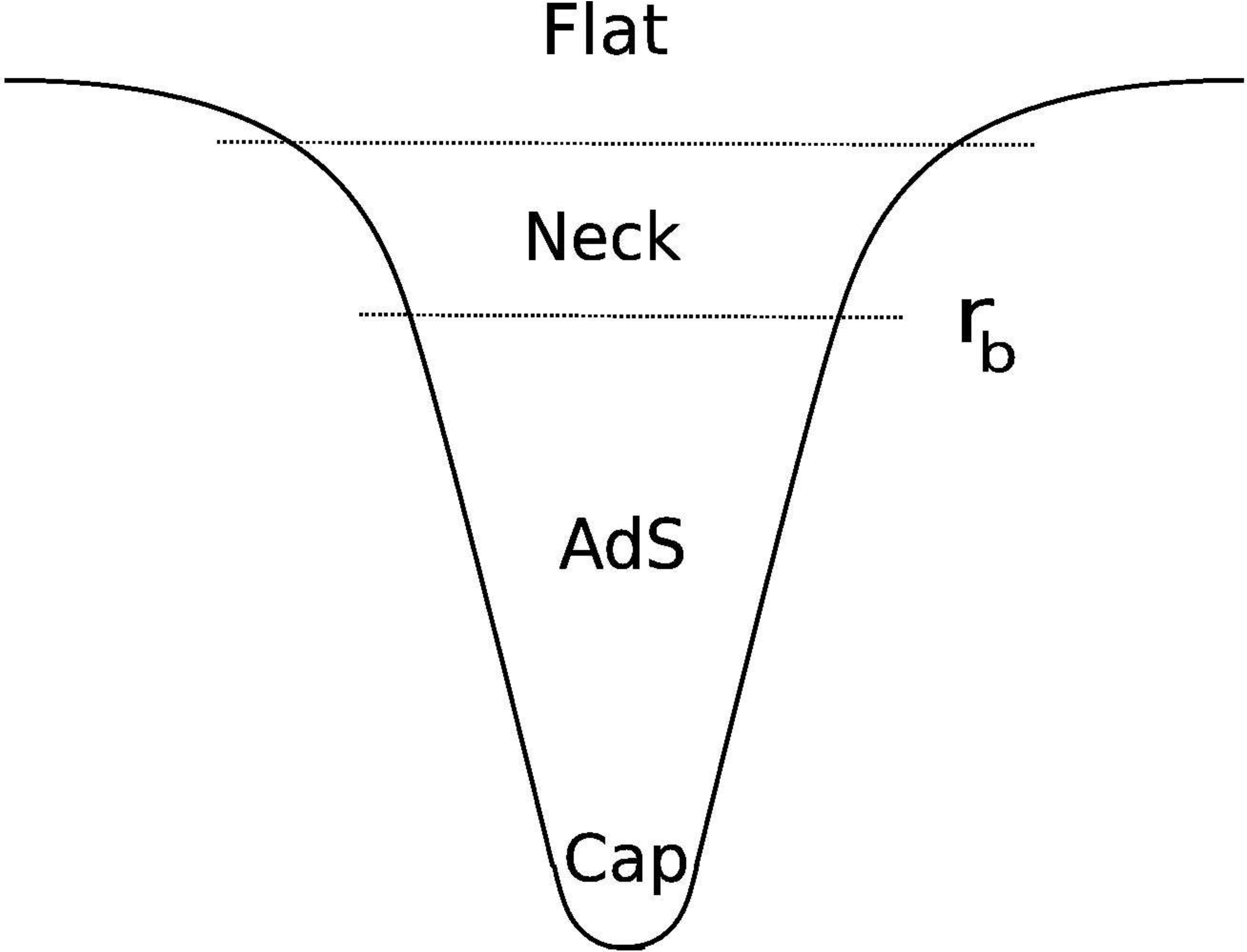}}
      \caption[An illustration of the geometry created by stacks of
      branes]{(a) The geometry of branes is flat at infinity, then we
        have a ``neck'', and further in the geometry takes the form
        $AdS_{d+1}\times S^p$. (b) Still further in, the geometry ends
        in a ``fuzzball cap'' whose structure is determined by the
        choice of microstate. For the simple state that we choose for
        the D1D5 system, the Cap and AdS regions together are just a
        part of global $AdS$. \label{fig:throats}}
\end{center}
\end{figure}

Consider the geometry traditionally written down for branes that have
a near-horizon $AdS_{d+1}\times S^p$ region. This geometry has the
form
\begin{equation}
ds^2=H^{-\frac{2}{d}} \left[-dt^2+\sum_{i=1}^{d-1} dy_idy_i\right] +
H^{\frac{2}{p-1}}\left[dr^2+r^2d\Omega_{p}^2\right].
\label{fullmetric}
\end{equation}
If there is only one kind of brane producing the metric (and hence only one length scale) the function $H$ is given by
\begin{equation}
H=1+ \frac{Q}{r^{p-1}}.
\label{caseone}
\end{equation}
The BPS black holes studied in string theory are constructed from $\mathcal B$ sets of mutually BPS branes.  In these cases $H$ is given by
\begin{equation}
H= \prod_{i=1}^\mathcal B \left( 1+ \frac{Q_i}{r^{p-1}} \right)^\frac{1}{\mathcal B}.
\end{equation}
(This reduces to (\ref{caseone}) for $\mathcal B=1$.) Let $Q_\text{max}$ be the
largest of the $Q_i$ and $Q_\text{min}$ the smallest.  It is
convenient to define the length scale
\begin{equation}
R_s = \left( \prod_{i=1}^\mathcal B Q_i \right)^\frac{1}{\mathcal B(p-1)}.
\end{equation}
For small $r$ the angular directions give a sphere with radius $R_s$. 

We picture such a geometry in Figure~\ref{fig:throats-a}. The geometry
has three regions:
\begin{enumerate}
\item \textit{The outer region:}\quad For large $r$,
\begin{equation}
r\gg Q_{\text{max}}^{\frac{1}{p-1}},
\end{equation}
we have essentially flat space.

\item \textit{The intermediate region:}\quad For smaller $r$ we find a ``neck,'' which we write as
\begin{equation}
C Q_{\text{min}}^{\frac{1}{{p-1}}}<r<D Q_{\text{max}}^{\frac{1}{p-1}}; 
    ~~~(C\ll 1, ~~D\gg 1). \label{neck}
\end{equation}
 
\item\label{ads-region} \textit{The inner region:}\quad For
\begin{equation}
 r<C Q_{\text{min}}^{\frac{1}{p-1}},
\end{equation}
we can replace the harmonic function by a power:
\begin{equation}
 H \rightarrow \left( \frac{R_s}{r} \right)^{p-1}.
\end{equation}
\end{enumerate}
The directions $y_i$ join up with $r$ to make an $AdS$ space, and the
angular directions become a sphere of constant radius:
\begin{equation}
ds^2\approx  \left[\left(\frac{r}{R_s}\right)^{\frac{2 (p-1)}{d}}
     \left(-dt^2+\sum_{i=1}^{d-1} dy_idy_i\right)
     +R_s^2 \frac{dr^2}{r^2}\right]+R_s^2 d\Omega_p^2,
\label{metricads}
\end{equation}
which is $AdS_{d+1}\times S^p$. We introduce a new radial
coordinate $\tilde{r}$,
\begin{equation}\label{scaledRadialCoordinate}
\left( \frac{\tilde r}{R_s} \right) = \left( \frac{r}{R_s} \right)^\frac{p-1}{d},
\end{equation}
which makes the AdS physics more apparent. In terms of this new radial
coordinate one has the inner region metric,
\begin{equation}
ds^2 \approx \left[
   \left(\frac{\tilde r}{R_s} \right)^2 \left( -dt^2 + \sum_{i=1}^{d-1} dy_i dy_i \right) 
 + \left(R_s \frac{d}{p-1} \right)^2 \frac{d {\tilde r}^2}{{\tilde r}^2} \right]
 + R_s^2 d\Omega_p^2. \label{metricadsScaled}
\end{equation}
We can scale $t, y_i$ to put the $AdS$ into standard form, but it is
more convenient to leave it as above, since the coordinates $t,
y_i$ are natural coordinates at infinity and we need to relate
the $AdS$ physics to physics at infinity.  The radius of $AdS_{d+1}$
and the sphere $S^p$ are given by
\begin{equation}
R_{\text{AdS}_{d+1}}=R_s \frac{d}{p-1}, \qquad R_{S^p} =R_s.
\end{equation}

In region~(iii), we can put a boundary at
\begin{equation}
\tilde{r}=\tilde{r}_b,
\end{equation}
which we regard as the boundary of the $AdS$ space. We can then
replace the space at $\tilde{r}<\tilde{r}_b$ by a dual CFT. Let us
note that $\tilde{r}_b$ plays the role of a UV cutoff for the dual
field theory, via the usual correspondence. Thus, traditional AdS/CFT
calculations are carried out only with the region
$\tilde{r}<\tilde{r}_b$. Our interest, however, is in the emission and
absorption of quanta between the $AdS$ region and asymptotic infinity.
Thus we need a formalism to couple quanta in region~(i) to the CFT.

Next, we note that this ``traditional'' $AdS$ geometry cannot be
completely right. In the case of the D1D5 system, we know that the
ground state has a large degeneracy $\sim
\exp[2\sqrt{2}\pi\sqrt{N_1N_5}]$. At sufficiently large $r$, all these
states have the form \eqref{fullmetric}. But at smaller $r$ these
states differ from each other. None of the states has a horizon;
instead each ends in a different ``fuzzball
cap''~\cite{bala-review,lm4,lm5,lmm,mss,gms1,gms2,st-1,st-2,st-3,
  st-4, bena-6, bena-7, bena-8, bena-9, ross, gimon-1, gimon-2,
  gimon-3, gimon-4, BDSMB, BSMB, fuzzballs-elementary,
  fuzzballs-structure, bena-review, skenderis-review}.  We can choose
special states where this cap is given by a classical geometry. In the
simplest case the cap is such that the entire region
$\tilde{r}<\tilde{r}_b$ has the geometry of \emph{global} $AdS_3\times
S^3$. We picture the full geometry for this state in
Figure~\ref{fig:throats-b}.

The geometry without the cap (Figure~\ref{fig:throats-a}) needs to be
supplemented with boundary conditions for the gravity fields at
$\tilde r=0$.  But since the actual states of the system (e.g.
Figure~\ref{fig:throats-b}) have ``caps,'' we do have a well defined
duality between the physics of the region $\tilde{r}<\tilde{r}_b$ and
the CFT at $\tilde{r}_b$.

\section{The Coupling between Gravity Fields and CFT Operators}\label{sec:AdS-CFT-norm}

Let us start with region~(iii), where we have the AdS--CFT duality
map. This map says that the partition function of the CFT, computed
with sources $\phi_b$, equals the partition function for gravity in
the AdS, with field values at the boundary equal to
$\phi_b$~\cite{witten}:
\begin{equation}\label{eq:prelim-coupling}
\int D[X]\, e^{-S_\text{CFT}[X]+\mu\int \drm^d y \sqrt{g_d} \phi_b(y) \mathcal{V}(y)}
    = \int D[\phi]\, e^{-S_\text{AdS}[\phi]}\bigg|_{\phi({\tilde r}_b)=\phi_b}.
\end{equation}
Here we have rotated to Euclidean spacetime, setting $t=y_{d}$.  The
symbol $\phi$ denotes fields in gravity, and $\hat{\mathcal{V}}$ are
CFT operators (depending on $X$) coupling to the gravity fields. Our
first task is to determine the coupling constant $\mu$, for the
normalizations of $\phi_b$ and $\hat{\mathcal{V}}$ that we choose.

Let us discuss the angular dependence in more detail. On $S^p$ the
fields $\phi(\tilde{r}, y, \Omega)$ can be decomposed into spherical harmonics,
\begin{equation}
\phi(\tilde{r}, y, \Omega) = \sum_{l, \vec{m}}\tilde{\phi}_{l, \vec{m}}(\tilde{r}, y) 
	Y_{l,\vec{m}}(\Omega).
\end{equation}
For the remainder of this discussion we consider a fixed $l$ mode.
Note that the reality of $\phi$ imposes the condition
\begin{equation}
\tilde{\phi}_{l, \vec{m}}^* = \tilde{\phi}_{l, -\vec{m}}.
\end{equation}
Thus if we expand field in the usual spherical harmonics then we
should write Equation~\eqref{eq:prelim-coupling} as
\begin{equation}\label{eq:ang-dep-adscft}
\int D[X]\exp\left(-S_\text{CFT}[X] 
	+ \mu \sum_{\vec{m}}\int\drm^d y\sqrt{g_d}\tilde{\phi}_{b,\vec{m}}\mathcal{V}_{\vec{m}}\right)
	= \int D[\phi]
	e^{-S_\text{AdS}}
		\bigg|_{\tilde{\phi}_{\vec{m}}(\tilde{r}_b, y)\equiv\tilde{\phi}_{b,\vec{m}}(y)}.
\end{equation}
Reality of the action requires
\begin{equation}
\hat{\mathcal{V}}_{\vec{m}}^\dg(y) = \hat{\mathcal{V}}_{-\vec{m}}(y).
\end{equation}

At leading order we can replace the gravity path integral by the
classical action evaluated with the given boundary values of the
gravity fields,
\begin{equation}
\int D[\phi] e^{-S_\text{AdS}[\phi]}\bigg|_{\phi(r_b)=\phi_b}=e^{-S_\text{AdS}[\phi^\text{cl.}]}.
\end{equation}
Then the 2-point function in the CFT is given by
\begin{equation}
\vev{\hat{\mathcal{V}}_{\vec{m}}(y_1)\hat{\mathcal{V}}_{\vec{m}'}(y_2)}
   =\frac{1}{\mu^2}\frac{\delta}{\delta \tilde{\phi}_{b,\vec{m}}(y_1)}
	\frac{\delta}{\delta \phi_{b,\vec{m}'}(y_2)}
		\big(-S_\text{AdS}\big).
\label{final}
\end{equation}

Let us now define the normalizations of $\phi$ and
$\hat{\mathcal{V}}$. We consider a minimal scalar field for
concreteness, though our computations should be extendable to other
supergravity fields with no difficulty. The gravity action is
\begin{equation}
S_\text{AdS} = \frac{1}{16\pi G_D} \int \drm^Dx \sqrt{g} 
   \left(\frac{1}{2}\pd\phi\pd\phi\right),
\end{equation}
where $D=d+1+p$ is the dimension of the spacetime ($D=10$ for string
theory, $D=11$ for M theory, and we have $D=6$ for the D1D5 system
after we reduce on a compact $T^4$). In region (iii), the spacetime
has the form
\begin{equation}
AdS_{d+1}\times S^p\times \mathcal{M}.
\end{equation}
For cases like the D1D5 system we have the additional compact
4-manifold $\mathcal{M}=T^4$ or $\mathcal{M}=K3$.  We take $\phi$ to
be a zero mode on $\mathcal{M}$ and dimensionally reduce on
$\mathcal{M}$, so that we again have a space $AdS_{3}\times S^3$, with
$G$ now the 6-d Newton's constant.

If the spherical harmonics are normalized such that,
\begin{equation}
\int \drm\Omega\, |Y_{l, \vec{m}}(\Omega)|^2=1,
\end{equation}
then dimensionally reducing on $S^p$ yields
\begin{equation}
S_\text{AdS}=\frac{R_s^p}{16\pi G_D}\sum_{\vec{m}}\int \drm^{d+1} y \sqrt{g_{d+1}}
   \left[\frac{1}{2} |\pd\tilde{\phi}_{\vec{m}}|^2 +\frac{1}{2} {m^2}|\tilde{\phi}_{\vec{m}}|^2\right].
\end{equation}
The $l$-dependent mass, $m$, comes from
\begin{equation}\label{mass}
m^2=\frac{\Lambda}{R_s^2}, \qquad \triangle_p Y(\Omega) = -\Lambda Y(\Omega), \qquad
	\Lambda =l(l+p-1).
\end{equation}

The CFT lives on the surface $\tilde r={\tilde r}_b$. The metric on this surface is
(from (\ref{metricadsScaled}))
\begin{equation}
ds^2=\left(\frac{{\tilde r}_b}{R_s}\right)^2\sum _{i=1}^{d}dy_idy_i.
\label{metricb}
\end{equation}
We choose the normalization of the operators $\hat{\mathcal{V}}$ by
requiring the 2-point function to have the short distance expansion
$\sim \frac{1}{(\text{distance})^{2\Delta}}$:
\begin{equation}
\vev{\hat{\mathcal{V}}_{\vec{m}} (y_1) \hat{\mathcal{V}}_{\vec{m}'}(y_2)}
	= \frac{\delta_{\vec{m} + \vec{m}',0}}
		{\left[(\frac{{\tilde r}_b}{R_s})|y_1-y_2|\right]^{2\Delta}}.
\label{first}
\end{equation}
Following~\cite{witten}, we define the boundary-to-bulk propagator
which gives the value of $\phi$ in the $AdS$ region given its value on
the boundary at $r_b$:
\begin{equation}
\tilde{\phi}_{\vec{m}}(\tilde r, y) = \int K(\tilde r,y; y')
		\tilde{\phi}_{b,\vec{m}}(y') \sqrt{g_d} \drm^d y',
\end{equation}
where $\sqrt{g_d}d^d y'$ is the volume element on the metric
(\ref{metricb}) on the boundary. We have
\begin{equation}
K(\tilde r,y;y') =  \frac{R_\text{AdS}^{2 \Delta-d}}{{\tilde r}_b^{\Delta} \pi^\frac{d}{2} } \frac{\Gamma(\Delta) }{\Gamma(\Delta- \frac{d}{2})}  \left[ \frac{\tilde r}{R_\text{AdS}^2 + \frac{{\tilde r}^2}{R_s^2}  |y-y'|^2} \right]^\Delta,
\label{kernel}
\end{equation}
where
\begin{equation}
\Delta=\frac{1}{2}(d+\sqrt{d^2+4 m^2 R_\text{AdS}^2}) = (l+p-1) \frac{d}{p-1}. \label{delta}
\end{equation}
We then have
\begin{equation}
\frac{\delta}{\delta \tilde{\phi}_{b,\vec{m}}(y_1)}
	\frac{\delta}{\delta\phi_{b,\vec{m}'}(y_2)}\big(-S_\text{AdS}\big)
  =-\frac{ R_s^p}{16\pi G_D}\delta_{\vec{m}+\vec{m}',0}
	\left(\frac{2\Delta-d}{\Delta}\right) 
	\pd_{\tilde r} K(\tilde r, y_1; y_2)\Big(\sqrt{g^{\tilde r\tilde r}}
		\Big|_{\tilde r=\tilde r_b}\Big ),
\label{second}
\end{equation}
where the extra factor $\frac{2\Delta-d}{\Delta}$ comes from taking
care with the limit $\tilde{r}_b\rightarrow\infty$ when using the
kernel~\eqref{kernel}~\cite{mathurfreedman}.

Putting (\ref{first}) and (\ref{second}) into (\ref{final}) we get
\begin{equation}
\mu=\Bigg[
\frac{R_s^{2\Delta-(d+1)+p}\left(\frac{d}{p-1}\right)^{2\Delta-(d+1)}}
     {16\pi G_D}
\frac{(2\Delta-d)}{\pi^{\frac{d}{2}}}
\frac{\Gamma(\Delta)}{\Gamma(\Delta-\frac{d}{2})}\Bigg]^{\frac{1}{2}}.
\end{equation}

\section{The Outer Region}\label{sec:outer}

The wave equation for the minimal scalar is
\begin{equation}
\square \phi=0.
\end{equation}
We write
\begin{equation}
\phi=h(r)Y(\Omega)e^{-iE t}e^{i\vec \lambda \cdot \vec y}, \label{WaveAnsatz}
\end{equation}
getting the solution
\begin{equation}
h=\frac{1}{r^\frac{p-1}{2}}\left[C_1\, J_{l+\frac{p-1}{2}}
    \left(\sqrt{E^2-\lambda^2} \,r\right)
     +C_2\, J_{-l-\frac{p-1}{2}}\left(\sqrt{E^2-\lambda^2} \,r\right)\right].
\end{equation}
Note that we can define a CFT in the $AdS$ region only if the
excitations in the $AdS$ region decouple to leading order from the
flat space part of the geometry \cite{maldacena}. Such an approximate
decoupling happens for quanta with energies $E\ll 1/R_s$: waves
incident from infinity with wavelengths much longer than the $AdS$
curvature scale almost completely reflect off the 'neck' region and
there is only a small probability of absorption into the $AdS$ part of
the geometry. Correspondingly, waves with such energies trapped in the
$AdS$ region have only a small rate of leakage to flat space. Thus we
work throughout this paper with the assumption
\begin{equation}
E\ll \frac{1}{R_s}.
\end{equation}
With this, we find that to leading order the wave in the outer region
has the $C_2\approx 0$ \cite{hottube}:
\begin{equation}
h\approx
C_1\frac{1}{\Gamma(l+\frac{p+1}{2})}
   \left[\frac{\sqrt{E^2-\lambda^2}}{2}\right]^{l+\frac{p-1}{2}}r^l.
\end{equation}

A general wave is a superposition of different modes of the
form~\eqref{WaveAnsatz}. We wish to extract a given spherical harmonic
from this wave, so that we can couple it to the appropriate vertex
operator of the CFT. Define
\begin{equation}\label{eq:diff-op}
\big[\pd^l \phi\big]^{l,\vec{m}} 
= Y^{k_1k_2\dots k_l}_{l,\vec{m}}
	\pd_{k_1}\pd_{k_2}\dots\pd_{k_l}\phi,
\end{equation}
where the above differential operator is normalized such
that\footnote{See Appendix~\ref{ap:spherical} for more details and some examples.}
\begin{equation}\label{eq:diff-op-norm}
Y^{k_1k_2\cdots k_l}_{l,\vec{m}}
	\pd_{k_1}\pd_{k_2}\dots\pd_{k_l}\big[r^{l'} Y_{l', \vec{m}'}(\Omega)\big]
= \delta_{ll'}\delta_{\vec{m}, \vec{m}'}.
\end{equation}
Thus the required angular component of $\phi$ at small $r$ satisfies
\begin{equation}
\phi\approx [\pd^l\phi]\Big|^{l,\vec{m}}_{r\rightarrow 0}~r^l Y_{l,\vec{m}}(\Omega).
\end{equation}

\section{The Intermediate Region}\label{sec:intermediate}

In the ``neck'' region we can set $E$ and $\lambda$ to zero in solving
the wave equation, since we assume that we are at low energies and
momenta so the wavelength is large compared to the size of the
intermediate region. Thus the $E$ and  $\lambda$ terms do not induce
oscillations of the waveform in the limited domain of the intermediate
region.

With this approximation we now have to solve the wave equation in the
intermediate region. From this solution, we need the following
information to construct our full solution. Suppose in the outer part
of the intermediate region $r \sim Q_{\text{max}}^\frac{1}{p-1}$ we
have the solution
\begin{equation}
\phi\approx r^l.
\end{equation}
Evolved to the inner part of the intermediate region $r\ll
Q^\frac{1}{p-1}$, we have a form given by $AdS$ physics:
\begin{equation}
\phi\approx b_l {\tilde r}^{\Delta-d}.
\end{equation}
These two numbers, $b_l, \Delta$, give the information we need about
the effect of the intermediate region on the wavefunction. $\Delta$ is
known from the CFT, while the number $b_l$ appears in our final
expression for the emission amplitude as representing the physics of
the intermediate region which connects the $AdS$ region to flat
infinity. In essence, $b_l$ is a tunneling coefficient telling us if
we start with a unit amount of $\phi$ in the outer region how much
makes it through to the inner region.

We now note that for the case that we work with, the minimally coupled
scalar, we can in fact write down the values of $b_l$ and $\Delta$.
The wave equation for a minimally coupled scalar in the background
\eqref{fullmetric} with the ansatz \eqref{WaveAnsatz} takes the form
\begin{equation}
H^\frac{2(d+p-1)}{d(p-1)} (E^2 - \lambda^2) r^2 h(r) + \frac{1}{r^{p-2}} \partial_r( r^p \partial_r h(r)) - l(l+p-1) h(r)=0.
\end{equation}
The term
\begin{equation}
H^\frac{2(d+p-1)}{d(p-1)} r^2
\end{equation}
is bounded in the ``neck'' region \eqref{neck}. Therefore, assuming
small $E, \lambda$, the wave equation in the neck is
\begin{equation}
\frac{1}{r^{p-2}} \partial_r( r^p \partial_r h(r)) - l(l+p-1) h(r)=0.
\end{equation}
This has the solution
\begin{equation}
h(r) = A r^l + B r^{-l-p+1}.
\end{equation} 
Thus we see that if we use the coordinate $r$ throughout the
intermediate region, then there is no change in the functional form of
$\phi$ as we pass through the intermediate region. We are interested
in the $r^l$ solution, so we set $B=0$. We must now write this in
terms of the coordinate $\tilde r$ appropriate for the $AdS$
region. First consider the case $d=p-1$ which holds for the D3 and
D1D5 cases. Then we see that
\begin{equation}
\tilde r=r, ~~~b_l=1.
\end{equation}
Now consider $d\ne p-1$. From~\eqref{scaledRadialCoordinate} and
\eqref{delta} we get
\begin{equation}
b_l=R_s^{l(1- \frac{d}{p-1})}.
\end{equation}

In general the scalars are not ``minimal;'' i.e., they can have
couplings to the gauge fields present in the geometry. Then $b_l$
needs to be computed by looking at the appropriate wave equation.
Examples of such non-minimal scalars are ``fixed'' scalars discussed
in~\cite{Callan:1996tv}.

To summarize, we find that the change of the waveform through the
intermediate region is given by $b_l, \Delta$. The wave at the
boundary $\tilde{r}=\tilde{r}_b$ is then
\begin{equation}
\tilde{\phi}_{b,\vec{m}}(y)=b_l ~\tilde{r}_b^{\Delta-d}~ [\pd^l\phi(y)]\Big|^{l,\vec{m}}_{r=0}.
\end{equation}

\section{The Interaction}\label{sec:interaction}

We can break the action of the full problem into three parts
\begin{equation}
S_\text{total} = S_\text{CFT} + S_\text{outer} + S_\text{int},
\end{equation}
where the contribution of the interaction between the CFT and the
outer asymptotically flat region, $S_\text{int}$, vanishes in the
strict decoupling limit. We work in the limit where the interaction is
small but nonvanishing, to first order in the interaction.

The coupling of the external wave \emph{at $r_b$} to the
CFT is given by the interaction
\begin{equation}
S^l_\text{int}= -\mu \sum_{\vec{m}} 
	\int \sqrt{g_d} ~\drm^dy\,  \tilde{\phi}_{b,\vec{m}}(y) \hat{\mathcal{V}}_{l,\vec{m}}(y).
\end{equation}
If we want to \emph{directly} couple to the modes in the \emph{outer
region}, then we can incorporate the intermediate region physics into
$S_\text{int}$ by writing
\begin{equation}\label{eq:general-S-int}
S^l_\text{int} = -c_l\sum_{\vec{m}}\int \sqrt{g_d} ~\drm^dy\,  [\pd^l\phi(y)]\Big|^{l,\vec{m}}_{r=0}~
   \hat{\mathcal{V}}_{l,\vec{m}}(y),
\end{equation}
where
\begin{equation}
c_l=\mu~ b_l~{\tilde r}_b^{\Delta-d}.
\label{coupling}
\end{equation}
This is the general action connecting modes in the AdS/CFT with the
modes in the asymptotically flat space. We focus specifically on
(first-order) emission processes, but one can consider more general
interactions (absorption, scattering, etc.).

\section{Emission}\label{sec:emission-formula}

Suppose we have an excited state in the ``cap'' region, $\ket{i}$, and
the vacuum in the outer region, $\vac_{\text{outer}}$.  Because of the
coupling~\eqref{coupling}, a particle can be emitted and escape to
infinity, changing the state in the cap to a lower-energy state
$\ket{f}$, and leaving a 1-particle state in the outer region,
$\ket{E,l,\vec{m}, \vec{\lambda}}_\text{outer}$.  We wish to compute
the rate for this emission, $\Gamma$. We can write the total amplitude
for this process as
\begin{equation}
\mathcal{A} = \bigg({}_\text{outer}\bra{E, l, \vec{m}, \vec{\lambda}}
     \bra{f}\bigg)\; iS_\text{int}\;\bigg(\ket{i}\vac_\text{outer}\bigg).
\end{equation}

We quantize the field $\phi$ in the outer region as
\begin{equation}\begin{split}
\hat{\phi} = \sqrt{\frac{16\pi G_D}{2V_y} }
  \sum_{\vec \lambda, l,\vec{m}}\int_0^\infty\drm E\,
  \tfrac{J_{l+\frac{p+1}{2}}(\sqrt{E^2-\lambda^2}~r)}{r^{\frac{p-1}{2}}}
  \bigg[&\hat{a}_{E,l, \vec{m},\lambda}
    Y_{l,\vec{m}}e^{i(\vec \lambda\cdot \vec  y - E t)}\\
  &+(\hat{a}_{E,l, \vec{m},\lambda})^\dg 
    Y_{l,\vec{m}}^*e^{-i(\vec \lambda \cdot\vec y - E t)}\bigg],
\end{split}\end{equation}
where
\begin{equation}
\com{\hat{a}_{E,l,\vec{m},\lambda}}
{(\hat{a}_{E',l',\vec{m}', \lambda'})^\dg}
= 
  \delta_{ll'}\delta_{\vec{m},\vec{m}'}
  \delta_{\lambda\lambda'}\delta(E-E').
\end{equation}

Using the asymptotic behavior,
\begin{equation}
J_\nu(z)\approx \frac{1}{\Gamma(\nu+1)}\Big ( \frac{z}{2}\Big)^\nu,
\end{equation}
we find that
\begin{multline}
[\pd^l\hat{\phi}]\Big|^{\vec{m}}_{r=0} = \sqrt{\frac{16\pi G_D}{2V_y}}
\sum_{\vec{\lambda}}\int_0^\infty\drm E\,
\frac{(\frac{\sqrt{E^2-\lambda^2}}{2})^{l+\frac{p-1}{2}}}
     {\Gamma(l+\frac{p+1}{2})}
    \bigg[\hat{a}_{E,l, \vec{m},\lambda}
    e^{i(\vec \lambda\cdot \vec  y - E t)}\\
  +(\hat{a}_{E,l, -\vec{m},\lambda})^\dg 
    e^{-i(\vec \lambda \cdot\vec y - E t)}\bigg].
\end{multline}
Using the coupling $c_l$ from (\ref{coupling}) we get the interaction
Lagrangian
\begin{multline}
S_\text{int} =  -\mu b_l {\tilde r}_b^{\Delta-d}\sqrt{\frac{16\pi G_D}{2V_y} }
\sum_{\vec \lambda, \vec{m}}\int\sqrt{g_d} \drm^d y\int_0^\infty\drm E\\    
\frac{(\frac{\sqrt{E^2-\lambda^2}}{2})^{l+\frac{p-1}{2}}}
     {\Gamma(l+\frac{p+1}{2})}
    \bigg[\hat{a}_{E,l, \vec{m},\lambda}e^{i(\vec \lambda\cdot \vec  y - E t)}
	  +(\hat{a}_{E,l,-\vec{m},\lambda})^\dg 
    e^{-i(\vec \lambda \cdot\vec y - E t)}\bigg]~\hat{\mathcal{V}}_{l,\vec{m}}(t,\vec y).
\end{multline}

We can pull out the $(t,\vec{y})$ dependence of the CFT part of the
amplitude by writing
\begin{equation}
\bra{f} \hat{\mathcal{V}}(t, \vec{y})\ket{i}=
     e^{-iE_0 t + i \vec \lambda_0\cdot \vec y}
         \bra{f}\hat{\mathcal{V}}(t=0, \vec y=0)\ket{i},
\end{equation}
where $E_0$ and $\lambda_0$ can be determined from the initial and
final states of the CFT, $\ket{i}$ and $\ket{f}$. We also work in the
case where the initial and final CFT states select out a single $l,m$
mode in the interaction, whose indices have been suppressed. The CFT
lives on a space of (coordinate) volume $V_y$. When computing CFT
correlators we work in a ``unit-sized'' space with volume
$(2\pi)^{d-1}$. Scaling the operator $\hat{\mathcal{V}}$ we have
\begin{equation}
\bra{f} \hat{\mathcal{V}}(t=0, \vec y=0)\ket{i}
     = \left[\frac{(2\pi)^{d-1}}
                  {(\frac{{\tilde r}_b}{R_s})^{d-1} V_y}
       \right]^\frac{\Delta}{d-1}
       \bra{f}\hat{\mathcal{V}}(t=0, \vec y=0)\ket{i}_\text{unit}.
\end{equation}
Rotating back to Lorentzian signature, the amplitude for emission of a
particle from an excited state of the CFT is
\begin{multline}
\mathcal{A}=-i \mu b_l {\tilde r}_b^{\Delta-d}\sqrt{\frac{16\pi G_D}{2V_y}}
\frac{(\frac{\sqrt{E^2-\lambda^2}}{2})^{l+\frac{p-1}{2}}}{\Gamma(l+\frac{p+1}{2})}
\left[ \frac{(2\pi)^{d-1}}{\big(\frac{{\tilde r}_b}{R_s}\big)^{d-1} V_y}\right]^\frac{\Delta}{d-1}
   \bra{f} \hat{\mathcal{V}}_{l,\vec{-m}}(t=0, \vec y=0)\ket{i}_\text{unit}
  \\
 \times
\int_0^T \drm t \int \big(\tfrac{{\tilde r}_b}{R_s}\big)^d 
   \drm^{d-1} y\, e^{-i(E_0-E)t}e^{i(\vec{\lambda}_0-\vec{\lambda})\cdot \vec y}.
\end{multline}
The amplitude gives the emission rate in a straightforward calculation:
\begin{calc}\label{eq:general-decay-rate}
\der{\Gamma}{E} &= \lim_{T\to\infty}\frac{|\mathcal{A}|^2}{T}\\
       &=(2\pi)^{2\Delta+1}\frac{R_s^{4\Delta-3d+p-1} 
         \left( \frac{d}{p-1}\right)^{2 \Delta -(d+1)}}{V_y^{\frac{2\Delta-d+1}{d-1}}}
\left[\frac{(2\Delta-d)}{2\pi^\frac{d}{2}}\frac{\Gamma(\Delta)}{\Gamma(\Delta-\frac{d}{2})}\right]
  |b_l|^2\left[\frac{1}{\Gamma(l+\frac{p+1}{2})}\right]^2\\
  &\qquad\times\left(\frac{{E^2-\lambda^2}}{4}\right)^{l+\frac{p-1}{2}}
   \Big|\bra{0}\hat{\mathcal{V}}_{l,-\vec{m}}(0)\ket{1}_\text{unit}\Big|^2\,
   \delta_{\vec{\lambda},\vec{\lambda}_0}\delta(E- E_0).
\end{calc}
From this expression, we see that in the strict decoupling limit where
$E R_s\rightarrow 0$ this rate vanishes as expected.

We have derived the above result for a general CFT and its
corresponding brane geometry.  In Chapter~\ref{ch:emission} we work
with the D1D5 system. For a minimal scalar in the D1D5 geometry we
have
\begin{equation}
d = 2, \qquad p = 3\qquad \Delta_\text{tot.} = l + 2\qquad b_l =1\qquad
R_s = (Q_1Q_5)^\frac{1}{4}\qquad V_y = 2\pi R.
\end{equation}
Plugging in, we reduce the decay rate formula to the form
\begin{equation}\label{eq:D1D5-decay-rate}
\der{\Gamma}{E} = 
	\frac{2\pi}{2^{2l+1}\,l!^2}\frac{(Q_1Q_5)^{l+1}}{R^{2l+3}}(E^2-\lambda^2)^{l+1}\,
	|\bra{0}\hat{\mathcal{V}}\ket{1}_\text{unit}|^2\,
	\delta_{\lambda,\lambda_0}\delta(E - E_0),
\end{equation}
which we use directly in Chapter~\ref{ch:emission}. Do not confuse the
radius of the $S^3$, $R_s$, with the radius of the boundary, $R$. Note
that most previous treatments would only have a proportionality, but
using our formalism we have completely fixed the equality.

\chapter{Emission from JMaRT Geometries}
\label{ch:emission}

In this chapter, we apply the formalism developed in
Chapter~\ref{ch:coupling} to the emission from a particular family of
smooth, horizon-free three-charge nonextremal geometries. In the
fuzzball proposal, they are candidate microstates of a three-charge
non-extremal black hole.  The particular family of geometries was
found by Jejjala, Madden, Ross, and Titchener~\cite{ross}, so they are
prosaically called JM(a)RT geometries. In~\cite{myers}, it was shown
that the JMaRT geometries have a classical instability associated with
the ergoregion in the geometry: one solves the classical wave equation
for a minimal scalar in these JMaRT background, and finds solutions
that are exponentially growing in time. Eventually, back-reaction
becomes important and the geometry decays.  Initially, this seemed
problematic for the fuzzball interpretation~\cite{myers, myers-2}:
black holes are only unstable \emph{semi}-classically due to Hawking
radiation---the JMaRT geometries decay much faster. In~\cite{cm1,
  cm2,cm3}, using heuristic CFT computations for some special cases,
it was argued that the ergoregion emission is the Hawking radiation
from this subset of microstates for the D1D5 black hole. In the CFT,
one observes that the states dual to the JMaRT geometry are
nongeneric, having many copies in the same state. The Hawking process,
then, gets a large Bose enhancement.

Here, we outline work done in~\cite{acm1, ac1}, which closes some of
the holes in~\cite{cm1, cm2, cm3}. We reproduce the \emph{entire}
spectrum and rate of emission for \emph{all} of the JMaRT geometries
using rigorous CFT calculations. We also note that the previous
calculations could only normalize the action of the vertex operator by
demanding agreement with Hawking radiation. Using the results in
Chapter~\ref{ch:coupling}, we can fix the coupling from the two-point
function in AdS, which makes the argument a bit more satisfying. One
of the main upshots of the calculations presented in this chapter is
that seemingly diverse gravitational processes are realized in
essentially the same way in the CFT via spectral flow and Hermitian
conjugation.

In Section~\ref{sec:jmart}, we briefly review the JMaRT geometries and
their ergoregion instability. In Section~\ref{sec:kappa-equals-1}, we
perform the CFT calculation that reproduces the emission for a subset
of the JMaRT geometries with $\kappa=1$. Using spectral flow, the
calculation is ultimately reduced to computing a two-point function of
the vertex operator. In Section~\ref{sec:kappa-not-1}, we generalize
to $\kappa>1$, which gives the emission spectra for all of the JMaRT
geometries. The calculation for $\kappa>1$ cannot be reduced to a
two-point function via spectral flow. To complete the calculation, it
is necessary to use the technology developed in~\cite{lm1, lm2} for
computing correlation functions of twist operators. The technique is
reviewed in Appendix~\ref{ap:twist-corr}.  Finally in
Section~\ref{sec:discuss-emission}, we interpret the results and
explain why the ergoregion emission is Hawking radiation from the
microstates. We also discuss what lessons one might hope to apply to
the fuzzball proposal in general.

\section{JMaRT Geometries}\label{sec:jmart}

For a fixed amount of D1 charge, D5 charge, and $S^1$-momentum,
~\cite{ross} found a three-parameter family of geometries that have
the following properties. At infinity they are asymptotically flat,
then as one goes radially inward one encounters a ``neck'' region.
After passing through the neck, one finds an AdS throat which
terminates in an ergoregion cap. The fuzzball proposal interprets
these smooth, horizonless geometries as classical approximations to
microstates of a black hole with the same mass and charges.

\begin{figure}[ht]
\begin{center}
\subfigure[~The gravity description.]{
\includegraphics[width=6cm]{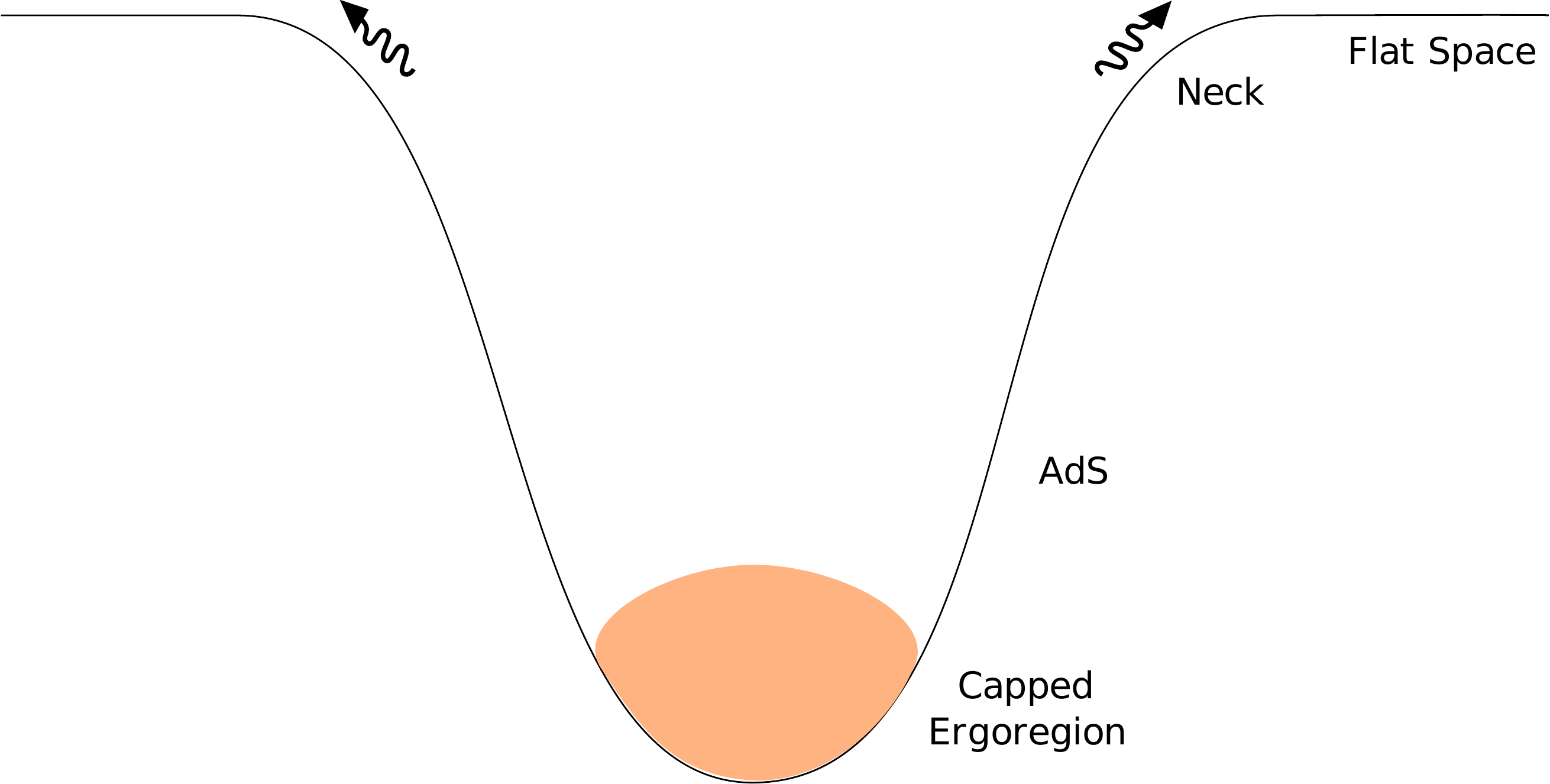}}
\hspace{30pt}
\subfigure[~The CFT description.]{
\includegraphics[width=6cm]{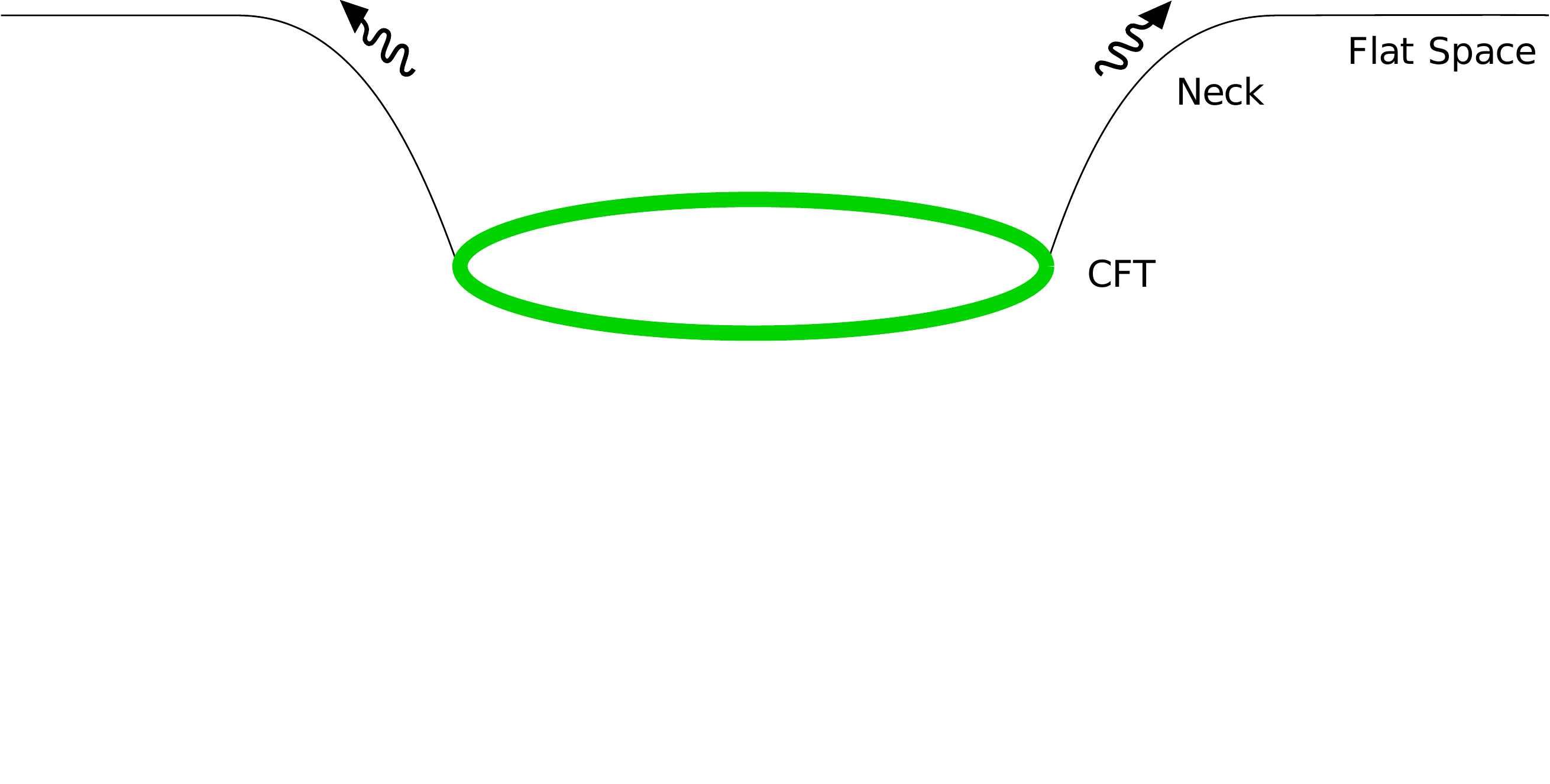}}
\caption[The gravity description of the emission process]{A depiction
  of the emission process we consider. In the cap region of the
  geometry, there is an ergoregion (shown in orange), which leads to
  the emission of particles into the flat space. In the CFT
  description particles are emitted by the CFT (shown in green)
  directly into the ``neck'' region of the
  geometry.}\label{fig:ergo-emission}
\end{center}
\end{figure}

The presence of ergoregions renders these geometries unstable
~\cite{myers,cm1,cm3}. The instability is exhibited by emission of
particles at infinity, with exponentially increasing flux, carrying
energy and angular momentum out of the geometry. In~\cite{cm1, cm2,
  cm3}, using heuristic CFT computations for some special cases, it
was argued that the ergoregion emission is the Hawking radiation from
this subset of microstates for the D1D5 black hole.
Figure~\ref{fig:ergo-emission} depicts the emission process in the
gravity and CFT descriptions.

These geometries are dual to CFT states parameterized by three
integers $n$, $\bar{n}$, and $\kappa$. In~\cite{acm1}, the spectrum
and rate of emission from the geometries with $\kappa=1$ was exactly
reproduced with a CFT computation. The parameter $\kappa$ (called $k$
in~\cite{cm3}), controls a conical defect in the geometry. For
$\kappa=1$, there is no conical defect or orbifold singularity.

\subsection{Features of the Geometries}

The D1D5 system we work with lives in a ten-dimensional geometry
compactified on $T^4\times S^1$. We wrap $N_5$ D5 branes on the full
compact space, $T^4\times S^1$; and we wrap $N_1$ D1 branes on the
circle, $S^1$. This gives rise to an $AdS_3\times S^3$ throat in the
noncompact space, which is dual to a two-dimensional
CFT~\cite{maldacena,swD1D5, lmD1D5,deBoerD1D5,
dijkgraafD1D5,frolov-1,frolov-2, Jevicki,David}. The core AdS region
has radius $(Q_1Q_5)^\frac{1}{4}$, where the D1 and D5 charges $Q_1$
and $Q_5$ are given by
\begin{equation}
Q_1 = \frac{g\Regge^3}{V}N_1 \qquad Q_5 = g\Regge N_5,
\end{equation}
with  string coupling $g$ and $T^4$ volume $(2\pi)^4V$.

The JMaRT solutions are found as special cases of nonextremal
three-charge black holes in~\cite{cveticyoum, cveticlarsen}. The three
charges come from the D1-branes wrapping $S^1$, the D5-branes wrapping
$T^4\times S^1$, and momentum on $S^1$. These solutions have
(string-frame) metric~\cite{ross}
\begin{equation}\begin{split}\label{eq:gen-metric}
ds^2 =& -\frac{f}{\sqrt{\tilde{H}_1\tilde{H}_5}}(dt^2 - dy^2) 
                                    + \frac{M}{\sqrt{\tilde{H}_1\tilde{H}_5}}(s_pdy - c_p dt)^2\\
      & + \sqrt{\tilde{H}_1\tilde{H}_5}\left(
                                \frac{r^2dr^2}{(r^2+a_1^2)(r^2 + a_2^2) - M r^2} + d\theta^2\right)\\
  & + \left(\sqrt{\tilde{H}_1\tilde{H}_5} - (a_2^2 - a_1^2)
                                   \frac{(\tilde{H}_1 + \tilde{H}_5 -f)\cos^2\theta}
                                        {\sqrt{\tilde{H}_1\tilde{H}_5}}\right) \cos^2\theta\,d\psi^2\\
      & + \left(\sqrt{\tilde{H}_1\tilde{H}_5} - (a_2^2 - a_1^2)
                                   \frac{(\tilde{H}_1 + \tilde{H}_5 -f)\sin^2\theta}
                                        {\sqrt{\tilde{H}_1\tilde{H}_5}}\right) \sin^2\theta\,d\phi^2\\
      & + \frac{M}{\sqrt{\tilde{H}_1\tilde{H}_5}}
                  \big(a_1 \cos^2\theta\, d\psi + a_2\sin^2\theta\,d\phi\big)^2\\
      & + \frac{2M\cos^2\theta}{\sqrt{\tilde{H}_1\tilde{H}_5}}
               \big[(a_1c_1c_5c_p - a_2s_1s_5s_p)dt + (a_2s_1s_5c_p - a_1c_1c_5s_p)dy\big]d\psi\\
      & + \frac{2M\sin^2\theta}{\sqrt{\tilde{H}_1\tilde{H}_5}}
               \big[(a_2c_1c_5c_p - a_1s_1s_5s_p)dt + (a_1s_1s_5c_p - a_2c_1c_5s_p)dy\big]d\phi\\
      & + \sqrt{\frac{\tilde{H}_1}{\tilde{H}_5}}\sum_{i=1}^4 dz_i^2,
\end{split}\end{equation}
where
\begin{equation}
\tilde{H}_i = f + M \sinh^2\delta_i\quad
f = r^2 + a_1^2\sin^2\theta + a_2^2\cos^2\theta\quad
c_i = \cosh\delta_i\quad
s_i = \sinh\delta_i\quad i=1,5,p.
\end{equation}
The D1 and D5 branes also source the RR 2-form, as expected. The
nonzero RR 2-form and dilaton may be found in~\cite{ross}. They are
not relevant for our purposes here. The metric in
Equation~\eqref{eq:gen-metric} is the ten-dimensional string frame
metric; however, if we omit the $T^4$ part, $dz_i^2$, then it is the
six-dimensional Einstein frame metric.  Our interest is mostly in this
six-dimensional space.

The constants $\delta_i$ determine the asymptotic charges via
\begin{equation}\begin{split}
Q_1 &= M \sinh\delta_1\cosh\delta_1 = \frac{M}{2}\sinh 2\delta_1\\
Q_5 &= M \sinh\delta_5\cosh\delta_5 = \frac{M}{2}\sinh 2\delta_5\\
Q_p &= M \sinh\delta_p\cosh\delta_p = \frac{M}{2}\sinh 2\delta_p.
\end{split}\end{equation}
For given asymptotic charges, $Q_i$, we can change the parameters
$a_1$, $a_2$, and $M$ subject to the above constraint. The geometry
has ADM mass and angular momentum given by
\begin{equation}\begin{split}
M_\text{ADM} &= \frac{M}{2}\big(\cosh 2\delta_1 + \cosh 2\delta_5 + \cosh 2\delta_p\big)\\
J_\psi &= -M\big(a_1 \cosh \delta_1 \cosh \delta_5\cosh \delta_p 
                       - a_2 \sinh \delta_1\sinh \delta_5 \sinh \delta_p\big)\\
J_\phi &= -M\big(a_2 \cosh \delta_1 \cosh \delta_5\cosh \delta_p 
                       - a_1 \sinh \delta_1\sinh \delta_5 \sinh \delta_p\big),
\end{split}\end{equation}
from which it is clear that we must choose $M\geq 0$~\cite{ross}.

\subsubsection{The Asymptotic Flat Space}

Let us first note that in the large $r$ limit the geometry takes the
form of $M^{4,1}\times S^1\times T^4$, as desired:
\begin{equation}
ds^2 \xrightarrow[r^2 \gg M]{} -dt^2  + dy^2 + dr^2 
                   + r^2\big(d\theta ^2 + \cos^2\theta\, d\psi^2 + \sin^2\theta \,d\phi^2\big)
                   + dz_i^2,
\end{equation}
where $y$ is the $S^1$ coordinate; $z_i$ are coordinates on the $T^4$;
and $\theta$, $\psi$ and $\phi$ are angular coordinates of $S^3$. We
can relate them to the usual Cartesian coordinates (at infinity) via
\begin{equation}\begin{split}
x_1 &= r\sin\theta\sin\psi\\
x_2 &= r\sin\theta\cos\psi\\
x_3 &= r\cos\theta\sin\phi\\
x_4 &= r\cos\theta\cos\phi.
\end{split}\end{equation}
We see then that $J_\psi$ and $J_\phi$ are the angular momenta
associated with rotation in the 1--2 and 3--4 planes, respectively.

\subsubsection{Horizon Removal}

The simplest way to ascertain whether the geometry has a horizon or
singularity is to examine the $rr$ component of the inverse
metric,
\begin{equation}
g^{rr} = \frac{1}{r^2\sqrt{\tilde{H}_1\tilde{H}_5}}
         \big[(r^2 + a_1^2)(r^2 + a_2^2) - M r^2\big]
\equiv
\frac{1}{r^2\sqrt{\tilde{H}_1\tilde{H}_5}}
         \big[(r^2 - r_+^2)(r^2 - r_-^2)\big].
\end{equation}
This vanishes for $r=r_\pm$, where
\begin{equation}
r^2_\pm = \frac{1}{2}\left[
(M - a_1^2 - a_2^2) \pm 
   \sqrt{(M-a_1^2 - a_2^2)^2- 4a_1^2a_2^2}\right].
\end{equation}
In order for $r^2_\pm$ to be real we must have
\begin{equation}
M \geq (a_1 + a_2)^2 \qquad \text{OR} \qquad M \leq (a_1-a_2)^2.
\end{equation}
Thus there are three different regions of parameter space that we can
consider~\cite{myers, ross}:
\begin{subequations}\begin{align}
& M \leq (a_1 - a_2)^2 &&\Longrightarrow & r_+^2 &< 0\label{eq:pr-jmrt}\\
& (a_1 - a_2)^2 < M < (a_1 + a_2)^2 &&\Longrightarrow & r_+^2 &\notin \re\\
& M \geq (a_1 + a_2)^2 &&\Longrightarrow & r_+^2 &> 0.
\end{align}\end{subequations}
We are interested in the first case, Equation~\eqref{eq:pr-jmrt},
where $r_+^2$ is negative. This case corresponds to the JMaRT
solution, which has no horizon. The second and third case give a naked
singularity and black hole, respectively.

Having $r^2 <0$ may seem strange, but we can in essence analytically
continue by defining a new radial coordinate
\begin{equation}
\rho^2 = r^2 - r_+^2.
\end{equation}
We can then look at what happens near $\rho^2 = 0$ or $r^2 = r_+^2$.
The determinant of the metric is proportional to
\begin{equation}
\det g \propto (r^2 - r_+^2)(r^2 - r_-^2),
\end{equation}
and thus vanishes at this point. This vanishing can signal either an
event horizon or a degeneration of the coordinate system, as happens
at the origin of a polar coordinate system. To have a fuzzball, we
want a horizonless solution and so we want the second case. This means
that at $r^2= r_+^2$ a circle in the geometry must pinch off to zero
size. This story will be familiar to anyone who has studied the
Euclidean Schwarzschild solution, for instance.  A general Killing
vector pointing in a circular direction can be written in the form
\begin{equation}
\xi = \pd_y - \alpha \pd_\psi - \beta \pd_\phi.
\end{equation}
So we must require that $\xi^2$ vanishes at $r^2 = r_+^2$ for some
choice of $\alpha$ and $\beta$. A necessary condition being that the
determinant of the metric at fixed $r$ and $t$,
$\det_{y\theta\phi\psi} g$, must vanish at $r^2=r_+^2$.

Working through these requirements, one finds~\cite{ross}
\begin{equation}\label{eq:M-fixed}
M = a_1^2 + a_2^2 - a_1a_2\frac{c_1^2c_5^2c_p^2 + s_1^2s_5^2s_p^2}{s_1c_1s_5c_5s_pc_p}
\quad\Longrightarrow\quad
r_+^2 = -a_1a_2\frac{s_1s_5s_p}{c_1c_5c_p},
\end{equation}
and the Killing vector that degenerates is given by
\begin{equation}
\alpha = - \frac{s_pc_p}{a_1\, c_1c_5c_p - a_2\, s_1 s_5 s_p}\qquad
\beta = - \frac{s_pc_p}{a_2\, c_1c_5c_p - a_1\, s_1s_5s_p}.
\end{equation}
This suggests that the natural angular coordinates in the interior of
the solution,
\begin{equation}
\tilde{\psi} = \psi + \alpha y\qquad \tilde{\phi} = \phi + \beta y,
\end{equation}
are shifted with respect to the natural angular coordinates at
infinity. We choose the $y$ coordinate to have range $y\in[0,2\pi R)$.
Thus, it would be natural to demand that the geometry be $2\pi
R$-periodic where the geometry pinches off, so that there is no
conical singularity; however, in string theory it is consistent to
consider orbifold singularities, since the worldsheet theory is
well-defined on these backgrounds. We therefore introduce an integer
parameter, $\kappa\in \ints^+$, and demand that the geometry be
$2\pi\kappa R$-periodic at $r^2 = r_+^2$. This
implies~\cite{ross}\footnote{We use hats to distinguish these
  parameters from variables we use later.}
\begin{equation}
\alpha R = -\hat{n}\qquad \beta R = \hat{m}\qquad
\hat{m},\hat{n}\in\ints,
\end{equation}
and
\begin{equation}
R = \frac{M}{\kappa\sqrt{a_1a_2}}
    \frac{s_1c_1s_5c_5\sqrt{s_1c_1s_5c_5s_pc_p}}
         {c_1^2c_5^2c_p^2 - s_1^2s_5^2s_p^2}.
\end{equation}
Thus for fixed $Q_1$, $Q_5$, and $R$, we have three integer parameters
that give distinct JMaRT geometries: $\kappa$, $\hat{m}$, and
$\hat{n}$. The requirement that $M$ be nonnegative implies that
$\hat{m}>\hat{n}\geq 0$, after making some choices that do not result
in loss of generality~\cite{ross}.

Let us note that the JMaRT solitons have mass and rotation outside of
the ``black hole bound.'' Thus, there is no nonextremal black hole
with the JMaRT conserved charges. This does not pose a major problem
for our interpretation for several reasons. Part of the fuzzball
proposal is identifying states in the CFT that are dual to geometries
(in the appropriate limits). We are certainly testing and exploring
that identification here. Within the CFT, there is no sharp
distinction between states that are in the black hole bound and states
that are not---they are all perfectly good states. Moreover, if one
works in a canonical ensemble, rather than a microcanonical ensemble,
then the CFT states considered here (and therefore the JMaRT
geometries) do enter as ``microstates'' in any calculation of
thermodynamic quantities.  It is expected based on this discussion,
however, that the JMaRT solutions will be nongeneric and may have
properties that differ significantly from more generic geometries.

\subsubsection{The Near-Horizon Geometry}

Of importance to our considerations is that the geometry has a
near-horizon ($\kappa$-orbifolded) $AdS_3\times S^3$. We take the
near-horizon limit by taking $Q_1,Q_5 \gg M, a_1^2, a_2^2$ and $r^2
\ll Q_1, Q_5$, which results in~\cite{ross}
\begin{equation}\begin{split}\label{eq:ross-ads}
ds^2 =& -\left(\frac{\rho^2}{L^2} + 1\right)d\tau^2
       + \left(\frac{\rho^2}{L^2} + 1\right)^{-1}d\rho^2
       + \rho^2d\varphi^2 \\
       &+ L^2\bigg[d\theta^2 
            + \sin^2\theta\left(d\phi + \hat{m}d\varphi 
                                      - \frac{\hat{n}}{L}d\tau\right)^2\\
       &\hspace{1.2cm} + \cos^2\theta\left(d\psi - \hat{n}d\varphi 
                                   + \frac{\hat{m}}{L}d\tau\right)^2\bigg],
\end{split}\end{equation}
where
\begin{equation}
L^2 = \sqrt{Q_1Q_5}\qquad
\varphi = \frac{y}{R}\qquad
\tau = \frac{t L}{R}\qquad
\rho^2 = \frac{R^2}{L^2}\left[r^2 + (M-a_1^2-a_2^2)s_p^2 + a_1a_2\sinh 2\delta_p\right].
\end{equation}
We see then that $\hat{m}$ and $\hat{n}$ give shifts to the angular
coordinates. This shifting of the angular coordinates corresponds to
spectral flow in the dual CFT.

\subsubsection{The Ergoregion}

The geometry also has an \emph{ergoregion}. If we start with a
timelike Killing vector at asymptotic infinity, it necessarily becomes
spacelike in the cap of the geometry---the lightcones tip over into an
angular direction. Let us note that this is different from an event
horizon, in which the lightcones tip over in a radial direction and
nothing can escape. In this case, if one is in the ergoregion, then
one must rotate with respect to the asymptotic flat space; however,
one can still exit an ergoregion.

The most general causal Killing field at infinity, can be written
as\footnote{Please do not confuse $\gamma$ here with the notation
  of~\cite{cm3}.}
\begin{equation}
 \zeta = \pd_t + \gamma\pd_y,
\end{equation}
for $\gamma^2 <1$. The norm may be written in the form
\begin{calc}
\zeta^2 &= g_{tt} + 2\gamma g_{ty} + \gamma^2 g_{yy}\\
        &= \frac{1}{\sqrt{\tilde{H}_1\tilde{H}_5}}
           \left[ -(1-\gamma^2)f +M(c_p - \gamma s_p)^2\right].
\end{calc}
The ``most'' timelike choice is $\gamma = s_p/c_p = \tanh \delta_p$,
but $\zeta$ still becomes spacelike for this choice in the region
defined by~\cite{ross}
\begin{equation}
\text{ergoregion:}\quad f = r^2 + a_1^2\sin^2\theta + a_2^2\cos^2\theta < M,
\end{equation}
where recall that $M$ is fixed by Equation~\eqref{eq:M-fixed}. Let us
emphasize that \emph{there is} a global time-like Killing vector and
hence no ergoregion within the asymptotic AdS geometry in
Equation~\eqref{eq:ross-ads}. It is only when comparing the
\emph{asymptotic flat space} notion of time and the interior space
notion of time that we see an ergoregion. The lightcones starting in
the asymptotic AdS do not tip over completely when going deep into the
cap.  This point is important since it means that we can only see the
physical effects of the ergoregion, namely ergoregion emission, if
\emph{we relax the strict decoupling limit} and re-attach the
asymptotic flat space.

\subsubsection{Identification in the CFT}

In order to perform a calculation with the CFT, we must identify the
CFT state dual to the JMaRT geometry. We can identify the weight and
charge of the state from the corresponding charges of the geometric
description. For instance, the energy of the state in the CFT, the sum
of the left and right conformal weight, is given by the (ADM) mass of
the geometry \emph{above extremality}, $\Delta M_{\text{ADM}}$. The
$S^1$-momentum gives the difference between the left and right
weights. Thus, we can identify $h$ and $\bar{h}$ of the state. The
angular momenta $J_\psi$ and $J_\phi$ give the left and right $J^3$
charges, $m$ and $\bar{m}$. This exercise gives~\cite{ross}
\begin{equation}\begin{aligned}\label{eq:CFT-id}
h &= N_1N_5\frac{1}{4\kappa^2}\left(\kappa^2 -1 + (\hat{m}+\hat{n})^2\right) &
m &= N_1N_5\frac{\hat{m} + \hat{n}}{2\kappa}\\
\bar{h} &= N_1N_5\frac{1}{4\kappa^2}\left(\kappa^2 -1 + (\hat{m}-\hat{n})^2\right) &
\bar{m} &= N_1N_5\frac{\hat{m} - \hat{n}}{2\kappa}.
\end{aligned}\end{equation}
The periodicity of the fermions is determined by the parity of
$\hat{m} + \hat{n}$: if the sum is odd, then we are in the R--R
sector; if the sum is even we are in the NS--NS sector~\cite{ross}. We
are interested in the R--R sector that is relevant for black hole
physics. 

Let us note the special case when $\hat{m} = \hat{n} + 1$ and $\kappa
= 1$. In this case, we see that $\bar{h} = \frac{1}{4}N_1N_5$ and
$\bar{m} = \frac{1}{2}N_1N_5$. The weight and charge of the right
sector uniquely determine the right part of the state to be the Ramond
vacuum with all of the ``base spins'' up. These are the three-charge
supersymmetric (extremal, BPS) geometries studied in~\cite{lunin,
  gms2, gimon-4}. In general the weight and charge of a state do not
suffice to identify it in the CFT; however, we can identify the JMaRT
states since the angular shifts in the boundary of the AdS region
correspond to spectral flow in the left and right sectors by $2n$
units on the left and $2\bar{n}$ units on the right, where
\begin{equation}
\frac{\hat{m} + \hat{n}}{\kappa} = 2n + \frac{1}{\kappa} \qquad
\frac{\hat{m} - \hat{n}}{\kappa} = 2\bar{n} + \frac{1}{\kappa}. 
\end{equation}
We find it more convenient to parameterize the geometries with $n$,
$\bar{n}$, and $\kappa$ for calculations in the CFT. When $n$ and
$\bar{n}$ both vanish, then $\hat{m} =1$ and $\hat{n}=0$ and both the
left and right sectors of the CFT are in the R ground state with all
base spins up. Thus, the JMaRT geometries correspond to an even number
of units of left and right spectral flow of the Ramond ground state.
The states where $n = 0$ or $\bar{n}=0$ are extremal and have no
ergoregion or ergoregion emission. Therefore, we see that the
existence of an ergoregion and therefore ergoregion emission is tied
to the nonextremality of the geometry. In~\cite{cm1, cm2, cm3}, it was
suggested that this may hold for more general nonextremal microstates:
perhaps generic nonextremal microstates do not have a global timelike
Killing vector. This would tie in nicely with the interpretation of
the ergoregion emission as the Hawking radiation process.

\subsection{Ergoregion Emission}

Ergoregions are usually encountered in geometries that also have a
horizon, i.e. black holes. For instance, outside the event horizon of
a Kerr black hole there is an ergoregion that has the same rotational
sense as the black hole. This is an example of frame-dragging. Because
there is no global timelike Killing vector, and therefore no global
notion of time-translation, one can have energies in the ergoregion
that are negative with respect to asymptotic infinity.

For ergoregions in the presence of a horizon, one gets the phenomenon
of superradiance. If one sends a wave at the ergoregion and black
hole, a greater amplitude wave can be reflected off. In this way, one
can mine energy and angular momentum out of the black hole geometry
until the ergoregion disappears. If one places a ``mirror'' outside
the ergoregion, then the reflected wave cannot vanish and one gets
repeated amplification until the mirror breaks. This is the ``black
hole bomb'' scenario~\cite{black-hole-bomb}.

Since the JMaRT geometries do not have a horizon, they are not
superradiant, but they suffer from another instability which we call
the ergoregion instability~\cite{friedman, cominsschutz, ashtekar,
  myers, myers-2}. The instability is associated with the ability to
form bound states in the negative energy region. This produces a large
classical instability to decay.  More specifically, if one solves the
classical wave equation of a minimal scalar in this background, one
finds modes with frequency,$\omega$, that have positive imaginary
part---the solution grows exponentially in time until back-reaction
becomes important and the geometry decays~\cite{myers, myers-2}. This
process happens very rapidly.  One can interpret this process as pair
production: negative energy particles with angular momentum canceling
out the ergoregion's condense in the ergoregion, while their positive
energy cousins carry energy and angular momentum out of the geometry
to asymptotic flat space~\cite{cm2}. 

Our goal in this chapter is to reproduce the spectrum and rate of the
ergoregion instability by using a CFT description. We work in the
large-charge limit where the AdS space is large and the the ergoregion
emission has been computed analytically via the method of matched
asymptotic expansions~\cite{myers}.  We also require the radius of
$S^1$ be large so that the AdS space is ``deep'' enough for the CFT to
be a good description. Previous calculations in~\cite{cm1, cm3} could
only reproduce the spectrum for certain special cases. The full
spectrum was found in~\cite{acm1, ac1}.

\section{Reproducing \texorpdfstring{$\kappa = 1$}{kappa = 1}  Emission using the Orbifold CFT}\label{sec:kappa-equals-1}

Early computations of radiation \cite{stromvafa, radiation-1,
  radiation-2, radiation-3, radiation-4, radiation-5} used the
somewhat heuristic picture of an ``effective string'' to describe the
D1D5 bound state.  We construct states and vertex operators in the
orbifold CFT, setting up notation and tools that allow us to compute
amplitudes with ease.  We apply these steps to compute the emission
rate of supergravity scalars from particular D1D5 states.  In
particular we can compute emissions in cases where it was unclear how
to proceed with the effective string model. The CFT amplitudes,
converted to radiation rates by our general formalism, show exact
agreement with the emission rates in the dual gravitational geometry.

Specifically, we perform the following steps:
\begin{enumerate}
\item As an example of our CFT techniques we consider minimal scalars
  in the D1D5 geometry. An example of such scalars is given by the
  graviton with indices along the compact torus directions. We
  construct the correctly normalized vertex operators for these
  scalars, which are obtained by starting with a twist operator in the
  CFT and dressing it with appropriate modes of the chiral algebra.

\item We use the notion of ``spectral flow'' to map states from the
  Ramond sector (which describes the D1D5 bound state) to the NS
  sector (which has the vacuum $\vac_{NS}$). This map helps us in two
  ways. First, it is not clear which states in the Ramond sector
  correspond to supergravity excitations (as opposed to string
  excitations). In the NS sector there \emph{is} a simple map: the
  supergravity excitations are given by chiral primaries and their
  descendants under the anomaly-free subalgebra of the chiral algebra.
  Thus the spectral flow map allows us to find the initial and final
  states of our emission process, given the quantum numbers of these
  states. Second, for the process of interest the final state turns
  out to be the \emph{vacuum} in the NS description; thus we do not
  need an explicit vertex operator insertion to create this state when
  computing amplitudes after spectral flowing to the NS sector.

\item We consider a simple decay process where an excited state of the
  CFT decays to the ground state and emits a supergravity quantum. We
  compute the CFT amplitude for this emission process. As mentioned
  above, the final state in the amplitude is nontrivial in the Ramond
  sector, but maps to the NS vacuum $\vac_{NS}$ under spectral flow.
  This converts a 3-point correlator in the Ramond sector to a 2-point
  correlator in the NS sector.  With this amplitude, we use the result
  in (i) to compute $\Gamma$, the rate of emission to flat space.

\item In the above computations we take the initial state to contain a
  few excitations above the ground state, and we compute the decay
  rate for these excitations. Alternatively, we can choose to start
  with the initial state having \emph{no} excitations in the NS
  sector, and then perform a spectral flow on this state. This
  spectral flow adds a large amount of energy to the state, giving a
  configuration which is described in the dual gravity by a
  \emph{nonextremal} classical metric~\cite{ross}. This metric is
  known to emit radiation by ergoregion emission~\cite{myers}, and
  in~\cite{cm1,cm2,cm3} it was shown that for a subset of possible
  supergravity emissions the CFT rate agreed exactly with the gravity
  emission rate.  We can now extend this agreement to all allowed
  emissions of supergravity quanta by using the orbifold CFT. It turns
  out that the simple decay process computed in steps~(iii) and (iv)
  can be used to give the emission rate from this highly excited
  initial state. This is because using spectral flow the initial state
  can be mapped to the vacuum, and in fact the entire amplitude maps
  under spectral flow to a time reversed version of the decay
  amplitude computed above.  We find exact agreement with the
  radiation rate in the dual gravity description.
\end{enumerate}

\subsection{The Initial State, the Final State, and the Vertex
  Operator}\label{sec:states-and-op}

Our main computation addresses the following physical process.
Consider a bound state of $N_1$ D1 branes and $N_5$ D5 branes, sitting
at the origin of asymptotically flat space.  As mentioned above, the
CFT describing this bound state is in the Ramond sector, which has a
number of degenerate ground states. We pick a particular Ramond ground
state. Instead of describing this choice directly in the Ramond
sector, we note that all Ramond sector ground states are obtained by
one unit of spectral flow from chiral primary states of the NS sector.
We pick the Ramond ground state that arises from the simplest chiral
primary: the NS vacuum $\vac_{NS}$. The gravity dual of this state can
be described as follows~\cite{bal, mm, lm3, lm4}: there is flat space
at infinity, then a ``neck,'' then an $AdS$ region, and then a
``cap,'' as pictured in Fig.~\ref{fig:throats-b}. While the structure
of the ``cap'' depends on the choice of Ramond ground state, in the
present case the structure is particularly simple: below $r_b$ in
Fig.~\ref{fig:throats-b} the geometry is a part of global $AdS_3\times
S^3$.

By itself such a D1D5 state is stable, and does not radiate energy. We
therefore add an excitation to the D1D5 brane state. In the
supergravity dual, the excitation we choose corresponds to adding a
supergravity quantum sitting in the ``cap.'' The supergravity field we
choose is a scalar $\phi_{ij}$, where $i,j=1, \dots 4$ are vector
indices valued in the compact $T^4$. These scalars arise from the
following fields:
\begin{enumerate}
\item A symmetric traceless matrix $h_{ij}$ with $i, j=1,\dots 4$
  giving the transverse traceless gravitons with indices in the $T^4$.

\item An antisymmetric matrix $B^{RR}_{ij}$ giving the components of
  the Ramond-Ramond $B$ field with indices in the torus.

\item The dilaton, which is a scalar in the full 10-dimensional theory.
\end{enumerate}
We can put all these scalars together into a $4\times 4$ matrix
$\phi_{ij}$, with the symmetric traceless part coming from $h_{ij}$,
the antisymmetric part from $B^{RR}_{ij}$ and the trace from the
dilaton. (Such a description was used for example in
\cite{radiation-3, lm4}. But we may need  to scale the above fields
by some function.  For instance, it is not the graviton, $h_{ij}$, but
$(H_5/H_1)^\frac{1}{4} h_{ij}$ which behaves as a minimal scalar in
the 6-d space obtained by dimensional reduction on $T^4$.

The supergravity particle is described by its angular momenta in the
$S^3$ directions given by $SU(2)_L\times SU(2)_R$ quantum numbers
$(\frac{l}{2}, m), (\frac{l}{2},\bar{m})$; and a ``radial quantum
number,'' $N$, where $N=0$ gives the lowest energy state with the
given angular momentum, and $N=1, 2, \dots$ give successively higher
energy states.

Adding this quantum to the $r<r_b$ region of the geometry corresponds
to making an excitation of the D1D5 CFT. Since we compute all
processes after spectral flowing to the NS sector, we should describe
this excitation in the NS sector. In the NS sector, the excitation is
a supersymmetry descendant of a chiral primary state, which is acted
on $N$ times with $L_{-1}$ to further raise the energy. We describe
the construction of this initial state in more detail below.

The process of interest is the emission of this supergravity particle
from the cap out to infinity. The final state is thus simple: in the
 Ramond sector description we return to the Ramond
ground state that we started with. In the spectral flowed NS sector
description that we compute with, the final state is just the NS
vacuum $\vac_{NS}$.

The emission is caused by the interaction Lagrangian in
Equation~\eqref{eq:general-S-int} which couples excitations in the CFT
to modes at infinity; the general structure of this coupling was
discussed in Chapter~\ref{ch:coupling}. We write down the vertex
operator $\hat{\mathcal{V}}$ which leads to the emission of the quanta
$\phi_{ij}$, and compute the emission amplitude
$\bra{f}\hat{\mathcal{V}}\ket{i}$.
  
We now describe the initial state, the final state, and the vertex
operator in detail.
 
\subsubsection{The Initial State in the NS Sector}

Let us first write the state, and then explain its structure. The left
and right parts of the state have similar forms, so we only write the
left part (indicated by the subscript $L$):
\begin{equation}
\ket{\phi_{N+1}^{\frac{l}{2},\frac{l}{2}-k}}_L^{A\dot{A}} 
      = \mathcal{C}_L
        L_{-1}^N(J_0^-)^k G^{-A}_{-\frac{1}{2}}
	\psi_{-\frac{1}{2}}^{+\dot{A}}\sigma^0_{l+1}\vac_{NS},
\end{equation}
where the normalization constant $\mathcal{C}_L$ is determined in
Appendix~\ref{ap:norm}:
\begin{equation}
\mathcal{C}_L = 
\sqrt{\frac{(l-k)!(l-\bar{k})!}{N!\bar{N}!(N+l+1)!(\bar{N}+l+1)!k!\bar{k}!(l+1)^2}}.
\end{equation}
The normalization is chosen such that
\begin{equation}
\mathop{\vphantom{a}}_{A\dot{A}}^L\hspace{-3pt}
\braket{\phi_{N+1}^{\frac{l}{2},\frac{l}{2}-k}|\phi_{N+1}^{\frac{l}{2},\frac{l}{2}-k}}_L^{B\dot{B}} = \delta_A^B\delta_{\dot{A}}^{\dot{B}}.
\end{equation}

Let us describe the structure of this state starting with the elements on the rightmost end:
\begin{enumerate}
\item We start with the NS vacuum $\vac_{NS}$. In this state each copy
  of the CFT is ``singly wound,'' and each copy is unexcited. In the
  supergravity dual, we have global AdS space with no particles in it.

\item We apply the chiral primary $\sigma^0_{l+1}$, thereby twisting
  together $l+1$ copies of the CFT into one ``multiply wound''
  component string. It also adds charge, so that we get a state with
\begin{equation}
  h=m=\bar h=\bar m= \frac{l}{2}.  
\end{equation} 
In the gravity dual, we have one supergravity quantum, with angular
momenta $(m, \bar m)$.

\item We act with $\psi_{-\frac{1}{2}}^{+\dot{A}}$. This increases
  both $h$ and $m$ by $\frac{1}{2}$, and so gives another chiral
  primary. We do the same for the right movers, so that overall the
  new state created is again bosonic. In the supergravity dual, it
  corresponds to a different bosonic quantum in the AdS.

\item We act with elements of the ``anomaly-free subalgebra'' of the
  chiral algebra:
\begin{enumerate}
\item The $G^{-A}_{-\frac{1}{2}}$ changes the chiral primary to a
  supersymmetry descendant of the chiral primary, corresponding to a
  different supergravity particle in the gravity dual. Again, because
  we apply this supersymmetry operator on both left and right movers,
  the new supergravity quantum is bosonic. We now find that the
  indices carried by this quantum are those corresponding to a minimal
  scalar with both indices along the $T^4$ in the gravity description.
  Thus we have finally arrived at the supergravity quantum that we
  wanted to consider.

\item The $(J_0^-)^k$ rotate the quantum in the $S^3$ directions.
  Before this rotation, the quantum numbers $(m, \bar m)$ were the
  highest allowed for the given supergravity particle state. The
  application of the $(J_0^-)^k, (\bar{J}_0^-)^k$ give us other
  members of the $SU(2)_L\times SU(2)_R$ multiplet.

\item The $L_{-1}^N$ move and boost the quantum around in the $AdS$,
  thus increasing its energy and momentum.
\end{enumerate}

\item Finally, we have the normalization constant. We derive this in
  detail in Appendix~\ref{ap:norm}. The final expression for the
  radiation rate involves factors appearing in this normalization.
\end{enumerate}

In the gravity description it is natural to write the field as
$\phi_{ij}$, with vector indices $ij$ of the internal symmetry group
$SO(4)_I$ of the $T^4$ directions. For CFT computations it is more
useful to use indices $A\dot A$ for $SU(2)_1\times SU(2)_2$, as we do
above. The conversion is achieved by
\begin{equation}
\ket{\phi_{N+1}^{\frac{l}{2},\frac{l}{2}-k}}_L^i
= \frac{1}{\sqrt{2}}(\sigma^i)^{A\dot{A}}\epsilon_{AB}\epsilon_{\dot{A}\dot{B}}
\ket{\phi_{N+1}^{\frac{l}{2},\frac{l}{2}-k}}_L^{B\dot{B}}. 
\end{equation}
We then have
\begin{equation}
\mathop{\vphantom{E}}_L^i\hspace{-3pt}\braket{\phi_{N+1}^{\frac{l}{2},\frac{l}{2}-k}|\phi_{N+1}^{\frac{l}{2},\frac{l}{2}-k}}_L^j
  = \delta^{ij}.
\end{equation}

\begin{sloppypar}
Similarly, one typically labels the angular momentum eigenstates in
terms of $(l, m_\psi, m_\phi)$, instead of $(l, m, \bar{m})$. The two
bases are related via
\begin{equation}\begin{split}
m_\psi &= -(m + \bar{m})\\ 
m_\phi &= m - \bar{m},
\end{split}\end{equation}
where the values on the right-hand side are the angular momenta of the
initial state in the NS sector.
\end{sloppypar}
 
\subsubsection{The Final State}

In the supergravity description the initial state had one quantum in
it. The emission process of interest leads to the emission of this
quantum. Thus the final state has no quanta, and in the NS description
is just the vacuum
\begin{equation}
\ket{f}=\vac_{NS}.
\end{equation}

\subsubsection{The Vertex Operator}

We need the vertex operator that emits the supergravity quantum
described by the initial state $\ket{i}$. Vertex operators describing
supergravity particles are given by chiral primaries and their
descendants under the anomaly-free part of the chiral algebra.

For the process of interest the emission vertex must have appropriate
charges to couple to the supergravity field under consideration.
Thus, one naturally concludes that the vertex operator has essentially
the same structure as the state $\ket{i}$, with two differences.
First, the operator has charges that are opposite to the charges
carried by the state. (We get a nonvanishing inner product between
$\ket{i}$ and the Hermitian \emph{conjugate} of $\ket{i}$.) Second,
the operator does not have the $L_{-1}$ modes present in the
description of the CFT state. This is because applying an $L_{-1}$
mode is equivalent to translating the location of the vertex
insertion, and we have already chosen the insertion to be the point
$(\sigma, \tau)$. Note that after applying the supercurrent to give
the operator the correct $SO(4)_I$ index structure, one finds that the
operator already has the correct weight to couple to a minimal scalar
in Equation~\eqref{eq:general-S-int} and form a scale invariant
action.

The vertex operator, then, is given by (we drop the hat on the vertex
operator from now on)
\begin{equation}\label{eq:vertex-op}
\widetilde{\mathcal{V}}^{A\dot{A}B\dot{B}}_{l, l-k-\bar{k}, k-\bar{k}}(\sigma, \tau)
	= \sqrt{\frac{(l-k)!(l-\bar{k})!}{(l+1)^2(l+1)!^2\,k!\,\bar{k}!}}
	(J^+_0)^k(\bar{J}^+_0)^{\bar{k}}
	G^{+A}_{-\frac{1}{2}}\psi^{-\dot{A}}_{-\frac{1}{2}}
	\bar{G}^{\dot{+}B}_{-\frac{1}{2}}\bar{\psi}^{\dot{-}\dot{B}}_{-\frac{1}{2}}
	\tilde{\sigma}^0_{l+1}(\sigma, \tau).
\end{equation}
The subscript on the vertex operator $\mathcal{V}_{l,m_\psi, m_\phi}$
are the $SO(4)_E$ angular momenta labels. Again the normalization is a
crucial part of the final amplitude, so we perform it in more detail
below.

Note that for $l=0$, the vertex operator reduces to $[\pd
X]^{A\dot{A}}[\pdb X]^{B\dot{B}}$, the old ``effective string''
coupling found by expanding the DBI action~\cite{radiation-3}.

We map the operator from the cylinder onto the complex plane via
\begin{equation}\label{eq:map-to-the-plane}
z=e^{\tau + i\sigma}\qquad
\bar{z} = e^{\tau - i\sigma}.
\end{equation}
The vertex operator has weight $\frac{l}{2}+1$
on both the left and the right, so we get
\begin{equation}\begin{split}
\widetilde{\mathcal{V}}^{A\dot{A}B\dot{B}}_{l, l-k-\bar{k}, k-\bar{k}}(\sigma, \tau) 
&= |z|^{l+2}
	\sqrt{\frac{(l-k)!(l-\bar{k})!}{(l+1)^2(l+1)!^2\,k!\,\bar{k}!}}\\
&\hspace{1in}\times\left((J^+_0)^k(\bar{J}^+_0)^{\bar{k}}
	G^{+A}_{-\frac{1}{2}}\psi^{-\dot{A}}_{-\frac{1}{2}}
	\bar{G}^{\dot{+}B}_{-\frac{1}{2}}\bar{\psi}^{\dot{-}\dot{B}}_{-\frac{1}{2}}
	\tilde{\sigma}^0_{l+1}(z,\bar{z})\right)_{z,\bar{z}}\\
&= |z|^{l+2}
	\mathcal{V}^{A\dot{A}B\dot{B}}_{l, l-k-\bar{k},k-\bar{k}}(z, \bar{z}).
\end{split}\end{equation}

The normalization of the vertex operator in the complex plane
is chosen such that
\begin{equation}\begin{split}
\vev{\mathcal{V}^{A\dot{A}B\dot{B}}_{l,-m_\psi, -m_\phi}(z)
\mathcal{V}^{C\dot{C}D\dot{D}}_{l,m_\psi, m_\phi}(0)}
 &= \frac{\epsilon^{AC}\epsilon^{\dot{A}\dot{C}}\epsilon^{BD}\epsilon^{\dot{B}\dot{D}}}
	{|z|^{l+2}}\\
\vev{\mathcal{V}^{ij}_{l,-m_\psi, -m_\phi}(z)
	\mathcal{V}^{kl}_{l,m_\psi, m_\phi}(0)} 
	&= \frac{\delta^{ik}\delta^{jl}}{|z|^{l+2}}.
\end{split}\end{equation}
Note that this is the normalization of the operator corresponding to
one particular way of permuting $l+1$ copies of the CFT. As mentioned
earlier, the actual vertex operator coupling to $\phi_{ij}$ is a
symmetrized sum over all possible ways of permuting $l+1$ copies from
the $N_1N_5$ available copies. We discuss the combinatorics of this
choice in Section~\ref{sec:combinatorics} below, and at that time note
the extra normalization factor which is needed to agree
with~(\ref{first}).

\subsection{Using Spectral Flow}\label{sec:use-spectral}

We wish to relate a CFT amplitude computed in the NS sector,
\begin{equation}
\mathcal{A}' = \bra{f'}\mathcal{V}(z, \bar{z})\ket{i'},
\end{equation}
to an amplitude in the Ramond sector, since the physical D1D5 system
has its fermions periodic around the $y$ circle.  In this section, we
show how to spectral flow~\cite{spectral,spectral-yu,vafa-warner} the
computation in the NS sector to the physical problem in the R sector.
Furthermore, we find that we can relate this NS sector computation to
a whole family of Ramond sector amplitudes, and each member of the
family describes a different physical emission process.

If spectral flowing the states $\ket{i'}$ and $\ket{f'}$ by $\alpha$
units is given by
\begin{equation}
\ket{i'}\mapsto \ket{i} = \mathcal{U}_\alpha\ket{i'}\qquad
\bra{f'}\mapsto \bra{f} = \bra{f'}\mathcal{U}_{-\alpha},
\end{equation}
then we can compute the amplitude in the Ramond sector by using
\begin{equation}
\mathcal{A}_\text{Ramond} = \bra{f} \mathcal{V}(z,\bar{z})\ket{i}
 			  = \big(\bra{f}\mathcal{U}_{\alpha}\big)
			     \big(\mathcal{U}_{-\alpha}\mathcal{V}\mathcal{U}_{\alpha}\big)
			     \big(\mathcal{U}_{-\alpha}\ket{i}\big)
			  = \bra{f'}\mathcal{V}'(z, \bar{z})\ket{i'}.
\end{equation}
Note that one finds $\mathcal{V}'$ by spectral flowing $\mathcal{V}$
by $-\alpha$ units.

We need to determine how the vertex operator transforms under
spectral flow. First, we demonstrate that the $G\psi$ part is
unaffected, since
\begin{calc}
\left(G^{+A}_{-\frac{1}{2}}\psi^{-\dot{A}}_{-\frac{1}{2}}\right)_z
	&= \oint_z\frac{\drm z_1}{2\pi i}\oint_z\frac{\drm z_2}{2\pi i} 
		\frac{G^{+A}(z_1)\psi^{-\dot{A}}(z_2)}{z_2-z}\\
	&= -\oint_z\frac{\drm z_1}{2\pi i}\oint_z\frac{\drm z_2}{2\pi i} 
		\frac{[\pd X(z_2)]^{\dot{A}A}}{(z_2-z)(z_1-z_2)}
\end{calc}
and the bosons are unaffected by spectral flow.

Therefore, we need only spectral flow the $k$ $J_0^+$'s and the chiral
primary. The effect of spectral flow by \emph{negative} $\alpha$ units
on chiral ($h=m$) and anti-chiral primaries ($h=-m$) is very simple:
\begin{equation}
\mathcal{O}'_\text{c.p.}(z) = z^{\alpha m}\mathcal{O}_\text{c.p.}(z)\qquad
\mathcal{O}'_\text{a.c.p.}(z) = z^{\alpha m}\mathcal{O}_\text{a.c.p.}(z).
\end{equation}
One can see this most directly after bosonizing the fermions; see 
Appendix~\ref{sec:spectral-flow} for details.

Under spectral flow by $-\alpha$ units the $J^\pm$ transform as
\begin{equation}
J^\pm(z) \mapsto z^{\pm\alpha}J^\pm(z),
\end{equation}
from which we see that
\begin{equation}
(J^+_0)_z = \oint_z\frac{\drm z'}{2\pi i} J^+(z')
\mapsto \oint_z \frac{\drm z'}{2\pi i} J^+(z') z'^{\alpha}
         = z^{\alpha}(J^+_0)_z +\alpha z^{\alpha-1}(J^+_1)_z + \dots.
\end{equation}
Only the first term contributes since the positive modes annihilate a
chiral primary. Therefore, we conclude that spectral flowing the
vertex operator by $-\alpha$ units has the effect of
\begin{equation}
\mathcal{V}'_{l,l-k-\bar{k}, k-\bar{k}}(z,\bar{z}) = z^{-\alpha(\frac{l}{2}-k)}
	\bar{z}^{-\bar{\alpha}(\frac{l}{2}-\bar{k})}
		\mathcal{V}_{l,l-k-\bar{k}, k-\bar{k}}(z,\bar{z}).
\end{equation}

Thus we observe that we can spectral flow the initial and final
states, keep the vertex operator unchanged, and compute the amplitude
\begin{equation} 
\mathcal{A}'=\bra{f'}\mathcal{V}(z,\bar{z})\ket{i'}.
\end{equation} 
The result we want, $\mathcal{A}_\text{Ramond}$, is then given by
\begin{equation}\label{eq:unspectral-flowed-V}
\mathcal{A}_\text{Ramond} = z^{-\alpha(\frac{l}{2}-k)}\bar{z}^{-\bar{\alpha}(\frac{l}{2}-\bar{k})}
	\bra{f'}\mathcal{V}(z,\bar{z})\ket{i'}
	= z^{-\alpha(\frac{l}{2}-k)}\bar{z}^{-\bar{\alpha}(\frac{l}{2}-\bar{k})}
		\mathcal{A}'.
\end{equation}
Here $\alpha$ is chosen to have a value that spectral flows from the
NS to the Ramond sector, but this can be achieved by \emph{any} odd
integral value of $\alpha$:
\begin{equation}
\alpha = (2n+1)\qquad \bar{\alpha} = (2\bar{n} + 1) \qquad n,\bar{n}\in\ints.
\end{equation}
For these values of $\alpha$ the initial and final states have weight and charge
\begin{equation}\begin{aligned}
h &= h' + (2n+1) m' + (2n+1)^2\frac{c_\text{tot.}}{24}\\
m &= m' + (2n+1)\frac{c_\text{tot.}}{12},
\end{aligned}\end{equation}
where $c_\text{tot.}$ is $c=6$ times the number of copies being
spectral flowed. A similar relation holds on the right sector.

In our present computation in the NS sector, we have
\begin{equation}\begin{aligned}
h'_i &= \frac{l}{2} + N + 1 \hspace{20pt}& h'_f &= 0\\
m'_i &= \frac{l}{2} - k                  & m'_f &= 0.
\end{aligned}\end{equation}
In the next section we look at the Ramond sector process for
$\alpha=\bar\alpha=1$. In this case the weights and charges of the
Ramond sector states are
\begin{equation}\begin{aligned}
h_i &= \frac{l}{2} + N + 1  + \left(\frac{l}{2} - k\right) + \frac{l+1}{4}
                     \hspace{20pt}& h_f &= (l+1)\frac{1}{4}\\
m_i &= \frac{l}{2} - k + \frac{l+1}{2}     & m_f &= (l+1)\frac{1}{2}.
\end{aligned}\end{equation}
We see that the final state has the weight and charge of the
``spin-up'' Ramond vacuum, while the initial state has the correct
weight and charge above the Ramond vacuum. Although the current
calculation is $\alpha = 1$, we leave $\alpha$ as an explicit
parameter in following calculations for later use and illustration.

In section~\ref{sec:nonextremal}, the full $\alpha$ and $\bar{\alpha}$
dependence is of physical interest, since how big $\alpha$ and
$\bar{\alpha}$ are roughly corresponds to how nonextremal the initial
state is.

\subsection{Evaluating the CFT Amplitude}\label{sec:CFT-evaluation}

Let us now compute the amplitude
\begin{equation}
{\mathcal{A}'}^{A\dot{A}}(\sigma, \tau)=\bra{f'}\widetilde{\mathcal{V}}(\sigma, \tau)\ket{i'}
                                = |z|^{l+2}\bra{f'}\mathcal{V}(z, \bar{z})\ket{i'}.
\end{equation}
We choose the charges of the initial state and the vertex operator so
that we get a nonvanishing amplitude. The nonvanishing amplitude is
\begin{equation}\begin{split}
{\mathcal{A}'}_L^{A\dot{A}} &= \frac{1}{\sqrt{2}}(\sigma^\ibar)_{B\dot{B}}\,
        z^{\frac{l}{2}+1}\mathop{\vphantom{a}}_{NS}\hspace{-3pt}\bvac
        \mathcal{V}^{A\dot{A}}_{L;l,k}(z)
\ket{\phi_{N+1}^{\frac{l}{2},\frac{l}{2}-k}}_L^{B\dot{B}}
\end{split}\end{equation}
where $A,\dot{A}$ and $\ibar$ are free indices. The $\ibar$ is the
index of the initial state excitation and $A,\dot{A}$ are the indices
on the vertex operator. 

Let us note that
\begin{equation}
\ket{\phi_{N+1}^{\frac{l}{2},\frac{l}{2}-k}}_L^{B\dot{B}}
 = (-1)^k
  \sqrt{\frac{(l+1)!}{N!(N+l+1)!}} 
  L_{-1}^N\mathcal{V}^{B\dot{B}}_{L;l, 2l-k}(0)\vac_{NS}
\end{equation}
and that the action of $L_{-1}$ on a primary
field $\mathcal{O}$ is
\begin{equation}
L_{-1}\mathcal{O}(0) = \oint\frac{\drm z}{2\pi i}T(z)\mathcal{O}(0)
  = \pd\mathcal{O}(0).
\end{equation}
Therefore, we may write
\begin{calc}
\vev{\mathcal{V}^{A\dot{A}}_{L;l,k}(z)\, 
     L_{-1}^N\mathcal{V}^{B\dot{B}}_{L;l, 2l-k}}
     = \epsilon^{AB}\epsilon^{\dot{A}\dot{B}}\lim_{v\to 0}
          \pd_v^N\frac{1}{(z-v)^{l+2}}
     = \frac{(N+l+1)!}{(l+1)!}\frac{1}{z^{\frac{l}{2} + N + 1}}
\end{calc}
and one finds the left amplitude reduces to the simple form
\begin{equation}\label{eq:NS-untwisting-amp}
{\mathcal{A}'}_L^{A\dot{A}} = (-1)^k\frac{1}{\sqrt{2}}(\sigma^\ibar)^{A\dot{A}}
   \frac{1}{z^{\frac{l}{2} + N+1}}
   \sqrt{\choose{N+l+1}{N}}.
\end{equation}

From Equation~\eqref{eq:unspectral-flowed-V}, we find that the left
part of the CFT amplitude in the Ramond sector is given by
\begin{calc}
\mathcal{A}_L^{A\dot{A}}  &= z^{-\alpha(\frac{l}{2} - k)}\mathcal{A}_L'\\
 &= (-1)^{k}\frac{1}{\sqrt{2}}(\sigma^\ibar)^{A\dot{A}}
    \frac{1}{z^{(1+\alpha)\frac{l}{2} -\alpha k +  N+1}}
    \sqrt{\choose{N+l+1}{N}}.
\end{calc}
Finally, converting back to $SO(4)$ indices for the vertex operator,
one gets in the Ramond sector
\begin{calc}\label{eq:final-L-amp}
\mathcal{A}_L^\ibar(z) &= \frac{1}{\sqrt{2}}
   (\sigma^\ibar)_{A\dot{A}}\mathcal{A}_L^{A\dot{A}}\\
   &= (-1)^{k+1}\frac{1}{z^{(1+\alpha)\frac{l}{2} -\alpha k +  N+1}}
    \sqrt{\choose{N+l+1}{N}},
\end{calc}
The free index $\ibar$ and a similar index from the right movers
$\jbar$ correspond to the indices $\phi_{ij}$ for the field coupling
to the emission vertex.

\subsection{Combinatorics}
\label{sec:combinatorics}

The full CFT has $N_1N_5$ copies of the basic $c=6$ CFT. In the above
section, we took a set of $l+1$ copies twisted together, and look at
an emission process where the emission vertex untwists these copies.
Now, we must put this computation in its full CFT context, by doing the
following:
\begin{enumerate}
\item We must compute the combinatorics of how we pick the particular
  way of twisting $l+1$ copies in the initial state from all $N_1N_5$
  copies.

\item We must similarly consider all the ways that the vertex operator
  can twist copies. This allows us to normalize the vertex operator in
  the full theory so that we reproduce (\ref{first}).

\item We can take the limit $N_1N_5\rightarrow\infty$ to get the
  ``classical limit'' of the D1D5 system; the result in this limit
  should agree with the computation in the dual supergravity theory.
\end{enumerate}

In fact we start with something a little more general. We assume that
the initial state has $\nu$ quanta of the same kind, and let the
emission process lead to the final state with $\nu-1$ quanta.  We then
observe a Bose enhancement of the emission amplitude by a factor
$\sqrt{\nu}$, which agrees with the enhancement observed in both CFT
and dual gravity computations in~\cite{cm1}.

\subsubsection{The Initial State}

We wish to have $\nu$ excitations, each of which involve twisting
together $l+1$ copies of the $c=6$ CFT. We can pick the needed copies
in any way from the full set of $N_1N_5$ copies, and because of the
orbifold symmetry between these copies the state must be a symmetrized
sum over these possibilities:
\begin{equation}\label{eq:comb-initial-state}
\ket{\Psi_\nu} = \mathcal{C}_\nu\bigg[\ket{\psi_\nu^1} + \ket{\psi_\nu^2} + \dots\bigg],
\end{equation}
where $\mathcal{C}_\nu$ is the overall normalization and each
$\ket{\psi^i_\nu}$ is individually normalized. To understand what we are
doing better, note that the state $\ket{\psi_\nu^1}$ can be written
schematically as
\begin{equation}
\ket{\psi_\nu^1} = \Ket{
	\big[12\cdots(l+1)\big]
	\big[(l+2)\cdots 2(l+1)\big]\cdots
	\big[\big(\nu(l+1)- l\big)\cdots\nu(l+1)\big]},
\end{equation}
where the numbers in the square brackets are indicating particular ways
of twisting individual strands corresponding to particular cycles of
the permutation group. For instance,
\begin{equation}
\ket{[1234]},
\end{equation}
indicates that we twist strand 1 into strand 2 into strand 3 into
strand 4 into strand 1 and leave strands 5 through $N_1N_5$ untwisted.

Our first task is to determine the number of terms in
Equation~\eqref{eq:comb-initial-state} and thereby its normalization
$\mathcal{C}_\nu$. To count the number of states we imagine
constructing one of these states and see how many choices we have
along the way. First, we choose $\nu(l+1)$ of the total $N_1N_5$
strands that are going to be twisted in some way. The remaining
strands are untwisted. Those $\nu(l+1)$ strands must now be broken
into sets of $l+1$. To do this, we first choose $l+1$ of the
$\nu(l+1)$, then the next set of $l+1$ from the remaining
$(\nu-1)(l+1)$, and so on. Note that $\ket{[12][34]}=\ket{[34][12]}$,
and so there is no sense in talking about the ``first'' set versus the
``second set.'' Therefore we should divide by the number of ways to
rearrange the $\nu$ sets of $l+1$. Finally, we should choose a
particular cycle for each set of $(l+1)$; since it does not matter
where we start on the final cycle, this gives a factor $l!$ for each
twisted cycle. Putting all of these factors together yields the number
of terms in Equation~\eqref{eq:comb-initial-state},
\begin{calc}
N_\text{terms}&= \choose{N_1N_5}{\nu(l+1)}\times
	\choose{\nu(l+1)}{l+1}\choose{(\nu-1)(l+1)}{l+1}\cdots
	\choose{l+1}{l+1}\times\frac{1}{\nu!}\times (l!)^\nu\\
	&= \frac{(N_1N_5)!}{(l+1)^\nu\,\nu![N_1N_5 - \nu(l+1)]!}.
\end{calc}
Without loss of generality, let us choose $\mathcal{C}_\nu$ to be
real, which gives
\begin{equation}
\mathcal{C}_\nu=\left[\frac{(N_1N_5)!}{(l+1)^\nu\,\nu![N_1N_5 - \nu(l+1)]!}\right]^{-\frac{1}{2}}.
\end{equation}

\subsubsection{The Final State}

The final state is simply $\ket{\Psi_{\nu-1}}$, with its corresponding
normalization $\mathcal{C}_{\nu-1}$.

\subsubsection{The Vertex Operator}

The vertex operator can twist together any $l+1$ copies of the CFT
with any $l+1$-cycle, and it should be written as a symmetrized sum
over these possibilities:
\begin{equation}
\mathcal{V}_\text{sym} = \mathcal{C}\sum_i\mathcal{V}_i.
\label{checktwo}
\end{equation}
Since the joined copies form a single long loop, the order of copies
matters but not which copy is the ``first one'' in the loop. Thus the
number of terms in the sum is
\begin{equation}
\choose{N_1N_5}{l+1}l! = \frac{(N_1N_5)!}{(l+1)[N_1N_5 - (l+1)]!}.
\end{equation}
This gives the normalization
\begin{equation}
\mathcal{C}=\left[\frac{(N_1N_5)!}{(l+1)[N_1N_5 - (l+1)]!}\right]^{-\frac{1}{2}}.
\end{equation}

\subsubsection{The Amplitude}

To compute the amplitude
\begin{equation}\label{eq:comb-amp}
\bra{\Psi_{\nu-1}}\mathcal{V}_\text{sym}\ket{\Psi_\nu},
\end{equation}
we have to count the different ways that terms in the initial state
can combine with terms in the vertex operator and terms in the final
state to produce a nonzero amplitude. For a given initial state term
$\ket{\psi_\nu^i}$, there are exactly $\nu$ vertex operators
$\mathcal{V}_i$ that can de-excite it into a final state. There is
only one final state that works, obviously. Thus the number of ways
that we can get a nonzero amplitude is simply
\[
\nu N_\text{terms} = \frac{\nu}{\mathcal{C}_\nu^2}.
\]
Let 
\begin{equation}
\bra{\psi_{\nu-1}^1}\mathcal{V}_1\ket{\psi_\nu^1}
\end{equation}
be the amplitude obtained by using only one allowed initial state $
\ket{\psi_\nu^1}$ from the set in Equation~\eqref{eq:comb-initial-state} and
one allowed vertex operator $\mathcal{V}_1$ from the set in
Equation~\eqref{checktwo}. Then we have
\begin{calc}
\bra{\Psi_{\nu-1}}\mathcal{V}_\text{sym.}\ket{\Psi_\nu}
	&= \mathcal{C}\mathcal{C}_\nu\mathcal{C}_{\nu-1}\cdot 
        \frac{\nu}{\mathcal{C}_\nu^2}
	   \bra{\psi_{\nu-1}^1}\mathcal{V}_1\ket{\psi_\nu^1}\\
	&= \sqrt{\nu}
        \sqrt{\frac{[N_1N_5-(\nu-1)(l+1)]![N_1N_5-(l+1)]!}
          {[N_1N_5-\nu(l+1)]!(N_1N_5)!}}
		\bra{\psi_{\nu-1}^1}\mathcal{V}_1\ket{\psi_\nu^1}.
\end{calc}

\subsubsection{The Large \texorpdfstring{$N_1N_5$}{N1N5} Limit}

We are ultimately interested in the limit of large $N_1N_5$. Then we have
\begin{equation}
\frac{[N_1N_5-(\nu-1)(l+1)]!}{[N_1N_5-\nu(l+1)]!}
	\longrightarrow
	(N_1N_5)^{l+1}
\hspace{20pt}
\frac{[N_1N_5-(l+1)]!}{(N_1N_5)!}
	\longrightarrow
	(N_1N_5)^{-(l+1)},
\end{equation}
which gives
\begin{equation}
\bra{\Psi_{\nu-1}}\mathcal{V}_\text{sym.}\ket{\Psi_\nu}
	\longrightarrow \sqrt{\nu}
	\bra{\psi_{\nu-1}^1}\mathcal{V}_1\ket{\psi_\nu^1}.
	\label{finalq}
\end{equation}
The prefactor $\sqrt{\nu}$ gives a ``Bose enhancement'' effect which
tells us that if we start with $\nu$ quanta, the amplitude to emit
another quantum is amplified by a factor $\sqrt{\nu}$ (compared to the
case when there was only one quantum). This gives an enhancement $\nu$
in the probability, which just tells us that if we start with $\nu$
quanta in the initial state, then the rate of emission is proportional
to $\nu$.

\subsection{The Rate of Emission}\label{sec:calc-rate-emission}

We now put together all the computations of the above sections to get
the emission rate for a quantum from the excited CFT state. We need to
do the following:
\begin{enumerate}
\item We use (\ref{finalq}) to relate the decay amplitude for one
  $(l+1)$-permutation to the amplitude with all the required
  symmetrizations put in
\begin{equation}
\bra{\Psi_{0}}\mathcal{V}_\text{sym.}\ket{\Psi_1}
	= \sqrt{\nu}\bra{\psi_{0}^1}\mathcal{V}_1\ket{\psi_1^1}.
\end{equation}

\item From Equation~\eqref{eq:final-L-amp}, we have 
  the decay amplitude for a given $l+1$-permutation (we put the
  right sector back in now):
\begin{calc}\label{eq:final-euc-amp}
\bra{\psi_{0}^1}\mathcal{V}_1\ket{\psi_1^1} &= \mathcal{A}^{\ibar\jbar}(z,\bar{z})\\ 
  &= (-1)^{k+\bar{k}}\sqrt{\choose{N+l+1}{N}\choose{\bar{N} + l+1}{\bar{N}}}
	 z^{-(\alpha +1)\frac{l}{2} +\alpha k - N - 1}
	\bar{z}^{-(\bar{\alpha} +1)\frac{l}{2} +\bar{\alpha} \bar{k} - \bar{N} - 1}.\\
\end{calc}
We now rotating back to Lorentzian signature and replacing $\tau,
\sigma$ by the physical $(t,y)$ coordinates. Note that we are still
working with a CFT with spatial section of ``unit size'' where the
spatial circle has length $(2\pi)$. The ``unit-sized'' amplitude is
thus
\begin{equation}\begin{split}\label{eq:physical-CFT-amp}
\mathcal{A}_\text{unit}^{\ibar\jbar}(t,y) &= (-1)^{k+\bar{k}}
	\sqrt{\choose{N+l+1}{N}\choose{\bar{N} + l+1}{\bar{N}}}\\
	 &\hspace{50pt} \times 
	e^{\frac{i}{R}\left(-(\alpha +1)\frac{l}{2} +\alpha k - N - 1\right)(y+t)}
	e^{-\frac{i}{R}\left(-(\bar{\alpha} +1)\frac{l}{2} +\bar{\alpha} \bar{k} 
		- \bar{N} - 1\right)(y-t)}.
\end{split}\end{equation}

\item From Equation~\eqref{eq:physical-CFT-amp} or by comparing the initial
and final states, we can read off
\begin{equation}\begin{split}
E_0 &= \frac{1}{R} \left[(\alpha + \bar{\alpha} + 2)\tfrac{l}{2} - \alpha k - \bar{\alpha}\bar{k}
	+ N + \bar{N} + 2\right]
	= \frac{1}{R}\left[2l - k - \bar{k} + N + \bar{N} + 2\right]\\
\lambda_0 &= \frac{1}{R}\left[-(\alpha -\bar{\alpha})\tfrac{l}{2} + \alpha k - \bar{\alpha}\bar{k}
	- N + \bar{N}\right]
	= \frac{1}{R}\left[k - \bar{k} - N + \bar{N}\right],
\end{split}\end{equation}
where we have set $\alpha = \bar{\alpha} = 1$ for the physical process
of interest. We also can determine the ``unit-sized'' amplitude with
the position dependence removed,
\begin{equation}
\bra{f}\mathcal{V}\ket{i}_\text{unit} = 
	\mathcal{A}^{\ibar\jbar}_\text{unit}(0,0) = 
	(-1)^{k+1}\sqrt{\nu}\sqrt{\choose{N+l+1}{N}\choose{\bar{N} + l+1}{\bar{N}}}.
\end{equation}
\end{enumerate}
Putting this into Equation~\eqref{eq:D1D5-decay-rate}, one finds the final
emission rate 
\begin{equation}
\der{\Gamma}{E} = \nu\frac{2\pi}{2^{2l+1}l!^2}\frac{(Q_1Q_5)^{l+1}}{R^{2l+3}}(E^2-\lambda^2)^{l+1}
	\choose{N+l+1}{N}\choose{\bar{N} + l+1}{\bar{N}}\delta_{\lambda, \lambda_0}\delta(E-E_0).
\end{equation}
This is the emission rate for one of $\nu$ excitations in the CFT to
de-excite and emit a supergravity particle with energy $E_0$,
$S^1$-momentum $\lambda_0$, and angular momentum
\begin{equation}\begin{split}\label{eq:emitted-ang-mom}
m_\psi &= -(m + \bar{m}) = -l + k + \bar{k}\\
m_\phi &= m - \bar{m} = \bar{k} - k.
\end{split}\end{equation}
The angular momentum can be read off from the angular momentum of the
NS sector initial state, or the difference in angular momentum between the
initial and final physical states.

The expression for the emission rate obtained above matches the one
obtained in \cite{gms2} where it is given in a slightly different
form. There the expression for a minimally coupled scalar to be
absorbed into the geometry and re-emerge is given in Equation~(5.34)
along with the time of travel in Equation~(5.33). The total
probability is the product of the probabilities to be absorbed and to
re-emerge, which are equal. Therefore, the rate of emission is the
square root of the total probability, with the energy and other
quantum numbers taking the corresponding values for excitations in the
background, divided by the time of travel. This expression is seen to
match the emission rate obtained above.

\subsection{Emission from Nonextremal Microstates}\label{sec:nonextremal}

From a physics point of view the emission computed above corresponds
to a very simple process. We take an extremal 2-charge D1D5
microstate, excite it by adding a quantum, and compute the rate at
which the state de-excites by emitting this quantum.

But this same computation can be slightly modified to obtain the
emission rate for a more interesting physical process. We start with a
nonextremal D1D5 microstate which has a large energy above
extremality. This particular microstate is obtained by taking an
extremal D1D5 microstate and performing a spectral flow on both the
left and right moving sectors. Such a spectral flow adds fermionic
excitations to \emph{every} component string. Thus we get a large
energy above extremality, not just the energy of one nonextremal
quantum as was the case with our earlier
computations~\cite{gms1,gms2,ross}.

This nonextremal state emits radiation, and we wish to compute the
rate of emission after $\nu$ quanta have been emitted. We again get
the ``Bose enhancement'' like that in (\ref{finalq}), so that the rate
of emission keeps increasing as more quanta are emitted. In~\cite{cm1}
it was shown that the resulting decay behavior is exactly the Hawking
radiation expected from this particular microstate. But the
computation of~\cite{cm1} was restricted to certain choices of spins
and excitation level $N=0$ for the emitted quantum; now we are able to
get a general expression for all values of spins and $N$.

\subsubsection{The CFT Process}

\begin{figure}[ht]
\begin{center}
\subfigure[]{\label{fig:spectral-a}
	\includegraphics[width=6cm]{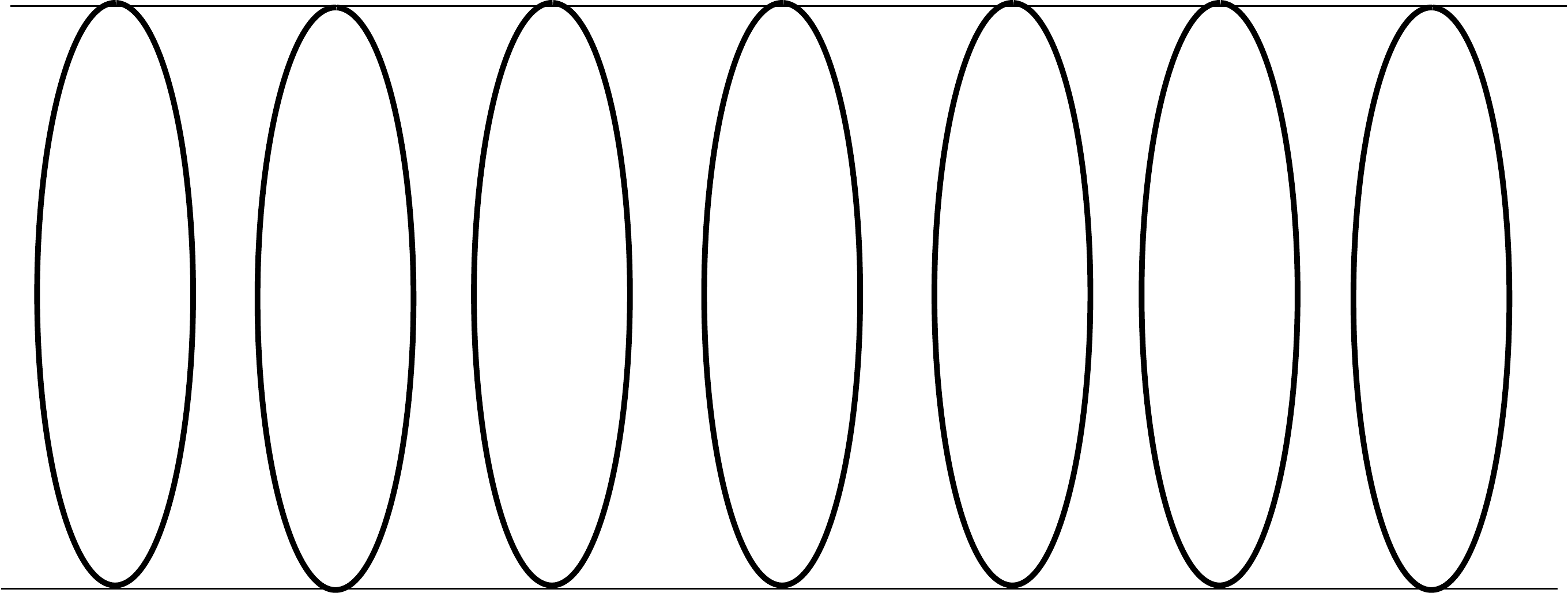}}
\raisebox{31pt}{$\xrightarrow{{\displaystyle \text{spectral flow}}}$}
\subfigure[]{\label{fig:spectral-b}
	\includegraphics[width=6cm]{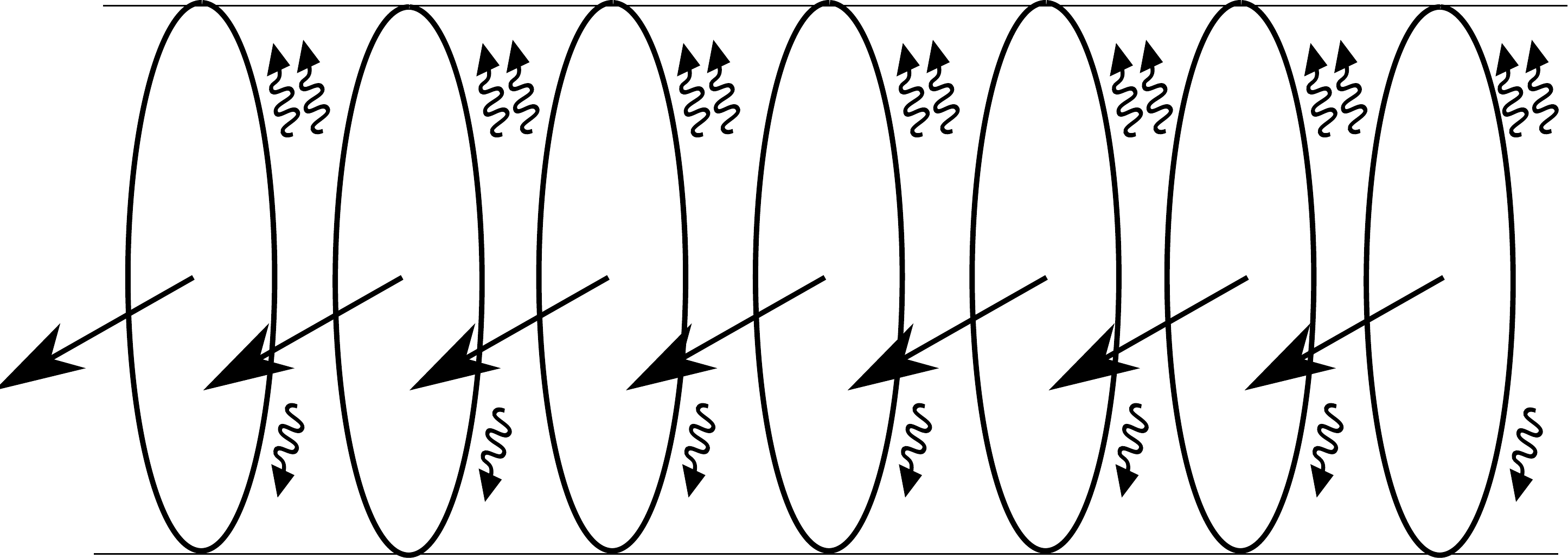}} 
      \caption[The effect of spectral flow on the NS vacuum]{(a) The
        NS vacuum state in the CFT and (b) the CFT state after
        spectral flow.  The arrows at the center of the circle
        indicate the ``base spin'' of component strings in the Ramond
        sector. The wavy arrows on top (bottom) of the strands
        represent fermionic excitations in the left (right)
        sector.\label{fig:SpectralFlow}}
\end{center}
\end{figure}

As discussed earlier, the physical D1D5 system is in the Ramond
sector. We can relate Ramond sector states to NS sector states by
spectral flow. Recall that under spectral flow the dimensions and
charges change as follows:
\begin{equation}\begin{split}
h' &= h + \alpha m+ \frac{c\alpha^2}{24}\\
m' &= m + \frac{c\alpha}{12}.
\end{split}\end{equation}
If we start with the NS vacuum $\vac_{NS}$ and spectra flow by
$\alpha=1$, we reach the Ramond vacuum state with $h=\frac{c}{24}$.
But we can also reach a Ramond state by spectral flow by $\alpha=3, 5,
\dots$, which are excited states with energy more than the energy of
the Ramond vacua. Let us take our initial state in the Ramond sector
to be the state obtained by spectral flow of $\vac_{NS}$ by
$\alpha=2n+1$ on the left and $\bar{\alpha} = 2\bar{n} + 1$ on the
right. The spectral flow adds fermions to the left and right sectors,
raising the level of the Fermi sea on both these sectors. Thus we get
an excited state of the D1D5 system, which we depict in
Fig.~\ref{fig:spectral-b}.

The vertex operator we have constructed can twist together $l+1$
copies of the CFT. In our earlier computation, we started with a set
of twisted copies, and the vertex operator ``untwisted'' these,
leading to a final state with no twists. This time the initial state
has all copies of the CFT ``untwisted,'' but these copies are all in
an excited state. The vertex operator can therefore twist together
$l+1$ copies, leading to a twisted component string in the final
state. Even though twisting a set of strings increases the energy,
this component string in the final state can have lower energy than
the strings in the initial state because of the fermionic excitations
present on the initial component strings. The energy difference
between the initial and final states is the energy of the emitted
supergravity particle.

Let us now set up the CFT computation needed for this process. We
observe that the amplitude can be obtained in a simple way from the
amplitude that we have already computed.

\subsubsection{The Initial State}

As before, we do all our computations in the NS sector. If we spectral
flow the starting state depicted in Fig.~\ref{fig:spectral-b} by
$-(2n+1)$ units, we arrive at the NS vacuum $\vac_{NS}$ depicted in
Fig.~\ref{fig:spectral-a}. It may appear that this vacuum state cannot
lead to any emission, but recall that we have used spectral flow only
as a technical trick; the actual initial state has a much higher
energy, and indeed leads to emission.

\begin{figure}[ht]
\begin{center}
\subfigure[~The initial state in the NS sector]{\label{fig:RossNSInitial}
	\includegraphics[width=7cm]{figures/NSNuEq1}}
\hspace{15pt}
\subfigure[~The final state in the NS sector]{\label{fig:RossNSFinal}
	\includegraphics[width=7cm]{figures/NSNuEq2}} \\
\subfigure[~The initial state in the Ramond sector]{\label{fig:RossRRInitial}
	\includegraphics[width=7cm]{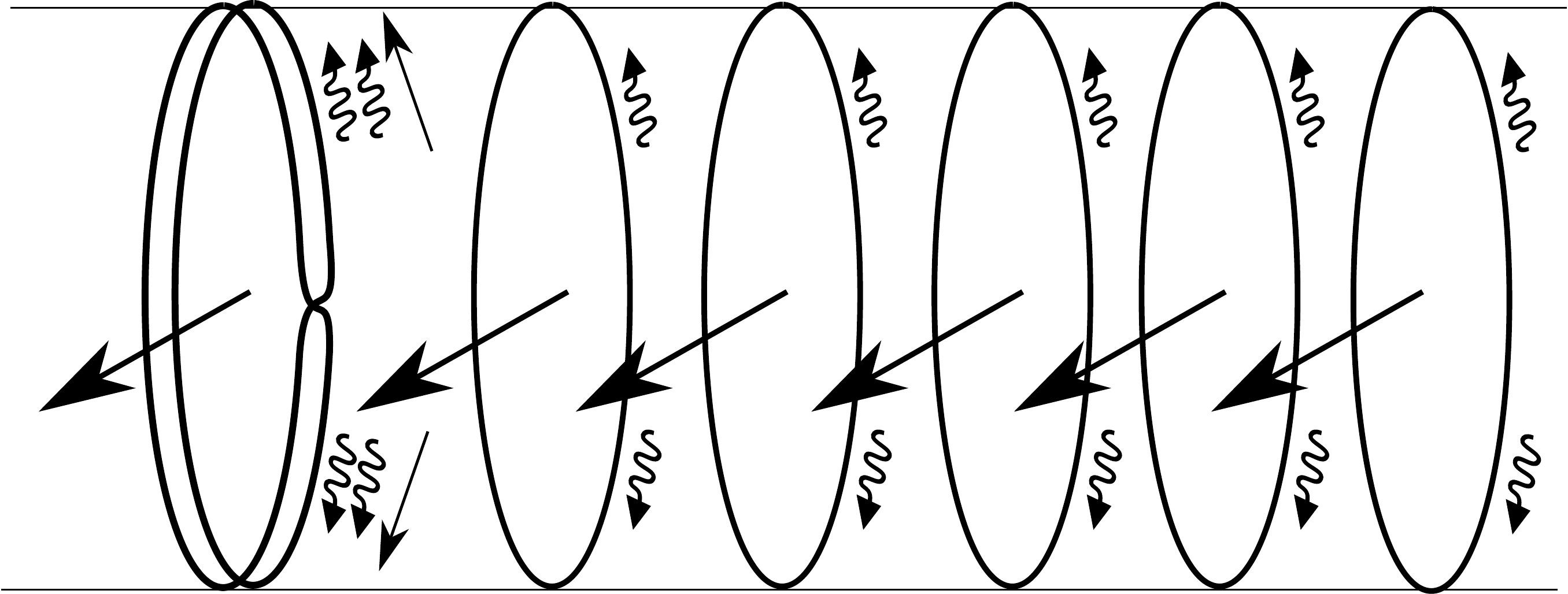}}
\hspace{15pt}
\subfigure[~The final state in the Ramond sector]{\label{fig:RossRRFinal}
  \includegraphics[width=7cm]{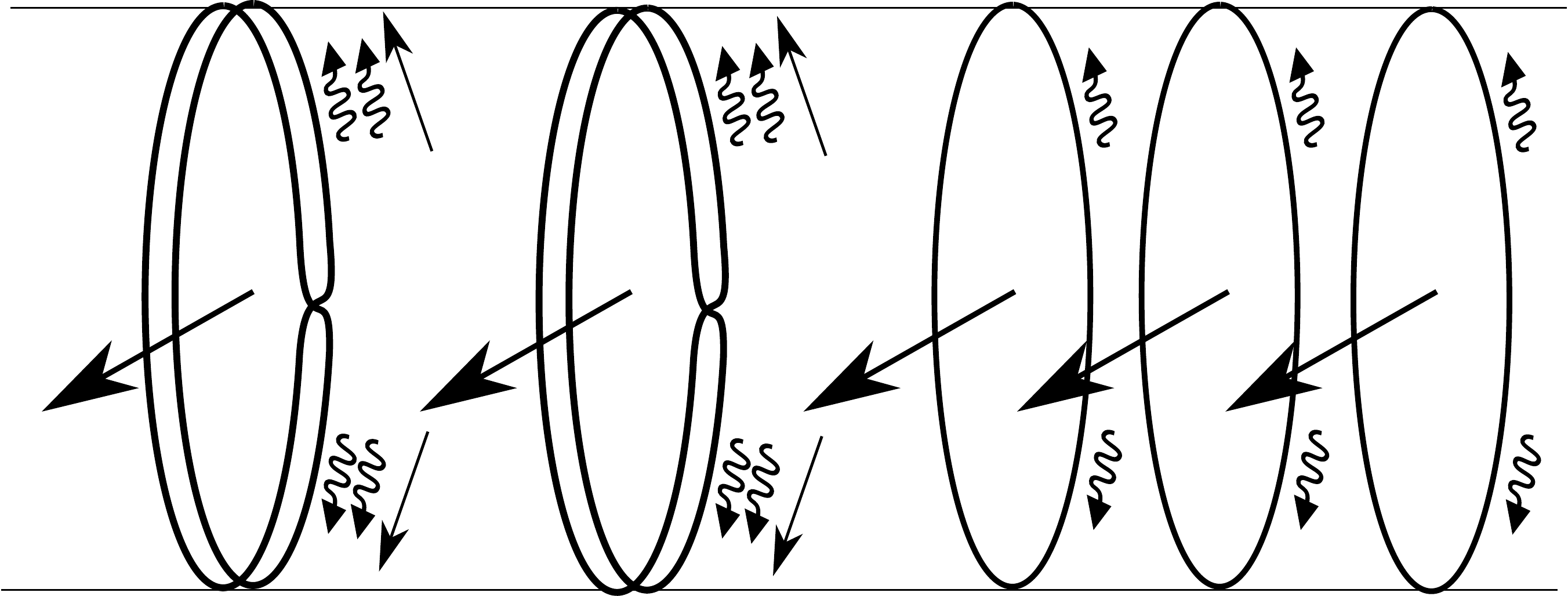}}
\caption[The initial and final states for nonextremal emission]{The
  initial and final states for the nonextremal emission process
  discussed in Section~\ref{sec:nonextremal}.  The specific case
  depicted is $\nu=2$ and $l=1$.\label{fig:RossDecay}}
\end{center}
\end{figure}

If we wanted to start with this state and proceed with the computation
we would set $\ket{i'}=\vac_{NS}$. But we instead look at a slightly
more general situation where $\nu-1$ quanta have already been emitted.
In this case, the initial state looks like the one depicted in
Fig.~\ref{fig:RossRRInitial}, where $\nu-1$ sets of $l+1$ copies have
already been twisted together.

This may look like a complicated initial state, but we look only at a
specific amplitude: the amplitude for emission of a further quantum of
the same kind as the quanta already present. This process therefore
requires us to take $l+1$ of the \emph{untwisted} copies of the CFT,
and use the vertex operator to twist them together. The other copies
of the CFT are unaffected by the vertex operator. Thus, for the
purposes of computing the amplitude, the initial state of the $l+1$
copies of interest is
\begin{equation}
\ket{i'}=\vac_{NS}.
\end{equation}

\begin{figure}[ht]
\begin{center}
\subfigure[~The initial state in the NS sector]{\label{fig:alphaNSInitial}
	\includegraphics[width=7cm]{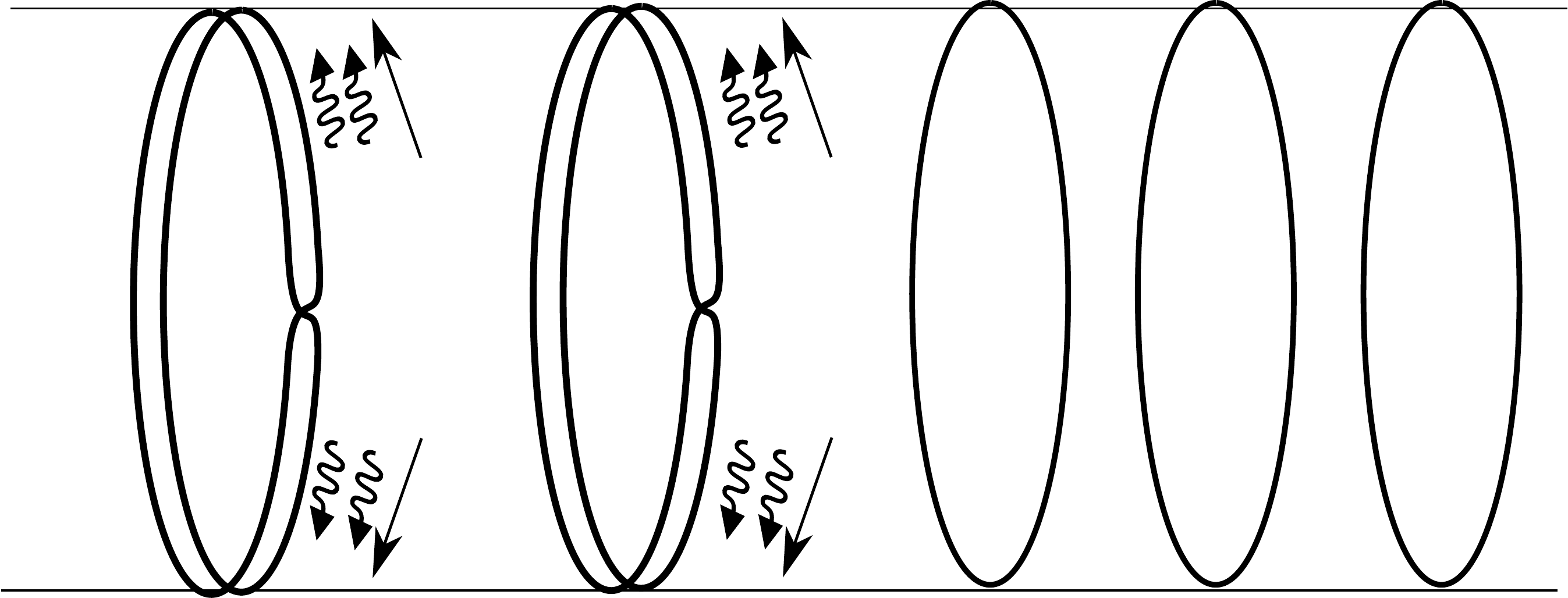}}
\hspace{15pt}
\subfigure[~The final state in the NS sector]{\label{fig:alphaNSFinal}
	\includegraphics[width=7cm]{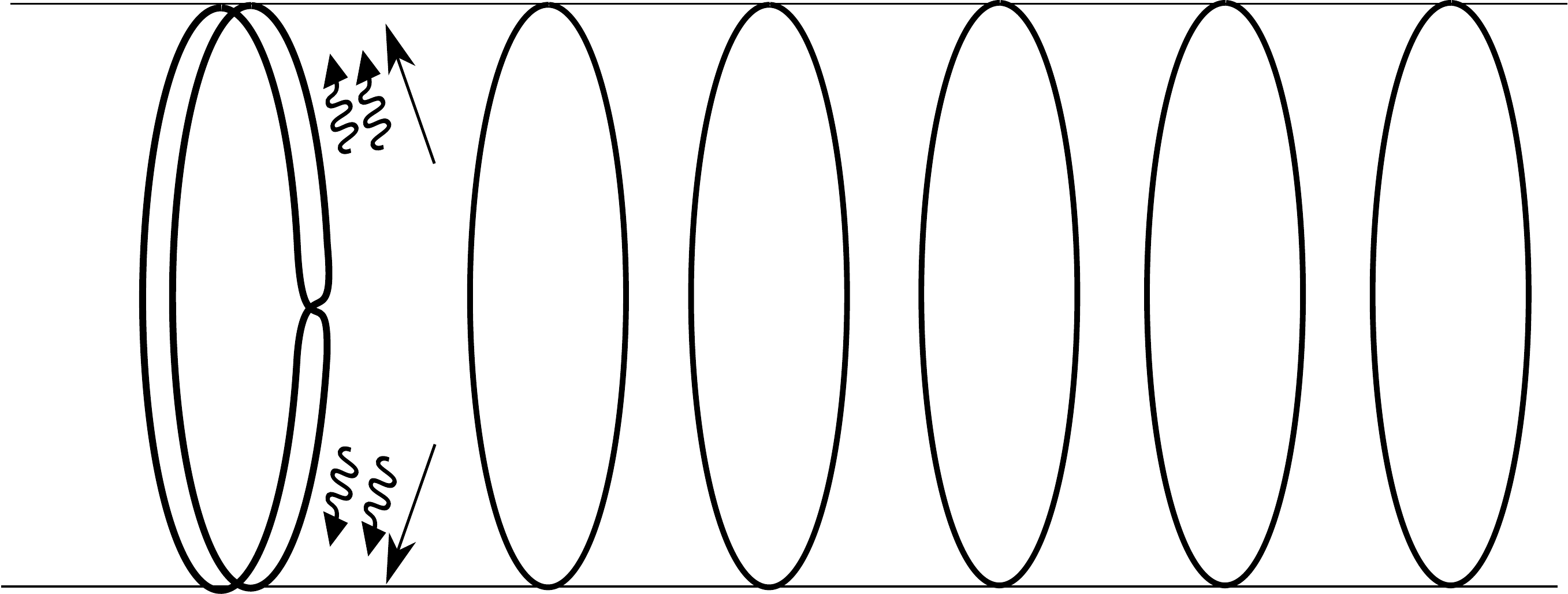}} \\
\subfigure[~The initial state in the Ramond sector]{\label{fig:alphaRRInitial}
	\includegraphics[width=7cm]{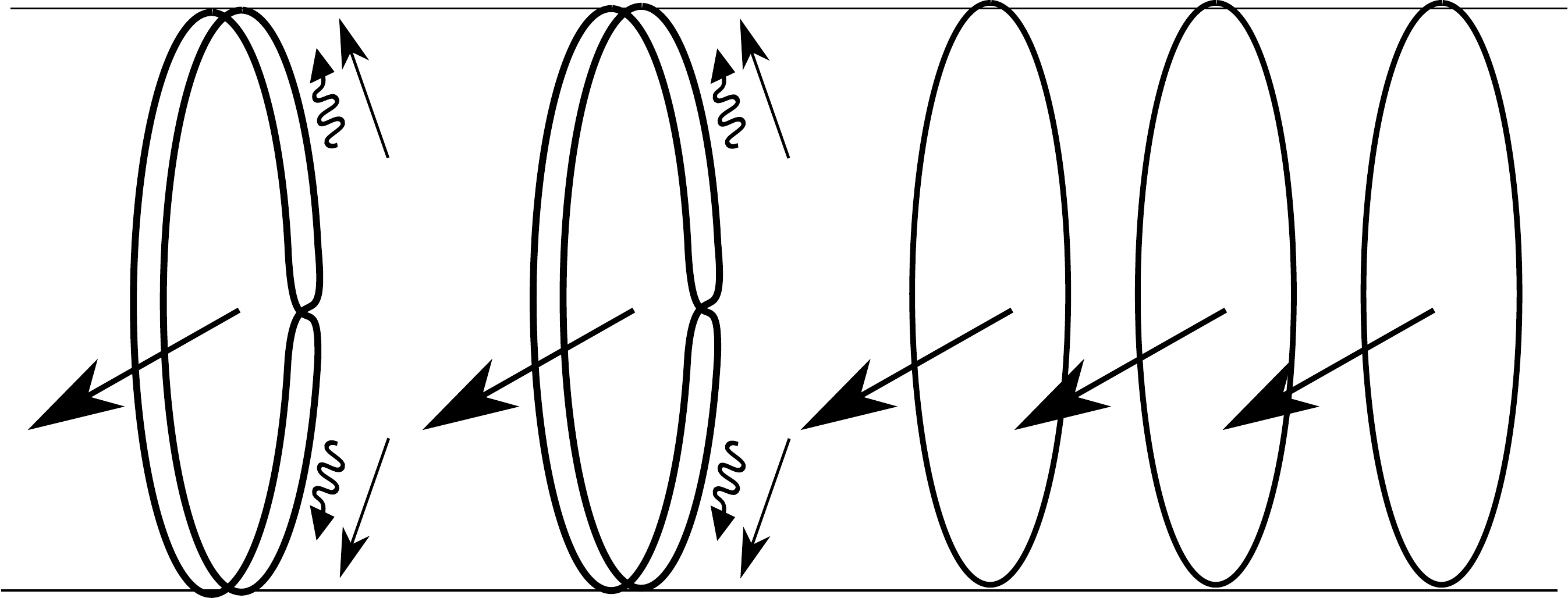}}
\hspace{15pt}
\subfigure[~The final state in the Ramond sector]{\label{fig:alphaRRFinal}
  \includegraphics[width=7cm]{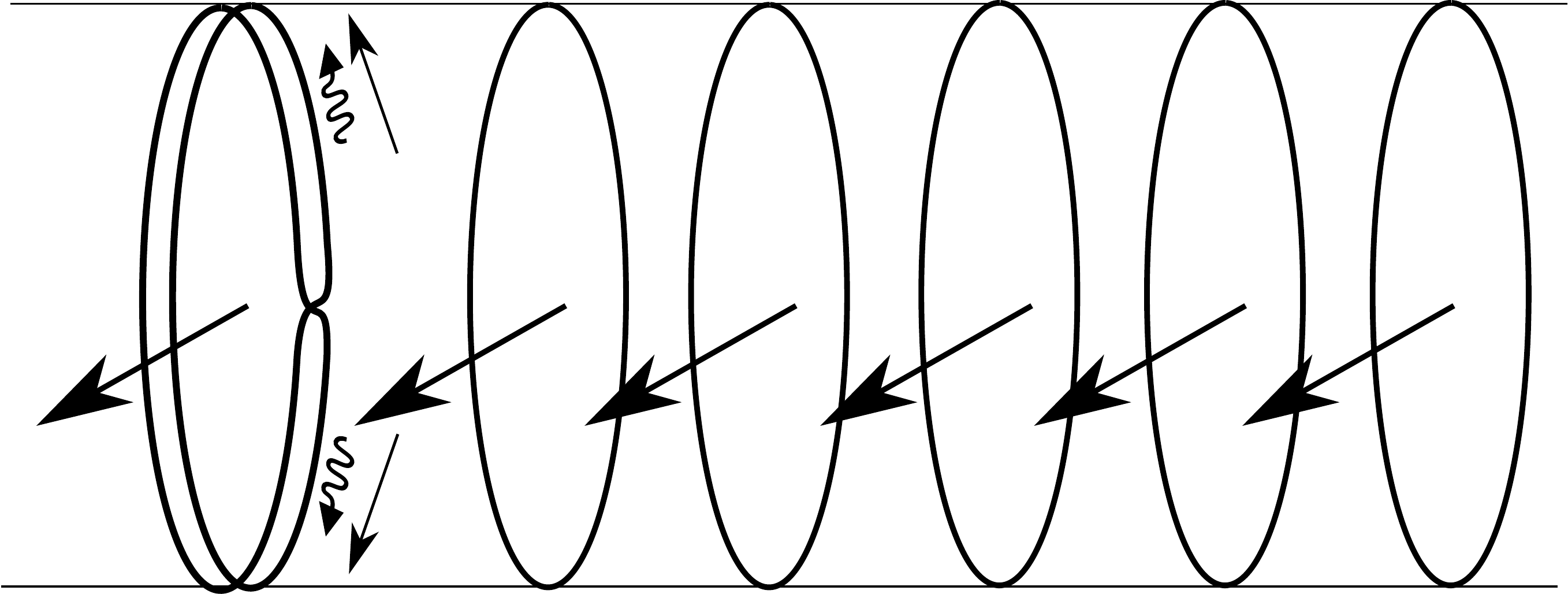}}
      \caption[The initial and final states for emission]{The initial
        and final states for the emission process discussed in
        Sections~\ref{sec:states-and-op},~\ref{sec:CFT-evaluation},
        and~\ref{sec:calc-rate-emission}. The pictures correspond to
        $\nu=2$ and $l=1$ emission.  The straight arrows pointing up
        (down) on the loops indicate bosonic excitations in the left
        (right) sector.\label{fig:AlphaDecay}}
\end{center}
\end{figure}

\subsubsection{The Final State}

The final state is determined by the fact that we are looking for the
amplitude to transition to a supergravity state, and we have a unique
supergravity excitation with given twist and angular quantum
numbers. Working again in the NS sector, arrived at by spectral flow
by $-(2n+1)$ units, we get
\begin{equation}
\ket{f'} = \ket{\phi_{N+1}^{\frac{l}{2},\frac{l}{2}-k}};
\end{equation}
the initial state of our previous calculation.

\subsubsection{The Vertex Operator}

The vertex operator is independent of the states it acts on. It is
completely determined by the supergravity scalar to which it couples
in Equation~\eqref{eq:general-S-int}.

We now see that the present process is similar to the amplitude we
computed earlier if we reverse the direction of time $\tau$. That is,
the initial state now is untwisted, while in our earlier computation
the \emph{final} state was untwisted. The vertex operator then leads
to a final twisted state, and since there is a unique supergravity
state with given quantum numbers, we can write down this state.

\subsubsection{Relating the Emission Amplitude to the Earlier Computed Amplitude}

We term the supergravity excitation emission in the previous sections
``untwisting'' emission since a twisted component string in the
initial state ``untwists'' under the action of the vertex operator and
leads to a final state with no twists. We call the emission of the
present section ``twisting'' emission since the initial state has no
twists, and the vertex operator leads to a twisted component string in
the final state.

By comparing the initial and final states of the two process, we
immediately see that the current NS sector twisting amplitude is
simply the Hermitian conjugate of the previous NS sector untwisting
amplitude,
\begin{equation}
{\mathcal{A}'}_{l, m_\psi, m_\phi}^\text{twisting}(t,y) 
	= \big[{\mathcal{A}'}_{l, -m_\psi, -m_\phi}^\text{untwisting}(-t,y)\big]^\dg,
\end{equation}
with flipped $SO(4)_E$ charges and reversed time. In the complex
plane, this statement becomes
\begin{equation}\label{eq:herm-conj-amp}
{\mathcal{A}'}^\text{twisting}_{l, m_\psi, m_\phi}(z, \bar{z}) = 
    \big[{\mathcal{A}'}^\text{untwisting}_{l, -m_\psi, -m_\phi}\big(\tfrac{1}{\bar{z}},\tfrac{1}{z}\big)\big]^\dg.
\end{equation}

To see the above relation explicitly, consider the Hermitian conjugate
of the previous, untwisting amplitude:
\begin{calc}
\big[{\mathcal{A}'}_{l,m_\psi, m_\phi}^\text{untwisting}(z,\bar{z})\big]^\dg 
	&= \left[|z|^{l+2}\bra{f'}\mathcal{V}_{l,m_\psi, m_\phi}
                (z, \bar{z})\ket{i'}\right]^\dg\\
	&= |z|^{l+2}\bra{i'}
            \big[\mathcal{V}_{l,m_\psi, m_\phi}(z,\bar{z})\big]^\dg\ket{f'}\\
	&= \frac{1}{|z|^{l+2}}
		\bra{i'}\mathcal{V}_{l,-m_\psi, -m_\phi}
                \big(\tfrac{1}{\bar{z}},\tfrac{1}{z}\big)\ket{f'},	
\end{calc}
where the $i'$ and $f'$ are from the previous calculation. The
amplitude we now wish to compute is (in terms of the previous
calculation's states)
\begin{equation}
{\mathcal{A}'}_{l,m_\psi, m_\phi}^\text{twisting}(z,\bar{z}) = |z|^{l+2}
                       \bra{i'}\mathcal{V}_{l,m_\psi, m_\phi}(z,\bar{z})\ket{f'}
\end{equation}
Comparing these two expressions, one arrives at
Equation~\eqref{eq:herm-conj-amp}.

From Equation~\eqref{eq:final-L-amp}, we have
\begin{equation}
{\mathcal{A}'}^\text{untwisting}(z, \bar{z})
	 = (-1)^{k+\bar{k}}
\sqrt{\choose{N+l+1}{N}\choose{\bar{N} + l+1}{\bar{N}}}
 z^{-\frac{l}{2} - N - 1}
\bar{z}^{-\frac{l}{2} - \bar{N} - 1},
\end{equation}
and so using Equation~\eqref{eq:herm-conj-amp} gives
\begin{equation}
{\mathcal{A}'}^\text{twisting} = (-1)^{k+\bar{k}}
	\sqrt{\choose{N+l+1}{N}\choose{\bar{N}+l+1}{\bar{N}}}
	z^{\frac{l}{2} + N + 1}\bar{z}^{\frac{l}{2} + \bar{N} +1}.
\end{equation}
Note that this amplitude is in the NS sector, and before we can use it
to get emission we have to spectral flow it to the Ramond sector.
Spectral flowing by $\alpha=2n+1$ units to the Ramond sector gives
\begin{calc}\label{eq:cft-twisting-amp}
\mathcal{A}^{(\alpha, \nu)}_\text{twisting} 
	&= z^{-\alpha(\frac{l}{2}-k)}\bar{z}^{-\bar{\alpha}(\frac{l}{2}-\bar{k})}
		\mathcal{A}'^{(\nu)}_\text{twisting}\\
	&= (-1)^{k+\bar{k}}\sqrt{\nu}\,
	\sqrt{\choose{N+l+1}{N}\choose{\bar{N}+l+1}{\bar{N}}}
		z^{\frac{l}{2} + N + 1- \alpha(\frac{l}{2}-k)}
	\bar{z}^{\frac{l}{2} + \bar{N} +1 - \bar{\alpha}(\frac{l}{2}-\bar{k})},
\end{calc}
where we have put the combinatoric factor in as well. Note that the
combinatorics work out the same as before since the combinatorics
cannot be affected by Hermitian conjugation; however, the
interpretation is different. The initial state starts with $\nu-1$
sets of $(l+1)$-twisted component strings, while the final state has
$\nu$ $(l+1)$-twisted component strings. Therefore, if at some initial
time all of the strands were untwisted and each twist corresponds to
an emitted supergravity particle, then the above is the amplitude for
the emission of the $\nu$th particle.

Comparing Equation~\eqref{eq:cft-twisting-amp} with
Equation~\eqref{eq:final-euc-amp}, we see that the amplitudes agree
except for the power of $z$, which is different because of the
different energies of the concerned states in the two processes. Thus
we can immediately write down the emission rate for the $\nu$th
particle from the cap into the flat space
\begin{equation}
\der{\Gamma}{E}
  = 
  \nu\frac{2\pi}{2^{2l+1}l!^2}\frac{(Q_1Q_5)^{l+1}}{R^{2l+3}}(E^2-\lambda^2)^{l+1}
	\choose{N+l+1}{N}\choose{\bar{N} + l+1}{\bar{N}}\delta_{\lambda, \lambda_0}\delta(E-E_0)
\end{equation}
where
\begin{equation}\begin{split}
E_0 &= \frac{1}{R}\left[(\alpha + \bar{\alpha} - 2)\tfrac{l}{2} 
  - \alpha k - \bar{\alpha}\bar{k} - N - \bar{N} - 2\right]\\
\lambda_0 &= \frac{1}{R}\left[-(\alpha - \bar{\alpha})\tfrac{l}{2} + \alpha k 
   - \bar{\alpha}\bar{k} + N - \bar{N}\right].
\end{split}\end{equation}
For sufficiently large $\alpha$ or $\bar{\alpha}$, $E_0$ is positive
and the physical process is emission and not absorption.  In taking the
hermitian conjugate we have flipped the the angular momentum of the
emitted particle from that in Equation~\eqref{eq:emitted-ang-mom};
therefore, the above emission rate is for
\begin{equation}
m_\psi =  l - k - \bar{k}\qquad m_\phi = k - \bar{k}.
\end{equation}

One can check that the emission rate above agrees with the emission
rate from the gravity dual \cite{myers,cm1}. Such a check was carried
out in \cite{cm1} only for states with excitation level $N=0$, because
it was not clear how to construct the initial state for $N>0$ in the
effective string description of the D1D5 bound state. With our present
construction of states and vertex operators in the orbifold CFT, we
can compute amplitudes for emission of supergravity quanta from all
initial states containing supergravity excitations.

\section{Reproducing \texorpdfstring{$\kappa > 1$}{kappa greater than one} Emission}\label{sec:kappa-not-1}

In this section, we extend the results by computing the emission from
a broader class of D1D5 states; those with $\kappa>1$. The above
calculation only reproduces the supergravity emission and spectrum for
the JMaRT geometries with $\kappa=1$.

The calculation may be broken into the following steps:
\begin{enumerate}
\item In Section~\ref{sec:physical-amp}, we set up the physical
  emission problem from the geometries of~\cite{ross} for $\kappa>1$,
  in the CFT language developed in~\cite{acm1}. Specifically, we
  describe the initial excited state and the final state that go into
  the CFT amplitude, which when plugged into
  Equation~\eqref{eq:D1D5-decay-rate} give the emission rate of the
  minimal scalars. Evaluating the CFT amplitude is a nontrivial
  exercise, to which the majority of Section~\ref{sec:kappa-not-1} is
  dedicated. The initial state is parameterized by three integers $n$,
  $\bar{n}$, and $\kappa$. The positive integer $\kappa$, called $k$
  in~\cite{cm3}, controls a conical defect in the AdS region. The
  spectrum and rate of emission found in the previous section are for
  $\kappa=1$. We extend those calculations to $\kappa>1$.

\item We do not directly compute the physical CFT amplitude of
  interest. Instead, we compute a CFT amplitude that does \emph{not}
  correspond to a physical gravitational process, and then map this
  ``unphysical'' amplitude onto the physical problem using spectral
  flow and hermitian conjugation. This route avoids some subtleties
  and allows us to use some results from the previous
  section (first derived in~\cite{acm1}) in the calculation. In
  Section~\ref{sec:unphysical-amp}, we precisely set up the unphysical
  CFT amplitude to be computed.

\item In Section~\ref{sec:specflow-hermconj}, we show how to relate
  the unphysical amplitude to the physical problem using spectral flow
  and Hermitian conjugation.

\item Before starting the calculation of the unphysical CFT amplitude,
  in Section~\ref{sec:method} we explain the specific method used to
  compute it . The unphysical amplitude, we explain, can be lifted to
  a covering space, where it becomes an amplitude computed in
  Section~\ref{sec:nonextremal}. The Jacobian factors that arise in
  mapping to the covering space, however, are highly nontrivial.
  Additionally there are some important combinatoric factors, which
  come in from symmetrizing over all $N_1N_5$ copies of the CFT.

\item In Section~\ref{sec:T}, we compute the Jacobian factors that
  arise from mapping the twist operators to the covering space. We use
  the methods for evaluating correlation functions of twist operators
  developed in~\cite{lm1}. For details, see
  Appendix~\ref{ap:twist-corr}.

\item In Section~\ref{sec:M}, we compute the Jacobian factors produced
  by the non-twist operator insertions in the unphysical amplitude.

\item In Section~\ref{sec:comb}, we compute the combinatoric factors
  that come from symmetrizing over all $N_1N_5$ copies of the CFT. The
  result simplifies in the large $N_1N_5$ limit that is physically
  relevant.

\item Finally in Section~\ref{sec:rate}, we use
  Section~\ref{sec:specflow-hermconj} to relate the computed
  unphysical amplitude to the final amplitude for emission. We then
  plug the amplitude into Equation~\eqref{eq:D1D5-decay-rate} to find
  the rate of emission. The explicit $\kappa$-dependence comes in the
  form of a power, $\kappa^{-2l-3}$, multiplying the rate for
  $\kappa=1$; the spectrum is also affected. The spectrum and rate
  exactly match the gravity calculation in~\cite{cm3}.
\end{enumerate}
We take some results from the previous section.

\subsection{Emission from \texorpdfstring{$\kappa$}{kappa}-orbifolded Geometries}\label{sec:physical-amp}

Below, we first describe the initial state of the physical CFT
problem, which is dual to the unperturbed background geometry. Then,
we roughly describe the final state of the physical CFT amplitude that
the initial state decays to. We do not precisely give these states
since we do not directly compute with them. In
Section~\ref{sec:unphysical-amp}, we give the precise states used in
the ``unphysical'' CFT amplitude, which in
Section~\ref{sec:specflow-hermconj} we relate to the physically
relevant states of this section.

\subsubsection{The CFT Initial State}

The physical initial state is the background geometry described
in~\cite{ross} and~\cite{cm3}. The decoupled AdS-part of the geometry
can be obtained by spectral flowing $\kappa$-orbifolded $AdS_3 \times
S^3$ by $\alpha = 2n+ \frac{1}{\kappa}$ units on the left-moving sector
and by $\bar{\alpha} = 2\bar{n}+ \frac{1}{\kappa}$ units on the
right-moving sector. The fractional spectral flow may seem strange;
however, it arises because of the conical defect. In
~\cite{mm,bal,gms2} geometries were constructed which had a decoupled
AdS part which can be understood in this context as spectral flow by
$\alpha=2n+\frac{1}{\kappa}$ and $\bar \alpha=\frac{1}{\kappa}$. However
these geometries are BPS and do not have an ergoregion and thus do not
radiate.  We discuss fractional spectral flow in the CFT context in
Section~\ref{sec:specflow}.

The $\kappa$-orbifolded $AdS_3$ is described, after a left and right
spectral flow by $\frac{1}{\kappa}$, as twisting the $N_1N_5$ strands
into $\kappa$-length component strings in the R sector. This geometry
is stable and does not emit anything. Performing further spectral
flow, however, adds fermionic excitations to all of the component
strings and allows for the possibility of emission.

\begin{figure}[ht]
\begin{center}
\subfigure[~The physical initial state.]{\label{phy-initial}
\includegraphics[width=6cm]{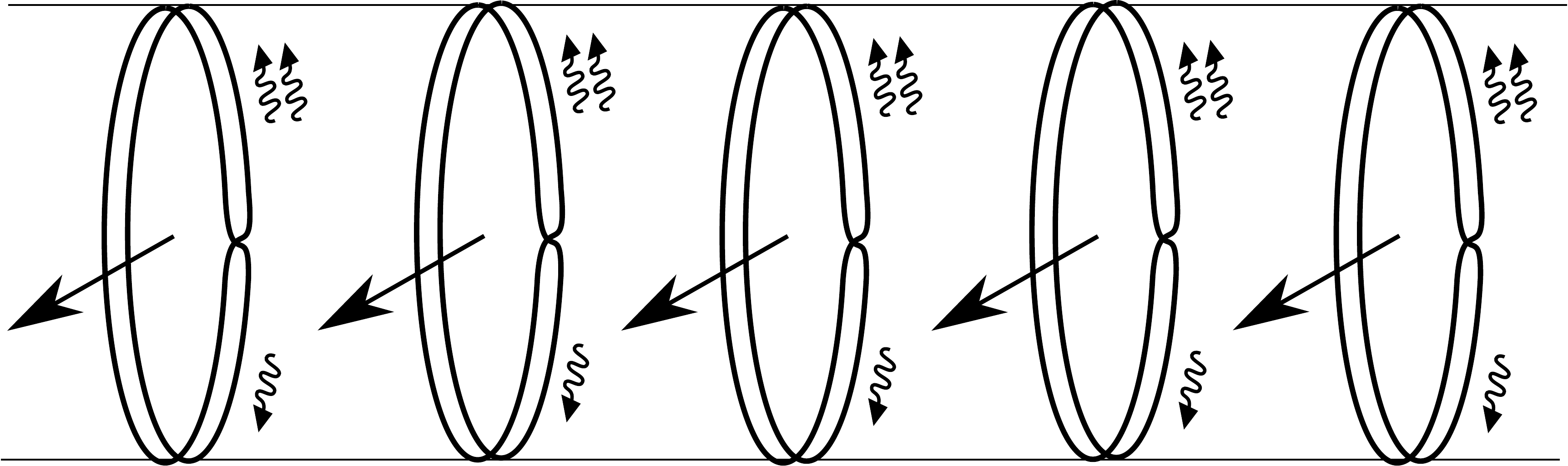}}
\hspace{30pt}
\subfigure[~The physical final state.]{\label{phy-final}
  \includegraphics[width=6cm]{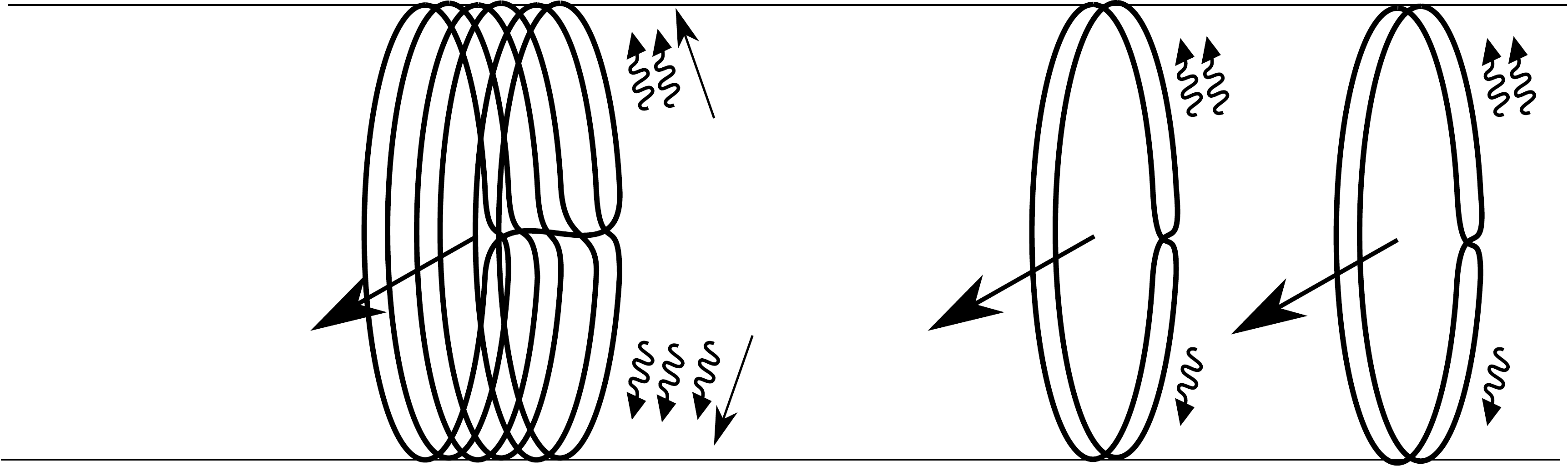}}
\caption[Initial and final states for $\kappa=2$.]{A depiction of the
  initial and final states of the physical amplitude for
  $\kappa=l=2$.}\label{fig:physical-states}
\end{center}
\end{figure}

The $\kappa$-twisted component string in the Ramond vacuum has weight
$\kappa/4$. The Ramond vacuum also has ``base spin'' coming from the
fermion zero modes. Let us start with the ``spin up'' $\kappa$-twisted
Ramond vacuum with weight and charge
\begin{equation}
h = \bar{h} = \frac{\kappa}{4} \qquad m = \bar{m} = \frac{1}{2}.
\end{equation}
We then add energy and charge to this state by spectral-flowing by
$2n$ units, where spectral flow by $\alpha$ units affects the weight
and charge of a state by~\cite{spectral}
\begin{equation}\begin{split}
h &\mapsto h + \alpha m + \frac{\alpha^2 c_\text{tot.}}{24}\\
m &\mapsto m + \frac{\alpha c_\text{tot.}}{12}.
\end{split}\end{equation}
Spectral flowing the $\kappa$-twisted Ramond vacuum by $2n$ units
corresponds to filling all fermion energy levels up to the ``Fermi sea
level'' $n \kappa$. Keep in mind that there are two complex fermions
which have spin up and spectral flow fills fermion levels with both
fermions. This gives the initial state's weight and charge~\cite{cm3}
\begin{equation}
h_i = \kappa\left(n^2 + \frac{n}{\kappa} + \frac{1}{4}\right) 
	\qquad m_i = n\kappa + \frac{1}{2}
\end{equation}
and similarly for the right sector replacing $n$ by $\bar{n}$. Compare
this to Equation~\eqref{eq:CFT-id}. This state is depicted in
Figure~\ref{phy-initial}.

\subsubsection{The CFT Final State}

In emitting a particle of angular momentum $(l, m_\psi, m_\phi)$ the
initial state is acted on by the supergravity vertex operator
$\mathcal{V}_{l, -m_\psi, -m_\phi}$ from
Equation~\eqref{eq:vertex-op}.

We consider the process where the $(l+1)$-twist operator acts on $l+1$
distinct component strings of the background forming one long
$\kappa(l+1)$-twisted component string. The final state, then, is in
the $\kappa(l+1)$-twisted R sector. In principle there are other ways
that the vertex operator may twist the initial state; however, we work
in the limit where $N_1N_5\gg \kappa$, in which case this process
dominates. Processes in which the vertex operator twists more than one
strand of the same $\kappa$-length component string have probabilities
suppressed by factors of $1/(N_1N_5)$.

The charge/angular momentum of the vertex operator
$\mathcal{V}_{l,l-k-\bar{k}, k-\bar{k}}$ is given by
\begin{equation}\begin{split}
m_v &= -\left(\tfrac{l}{2} -k\right)\hspace{50pt} 
	m^\text{vertex}_\psi = -m_\psi = l-k-\bar{k}\\
\bar{m}_v &= -\left(\tfrac{l}{2} - \bar{k}\right)\hspace{50pt} 
	m^\text{vertex}_\phi = -m_\phi = k-\bar{k}.
\end{split}\end{equation}
From the charge of the vertex operator and the initial state it is
easy to deduce the charge of the final state:
\begin{equation}
m_f = (l+1)\left(n\kappa + \tfrac{1}{2}\right) - \left(\tfrac{l}{2} - k\right)
  = (l+1)n\kappa + k + \frac{1}{2}.
\end{equation}

There can be different final states with the same charge corresponding
to different harmonics of the emitted quanta. Since we derive the
amplitude using spectral flowed states it is not important to write
down the weights of the states here. However they are found easily by
using the charge and weight of the spectral flowed states. The final
state, which we call $\ket{f}$, is shown in Figure~\ref{phy-final}.

\subsubsection{The Physical CFT Amplitude}

From the initial and final states, the CFT amplitude which we use to
find the emission spectrum and rate is given by
\begin{equation}\label{eq:cylinder-amp}
\mathcal{A}_{l, m_\psi, m_\phi} 
	= \bra{f}\widetilde{\mathcal{V}}_{l, m_\psi, m_\phi}(\sigma, \tau)\ket{i}.
\end{equation}
We prefer to calculate on the complex plane with $z$ coordinates
instead of on the cylinder. Mapping the vertex operator to the plane
using Equation~\eqref{eq:map-to-the-plane} gives a Jacobian factor,
$|z|^{l+2}$, from its conformal weight. Thus, the physical CFT
amplitude that we ultimately wish to find is written as
\begin{equation}
\mathcal{A}_{l,m_\psi, m_\phi}(z) 
	= |z|^{l+2}\bra{f}\mathcal{V}_{l,m_\psi, m_\phi}(z, \bar{z})\ket{i}.
\end{equation}
The majority of the calculation is expended in finding this amplitude.
The tilde on the vertex operator in Equation~\eqref{eq:cylinder-amp}
is to distinguish it from the operator in the complex plane.

\subsection{The ``Unphysical'' Amplitude}\label{sec:unphysical-amp}

We do not directly compute the amplitude of interest with the initial
and final states described above. Instead we compute an amplitude,
described below, that is related to the physical problem by spectral
flow and Hermitian conjugation. In this other amplitude, the
calculation is simpler and we have a better understanding of what the
initial and final states should be. In
Section~\ref{sec:kappa-equals-1}, the same technique was used.

In this section we describe the initial and final states of the
unphysical amplitude, $\ket{i'}$ and $\ket{f'}$, that we use to
calculate
\begin{equation}
\mathcal{A}' = \bra{f'}\mathcal{V}(z)\ket{i'}.
\end{equation}
Later, we relate this process to the physical problem by spectral flow
and Hermitian conjugation. Before we can give the states, we must
describe the notation we use, specifically addressing the fermions'
periodicity.

\subsubsection{Fermion Periodicities}

First, let us emphasize what the R and NS sectors \emph{mean} in the
twisted sector.  We define two parameters $\beta_+$ and $\beta_-$, which
specify the periodicity of the fermions in the theory:
\begin{equation}
\psi^{\pm \dot{A}}\big(ze^{2\pi i}\big) = e^{i\pi\beta_\pm}\psi^{\pm\dot{A}}(z).
\end{equation}
Obviously, $\beta_\pm$ are only defined modulo two under addition.

The R sector means $\beta_\pm\equiv 1$, whereas the NS sector means
$\beta_\pm\equiv 0$. Spectral flow by $\alpha$ units has the effect of
taking
\begin{equation}\label{eq:spectral-eta}
\beta_\pm \mapsto \beta_\pm \pm \alpha.
\end{equation}

In the $p$-twisted sector, the boundary conditions are
\begin{equation}
\psi^{\pm\dot{A}}_{(j)}\big(ze^{2\pi i}\big) = e^{i\pi \beta_\pm}\psi^{\pm\dot{A}}_{(j+1)}(z),
\end{equation}
where $(j)$ indexes the different copies of the target space. This
implies that
\begin{equation}\label{eq:base-twisted-eta}
\psi^{\pm\dot{A}}_{(j)}\big(z e^{2p\pi i}\big) = e^{ip\pi\beta_\pm}\psi^{\pm\dot{A}}_{(j)}(z).
\end{equation}
In the base space, over each point $z$, there are $p$ different copies
of each field. To calculate, it is convenient to map to a covering
space where these $p$ copies become a single-valued field. When one
goes to a covering space the periodicity of the total single-valued
field $\Psi(t)$, then is given by
\begin{equation}
\Psi^{\pm\dot{A}}\big(te^{2\pi i}\big) = \exp\left[i\big(p\beta_\pm + (p-1)\big)\pi\right]
		\Psi^{\pm\dot{A}}(t),
\end{equation}
where the extra factor of $(p-1)$ in the exponent comes from the
Jacobian of the weight-half fermion, under a map of the form $z\propto
t^p$. We label the periodicity in the cover by
\begin{equation}\label{eq:cover-eta}
\beta^{(\mathrm{cover})}_\pm = p(\beta_\pm+1) -1.
\end{equation}
In the untwisted sector, we use the natural definition of the
fermions, $\psi(z)$, as being periodic, without branch cut. Thus,
application of a spin field operator is necessary to give antiperiodic
boundary conditions. In the twisted sector, the periodicity of
fermions in the base space is neither ``naturally'' periodic nor
antiperiodic since there is a hole and branch cut from the twist
operator; however, in the covering space, the fields are, again,
naturally periodic.

We denote ``bare twists'' which insert the identity in the cover as
\begin{equation}
\sigma_p(z, \bar{z}) \xrightarrow[\quad\text{cover}\quad]{\text{to the}} \id (t,\bar{t})
\qquad h = \bar{h} = \Delta_p = \frac{c}{24}\left(p - \frac{1}{p}\right).
\end{equation}
After the above discussion, we see that one should always use
\begin{equation}
(\text{NS Sector}) \,\Longrightarrow \, 
\begin{cases}
\sigma_p(z,\bar{z}) & \text{$p$ odd}\\
S_p^\alpha(z)\bar{S}_p^{\dot{\alpha}}(\bar{z})\sigma_p(z,\bar{z}) & \text{$p$ even}
\end{cases}, \qquad
(\text{R Sector}) \,\Longrightarrow \, 
S_p^\alpha(z)\bar{S}_p^{\dot{\alpha}}(\bar{z})\sigma_p(z, \bar{z}).
\end{equation}
The $S_p$'s indicate that one should insert a spin field in the
$p$-fold covering space at the image of the point $z$. In the above
statement, one could equally well choose $SU(2)_2$ indices instead of
$SU(2)_{L/R}$ indices for the spin fields.

The CFT amplitude we compute uses the bare twists, which for odd twist
order correspond to the NS sector and for even twist order correspond
to neither the NS nor the R sector. Even though this amplitude is not
physically relevant, we can use spectral flow to relate it to the R
sector process of physical interest. Below, we give the initial and
final states of the unphysical amplitude, starting with the simpler
final state.

\begin{figure}[ht]
\begin{center}
\subfigure[~The initial state of the unphysical amplitude.]
{\label{fig:unphy-in}
\includegraphics[width=6cm]{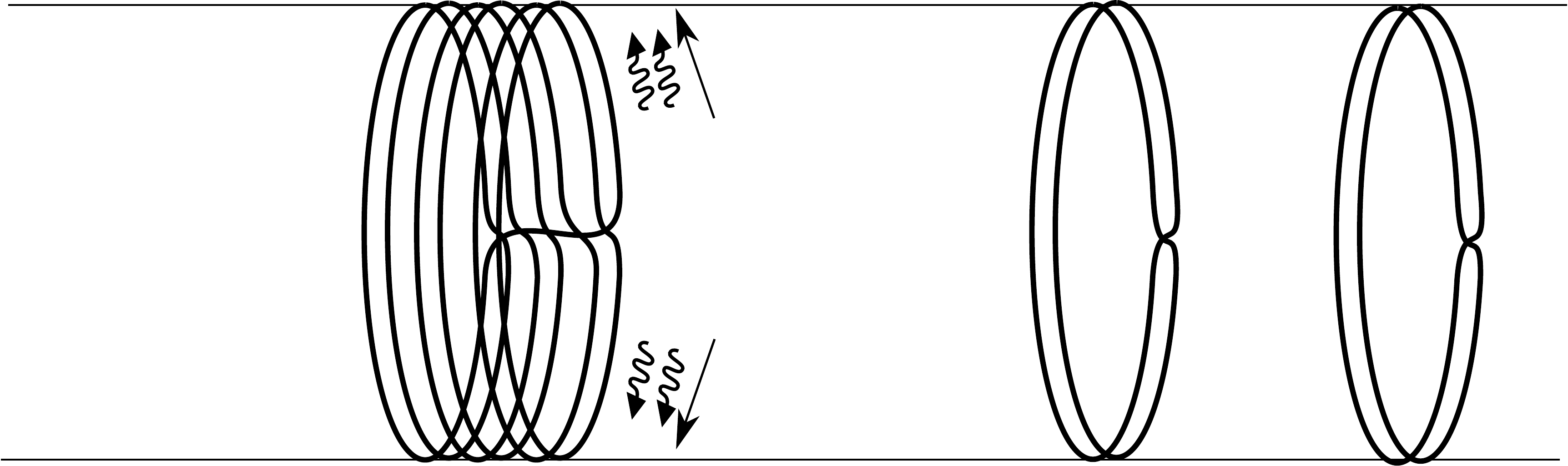}}
\hspace{30pt}
\subfigure[~The final state of the unphysical amplitude.]
{\label{fig:unphy-fin}
\includegraphics[width=6cm]{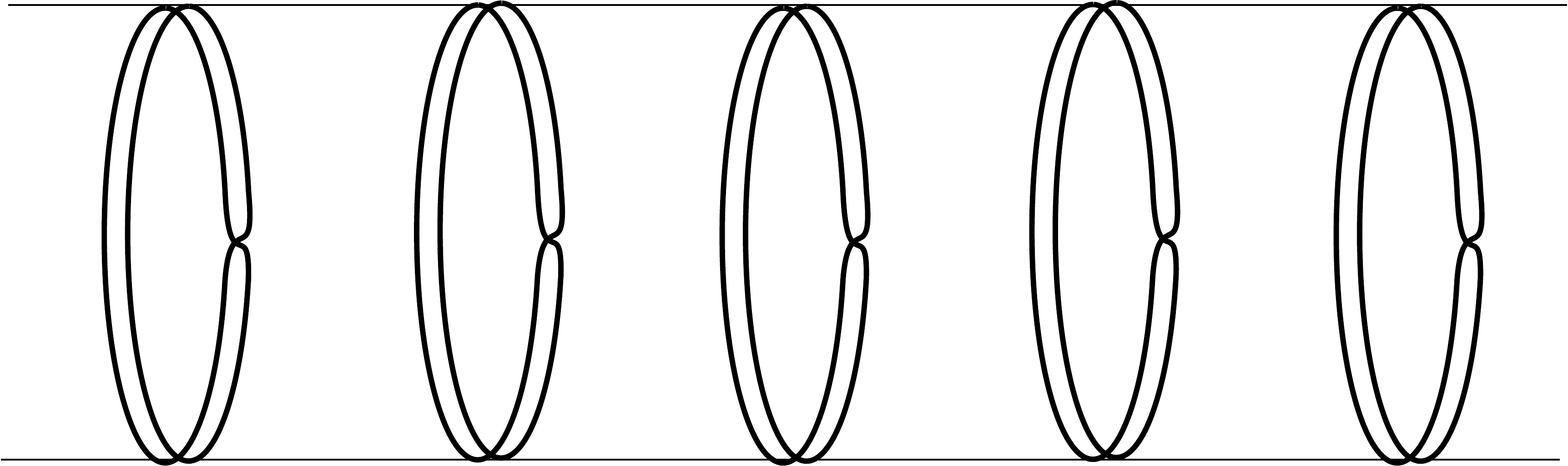}}
\caption[Initial and final states of the ``unphysical'' amplitude]{The
  initial and final states of the unphysical amplitude for
  $\kappa=l=2$. The initial state spectral flows to the physical final
  state, and the final state spectral flows to the physical initial
  state.}\label{fig:unphysical-states}
\end{center}
\end{figure}

\subsubsection{The Final State}

The \emph{final} state\footnote{Here, we just discuss the
$\kappa(l+1)$ strands that are involved in the nontrivial part of the
correlator. Later, we introduce the combinatoric factors which result
from symmetrizing over all $N_1N_5$ copies of the CFT.} spectral flows
to the \emph{initial} state of the physical problem. All of the
excitations and spin fields of the physical state can be acquired via
spectral flow, so we use bare twists for the unphysical amplitude
final state:
\begin{equation}
\ket{f'} = \underbrace{\sigma_\kappa\sigma_\kappa\cdots\sigma_\kappa}_{l+1}\vac_{NS}.
\end{equation}
This state has weight and charge
\begin{equation}
h_f = \bar{h}_f = \frac{1}{4}(l+1)\left(\kappa - \frac{1}{\kappa}\right)\qquad
m_f = \bar{m}_f = 0.
\end{equation}
Note that for odd $\kappa$ this state corresponds to the NS sector,
but for $\kappa$ even it corresponds to neither the NS nor the R
sector. This state, which we call $\ket{i}$, is shown in
Figure~\ref{fig:unphy-fin}.

\subsubsection{The Initial State}

The \emph{initial} state spectral flows to the physical final state,
and therefore consists of excitations in the $\kappa(l+1)$-twisted
sector, as shown in Figure~\ref{fig:unphy-in}. Below, modulo
normalization, we give the left part of the state
\begin{smalleq}\begin{equation}
\ket{i'} = 
\big(J_0^-\big)^k
G^{- A}_{-\frac{1}{2\kappa}}\big(L_{-\frac{1}{\kappa}}\big)^N \psi^{+\dot{A}}_{-\frac{1}{2\kappa}}
\begin{cases}
	J^+_{-\frac{l-1}{\kappa(l+1)}}J^+_{-\frac{l-3}{\kappa(l+1)}}
		\cdots
	J^+_{-\frac{1}{\kappa(l+1)}} \sigma_{\kappa(l+1)}\vac_{NS} &
		\text{$(l+1)$ odd}\\
		J^+_{-\frac{l-1}{\kappa(l+1)}}J^+_{-\frac{l-3}{\kappa(l+1)}}
			\cdots
		J^+_{-\frac{2}{\kappa(l+1)}}
	S^+_{\kappa(l+1)}\sigma_{\kappa(l+1)}\vac_{NS}  &  \text{$(l+1)$ even}.
\end{cases}
\end{equation}\end{smalleq}
The normalization of the state is $\kappa$-dependent, which plays an
important role in the calculation as discussed in
Section~\ref{sec:Mi}.

This state has weight and charge
\begin{equation}
h_i = \frac{1}{\kappa}\left(\frac{l}{2} + N + 1\right) 
		+ \frac{1}{4}(l+1)\left(\kappa - \frac{1}{\kappa}\right)\qquad
m_i = \frac{l}{2} - k,
\end{equation}
and similarly for the right sector. We conjecture the above form for
the initial state based on the $\kappa=1$ case in
Section~\ref{sec:kappa-equals-1} and work done in~\cite{cm3} for
$\kappa>1$. The excited initial state should be dual to a supergravity
excitation of the orbifolded-AdS background. In global AdS, one
identifies the supergravity duals as the descendants of chiral primary
states under the anomaly-free subalgebra; here, we propose the above
modification for the $\kappa$-orbifolded background. This is
motivated, in part, by the action of the vertex operator for
absorption of supergravity particles on $\kappa$-orbifolded AdS.

\subsubsection{The Amplitude}

The amplitude we compute, then, is
\begin{equation}
{\mathcal{A}'}^\kappa = \bra{f'}\mathcal{V}_{l, l- k-\bar{k}, k-\bar{k}}(z_2)\ket{i'}.
\end{equation}
Note that if we map to the $\kappa$-cover, then these states are the
initial and final states of the $\kappa=1$ calculation in
Section~\ref{sec:kappa-equals-1}.  In mapping to a $\kappa$-covering
space the final state's twist operators are removed, and the initial
state's twist operator becomes an $(l+1)$-twist. The $J^+$ modes
acting on the initial state, along with the spin field for even $l+1$,
act on the $(l+1)$-twist in the $\kappa$-cover in precisely the
correct way to form the lowest weight chiral primary twist operator
of~\cite{lm2}:
\begin{smalleq}\begin{multline}
\hspace{-4pt}J^+_{-\frac{l-1}{\kappa(l+1)}}\hspace{-1pt}J^+_{-\frac{l-3}{\kappa(l+1)}}\hspace{-2pt}
		\cdots
\begin{cases}
	J^+_{-\frac{1}{\kappa(l+1)}} \sigma_{\kappa(l+1)}\\
		J^+_{-\frac{2}{\kappa(l+1)}}
	S^+_{\kappa(l+1)}\sigma_{\kappa(l+1)}
\end{cases}\hspace{-16pt}
 \xrightarrow[\text{$\kappa$-cover}]{\text{to}}
J^+_{-\frac{l-1}{l+1}}\hspace{-1pt}J^+_{-\frac{l-3}{l+1}}\hspace{-2pt}\cdots
\begin{cases}
	J^+_{-\frac{1}{l+1}} \sigma_{l+1}\\
	J^+_{-\frac{2}{l+1}}
	S^+_{l+1}\sigma_{l+1}\\
\end{cases}\hspace{-15pt}
= \sigma^0_{l+1}.
\end{multline}\end{smalleq}
We follow the notation of~\cite{acm1} by denoting the lowest weight
chiral primary ${(l+1)}$-twist operator, introduced in~\cite{lm2}, as
$\sigma^0_{l+1}$. The above equation does not include Jacobian
factors, computed in Sections~\ref{sec:T} and~\ref{sec:M}, which are
introduced in going to the $\kappa$-covering space.

The specific covering space to which we map preserves the form of the
$(l+1)$-twist vertex operator. The fact that going to a $\kappa$
covering space gives the $\kappa=1$ amplitude along with the gravity
description of the emission process, is precisely the motivation for
introducing the specific form of the state
$\ket{i'}$. 

\subsection{Relating the Computed CFT Amplitude to the Physical Problem}\label{sec:specflow-hermconj}

There are two steps needed to map the physical problem onto the
unphysical CFT computation. First, we use spectral flow to map the
physical states, $\ket{i}$ and $\ket{f}$, to the ``primed states,''
$\ket{f'}$ and $\ket{i'}$.  Second, we use Hermitian conjugation to
reverse the initial and final primed states.

\subsubsection{Using Spectral Flow}\label{sec:specflow}

We wish to relate a CFT amplitude computed with the unphysical
``primed states,''
\begin{equation}
\mathcal{A}' = \bra{f'}\mathcal{V}(z, \bar{z})\ket{i'},
\end{equation}
to the physical amplitude in the Ramond sector.  In this section, we
show how to spectral flow~\cite{spectral,spectral-yu,vafa-warner} the
physical problem in the R sector to the actual CFT amplitude we
compute.

If spectral flowing the states $\ket{i'}$ and $\ket{f'}$ by $\alpha$
units is given by
\begin{equation}
\ket{f'}\mapsto \ket{i} = \mathcal{U}_\alpha\ket{f'}\qquad
\bra{i'}\mapsto \bra{f} = \bra{i'}\mathcal{U}_{-\alpha},
\end{equation}
then we can compute the physical Ramond sector amplitude for emission
of a particle with angular momentum $(l, m_\psi, m_\phi)$ by using
\begin{calc}
\mathcal{A}_{l, m_\psi, m_\phi} &= 
	|z|^{l+2}\bra{f} \mathcal{V}_{l, -m_\psi, -m_\phi}(z,\bar{z})\ket{i}\\
 	  &= |z|^{l+2}\big(\bra{f}\mathcal{U}_{\alpha}\big)
	     \big(\mathcal{U}_{-\alpha}\mathcal{V}_{l, -m_\psi, -m_\phi}\mathcal{U}_{\alpha}\big)
			     \big(\mathcal{U}_{-\alpha}\ket{i}\big)\\
	  &= |z|^{l+2}\bra{i'}\mathcal{V}'_{l, -m_\psi, -m_\phi}(z, \bar{z})\ket{f'}.
\end{calc}
Note that one finds $\mathcal{V}'$ by spectral flowing $\mathcal{V}$
by $-\alpha$ units.

In~\cite{acm1}, it was shown that the vertex operator transforms under
spectral flow as
\begin{equation}
\mathcal{V}'_{l,l-k-\bar{k}, k-\bar{k}}(z,\bar{z}) = z^{-\alpha(\frac{l}{2}-k)}
	\bar{z}^{-\bar{\alpha}(\frac{l}{2}-\bar{k})}
		\mathcal{V}_{l,l-k-\bar{k}, k-\bar{k}}(z,\bar{z}).
\end{equation}
We can use the vertex operator's transformation to write
\begin{equation}\label{eq:spectral-flow-A}
\mathcal{A}_{l, k+\bar{k} -l, \bar{k} - k} 
	= z^{-\alpha(\frac{l}{2}-k)}\bar{z}^{-\bar{\alpha}(\frac{l}{2}-\bar{k})}\,
	|z|^{l+2}\bra{i'}\mathcal{V}_{l,l-k-\bar{k}, k-\bar{k}}(z,\bar{z})\ket{f'}.
\end{equation}
Note that the primed states are reversed from what one would like.

\begin{sloppypar}
The spectral flow parameter, $\alpha$, is chosen to have a value that
spectral flows the primed states to the physical states, which is
achieved by
\begin{equation}
\alpha = 2n+\frac{1}{\kappa}\qquad \bar{\alpha} = 2\bar{n} + \frac{1}{\kappa} 
	\qquad n,\bar{n}\in\ints.
\end{equation}
Spectral flowing by non-integer units may seem strange, however, one
can show that this results from the peculiar bare twist
operators. Consider spectral flowing the bare twist $\sigma_\kappa$ by
$\frac{1}{\kappa}$ units, and recall that under spectral flow by
$\alpha$ units
\begin{equation}
\beta_\pm \mapsto \beta_\pm \pm \alpha,
\end{equation}
and
\begin{equation}
\beta^{(\text{cover})}_\pm = \kappa(\beta_\pm + 1) - 1.
\end{equation}
The bare twist has $\beta_\pm^{(\text{cover})}=0$, which after
spectral flowing by $1/\kappa$ units becomes
${\beta_\pm^{(\text{cover})}=1}$ consistent with the R boundary
conditions in the base space. Similarly, one finds that that the
fermion periodicity of $\ket{i'}$ becomes correct for the R sector
after spectral flowing by $1/\kappa$ units. After spectral flowing by
$1/\kappa$ units to get to the R sector, we spectral flow by an even
number of units to build up the fermionic excitations of the state
$\ket{i}$.
\end{sloppypar}

\subsubsection{Hermitian Conjugation}

Having related the physical amplitude to
\begin{equation}
\bra{i'}\mathcal{V}_{l, m_\psi, m_\phi}(z,\bar{z})\ket{f'},
\end{equation}
we now wish to switch the initial and final primed states by Hermitian
conjugation.

On the cylinder, the vertex operator Hermitian conjugates as
\begin{equation}
\left[\widetilde{\mathcal{V}}_{l, m_\psi, m_\phi}(\tau,\sigma)\right]^\dg 
	=  \widetilde{\mathcal{V}}_{l, -m_\psi, -m_\phi}(-\tau, \sigma),
\end{equation}
and therefore
\begin{equation}
\bra{i'}\widetilde{\mathcal{V}}_{l, m_\psi, m_\phi}(\tau,\sigma)\ket{f'}
 = \left[\bra{f'} \widetilde{\mathcal{V}}_{l,-m_\psi, -m_\phi}(-\tau, \sigma)\ket{i'}\right]^\dg.
\end{equation}
In the complex plane, this statement translates to
\begin{equation}
|z|^{l+2}\bra{i'}\mathcal{V}_{l, m_\psi, m_\phi}(z, \bar{z})\ket{f'}
  = |z|^{-(l+2)}\left[\bra{f'}
	\mathcal{V}_{l, -m_\psi, -m_\phi}\left(\tfrac{1}{z}, \tfrac{1}{\bar{z}}\right)
	\ket{i'}\right]^\dg.
\end{equation}

Applying this result to Equation~\eqref{eq:spectral-flow-A}, we may
write the physical CFT amplitude for emission of a particle with
angular momentum $(l, k+\bar{k}, \bar{k}-k)$ as
\begin{calc}\label{eq:unphysical-to-physical}
\mathcal{A}_{l, k+\bar{k}-l, \bar{k}-k} 
	&= z^{-\alpha(\frac{l}{2} - k)}\bar{z}^{-\bar{\alpha}(\frac{l}{2} - \bar{k})}
		|z|^{-(l+2)} 
	\left[\bra{f'}
	\mathcal{V}_{l, k+\bar{k}-l, \bar{k}-k}\left(\tfrac{1}{z}, \tfrac{1}{\bar{z}}\right)
	\ket{i'}\right]^\dg\\
	&= z^{-\frac{l}{2}-1-\alpha(\frac{l}{2} - k)}
		\bar{z}^{-\frac{l}{2} -1 -\bar{\alpha}(\frac{l}{2} - \bar{k})}
	\left[\mathcal{A}'_{l, k+\bar{k}-l, \bar{k}-k} 
		\left(\tfrac{1}{z}, \tfrac{1}{\bar{z}}\right)\right]^\dg.
\end{calc}
Thus, we compute the unphysical amplitude $\mathcal{A}'(z, \bar{z})$,
and use the above relation to find the physical amplitude
$\mathcal{A}$. The unphysical amplitude $\mathcal{A}'$ is independent
of $k$ and $\bar{k}$, so we suppress the subscripts.

\subsection{Method of Computation}\label{sec:method}

The way we compute the CFT amplitude ${\mathcal{A}'}^\kappa$ is by
noting that if one were to lift to the $\kappa$-cover, the amplitude
becomes ${\mathcal{A}'}^{\kappa=1}$ computed in
Section~\ref{sec:nonextremal}. Specifically, if we transform to a
coordinate $u$ via the map
\begin{equation}
z = b u^\kappa\qquad b = \frac{z_2}{u_2^\kappa},
\end{equation}
where $u_2$ is the image of the point $z_2$; then, the operators left
in the correlator are exactly those for ${\mathcal{A}'}^{\kappa=1}$.
In lifting to the cover, however, we do get some nontrivial ``Jacobian
factors'' from the various operators in the correlator transforming
under the map. Therefore, we write
\begin{equation}
{\mathcal{A}'}^\kappa 
	= \left(\frac{{\mathcal{A}'}^\kappa}{{\mathcal{A}'}^{\kappa=1}}\right)
		{\mathcal{A}'}^{\kappa=1}
	= TM{\mathcal{A}'}^{\kappa=1},
\end{equation}
where we have written the Jacobian factors as the product of two
contributions: one from the twists, $T$, and one from the modes, $M$.

\begin{figure}[ht]
\begin{center}
\includegraphics[width=4cm]{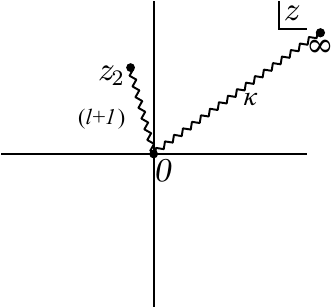}
\hspace{20pt}
\raisebox{1.8cm}{$\xrightarrow[]{\hspace{14pt}z\,=\,b u^\kappa\hspace{10pt}}$}
\hspace{20pt}
\includegraphics[width=4cm]{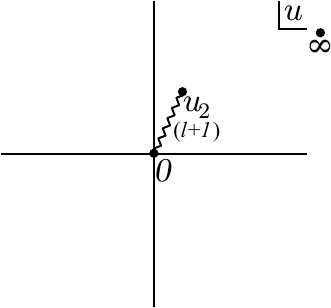}
\caption[The branch cuts created by the twist operators]{In the
  $z$-plane, there is a twist operator of order $\kappa(l+1)$ at the
  origin, a twist of order $l+1$ at a point we call $z_2$, and $l+1$
  $\kappa$-twists at infinity. The diagram on the left depicts the
  branch cuts connecting the three points. The number next to the
  branch cut depicts its ``order.'' One can remove the $l+1$ branch
  cuts of order $\kappa$ between the origin and infinity by mapping to
  the $u$-coordinate, as shown in the figure on the
  right.}\label{fig:branch-cuts}
\end{center}
\end{figure}

We begin with the lengthier calculation of the nontrivial factor $T$
that comes from the twist operators transforming. One can see that $T$
is given by the ratio
\begin{equation}\label{eq:T-def}
T = \lim_{z_3\to\infty}
    |z_3|^{4(l+1)\Delta_\kappa}
  \frac{\vev{\sigma_\kappa\dots\sigma_\kappa(z_3)\sigma_{l+1}(z_2)\sigma_{\kappa(l+1)}(0)}}
	{\vev{\sigma_{l+1}(u_2)\sigma_{l+1}(0)}},
\end{equation}
where all of the above twists are ``bare'' twists which give \emph{no}
insertions in total $\kappa(l+1)$-covering space. The $z_3$ prefactor
ensures that in the limit as $z_3\to\infty$ the $l+1$ twists at
infinity make a normalized bra.

After computing $T$, we compute the mode Jacobian factor $M$, which is
given by the product of all of the Jacobian factors that the modes
circling the origin and the point $z_2$ acquire. 

\subsection{Computing the Twist Jacobian Factor \texorpdfstring{$T$}{T}}\label{sec:T}

Since the \emph{effect} of the twist operators is inherently nonlocal,
they do not transform in a simple local way when lifting to a covering
space. The simplest way to compute the numerator of
Equation~\eqref{eq:T-def}, is by lifting to the total covering space
and then computing the Liouville action that comes in conformally
transforming the induced metric on the covering space to a fiducial
metric. This is the method developed in~\cite{lm1}. Recently,
\cite{rastelli1, rastelli2} developed new technology for correlators
of $S_{N_1N_5}$-twist operators; however, we do not use those
techniques here.

The denominator of Equation~\eqref{eq:T-def} is just the two-point
function of some normalized twist operators, and so we know the
answer:
\begin{equation}
\vev{\sigma_{l+1}(u_2)\sigma_{l+1}(0)} = \frac{1}{|u_2|^{\frac{c}{6}\left(l+1 - \frac{1}{l+1}\right)}}
	= \frac{1}{|u_2|^{4\Delta_{l+1}}}.
\end{equation}

From the weights of the twist operators and $SL(2,\co)$ invariance of
the correlator, we know that any three-point function of quasi-primary
fields must be given by
\begin{equation}\begin{split}
&\vev{\mathcal{O}_3(z_3)\,\mathcal{O}_2(z_2)\,
	\mathcal{O}_1(z_1)}\\
&\qquad= \frac{|C_{123}|^2}{
	\left(|z_1 - z_2|^{\Delta_1 + \Delta_2 - \Delta_3}
	       |z_2 - z_3|^{\Delta_2 + \Delta_3 - \Delta_1}
	       |z_3 - z_1|^{\Delta_3 + \Delta_1 - \Delta_2}\right)^2}.
\end{split}\end{equation}
In the present case where $z_3 = \infty$ and $z_1 = 0$, this gives
(after regularizing the divergence which corresponds to correctly
normalizing the final state)
\begin{equation}\label{eq:three-point-function}
\lim_{z_3\to\infty}|z_3|^{4(l+1)\Delta_\kappa}
\vev{\sigma_\kappa\dots\sigma_\kappa(z_3)\sigma_{l+1}(z_2)\sigma_{\kappa(l+1)}(0)}
 = \frac{|C|^2}{|z_2|^{2(\Delta_{\kappa(l+1)} + \Delta_{l+1} - (l+1)\Delta_\kappa)}}.
\end{equation}
It is the constant $C$ which is nontrivial and is determined in the
calculation that follows. From the above, one sees that the twist
Jacobian factor is completely determined by the constant $C$:
\begin{equation}\label{eq:C-gives-T}
T = |C|^2 \frac{|u_2|^{4\Delta_{l+1}}}
	{|z_2|^{2(\Delta_{\kappa(l+1)} + \Delta_{l+1} - (l+1)\Delta_\kappa)}}.
\end{equation}

To compute $C$ in Equation~\eqref{eq:three-point-function} and thereby
the twist Jacobian factor, $T$, we undertake the following steps:
\begin{enumerate}
\item We find the map to the total covering space, $z=z(t)$, that has the required
  properties.
\item We compute the Liouville action that comes from
  conformally mapping the induced metric of the total covering space
  to the fiducial form in Equation~\eqref{eq:covering-metric}.
\item The above correlators are all of normalized twist operators. The
  normalization is such that the two-point function has unit
  correlator at unit separation, as determined in~\cite{lm1}. We put
  in the normalization factors, which cancel out the
  $\veps$-dependence of the correlator.
\item Finally, we put all of the pieces together to determine $C$,
  and thereby $T$.
\end{enumerate}

\subsubsection{The Map to the Total Covering Space}

The first step in the calculation of $T$, then, is to find the map to
the total covering space. We have an $SL(2,\co)$ symmetry of the
covering space that allows us to fix three points. We fix the image of
the origin of the $z$-plane to the origin of the $t$-plane, the image
of $z_2$\footnote{We should point out that there are other images of
$z_2$, but only $t=1$ corresponds to where the vertex operator
acts. The fact that there are other images of $z_2$ corresponds to the
fact that the vertex operator only acts on $l+1$ strands of the
$\kappa(l+1)$ initial strands.} to the point $t=1$, and one image of
$z=\infty$ to $t=\infty$. Having expended the $SL(2,\co)$ symmetry, we
do not have freedom to fix the locations of the other images of
infinity in the covering space. We call the other images of infinity
$t_\infty^{(j)}$, where $j=1,\dots, l$.

Thus, we require our map to behave as
\begin{equation}\begin{aligned}
z &\sim t^{\kappa(l+1)}  & z&\approx 0,\; t\approx 0\\
z -z_2 &\sim  (t-1)^{l+1} &  z &\approx z_2,\; t\approx 1\\
z &\sim t^\kappa  & z &\to \infty,\; t\to \infty\\
z &\sim (t-t_\infty^{(j)})^{-\kappa} & z &\to\infty,\; t\to t_\infty^{(j)},
\end{aligned}\end{equation}
and be regular at all other points.  Here, $\sim$ means proportional
to at leading order.  Generically, one expects $\kappa(l+1)$ images of
infinity. However, in this case there should only be $l+1$ images of infinity since there
are only $l+1$ disconnected strings at $z\to\infty$. A priori, we do
not know the $t_\infty^{(j)}$; this is part of the output in finding
the map.

From the above conditions and requiring regularity everywhere but the
above special points, we see that $\der{z}{t}$ may have zeroes only at
$t=0$ (of degree $\kappa(l+1) -1$) and $t=1$ (of degree $l$). The
meromorphic map we want can be written as the ratio of two
polynomials,
\begin{equation}
z = \frac{f_1(t)}{f_2(t)},
\end{equation}
where $f_1$ must be of degree $\kappa(l+1)$ and $f_2$ must be of
degree $\kappa l$, from the number of sheets and the behavior at
infinity. Furthermore, near the origin we know that $f_1$ must behave
as $t^{\kappa(l+1)}$ and $f_2$ a constant. Thus, we already have
determined that
\begin{equation}
f_1 \propto t^{\kappa(l+1)}.
\end{equation}
From familiarity with the two-point function map in~\cite{lm1} and
consideration of the above requirements, one can immediately write
down the correct map:
\begin{equation}\label{eq:themap}
z = z_2 \frac{t^{\kappa(l+1)}}{\big[t^{l+1}-(t-1)^{l+1}\big]^\kappa}.
\end{equation}
Note that when $\kappa=1$, the map reduces to the usual two-point map
between $0$ and $z_2$; and when $l=0$, the map reduces to
\begin{equation}
z = z_2 t^\kappa,
\end{equation}
which is the correct map for a two-point function between $0$ and
infinity. Furthermore, that the map is simply raising the two-point
function map to the $\kappa$ power makes sense, since when one goes to
the $\kappa$-cover the correlator should become just the order $(l+1)$
two-point function as depicted in Figure~\ref{fig:branch-cuts}. The
derivative of the map is
\begin{equation}
\der{z}{t} = \frac{\kappa(l+1) z_2}{\big[t^{l+1} - (t-1)^{l+1}\big]^{\kappa+1}}t^{\kappa(l+1)-1}
	(t-1)^l,
\end{equation}
which has zeroes at the correct points. One can see from the map that
it only has $l+1$ distinct images of infinity, as required.

The behavior of the map near the irregular points is needed for the
computation. Near the origin
\begin{equation}
z \approx (-1)^{\kappa l} z_2 t^{\kappa (l+1)}\qquad 
\der{z}{t}\approx (-1)^{\kappa l}\kappa(l+1) z_2t^{\kappa(l+1) -1}.
\end{equation}
Near $t=1$,
\begin{equation}
z\approx z_2 + z_2 \kappa (t-1)^{l+1}\qquad 
\der{z}{t}\approx z_2 \kappa(l+1) (t-1)^l.
\end{equation}
For large $t$,
\begin{equation}
z \approx \frac{z_2}{(l+1)^\kappa}t^\kappa\qquad
\der{z}{t} \approx \frac{z_2\kappa}{(l+1)^\kappa} t^{\kappa -1}.
\end{equation}

The images of infinity at finite $t$ are the same as for the $l+1$
order two-point map, that is
\begin{equation}
t_\infty^{(j)} = \frac{1}{1-\alpha_j}\qquad (\alpha_j)^{l+1}=1,
\end{equation}
where the $\alpha_j$ are the $(l+1)$ roots of unity:
\begin{equation}
\alpha_j = e^{2\pi i\frac{j}{l+1}}.
\end{equation} 
Note that $\alpha_0=1$, which corresponds to $t\to\infty$; the case we
already covered. Therefore, the images of infinity at finite $t$
correspond to $j=1,\dots, l$. If we let
\begin{equation}
t = \frac{1}{1-\alpha_j} + x,
\end{equation}
for small $x\in\co$, then near the images of infinity the map behaves
as
\begin{equation}
z \approx (-1)^\kappa z_2 \left[\frac{\alpha_j}{(l+1)(1-\alpha_j)^2} \frac{1}{x}\right]^\kappa.
\end{equation}
Plugging back in with $t$, one finds the behavior of the map near the
finite images of infinity:
\begin{equation}\label{eq:behavior-finite-images}
\begin{split}
z &\approx (-1)^\kappa
	z_2\left[\frac{\alpha_j}{(l+1)(1-\alpha_j)^2}\right]^\kappa
	\frac{1}{(t-t_\infty^{(j)})^\kappa}\\
\der{z}{t} &\approx (-1)^{\kappa+1}
	\kappa z_2 \left[\frac{\alpha_j}{(l+1)(1-\alpha_j)^2}\right]^\kappa
	\frac{1}{(t-t_\infty^{(j)})^{\kappa+1}}.
\end{split}
\end{equation}
The map is regular at all other points not considered above.

\subsubsection{The Twist Jacobian Factor \texorpdfstring{$T$}{T}}\label{sec:norm}

In Appendix~\ref{ap:twist-corr}, we use the local behavior of the map
given above to compute the various contributions to the Liouville
action. In the notation used there, the quantities which define the
correlator are
\begin{equation}\begin{gathered}
M = 2    \qquad F = N = l+1 \qquad s = \kappa(l+1)\\
z_1 = 0  \qquad z_2 = z_2\\
t_1 = 0  \qquad t_2 = 1\\ 
p_1 = \kappa(l+1) \qquad p_2 = l+1 \qquad q_j \equiv \kappa,
\end{gathered}\end{equation}
and the local behavior of the map one needs is encoded in the
coefficients
\begin{equation}\begin{aligned}
  a_1&=(-1)^{\kappa l} z_2\\
  a_2&= z_2\kappa\\
  b_0&=\frac{z_2}{(l+1)^\kappa}\\
  b_j&=(-1)^\kappa z_2\left[\frac{\alpha_j}{(l+1)(1-\alpha_j)^2}\right]^\kappa
\qquad
  j = 1, \dots, l.
\end{aligned}\end{equation}
Plugging in with these values, we find that~\cite{ac1} 
\begin{equation}
Z = \delta^{4(l+1)\Delta_\kappa} \kappa^{-\frac{c}{12}\frac{l(l+2)}{l+1}},
\end{equation}
where we use the identity
\begin{equation}
\prod_{j=1}^l\alpha_j^l(1-\alpha_j)^2 = (l+1)^2.
\end{equation}

We normalize the twists at infinity in the same way as those in the
finite $z$-plane: at unit separation in their local coordinates they
should have unit correlator with themselves. From
Appendix~\ref{ap:twist-corr}, we see that if we normalize the twists
at infinity in this way, then~\cite{ac1}
\begin{equation}
\vev{\sigma_\kappa(\infty)\sigma_\kappa(0)}_{\delta} = \delta^{4\Delta_\kappa}.
\end{equation}
To ensure that the final state corresponds to a normalized bra, then,
we should use
\begin{equation}
\lim_{z_3\to\infty}|z_3|^{4\Delta_\kappa}\sigma_\kappa(z_3) 
  = \delta^{-4\Delta_\kappa}\sigma_\kappa(\infty).
\end{equation}
Thus, the $\delta$ dependence cancels out.  

The constant $C$, then, is given as a power of $\kappa$:
\begin{equation}
|C|^2 = \kappa^{-\frac{c}{12}\frac{l(l+2)}{l+1}}.
\end{equation}
The power of $\kappa$ may seem mysterious, until one realizes that
\[
\Delta_{l+1} = \frac{c}{24}\left(l+1 - \frac{1}{l+1}\right)
	= \frac{1}{4}\frac{l(l+2)}{l+1},
\]
and so we write
\begin{equation}
|C|^2 = \kappa^{-2\Delta_{l+1}}.
\end{equation}
Plugging in the value of $C$ into Equation~\eqref{eq:C-gives-T}, one
finds the twist Jacobian factor,
\begin{equation}
T = \kappa^{-2\Delta_{l+1}}\frac{|u_2|^{4\Delta_{l+1}}}
	{|z_2|^{2\frac{\kappa+1}{\kappa}\Delta_{l+1}}}.
\end{equation}
Note that this Jacobian factor cannot naturally be thought of as the
product of Jacobian factors from the different twist operators. This
fact makes correlation functions of twist operators nontrivial.

\subsection{Computing the Mode Jacobian Factor \texorpdfstring{$M$}{M}}\label{sec:M}

To compute the mode Jacobian factor, $M$, we first illustrate how
modes transform under the map $z=bu^\kappa$. Then, we compute the mode
Jacobian factor from the initial state modes, followed by the mode
Jacobian factor from the vertex operator modes. The final state does
not have any modes, and therefore does not contribute to $M$. In
discussing the initial state, we find that we must also consider the
$\kappa$-dependence of the normalization of the state.

\subsubsection{Transformation Law for Modes}

First, let us examine how the different modes transform under the map,
\begin{equation}
z = b u^\kappa.
\end{equation}
Modes that circle the origin get opened up by a factor of $\kappa$,
and also get a Jacobian factor. They transform as
\begin{equation}\label{eq:mode-trans}
\mathcal{O}_m^{(z)} = b^m \kappa^{1-\Delta}\mathcal{O}^{(u)}_{m\kappa},
\end{equation}
where $\Delta$ is the weight of the \emph{field} $\mathcal{O}(z)$. One
can see this from the definition of modes in the $\kappa(l+1)$-twisted
sector~\cite{lm2}:
\begin{equation}
\mathcal{O}_{\frac{m}{\kappa(l+1)}}^{(z)} = \oint\frac{\drm z}{2\pi i}\sum_{j=1}^{\kappa(l+1)}
\mathcal{O}_{(j)}(z)e^{2\pi i\frac{m}{\kappa(l+1)}(j-1)}z^{\Delta - 1 + \frac{m}{\kappa(l+1)}}.
\end{equation}
The subscripted parenthetical index on the field $\mathcal{O}_{(j)}$
denotes the copy the field lives on.

Modes which circle $z_2$, where there is no branching point, transform
as
\begin{equation}
\mathcal{O}^{(z)}_m = \left(\der{z}{u}\bigg|_{u_2,z_2}\right)^m\mathcal{O}_m^{(u)}.
\end{equation}
Near the point $u_2$, the map behaves as
\begin{calc}
z &\approx z_2 + \kappa b u_2^{\kappa-1}(u-u_2) \\
  &= z_2 + \kappa \frac{z_2}{u_2}(u-u_2).
\end{calc}
Therefore, we see that, under this map, modes which circle the point
$z_2$ transform as
\begin{equation}
\mathcal{O}^{(z)}_{m} = \left(\kappa \frac{z_2}{u_2}\right)^m\mathcal{O}_m^{(u)}.
\end{equation}

Since the spectral flowed final state does not have any modes, we can
write $M$ as a product of Jacobian factors from the initial state and
Jacobian factors from the vertex operator,
\begin{equation}
M = M_i M_v.
\end{equation}

\subsubsection{The Mode Jacobian Factor from the Initial State}\label{sec:Mi}

For the initial state, we need to be a little careful about the
normalization. Consider an initial state defined in the base space as
\begin{equation}\label{eq:state-base}
\ket{\psi^{(z)}} = \mathcal{C}\mathcal{O}_m^{(z)}\ket{\sigma_{\kappa(l+1)}}.
\end{equation}
The norm of this state is given by
\begin{calc}
\braket{\psi^{(z)}|\psi^{(z)}} 
	&= |\mathcal{C}|^2\bra{\sigma_{\kappa(l+1)}}
		\mathcal{O}^{\dg(z)}_{-m}\mathcal{O}^{(z)}_m\ket{\sigma_{\kappa(l+1)}}\\
	&= |\mathcal{C}|^2\bra{\sigma_{l+1}}
		\left(b^{-m}\kappa^{1-\Delta}\mathcal{O}^{\dg(u)}_{-m\kappa}\right)
		\left(b^m\kappa^{1-\Delta}\mathcal{O}^{(u)}_{m\kappa}\right)\ket{\sigma_{l+1}}\\
	&= |\mathcal{C}|^2\kappa^{2(1-\Delta)}\bra{\sigma_{l+1}}\mathcal{O}^{\dg(u)}_{-m\kappa}
		\mathcal{O}^{(u)}_{m\kappa}\ket{\sigma_{l+1}}.
\end{calc}

On the other hand, suppose one were to define the state directly in
the cover as
\begin{equation}
\ket{\psi^{(u)}} = \mathcal{C}_\text{cover}\mathcal{O}_{m\kappa}^{(u)}\ket{\sigma_{l+1}},
\end{equation}
then its norm is given by
\begin{equation}
\braket{\psi^{(u)}|\psi^{(u)}} = |\mathcal{C}_\text{cover}|^2
\bra{\sigma_{l+1}}\mathcal{O}^{\dg(u)}_{-m\kappa}\mathcal{O}^{(u)}_{m\kappa}\ket{\sigma_{l+1}}.
\end{equation}
Requiring that both $\ket{\psi^{(z)}}$ and $\ket{\psi^{(u)}}$ be
normalized, means
\begin{equation}
\mathcal{C} = \frac{\mathcal{C}_\text{cover}}{\kappa^{1-\Delta}}.
\end{equation}
Thus, using Equation~\eqref{eq:mode-trans} with
Equation~\eqref{eq:state-base}, one finds
\begin{equation}
\ket{\psi^{(z)}} = b^m\ket{\psi^{(u)}}.
\end{equation}
From this we see that the $\kappa$ factor in
Equation~\eqref{eq:mode-trans} gets cancelled out by the $\kappa$s one
gets in normalizing the state in the covering space. This argument
easily generalizes to all of the modes acting in the initial state.

The only contributions to $M_i$, therefore, come from the factors
$b^m$. Because of the way they come in, we see that the the total
factor we get (from the left) is
\begin{equation}
b^{-h^\text{modes}_i},
\end{equation}
where $h^\text{modes}_i$ is the weight of all of the modes in the
initial state. This is simply the total weight of the initial state
minus the weight of the bare twist:
\begin{calc}
h^\text{modes}_i &= h'_{i} - \Delta_{\kappa(l+1)}\\
	&= \frac{l+1}{4}\left(\cancel{\kappa}-\frac{1}{\kappa}\right)
		+\frac{1}{\kappa}\left(\frac{l}{2} + 1\right) + \frac{N}{\kappa}
		-\frac{1}{4}\left(\cancel{\kappa(l+1)} - \frac{1}{\kappa(l+1)}\right)\\
	&= \frac{1}{\kappa}\left[
	\frac{l}{2} +N + 1 - \Delta_{l+1}
		\right].
\end{calc}

Therefore, the total contribution (including both the left and right
sectors) to the mode Jacobian factor from the initial state is given
by
\begin{calc}
M_i &= |b|^{-\frac{2}{\kappa}\left[\frac{l}{2} + 1 - \Delta_{l+1}\right]}
	b^{-\frac{N}{\kappa}}\bar{b}^{-\frac{\bar{N}}{\kappa}}\\
     &= \frac{|u_2|^{2\left[\frac{l}{2} + 1 - \Delta_{l+1}\right]}}
	{|z_2|^{\frac{2}{\kappa}\left[\frac{l}{2} + 1 - \Delta_{l+1}\right]}}
        \frac{u_2^N\bar{u}_2^{\bar{N}}}{z_2^\frac{N}{\kappa}\bar{z}_2^\frac{\bar{N}}{\kappa}}.
\end{calc}

\subsubsection{The Mode Jacobian Factor from the Vertex Operator}

The vertex operator, we can see from
\[
\mathcal{O}^{(z)}_{m} = \left(\kappa \frac{z_2}{u_2}\right)^m\mathcal{O}_m^{(t)},
\]
contributes a factor of
\begin{equation}
M_v = \left(\kappa\tfrac{z_2}{u_2}\right)^{-h^\text{modes}_v}
      \left(\kappa\tfrac{\bar{z}_2}{\bar{u}_2}\right)^{-\bar{h}^\text{modes}_v},
\end{equation}
where $h^\text{modes}_v$ is the weight of the vertex operator less the
weight of the bare $(l+1)$-twist, that is,
\begin{equation}
h^\text{modes}_v = \bar{h}^\text{modes}_v
 = \frac{l}{2} + 1 - \Delta_{l+1}.
\end{equation}
Therefore, we see that the contribution to the mode Jacobian factor
from the vertex operator is given by
\begin{equation}
M_v = \kappa^{2\Delta_{l+1} - (l+ 2)}\left(\frac{|u_2|}{|z_2|}\right)^{l+2 - 2\Delta_{l+1}}.
\end{equation}

\subsection{The CFT Amplitude}\label{sec:A-prime}

Multiplying the mode Jacobian factor,
\begin{equation}
M = M_iM_v = \kappa^{2\Delta_{l+1} - (l+2)}\frac{|u_2|^{2(l+2) - 4\Delta_{l+1}}}
	{|z_2|^{\frac{\kappa+1}{\kappa}[l+2 - 2\Delta_{l+1}]}}
	\frac{u_2^N\bar{u}_2^{\bar{N}}}{z_2^\frac{N}{\kappa}\bar{z}_2^\frac{\bar{N}}{\kappa}},
\end{equation}
with the twist Jacobian factor,
\[
T = \kappa^{-2\Delta_{l+1}}\frac{|u_2|^{4\Delta_{l+1}}}
	{|z_2|^{2\frac{\kappa+1}{\kappa}\Delta_{l+1}}},
\]
one finds
\begin{equation}
TM = \kappa^{-(l+2)}\frac{|u_2|^{2(l+2)}}{|z_2|^{\frac{\kappa+1}{\kappa}(l+2)}}
     \frac{u_2^N\bar{u}_2^{\bar{N}}}{z_2^\frac{N}{\kappa}\bar{z}_2^\frac{\bar{N}}{\kappa}}.
\end{equation}

Finally, from Section~\ref{sec:kappa-equals-1} we take the $\kappa=1$
CFT amplitude (before spectral flow and \emph{without Jacobian
  prefactor $|z|^{l+2}$ from mapping from the cylinder to the complex
  plane}), which can be deduced from Equation~\eqref{eq:final-L-amp}:
\begin{equation}\label{eq:kappa1amp}
{\mathcal{A}'}^{\kappa=1} = (-1)^{k+\bar{k}}
	\sqrt{\choose{N+l+1}{N}\choose{\bar{N} + l+1}{\bar{N}}}
	\frac{1}{|u_2|^{2(l+2)}u_2^N\bar{u}_2^{\bar{N}}}.
\end{equation}
We use the Jacobian factor $TM$ to find
\begin{equation}\label{eq:final-primed-amp}
{\mathcal{A}'}^\kappa(z_2, \bar{z}_2) = \kappa^{-(l+2)}
	\sqrt{\choose{N+l+1}{N}\choose{\bar{N} + l+1}{\bar{N}}}
	\frac{1}{|z_2|^{\frac{\kappa+1}{\kappa}(l+2)}
		z_2^\frac{N}{\kappa}\bar{z}_2^\frac{\bar{N}}{\kappa}},
\end{equation}
which reduces to Equation~\eqref{eq:kappa1amp} for $\kappa=1$. 


\subsection{Combinatorics}\label{sec:comb}

As in Section~\ref{sec:kappa-equals-1}, we want to consider an initial
state with $\nu$ excitations that de-excites into a final states with
$\nu-1$ excitations (shown in Figure~\ref{fig:combinatorics-states});
however, the background geometry now consists of $\kappa$-twisted
component strings. We assume, therefore, that $\kappa$ divides
$N_1N_5$. Because the theory is orbifolded by $S_{N_1N_5}$, the
initial state, the final state, and the vertex operator must all be
symmetrized over the $N_1N_5$ copies.

We relate the full symmetrized states and operator to the amplitude
computed above, in which we do not worry about these combinatoric
factors.  The combinatorics for this problem are quite similar to and
for $\kappa=1$ reduce to the combinatorics in
Section~\ref{sec:combinatorics}.

\subsubsection{The Initial State}

We write the initial state as a sum over all the symmetric
permutations:
\begin{equation}\label{eq:sym-nu-sum}
\ket{\Psi_\nu} = \mathcal{C}_\nu\bigg[\ket{\psi_\nu^1} + \ket{\psi_\nu^2} + \dots\bigg],
\end{equation}
where $\mathcal{C}_\nu$ is the overall normalization and each
$\ket{\psi^i_\nu}$ is individually normalized. The initial state for
$\nu=2$, $\kappa=2$, and $l=2$ is shown in
Figure~\ref{fig:unphy-comb-in}.

To understand what we are doing better, note that the state
$\ket{\psi_{\nu=2}^1}$ with $\kappa =2$ and $l+1=3$ can be written
schematically as
\begin{equation}\label{eq:schematic-nu-state}
\ket{\psi_\nu^1} = \Ket{
	\big[1 \cdots 6\big]
	\big[7 \cdots 12\big]
	\big[12,13]\big[14,15]\cdots}
\end{equation}
where the numbers in the square brackets are indicating particular ways
of twisting individual strands corresponding to particular cycles of
the permutation group. For instance,
\begin{equation}
\ket{[1234]},
\end{equation}
indicates that we twist strand 1 into strand 2 into strand 3 into
strand 4 into strand 1 and leave strands 5 through $N_1N_5$ untwisted.
In Equation~\eqref{eq:schematic-nu-state}, the first two $6$-twists
are the two $\kappa(l+1)$-twists which indicate the two
excitations; the remaining $(\kappa=2)$-twists correspond to the
background.

To determine the normalization $\mathcal{C}_\nu$, we need to count the
number of distinct terms in Equation~\eqref{eq:sym-nu-sum}. To count
the number of states, we imagine constructing one of the terms and
count how many choices we have. First, of the $N_1N_5$ strands we must
pick $\nu\kappa(l+1)$ of them to make into the $\nu$ excited states.
Next, we must break the $\nu\kappa(l+1)$ strands into sets of
$\kappa(l+1)$ to be twisted together. Then, we must pick a way of
twisting together each set of $\kappa(l+1)$. Finally, we must make
similar choices for the $N_1N_5 - \nu\kappa(l+1)$ strands that are
broken into the $\kappa$-twists of the background. Putting these
combinatoric factors together, one finds
\begin{calc}
N_\text{terms} &= \choose{N_1N_5}{\nu\kappa(l+1)}\\
 &\qquad \times \choose{\nu\kappa(l+1)}{\kappa(l+1)}\choose{(\nu-1)\kappa(l+1)}{\kappa(l+1)}
                   \cdots \choose{\kappa(l+1)}{\kappa(l+1)}\frac{1}{\nu!}\times
                   \big([\kappa(l+1)-1]!\big)^\nu\\
 &\qquad \times \choose{N_1N_5 - \nu\kappa(l+1)}{\kappa}
                \cdots \choose{\kappa}{\kappa}
                \frac{1}{\left[\frac{N_1N_5}{\kappa} - \nu(l+1)\right]!}
                \times \big((\kappa -1)!\big)^{\frac{N_1N_5}{\kappa} - \nu(l+1)}\\
  &= \frac{(N_1N_5)!}{[\kappa(l+1)]^\nu\, \kappa^{\frac{N_1N_5}{\kappa} - \nu(l+1)}\,
                   \left(\frac{N_1N_5}{\kappa} - \nu(l+1)\right)!\,\nu!}.
\end{calc}
Choosing $\mathcal{C}_\nu$ to be real, one finds
\begin{equation}
\mathcal{C}_\nu = \frac{1}{\sqrt{N_\text{terms}}}
                = \left[
                  \frac{(N_1N_5)!}{[\kappa(l+1)]^\nu\, \kappa^{\frac{N_1N_5}{\kappa} - \nu(l+1)}\,
                   \left(\frac{N_1N_5}{\kappa} - \nu(l+1)\right)!\,\nu!}
                  \right]^{-\frac{1}{2}}.
\end{equation}

\subsubsection{The Final State}

\begin{figure}[t]
\begin{center}
\subfigure[~The initial state with $\nu=2$ excitations.]
{\label{fig:unphy-comb-in}
  \includegraphics[width=6cm]{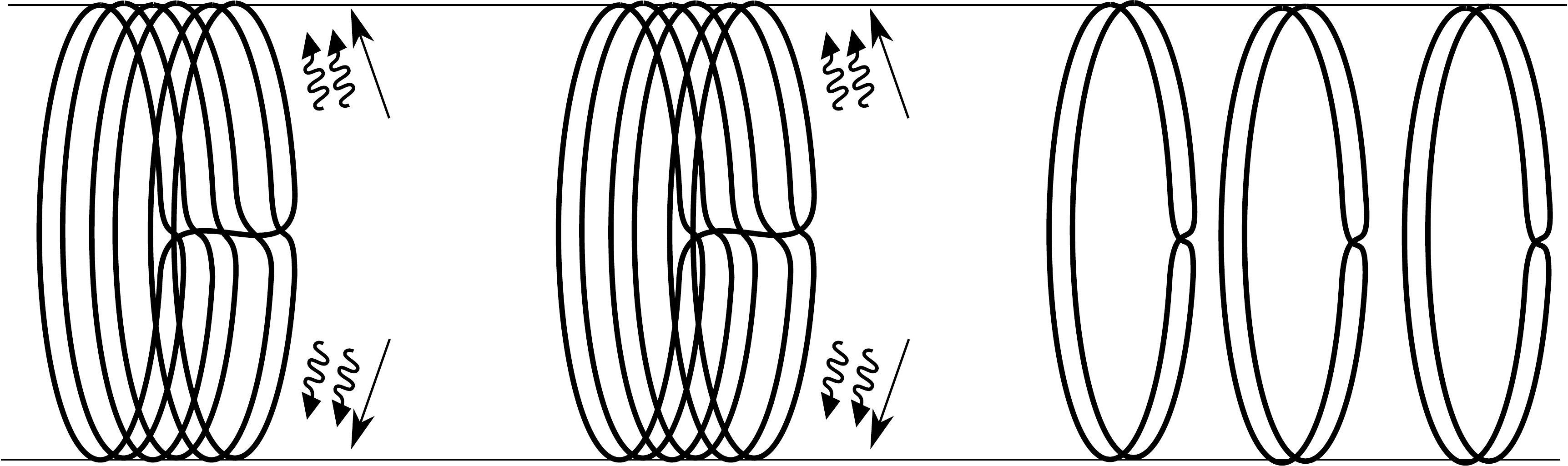}}
\hspace{30pt} \subfigure[~The final state with $\nu-1=1$ excitations.]
{\label{fig:unphy-comb-fin}
  \includegraphics[width=6cm]{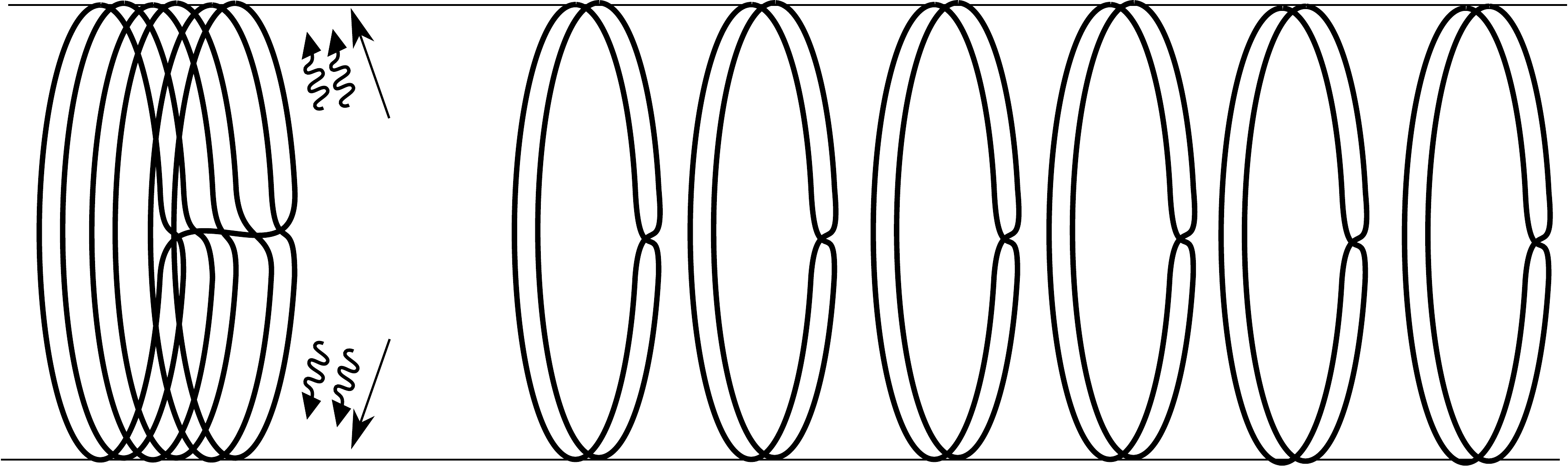}}
\caption[The initial and final states of the ``unphysical'' amplitude
for $\nu=2$ excitations]{The initial and final states of the
  unphysical amplitude with some excitations present in the
  background. An $l=2$ transition from $\nu=2$ excitations to $1$
  excitation of the $\kappa=2$ background is
  shown.}\label{fig:combinatorics-states}
\end{center}
\end{figure}

The final state is simply $\ket{\Psi_{\nu-1}}$ with its corresponding
normalization $\mathcal{C}_{\nu-1}$. This state is shown for $\nu=2$,
$\kappa=2$, and $l=2$ in Figure~\ref{fig:unphy-comb-fin}.

\subsubsection{The Vertex Operator}

The normalization of the \emph{full} vertex operator is determined in
Section~\ref{sec:combinatorics}. The full symmetrized vertex operator
is written
\begin{equation}
\mathcal{V}_\text{sym} = \mathcal{C}\sum_i\mathcal{V}_i,
\end{equation}
where
\begin{equation}
\mathcal{C}=\left[\frac{(N_1N_5)!}{(l+1)[N_1N_5 - (l+1)]!}\right]^{-\frac{1}{2}}.
\end{equation}

\subsubsection{The Amplitude}

The full amplitude of interest is
\begin{equation}
\bra{\Psi_{\nu-1}}\mathcal{V}_\text{sym}\ket{\Psi_\nu},
\end{equation}
which we wish to relate to a ``single'' unsymmetrized amplitude.
Having the normalization of the initial state, the final state, and
the vertex operator, all that remains is to determine which terms in
the symmetrized states and operator combine to give a nonzero
amplitude. For each initial state there are $\nu$ excitations that a
term in the vertex operator can de-excite. We are left with the
question, how many $(l+1)$-twists can untwist a given
$\kappa(l+1)$-twist into $l+1$ $\kappa$-twists?  In fact, there are
$\kappa$ such twist operators, which brings the number of nonzero
amplitudes to
\begin{equation}\label{eq:number-amplitudes}
N_\text{terms}\, \nu\kappa  = \kappa\frac{\nu}{\mathcal{C}_\nu^2}.
\end{equation}

To illustrate, consider the initial state in
Equation~\eqref{eq:schematic-nu-state} for $\nu=2$, $\kappa=2$, and
$l+1=3$. First, there are two different excitations to untwist.
Consider untwisting the first excitation $[123456]$. There are exactly
two $3$-twist vertex operators that untwist the excitation into three
sets of $2$-twists: $[531]$ and $[642]$. The $[531]$ vertex operator
breaks the initial state into $[12][34][56]$, while the $[642]$ vertex
operator breaks the initial state into $[61][23][45]$.

Using Equation~\eqref{eq:number-amplitudes} and the normalizations, we
can relate the total symmetrized amplitude to the amplitude computed
with only one term from the initial state, the final state, and the
vertex operator:
\begin{calc}\label{eq:full-comb-factor}
\bra{\Psi_{\nu-1}} \mathcal{V}_\text{sym} \ket{\Psi_{\nu}}
  &= \mathcal{C}_{\nu-1}\mathcal{C}\mathcal{C}_\nu\, \kappa\frac{\nu}{\mathcal{C}_\nu^2}
     \bra{\psi^1_{\nu-1}}\mathcal{V}_1 \ket{\psi^1_{\nu}}\\
  &= \sqrt{\kappa^{l+2}\nu
            \frac{\left[\frac{N_1N_5}{\kappa}-(\nu-1)(l+1)\right]!}
                 {\left[\frac{N_1N_5}{\kappa}-\nu(l+1)\right]!}
            \frac{\left[N_1N_5 - (l+1)\right]!}{(N_1N_5)!}}     
\bra{\psi^1_{\nu-1}}\mathcal{V}_1 \ket{\psi^1_{\nu}}.
\end{calc}

\subsubsection{The Large \texorpdfstring{$N_1N_5$}{N1N5} Limit}

While the expression in Equation~\eqref{eq:full-comb-factor} is
complicated, it simplifies considerably in the large $N_1N_5$ limit of
ultimate interest. We take both $N_1N_5$ and $N_1N_5/\kappa$ to be
large,
\begin{equation}\label{eq:comb-factors-limit}
\frac{\left[\frac{N_1N_5}{\kappa}-(\nu-1)(l+1)\right]!}
                 {\left[\frac{N_1N_5}{\kappa}-\nu(l+1)\right]!}
                \longrightarrow \left(\frac{N_1N_5}{\kappa}\right)^{(l+1)}
\hspace{20pt}
\frac{[N_1N_5-(l+1)]!}{(N_1N_5)!}
	\longrightarrow
	(N_1N_5)^{-(l+1)},
\end{equation}
in which case Equation~\eqref{eq:full-comb-factor} reduces to
\begin{equation}\label{eq:large-N-comb}
\bra{\Psi_{\nu-1}} \mathcal{V}_\text{sym} \ket{\Psi_{\nu}}
  = \sqrt{\kappa\nu}\, \bra{\psi^1_{\nu-1}}\mathcal{V}_1 \ket{\psi^1_{\nu}}.
\end{equation}
The $\sqrt{\nu}$ prefactor can be thought of as a ``Bose enhancement''
effect.

\subsection{The Rate of Emission}\label{sec:rate}

We now can put all of the pieces together to find the spectrum and
rate of emission. Plugging Equation~\eqref{eq:final-primed-amp} into
Equation~\eqref{eq:unphysical-to-physical} along with the combinatoric
factors from Equation~\eqref{eq:large-N-comb}, one finds the amplitude
for emission with angular momentum $(l, k+\bar{k}-l, \bar{k}-k)$
\begin{multline}
\mathcal{A}_{l, k+\bar{k}-l, \bar{k}-k} = \sqrt{\nu\kappa}\,\kappa^{-(l+2)}
 	\sqrt{\choose{N+l+1}{N}\choose{\bar{N} + l+1}{\bar{N}}}\\
	\times z^{\frac{1}{\kappa}(\frac{l}{2} + N + 1)-\alpha(\frac{l}{2}-k)}
   \bar{z}^{\frac{1}{\kappa}(\frac{l}{2} + \bar{N} + 1) - \bar{\alpha}(\frac{l}{2} - \bar{k})}.
\end{multline}
Plugging back in with the physical cylindrical coordinates,
\begin{equation}
z = e^{\frac{i}{R}(y+t)}\qquad \bar{z} = e^{-\frac{i}{R}(y-t)},
\end{equation}
we can read off the spectrum for emission,
\begin{equation}\begin{split}\label{eq:final-spectrum}
E_0 &= \frac{1}{\kappa R}\left[(\kappa\alpha + \kappa\bar{\alpha} - 2)\tfrac{l}{2}
		- \kappa(\alpha k + \bar{\alpha}\bar{k}) - N - \bar{N} - 2\right]\\
\lambda_0 &= \frac{1}{\kappa R}
	\left[-\kappa(\alpha - \bar{\alpha})\tfrac{l}{2} + \kappa(\alpha k - \bar{\alpha}\bar{k})
	 + N - \bar{N}\right],
\end{split}\end{equation}
where recall that $\alpha = 2n + 1/\kappa$ and
$\bar{\alpha}=2\bar{n}+1/\kappa$.  For there to be emission the energy
of the emitted particle must be positive, meaning that $E_0>0$.

\begin{sloppypar}
The unit amplitude with the $(\sigma, \tau)$ dependence removed is
\begin{equation}
\mathcal{A}_\text{unit}(0) = \sqrt{\nu} \kappa^{-l -\frac{3}{2}}
 	\sqrt{\choose{N+l+1}{N}\choose{\bar{N} + l+1}{\bar{N}}}.
\end{equation}
Section~\ref{sec:comb}, where $\nu$ is defined, calculates the
combinatorics for the \emph{unphysical} amplitude. For the
\emph{physical} amplitude, the combinatorics are the same except that
the emission process is a transition from $\nu-1$ to $\nu$
$\kappa(l+1)$-twists. Thus, the above is the amplitude for emission of
the $\nu$th particle. Plugging into
Equation~\eqref{eq:D1D5-decay-rate}, one finds the emission rate for
the $\nu$th particle,
\begin{equation}\label{eq:final-rate}
\der{\Gamma}{E} = \nu\kappa^{-2l-3}
	\frac{2\pi}{2^{2l+1}\,l!^2}\frac{(Q_1Q_5)^{l+1}}{R^{2l+3}}(E^2-\lambda^2)^{l+1}\,
	\choose{N+l+1}{N}\choose{\bar{N} + l+1}{\bar{N}}
	\delta_{\lambda,\lambda_0}\delta(E - E_0),
\end{equation}
with energy and momentum in Equation~\eqref{eq:final-spectrum} and
angular momentum $(l, k+\bar{k}-l,\bar{k}-k)$. This answer exactly
matches the gravity calculation in~\cite{cm3}.
\end{sloppypar}

\section{Discussion}\label{sec:discuss-emission}

We reproduced the full spectrum and emission rate of supergravity
minimal scalars from the geometries found in~\cite{ross} by using a
CFT formalism. In~\cite{cm1,cm2,cm3}, using a heuristic picture of the
CFT process the spectrum and rate was found, but only for special
cases ($N=\bar{N}=0$ and $k=\bar{k}=0$) and the normalization of the
heuristic vertex operator was determined only indirectly.  In
Section~\ref{sec:kappa-equals-1}, the full rate and spectrum was found
as a rigorous calculation for $\kappa=1$ with a vertex operator whose
coupling to flat space and normalization was determined directly from
the AdS--CFT correspondence.

The new feature of Section~\ref{sec:kappa-not-1} is the
$\kappa$-dependence. Using the old effective string description, the
$\kappa$-dependence in Equations~\eqref{eq:final-spectrum}
and~\eqref{eq:final-rate} could have been guessed via a heuristic
argument that we now describe.  Taking higher values of $\kappa$
corresponds to twisting the background strings by $\kappa$. The
process, then, looks the same as for $\kappa=1$ but taking $R \mapsto
\kappa R$. This reproduces the explicit $\kappa$ dependence of
Equation~\eqref{eq:final-rate}, but the spectrum is slightly more
complicated. In the spectrum, we also should take $R \mapsto \kappa
R$, since the energy level spacing becomes reduced by a factor of
$\kappa$; however, things are complicated by the parameters $n$ and
$\bar{n}$. Why does the $n$ and $\bar{n}$ part of the spectrum get
multiplied by $\kappa$ with respect to the rest of the contributions
to the energy? The parameters $n$ and $\bar{n}$ control the Fermi
level of the physical initial state. In the $\kappa$-cover, the
fermions fill up to energy level $\kappa n$ and not $n$; thus, the
extra factor of $\kappa$.

With the final answer and CFT computation in front of us, this
heuristic argument seems compelling indeed; however, we argue that it
is not completely obvious a priori that the effective string reasoning
works for this calculation. Certainly, the way that the
$\kappa$-dependence works out in the rigorous CFT calculation seems
quite nontrivial and nonobvious. That the action of the vertex
operator so simply gets ``divided by'' $\kappa$ does not seem obvious.
In any case, one of the goals in this chapter is to demonstrate the
formalism in Chapter~\ref{ch:coupling} and put the reasoning
in~\cite{cm1, cm2, cm3} on a firmer footing. Along the way, we found
(conjectured?) the form of supergravity excitations on the
orbifolded-AdS background; it would be nice to better understand the
identification of the supergravity multiplet in orbifolded-AdS.

That the rate of emission computed in this class of geometries can be
reproduced via a CFT calculation may seem insignificant in the face of
so many AdS--CFT successes. There are two reasons why this calculation
is of interest. First, most AdS--CFT calculations use a gravity
calculation in the AdS to compute a CFT correlator, whereas we use the
methods of Chapter~\ref{ch:coupling} to compute the rate of emission
\emph{out of the} CFT or AdS. This demonstrates the formalism of
Chapter~\ref{ch:coupling}~\cite{acm1}, which hearkens back to and puts
a firmer footing on the ``effective string'' calculations
of~\cite{stromvafa,radiation-1,radiation-2,radiation-3,radiation-4,
  radiation-5}. Second, as explained in~\cite{cm1, cm2, cm3}, the CFT
calculation justifies interpreting the ergoregion emission as Hawking
radiation from these, albeit nongeneric, microstates of a black hole.
This may help us better understand black holes in string theory and
thereby quantum gravity.

If one computes the CFT amplitude for emission from a thermally
distributed initial state and sums over all possible final states,
then one reproduces the full Hawking radiation spectrum. The
calculation was performed in~\cite{cv}, using precisely the formalism
developed in Chapter~\ref{ch:coupling}. The usual thermal field theory
technique of compactifying time was used, putting the CFT on a torus.
The answer precisely matched the gravity calculation. The important
point being that it is the \emph{same} vertex operator $\mathcal{V}$
in the CFT for Hawking radiation and for the ergoregion emission
described above.  The only difference being whether one considers a
particular nongeneric state of the CFT, or a thermal distribution.
This strongly suggests that the naive nonextremal black hole and its
Hawking radiation is the result of thermally averaging over
microstates. Some of which, at least, allow a geometric description.

All of the calculation were performed on the ``orbifold point'' of the
D1D5 CFT. The emission process described in the gravity side should be
dual to the CFT off of the orbifold point, and so it is an open
question why this and many other calculations on the orbifold point so
accurately reproduce the gravitational physics. Although, some
calculations should agree due to established non-renormalization
theorems~\cite{deBoer-attractor, dmw02}. In Chapter~\ref{ch:orbifold},
we discuss how to move off of the orbifold point.
\chapter{Moving off the Orbifold Point}
\label{ch:orbifold}

As discussed in Chapter~\ref{ch:d1d5}, the point in moduli space where
the orbifold CFT is a good description does not coincide with points
in moduli space where supergravity is a good description. If one wants
to understand black holes better, then one needs a CFT description at
points in moduli space where one also has black hole physics. In
Section~\ref{sec:marg-def}, we introduced the marginal deformations of
the orbifold CFT that move the CFT in its 20-dimensional moduli space.
Of the 20 marginal deformations, we are interested in the 4 twist
deformations that correspond to blow-up modes of the orbifolded target
space. While the supergravity is far from the orbifold point in moduli
space, we take a perturbative approach and study the effect of a
single application of the deformation operator. This work was
performed in~\cite{acm2, acm3, ac2}.  It is hoped that these results
will provide some qualitative and maybe even quantitative answers to
questions about black holes. Most of the analysis is deferred to later
work.

In Section~\ref{sec:the-def-op}, we recall the structure of the
marginal deformation operator of interest and make its structure more
explicit. In Section~\ref{sec:app-to-vac}, we find the state created
by the application of the deformation operator to the vacuum. The
state is a ``squeezed state'' with a long power-law
``tail''~\cite{acm2}. In Section~\ref{sec:excited}, we show how to
calculate the state created by the marginal deformation on excited
states~\cite{acm3}. In Section~\ref{sec:intertwining}, we show an
alternative method for computing the state created by the marginal
deformation on excited states. This alternative method may be
preferable because it comes in the form of intertwining relations for
individual modes~\cite{ac2}.  These can loosely be thought of as
Bogolyubov coefficients for bosonic and fermionic modes.
Unfortunately, this second method suffers from ambiguous
multidimensional series---the ambiguity is resolved via a physically
motivated prescription. In Section~\ref{sec:example} we use the
results from Section~\ref{sec:intertwining} in two different ways,
demonstrating the techniques. In Section~\ref{sec:orbifold-future}, we
briefly recapitulate the chapter, and outline some future directions
for which these results may be useful.

Before proceeding, let us alert the careful reader to some notational
changes made in this chapter. Since the entire discussion focuses on
going from two singly-wound copies to a single doubly-wound copy, we
find it convenient to use parenthetical superscripts to indicate the
copy number ``before the twist'' and no corresponding notation for
modes and fields ``after the twist.'' Furthermore, modes in the
twisted sector are defined with a factor of 2 with respect to previous
chapters. So $\alpha_{n}$ in this chapter corresponds to
$\alpha_{\frac{n}{2}}$ in previous chapters.

\section{The Deformation Operator}\label{sec:the-def-op}

The deformation operator of interest, $\mathcal{T}$, is introduced in
Section~\ref{sec:marg-def}. Recall that the deformation is a singlet
under $SU(2)_L\times SU(2)_R$.  To obtain such a singlet we apply
$G^\mp_{A, -\frac{1}{2}}$ to $\sigma_2^\pm$. A singlet under
$SU(2)_L\times SU(2)_R$ can be made as (writing both left and right
moving sectors)~\cite{gava-narain, gomis02, dmw99}
\begin{equation}
\mathcal{T}_{AB}\propto\frac{1}{4}
 \epsilon_{\alpha\beta}\epsilon_{\dot{\alpha}\dot{\beta}}  
\Big[\int_{w_0} \frac{\drm w}{2\pi i} G^\alpha_{A}(w)\Big]
\Big[\int_{\bar w_0}\frac{\drm\bar w}{2\pi i}{\bar G}^{\dot\alpha}_{B}(\bar w)\Big]
 \sigma_2^{\beta\dot\beta}\label{deformation}
\end{equation}
In Section~\ref{sec:short}, we demonstrate that $G^{-}_{-\frac{1}{2}}$
acting on a chiral primary gives the top member of $SU(2)_L$. In this
case, that means that $\mathcal{T}_{\dot{A}\dot{B}}$ is a singlet---a
fact pointed out in Section~\ref{sec:marg-def}. Thus we can write the
deformation operator as (we choose its normalization at this stage)
\begin{equation}
\mathcal{T}_{AB}=
\Big[\int_{w_0} \frac{\drm w}{2\pi i} G^-_{A} (w)\Big]
\Big[\int_{\bar w_0} \frac{\drm\bar w}{2\pi i}\bar{G}^-_{B}(\bar w)\Big]
   \sigma_2^{++}(w_0)
\end{equation}
The normalization of $\sigma_2^{++}$ is specified below. The indices
$\dot{A}$ and $\dot B$ can be contracted to rewrite the above four
operators as a singlet and a triplet of $SU(2)_1$.\footnote{Since we
  can write the deformation operator in terms of $G^-\sigma_2^+$ or in
  terms of $G^+\sigma_2^-$, it provides a good check on the results
  that we get the same final state each way.}

\begin{figure}[ht]
\begin{center}
\includegraphics[width=4cm]{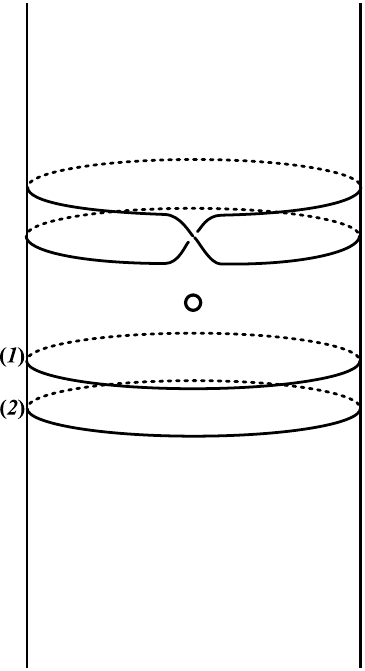}
\caption[The effect of the twisting part of the deformation]{The
  effect of the twist contained in the deformation operator: two
  circles at earlier times get joined into one circle after the
  insertion of the twist.}\label{2-twist}
\end{center}
\end{figure}

\subsection{Normalization of \texorpdfstring{$\sigma_2^+$}{2-twists}}\label{normsection}

Now we describe the normalization of $\sigma_2^{++}$. Let us focus on
the left moving part of the operator, which we denote by $\sigma_2^+$.
Let the conjugate operator be called $\sigma_{2, +}$, and normalize
these operators so that they have the OPE
\begin{equation}
\sigma_{2,+}(z')\sigma_2^{+}(z)\sim \frac{1}{z'-z}
\end{equation}
On the cylinder this implies
\begin{equation}
  \lim_{\tau'\to\infty} 
  \lim_{\tau\to-\infty} 
{}^{(1)}\bra{0} {}^{(2)}\langle 0|e^{\frac{1}{2}(\tau'+i\sigma)}
           \sigma_{2,+}(\tau'+i\sigma)e^{-\frac{1}{2}(\tau+i\sigma)}
           \sigma_2^+(\tau+i\sigma)|0\rangle^{(1)} |0\rangle^{(2)} = 1
\label{qwone}
\end{equation}
where $|0\rangle^{(1)}|0\rangle^{(2)}$ is the (untwisted) NS vacuum.

Let us perform a spectral flow~\eqref{eq:spectral} with parameter
$\alpha=-1$. This changes the untwisted NS vacuum as
\begin{equation}
\ket{0}^{(1)}\ket{0}^{(2)}\to \ket{0_R^{-}}^{(1)} \ket{0_R^{-}}^{(2)}
\end{equation}
We also have to ask what happens to the twist operator
$\sigma_2^+(w_0)$ under this spectral flow. As discussed in
Section~\ref{sec:use-spectral}, the action of spectral flow is simple
for operators where the fermion content can be expressed as a simple
exponential in the language where we bosonize the fermions.  For such
operators with charge $j$, spectral flow with parameter $\alpha$ leads
to a multiplicative factor~\cite{acm1}
\begin{equation}
\mathcal{O}_j(w)\to e^{-\alpha j w} \mathcal{O}_j(w)
\label{spectralformula}
\end{equation}
The operator $\sigma_2^+$ is indeed of this simple form~\cite{lm2}, so
we just get a multiplicative factor under spectral flow.  Its charge
is $q=\frac{1}{2}$, so we get
\begin{equation}
\sigma_2^+(w)\to e^{-\alpha \frac{1}{2} w} \sigma_2^+(w_0)
   =e^{\frac{1}{2} (\tau+i\sigma)} \sigma_2^+(\tau+i\sigma)
\end{equation}
The operator $\sigma_{2,+}$ has charge $-\frac{1}{2}$, so we get
\begin{equation}
\sigma_{2,+}(w')\to e^{\alpha \frac{1}{2} w'} \sigma_{2,+}(w')
  =e^{ -\frac{1}{2} (\tau'+i\sigma)} \sigma_{2,+}(\tau'+i\sigma)
\end{equation}
Thus under spectral flow the relation (\ref{qwone}) gives
\begin{equation}
\lim_{\tau'\to\infty} \lim_{\tau\to-\infty}{}^{(1)}\langle 0_{R,-}| {}^{(2)}\langle 0_{R,-}|~
\sigma_{2,+}(\tau'+i\sigma)\sigma_2^+(\tau+i\sigma)~|0_R^-\rangle^{(1)} |0_R^-\rangle^{(2)}=1
\label{qwonep}
\end{equation}
We write
\begin{equation}
|0^-_R\rangle\equiv \lim_{\tau\to-\infty}\sigma_2^+(\tau+i\sigma)~|0_R^-\rangle^{(1)} |0_R^-\rangle^{(2)}
\label{qwtwo}
\end{equation}
This is one of the two Ramond vacua of the CFT on the double circle.
The spin of this vacuum is $-\frac{1}{2}$, arising from the spin
$-\frac{1}{2}$ on each of the two initial Ramond vacua before twisting
and the spin $\frac{1}{2}$ of the twist operator $\sigma_2^+$.
Similarly, we write
\begin{equation}
\langle 0_{R,-}|\equiv \lim_{\tau'\to\infty} {}^{(1)}\langle 0_{R,-}| {}^{(2)}\langle 0_{R,-}|~
\sigma_{2,+}(\tau'+i\sigma)
\label{qsone}
\end{equation}
From the relation (\ref{qwonep}) we have
\begin{equation}
\langle 0_{R,-}|0^-_R\rangle=1
\label{qstwo}
\end{equation}
The relation (\ref{qwtwo}) implies that if we insert $\sigma_2^+$ at a
general point $w$, then we get a state of the form
\begin{equation}
\sigma_2^+(w)|0_R^-\rangle^{(1)} |0_R^-\rangle^{(2)}=|0_R^-\rangle+\dots
\label{qwthree}
\end{equation}
where the coefficient of the vacuum on the right-hand side (RHS) is
unity, and the `$\dots$' represent excited states of the CFT on the
doubly wound circle. We will use this relation below.

\section{Applying the Deformation Operator to the Vacuum}\label{sec:app-to-vac}

\begin{figure}[ht]
\begin{center}
\includegraphics[width=4cm]{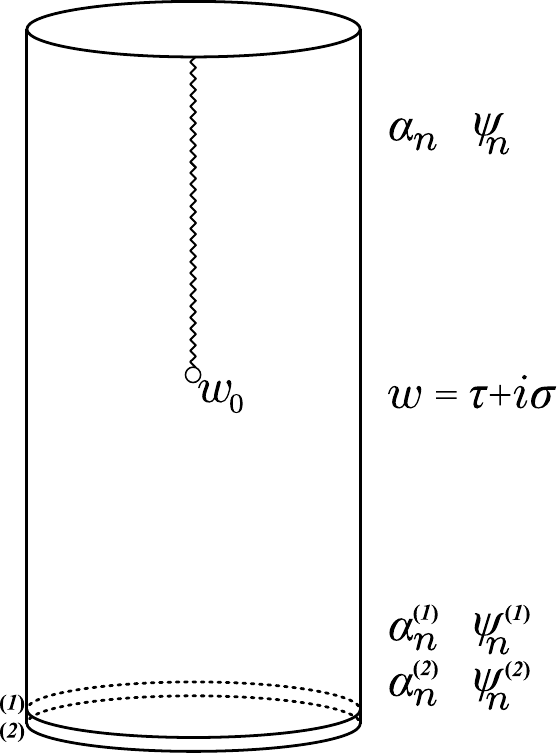}
\caption[The deformation operator on the cylinder]{Before the twist
  insertion we have boson and fermion modes on two copies of the $c=6$
  CFT. These modes are labeled with superscripts $(1)$ and $(2)$,
  respectively. The twist inserted at $w_0$ joins these to one copy
  for $\tau>\tau_0$; the modes here do not carry a superscript. The
  branch cut above $w_0$ indicates that we have two sets of fields at
  any given $\sigma$; these two sets go smoothly into each other as we
  go around the cylinder, giving a continuous field on a doubly wound
  circle.}\label{two}
\end{center}
\end{figure}

Consider the amplitude depicted in Figure~\ref{two}. Let us write down
all the states and operators in this amplitude. In the initial state,
we have two component strings. Since each is in the Ramond sector, we
have to choose one of the Ramond ground states. We choose
\begin{equation} \label{initial}
\ket{\Psi}_i=\ket{0_R^{--}}^{(1)}\ket{0_R^{--}}^{(2)},
\end{equation}
as the initial state. To this state we apply the deformation operator
at some point $w_0$ to arrive at the ``final state:'' 
\begin{equation}
\ket{\Psi}_f = \mathcal{T}_{\dot A\dot B}(w_0)\ket{\Psi}_i
     = \Big[\int_{w_0} \frac{\drm w}{2\pi i} G^-_{A}(w)\Big]
       \Big[\int_{\bar w_0} \frac{\drm\bar{w}}{2\pi i}\bar{G}^-_{B}(\bar{w})\Big]
       \,\sigma_2^{++}(w_0) |0_R^{--}\rangle^{(1)}|0_R^{--}\rangle^{(2)}
\label{pone}
\end{equation}
The final state contains one component string with winding number $2$,
since the deformation operator contains the twist $\sigma_2$. From
this stage on, we will write only the left moving part of the state,
and join it up with the right moving part at the end. Thus we write
\begin{equation}
|\Psi\rangle_f=|\psi\rangle|\bar\psi\rangle
\end{equation}
and work with $|\psi\rangle$ in what follows.

\subsection{Outline of the Computation}\label{outline}

Let us outline our steps for computing $|\psi\rangle$.

\begin{enumerate}
\item The essence of the computation lies in the nature of the
  deformation operator. This operator is given by a supercharge acting
  on the twist operator $\sigma_2^+$.  This supercharge is given by a
  contour integral of $G^-_{A}$ around the twist insertion. We
  first deform this contour to a pair of contours: one above and one
  below the insertion. These contours give zero modes of the
  supercurrent on the states before and after the twist insertion. We
  handle these zero modes at the end, and focus first on the state
  produced by just the twist insertion $\sigma_2^+$; we call this
  state $|\chi\rangle$.

\item Let us now look at the nature of the twist operator for bosonic
  fields. As we circle the twist, the two copies of the boson go into
  each other. The twist operator is defined by cutting a small hole
  around its insertion $w_0$, and taking boundary condition at the
  edge of this hole given by filling the hole in the {\it covering
    space} with a disc; i.e. there are no operator insertions in this
  covering space and we have just the vacuum state \cite{lm1}. To use
  this structure of the twist operator, we first map the cylinder to
  the plane via $z=e^w$, and then pass to the covering space $t$ by
  the map $z=z_0+t^2$ (here $z_0=e^{w_0}$ is the location of the
  twist). The small hole cut out on the cylinder around $w_0$ becomes
  a small hole around $t=0$. Since the boundary condition on the edge
  of this hole is generated by filling this hole by a disc, we get
  just the vacuum state at the origin in the $t$ plane. This
  observation takes into account the entire effect of the twist on the
  bosons.

\item On the cylinder we can specify the initial state of the system
  on the two circles at $\tau\to-\infty$ corresponding to the two
  copies of the $c=6$ CFT. On the $t$ plane these circles map to
  punctures at $t=\pm z_0^\frac{1}{2}\equiv \pm ia$. Since we have taken no
  bosonic excitations in our initial state, the bosonic part of the
  states at these punctures is just the vacuum, and we can close these
  punctures smoothly, just like the hole at $t=0$. Thus we have no
  insertions anywhere in the $t$ plane.

\item Our goal is to find the state at a circle $\tau\to\infty$ on the
  cylinder. But this circle maps to the circle $|t|=\infty$ on the $t$
  plane. Thus what we need is the state in the $t$ plane at infinity.
  But since there are no insertions anywhere on the $t$ plane, this
  state is just the $t$ plane vacuum $|0\rangle_t$. One might think
  that this means there are no excitations in the final state, but
  this is not the case: the vacuum on the $t$ plane is killed by
  positive energy modes defined with respect to the $t$ coordinate,
  and these will map to a linear combination of positive and negative
  energy modes in the original cylinder coordinate $w$. Thus all we
  have to do is express the state $|0\rangle_t$ in terms of the modes
  on the cylinder, and we would have obtained the bosonic part of the
  state arising from the twist insertion.

\item Let us now ask if we can guess the nature of this state in terms
  of the modes on the cylinder. In the treatment of quantum fields on
  curved space we often come across a situation where we have to
  express the vacuum with respect to one set of field modes in terms
  of operators defined with respect to another set of field modes. The
  state with respect to the latter modes has the form of an
  exponential of a quadratic, i.e. of the form
  $e^{\gamma_{mn}a^\dagger_m a^\dagger_n}|0\rangle$. The essential
  reason for getting this form for the state can be traced to the fact
  that free fields have a quadratic action, and if there are no
  operator insertions anywhere then in all descriptions of the state
  we can only observe exponentials of a quadratic form.

  For our problem, we make the ansatz that the state $\ket{\chi}$ has
  a similar exponential form. We find the $\gamma_{mn}$ by computing
  the inner product of the state with a state containing a pair of
  operator modes. In Appendix~\ref{ap:exp-ansatz} we prove that this
  exponential ansatz is indeed correct to all orders. A state of this
  form is frequently called a squeezed state in atomic physics.

\item Let us now ask if similar arguments can be applied to the
  fermionic field. The initial state on the cylinder has Ramond vacua
  for the two copies of the CFT. If we map to the $t$ plane these
  would give nontrivial states at $t=\pm ia$. Thus we first perform a
  spectral flow on the cylinder, which maps the problem to one where
  these Ramond vacua map to NS vacua at $\tau\to-\infty$ on the
  cylinder. These NS vacua will map to NS vacua at $t=\pm ia$, so
  there will be no operator insertions in the $t$ plane at these
  points. We should also check the effect of this spectral flow on the
  twist $\sigma_2^+(w_0)$.  From (\ref{spectralformula}) we find that
  $\sigma_2^+(w_0)$ will change by just a multiplicative constant.
  This constant will not matter at the end since we know the
  normalization of our final state by (\ref{qwthree}).

  We can now pass to the covering space $t$. We must now ask for the
  state at the edge of the hole around $t=0$. One finds that the
  fermions in the $t$ plane are antiperiodic around $t=0$ \cite{lm2}.
  Thus the state given by the operator $\sigma_2^+$ corresponds to
  having the positive spin Ramond vacuum $|0^+_R\rangle_t$.  As it
  stands this implies that we have a nontrivial state at $t=0$, but we
  perform another spectral flow, this time in the $t$ plane. Under
  this second spectral flow the Ramond vacuum $|0^+_R\rangle_t$ maps
  to the NS vacuum of the $t$ plane $|0\rangle_t$. We take the
  normalization
  \begin{equation} {}_t\langle 0|0\rangle _t=1
    \label{qsthree}
  \end{equation}
  for this NS vacuum. At this stage we have indeed no insertions
  anywhere in the $t$ plane, and the state at $t=\infty$ is just the
  $t$ plane vacuum for the fermions. Since all coordinate maps and
  spectral flows were linear in the operator modes, we again expect
  the state to be given by the exponential of a bilinear in fermion
  modes. We write such an ansatz, and find the coefficients in the
  exponential.

\item We can summarize the above discussion in the following general
  relation
  \begin{equation}
   \bra{0_{R,-}}\Big(\mathcal{O}_1 \mathcal{O}_2 \dots \mathcal{O}_n\Big) 
                      \sigma_2^+(w_0)\ket{0_R^-}^{(1)} \ket{0_R^-}^{(2)}
    ={}_t\bra{0}\Big(\mathcal{O}'_1, \mathcal{O}'_2, \dots \mathcal{O}'_n\Big)\ket{0}_t
    \label{master}
  \end{equation}
  The state $\bra{0_{R,-}}$ is defined in (\ref{qsone}).
  $\mathcal{O}_i$ are any operators inserted after the twist insertion
  (we will need to insert boson and fermion modes in finding the
  coefficients $\gamma_{mn}$). On the RHS, the operators
  $\mathcal{O}'_i$ are obtained by mapping the operators
  $\mathcal{O}_i$ through all coordinate changes and spectral flows
  till we reach the $t$ plane with the NS vacuum at $t=0$. The
  normalizations (\ref{qstwo}) and (\ref{qsthree}) tell us that there
  is no extra constant relating the left-hand side (LHS) to the
  right-hand side (RHS) of (\ref{master}).

\item After obtaining the state $|\chi\rangle$ generated by the action
  of the twist $\sigma_2^+$ on
  $|0_R^-\rangle^{(1)}|0_R^-\rangle^{(2)}$, we act with the zero mode
  of the supercurrent to obtain the final state $|\psi\rangle$
  obtained by the action of the full deformation operator on
  $|0_R^-\rangle^{(1)}|0_R^-\rangle^{(2)}$.
\end{enumerate}

\subsection{Mode Expansions on the Cylinder}

Let us give the mode expansions on the cylinder, where the CFT is
defined. Below the twist insertion, $\tau < \tau_0$, the fields are
$2\pi$-periodic and we define the modes as
\begin{subequations}\begin{align}\label{eq:cylinder-modes-before}
\alpha^{(i)}_{A\dot A, n}&= 
    \int_{\sigma=0}^{2\pi} \frac{\drm w}{2\pi}\pd_w X^{(i)}_{A\dot A}(w) e^{nw} 
                \qquad i=1,2\\
\psi^{(i)\alpha \dot{A}}_n &=  
\int_{\sigma=0}^{2\pi}\frac{\drm w}{2\pi i} \psi^{(i)\alpha \dot{A}}(w) e^{nw} \qquad i=1,2
\end{align}\end{subequations}
which gives the inverse relationship,
\begin{subequations}\begin{align}
\pd_w X^{(i)}_{A\dot A}(w) &=
   -i \sum_n \alpha^{(i)}_{A\dot A, n} e^{-nw}\qquad i=1,2\label{pfthree}\\
\psi^{(i)\alpha \dot{A}}(w) &= \sum_n \psi^{(i)\alpha\dot A}_n e^{-n w}\qquad i=1,2.
\end{align}\end{subequations}
The commutation relations are
\begin{subequations}\begin{align}
\com{\alpha^{(i)}_{A\dot A, m}}{\alpha^{(j)}_{B\dot B, n}} &=
                -\epsilon_{AB}\epsilon_{\dot A\dot B}\delta^{ij} m \delta_{m+n,0}\\
\ac{\psi^{(i)\alpha\dot A}_m}{\psi^{(j)\beta\dot B}_n} &= 
                -\epsilon^{\alpha\beta}\epsilon^{AB}\delta^{ij}\delta_{m+n,0}
\end{align}\end{subequations}

Above the twist, $\tau > \tau_0$, we have a doubly twisted circle and
the $4\pi$-periodic. We have a choice of normalization in how we
define modes on the doubly wound circle, and we take 
\begin{subequations}\begin{align}\label{qaone}
\alpha_{A\dot{A}, n} &= 
   \int_{\sigma=0}^{4\pi}\frac{\drm w}{2\pi} \pd_w X_{A\dot A}(w)e^{\frac{n}{2}w}\\
\psi^{\alpha\dot A}_n &= \int_{\sigma=0}^{4\pi}\frac{\drm w}{2\pi i} \psi^{\alpha\dot A}(w) e^{\frac{n}{2} w}
\end{align}\end{subequations}
Taking the normalizations as above, one finds 
\begin{subequations}\begin{align}
\pd_w X_{A\dot{A}}(w) &=-\frac{1}{2} i \sum_n \alpha_{A\dot A, n} e^{-\frac{n}{2}w} \\
\psi^{\alpha\dot A}(w)    &= \frac{1}{2}   \sum_n \psi^{\alpha\dot A}_n e^{-\frac{n}{2}w}.
\end{align}\end{subequations}
Note the factor of $\frac{1}{2}$ that appears in these equations.  The
commutation relations turn out to be 
\begin{subequations}\begin{align}
\com{\alpha_{A\dot A, m}}{\alpha_{B\dot{B}, n}}
   &= -\epsilon_{AB}\epsilon_{\dot A\dot B} m \delta_{m+n,0}\label{bcommtwist}\\
 \ac{\psi^{\alpha\dot A}_m}{\psi^{\beta\dot  B}_n}
   &= -2\epsilon^{\alpha\beta}\epsilon^{\dot A\dot B}\delta_{m+n,0} \label{fcommtwist}.
\end{align}\end{subequations} 
Again note the factor of $2$ in the fermion relation. The difference
between the boson and the fermion cases arises from the fact that they
have different scaling dimensions.

\subsection{The \texorpdfstring{$G^-_{A, -\frac{1}{2}}$}{G} Operator}\label{sec:G-contour}

\begin{figure}[ht]
\begin{center}
\includegraphics[width=8cm]{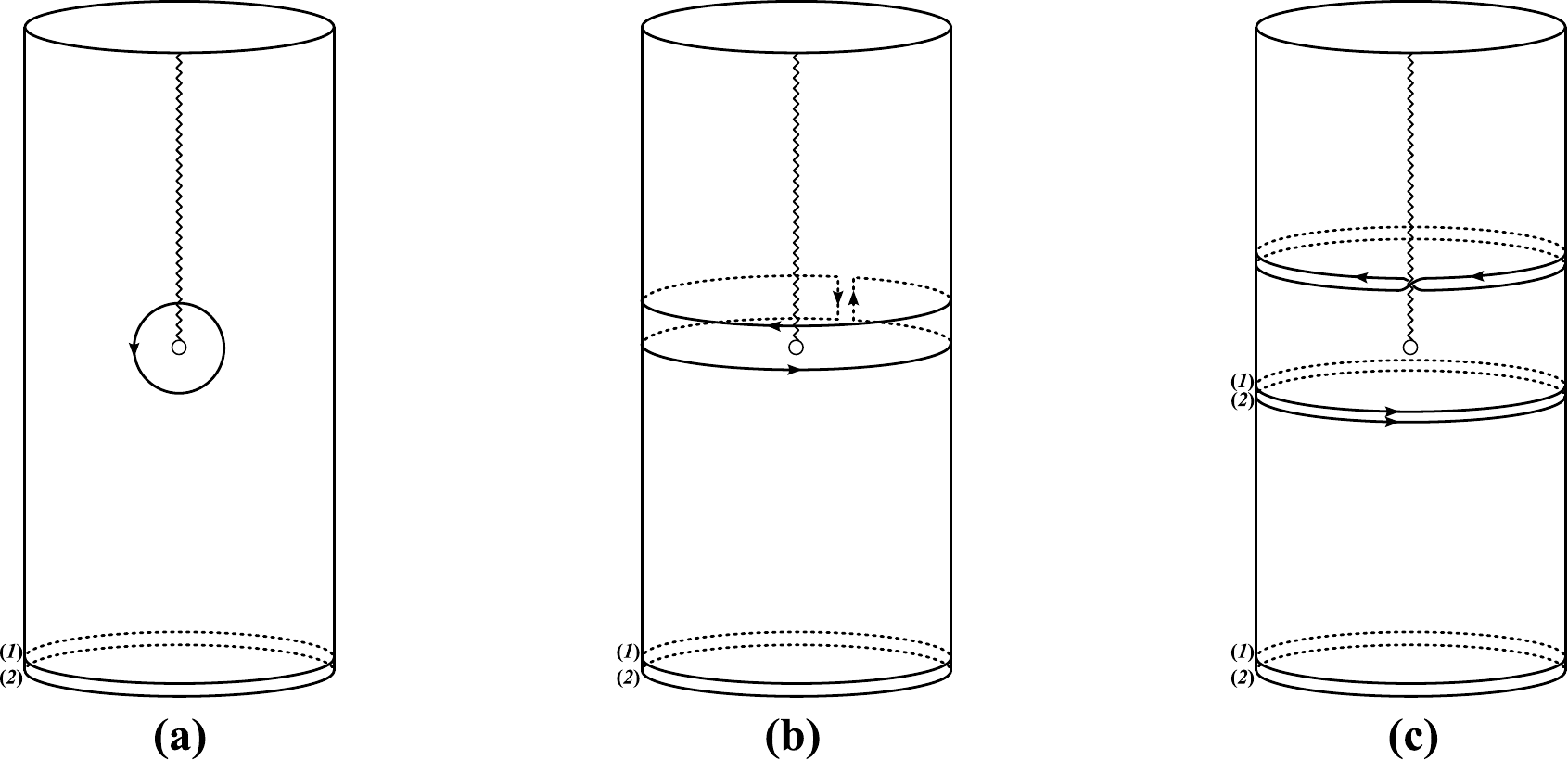}
\caption[Manipulating the supercharge contour]{(a) The supercharge in
  the deformation operator is given by integrating $G^-_{ A}$
  around the insertion at $w_0$. (b) We can stretch this contour as
  shown, so that we get a part above the insertion and a part below,
  joined by vertical segments where the contributions cancel. (c) The
  part above the insertion gives the zero mode of the supercharge on
  the doubly wound circle, while the parts below give the sum of this
  zero mode for each of the two initial copies of the
  CFT.}\label{G-contour}
\end{center}
\end{figure}

Let us first put the $\int_{w_0} \frac{\drm w}{2\pi i} G^-_{A}(w)$ operator in (\ref{pone}) in a more convenient form. The
contour in this operator runs circles the insertion $w_0$
(Figure~\ref{G-contour}(a)). We can stretch this to a contour that
runs around the rectangle shown in figure~\ref{G-contour}(b). The
vertical sides of the contour cancel out.  We can thus break the
contour into a part above the insertion and a part below the insertion
(Figure~\ref{G-contour}(c)). The lower leg gives
\begin{equation}
\int_{w=\tau_0-\epsilon}^{\tau_0-\epsilon+2\pi i}\frac{\drm w}{2\pi i}G^-_{A}(w)
   =\int_{w=\tau_0-\epsilon}^{\tau_0-\epsilon+2\pi i}\frac{\drm w}{2\pi i}
    [G^{(1)-}_{A}(w)+G^{(2)-}_{A}(w)]
   \equiv G^{(1)-}_{A, 0}+G^{(2)-}_{A, 0}
\end{equation}
The upper leg gives
\begin{equation}
-\int_{w=\tau_0+\epsilon}^{\tau_0+\epsilon+4\pi i}\frac{\drm w}{2\pi i}G^-_{A}(w)
        \equiv -G^-_{A, 0}
\end{equation}
where we note that the two copies of the CFT have linked into one  copy on a doubly wound circle, and we just get the zero mode of $G^-_{A}$ on this single copy.

Note that
\begin{equation}
G^{(i)-}_{A, 0}|0_R^{--}\rangle^{(i)}=0, \qquad i=1,2
\end{equation}
since the $\alpha$ index of $|0_R^{\alpha -}\rangle^{(i)}$ forms a
doublet under $SU(2)_L$, and we cannot further lower the spin of
$|0_R^{--}\rangle^{(i)}$ without increasing the energy level. Thus the
lower contour gives nothing, and we have
\begin{equation}
|\psi\rangle = -G^-_{A, 0}\sigma_2^{+}(w_0)|0_R^{-}\rangle^{(1)} |0_R^{-}\rangle^{(2)}.
\label{ptwo}
\end{equation}
We find it convenient to write this as
\begin{equation}
|\psi\rangle=-G^-_{A, 0}|\chi\rangle,
\end{equation}
where
\begin{equation}\label{eq:chiq}
|\chi\rangle=\sigma_2^{+}(w_0)|0_R^{-}\rangle^{(1)} |0_R^{-}\rangle^{(2)}.
\end{equation}

\subsection{Ansatz for \texorpdfstring{$|\chi\rangle$}{chi}}

Let us leave aside the action of $G^-_{A, 0}$ for the moment, and
consider the state $|\chi\rangle$.

The spin of $|\chi\rangle$ is $-\frac{1}{2}$, since each of the two
Ramond ground states have spin $-\frac{1}{2}$ and $\sigma_2^+$ has
spin $\frac{1}{2}$. Further, the fermions on the double circle
produced after the twist are periodic after we go around the double
circle; this follows by noting that these fermions were periodic on
each of the two copies before the twist, and we have simply cut and
rejoined the copies into one long loop. Thus the state $|\chi\rangle$
can be considered as built up by adding excitations (with total charge
zero) to the Ramond vacuum $|0^-_R\rangle$ of the doubly twisted
theory (we assume that this vacuum is normalized to unit norm)
\begin{equation}
|\chi\rangle=\prod \{ \alpha_{C\dot C, m_i}\} \prod \{ \psi^{\beta}_{\dot D, n_j}\} |0^-_R\rangle
\label{pfour}
\end{equation}
As discussed in section (\ref{outline}), the state $|\chi\rangle$
should have the form of an exponential in the boson and fermion
creation operators. Let us write down the ansatz and then explain some
of its points. In Appendix~\ref{ap:exp-ansatz} we show that for the
bosonic case this ansatz is correct to all orders in the number of
excitations; the fermionic case can be done in a similar way. We 
write
\begin{equation}\label{pfive}
\ket{\chi} = 
\exp\left[-\frac{1}{2}\sum_{ m\ge 1, n\ge 1}\gamma^B_{mn}
        \epsilon^{AB}\epsilon^{\dot{A}\dot{B}}
                \alpha_{A\dot{A},-m}\alpha_{B\dot{B},-n}
+\sum_{m\ge 0,n\ge 1}\gamma^F_{mn}\epsilon_{\dot{A}\dot{B}}
               \psi^{+\dot{A}}_{-m}\psi^{-\dot{B}}_{-n}\right]\ket{0^-_R}
\end{equation}
Below we define more precisely the modes $\alpha_{A\dot A, n},
\psi^{\alpha\dot A}_n$ on the double circle. For now, let us note some points
about the above expression:
\begin{enumerate}
\item From eq. (\ref{qwthree}) we see that there will not be any
  additional multiplicative constant on the RHS; the coefficient of
  first term obtained by expanding the exponential (i.e. the number
  unity) is set by (\ref{qwthree}).

\item The initial state $|0^-_R\rangle$ is a singlet of $SO(4)_I$, the
  symmetry group in the torus directions. The operator $\sigma^+_2$ is
  a singlet under this group also. Thus the state $|\chi\rangle$ will
  have to be a singlet under this group. Thus we have written the
  ansatz in a way that the $A$ and $\dot A$ indices of $\alpha_{A\dot
    A}$ and $\psi^{\alpha\dot A}$ are grouped to make singlets under
  $SO(4)_I$.

\item We have
  \begin{equation}
    \alpha_{A\dot A, 0}|0^-_R\rangle=0
  \end{equation}
  since there is no momentum in the state. Thus the mode summations
  for the bosons start with $m,n=1$ and not with $m,n=0$.

\item We have
  \begin{equation}
    \psi_0^{-\dot A}|0^-_R\rangle=0
  \end{equation}
  Thus the sum over fermion modes starts with $n=1$ for $\psi^{--},
  \psi^{-+}$ and with $m=0$ for $\psi^{++}, \psi^{+-}$.  By writing
  modes this way we remove a normal ordering term that can arise from
  the anticommutation of zero modes; such a contribution would then
  have to be reabsorbed in an overall normalization constant in front
  of the exponential. We will find later that the value $m=0$ does not
  occur either because the $\gamma^F_{mn}$ vanish for that case; in
  fact we will find that $\gamma^B_{mn}$ and $\gamma^F_{mn}$ are
  nonzero only for $m,n$ odd.
\end{enumerate}

\subsection{The First Spectral Flow}

Let us continue to work with the state $\ket{\chi}$, and return to
$\ket{\psi}$ at the end. First, we perform a spectral
flow~\eqref{eq:spectral} with parameter $\alpha=1$. Let the resulting
state be called $\ket{\chi}_{\alpha=1}$. This spectral flow has the
following effects:
\begin{enumerate}
\item The Ramond ground states $\ket{0_R^{-}}^{(i)}$, $i=1,2$ change
  to NS vacua
  \begin{equation}
    |0_R^{-}\rangle^{(1)}\to |0\rangle^{(1)}\qquad
    |0_R^{-}\rangle^{(2)}\to |0\rangle^{(2)}.
  \end{equation}

\item The operator $\sigma_2^+$ changes by a phase which depends on
  its charge $q$; this charge is $q=\frac{1}{2}$.  So we get
  \begin{equation} 
    \sigma_2^+(w_0)\to e^{-\frac{1}{2} w_0} \sigma_2^+(w_0)
  \end{equation}
  Thus we get
  \begin{equation}
    |\chi\rangle_{\alpha=1}=e^{-\frac{1}{2}w_0}\sigma_2^+(w_0)|0\rangle^{(1)} |0\rangle^{(2)}
  \end{equation}

\item Modes of bosonic operators are not affected. The fermionic field
  changes as follows
  \begin{equation}\label{pseven}
    \psi^{+ \dot A}(w)\to e^{-\frac{1}{2} w} \psi^{+\dot A}(w), \qquad
    \psi^{-\dot A}(w)\to e^{\frac{1}{2} w} \psi^{-\dot A}(w)
  \end{equation}
Thus fermionic modes change as follows
\begin{subequations}\begin{align}\label{psix}
\psi^{(i)\pm\dot A}_n &\to \int_{\sigma=0}^{2\pi}\frac{\drm w}{2\pi i} 
                    \psi^{(i)\pm\dot A}(w) e^{(n\mp\frac{1}{2})w}, \qquad i=1,2\\
\psi^{\pm\dot A}_n &\to \int_{\sigma=0}^{4\pi}\frac{\drm w}{2\pi i} \psi^{\pm\dot A}(w) 
                    e^{\frac{(n\mp 1)}{2} w}.
\end{align}\end{subequations}

\item For $\tau>\tau_0$, we have one copy of the CFT on a doubly long
  circle. Before the spectral flow, the state in this region was built
  by applying excitations to $|0^-_R\rangle$ in Equation\eqref{pfour}.
  Under the spectral flow we have
  \begin{equation}
    |0^-_R\rangle\to |0^+_R\rangle.
  \end{equation}
\end{enumerate}

\subsection{Mode Expansions on the \texorpdfstring{$z$}{z} Plane}

We wish to go to a covering space which will allow us to see
explicitly the action of the twist operator. As a preparatory step, it
is convenient to map the cylinder with coordinate $w$ to the plane
with coordinate $z$
\begin{equation}
z=e^w
\end{equation}
Under this map the operator modes change as follows. Before the
insertion of the twist ($|z|<e^{\tau_0}$) we have
\begin{subequations}\begin{align}
\alpha^{(i)}_{A\dot A, n} &\to \int_{z=0}\frac{\drm z}{2\pi} \pd_z X^{(i)}_{A\dot A}(z) z^n, 
                  \qquad i=1,2\\
\psi^{(i)+\dot A}_n &\to \int_{z=0}\frac{\drm z}{2\pi i} \psi^{(i)+\dot A}(z) z^{n-1},
                  \qquad i=1,2\\
\psi^{(i)-\dot A}_n &\to \int_{z=0}\frac{\drm z}{2\pi i} \psi^{(i)-\dot A}(z) z^{n}, 
                  \qquad i=1,2
\end{align}\end{subequations}
After the twist ($|z|>e^{\tau_0}$) we have
\begin{subequations}\begin{align}
\alpha_{A\dot A, n} &\to \int_{z=\infty}\frac{\drm z}{2\pi} \pd_z X_{A\dot A}(z) z^{\frac{n}{2}}\\
\psi^{+\dot A}_n &\to \int_{z=\infty}\frac{\drm z}{2\pi i} \psi^{+\dot A}(z) z^{\frac{n-2}{2}}\\
\psi^{-\dot A}_n &\to \int_{z=\infty}\frac{\drm z}{2\pi i} \psi^{-\dot A}(z) z^{\frac{n}{2}}
\end{align}\end{subequations}

\subsection{Mapping to the Covering Space}

\begin{figure}[ht]
\begin{center}
\includegraphics[width=5cm]{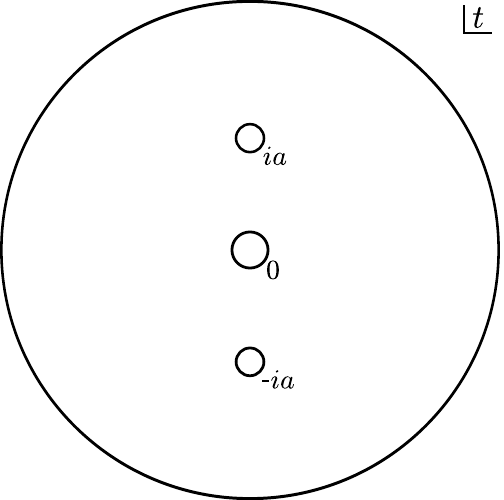}
\caption[The $t$-plane for a $\sigma_2(z_0)$]{The $z$ plane is mapped
  to the covering space -- the $t$ plane -- by the map $z=z_0+t^2$.
  The point $z=0$ corresponds to $\tau\to -\infty$ on the cylinder,
  and the two copies of the CFT there correspond to the points $t=\pm
  ia$. The location of the twist operator maps to $t=0$. The top the
  cylinder $\tau\to\infty$ maps to $t\to \infty$. After all maps and
  spectral flows, we have the NS vacuum at $t=0, \pm ia$, and so we
  can smoothly close all these punctures. The state $|\chi\rangle$ is
  thus just the $t$ plane vacuum; we must write this in terms of the
  original cylinder modes and apply the supercharge to get the final
  state $|\psi\rangle$.}\label{t-plane}
\end{center}
\end{figure}

We pass to the cover of the $z$ plane via the map
\begin{equation}
z=z_0+t^2
\end{equation}
where
\begin{equation}
z_0=e^{w_0}\equiv a^2
\end{equation}
Thus the map from the $z$ plane to the $t$ plane has second order
branch points at $z=z_0$ (the location of the twist $\sigma_2^+$) and
at infinity (corresponding to the top of the cylinder, where we can
imagine the dual twist $\sigma_{2,+}$ being placed). Under this map we
have the following changes:
\begin{enumerate}
\item Consider the circle on the cylinder at $\tau\to -\infty$; this is
  the location of the initial states on the cylinder. This circle maps
  to $z=0$, and then to $t=\pm ia$ on the $t$ plane. There were two
  copies of the CFT at $\tau\to-\infty$ on the cylinder, and the
  initial state of one copy (copy $(1)$) will map to the point $t=ia$
  and the state of the other copy (copy $(2)$) will map to $t=-ia$.

  Note that in our present problem the states of these copies, were
  $|0^-_R\rangle^{(i)}$ to start with, which became NS vacua
  $|0\rangle^{(i)}$ after the first spectral flow. Now when we map
  them to the $t$ plane we find that there is just the $t$ plane
  vacuum at the points $\pm ia$, so we may smoothly close the
  punctures at these points with no insertions at the puncture.

\item The location of the twist insertion $\sigma_2^+$ maps to $t=0$.
  At this location we have the state $|0^+_R\rangle_t$, the spin up
  Ramond vacuum of the $t$ plane.

\item The operator modes before the twist become
\begin{subequations}\begin{align}\label{eq:t-modes-before}
\alpha^{(1, 2)}_{A\dot A, n} &\to  \int_{t=\pm ia}\frac{\drm t}{2\pi} \pd_t X_{A\dot A}(t) (z_0+t^2)^n\\
\psi^{(1, 2)+\dot A}_n &\to  \int_{t=\pm ia}\frac{\drm t}{2\pi i}
                                  \psi^{+\dot A}(t) (z_0+t^2)^{n-1}\sqrt{2t}\\ 
\psi^{(1, 2)-\dot A}_n &\to  \int_{t=\pm ia}\frac{\drm t}{2\pi i} \psi^{-\dot A}(t) (z_0+t^2)^{n} \sqrt{2t}
\end{align}\end{subequations}
After the twist we have
\begin{subequations}\begin{align}\label{eq:t-modes-after}
\alpha_{A\dot A, n} &\to \int_{t=\infty}\frac{\drm t}{2\pi} 
                     \pd_t X_{A\dot A}(t) (z_0+t^2)^{\frac{n}{2}}\\
\psi^{+\dot A}_n &\to \int_{t=\infty}\frac{\drm t}{2\pi i}\psi^{+\dot A}(t) (z_0+t^2)^{\frac{n-2}{2}} \sqrt{2t}\\
\psi^{-\dot A}_n &\to \int_{t=\infty}\frac{\drm t}{2\pi i}\psi^{-\dot A}(t) (z_0+t^2)^{\frac{n}{2}} \sqrt{2t}
\end{align}\end{subequations}
\end{enumerate}

\subsection{The Second Spectral Flow}

We have now mapped the problem to the $t$ plane, where we have found a
state $|0^+_R\rangle_t$ at $t=0$, and no other insertions anywhere.
Let us perform a spectral flow with $\alpha=-1$ in the $t$ plane. This
has the following effects
\begin{enumerate}
\item The Ramond ground state at $t=0$ maps to the NS vacuum in the
  $t$ plane,
  \begin{equation}
    |0^+_R\rangle_t\to |0\rangle_t.
  \end{equation}

\item The operator modes change as follows. The bosons are not
  affected. The fermion field changes as
  \begin{equation}
    \psi^{\pm\dot A}(t) \to t^{\pm \frac{1}{2}} \psi^{\pm\dot A}(t).
  \end{equation}
\end{enumerate}
Before the twist we get the modes
\begin{subequations}\begin{align}
\alpha^{(1, 2)}_{A\dot A, n} &\to
       \int_{t=\pm ia}\frac{\drm t}{2\pi} \pd_t X_{A\dot A}(t) (z_0+t^2)^n\\
\psi^{(1,2)+\dot A}_n &\to \sqrt{2} \int_{t= \pm ia}\frac{\drm t}{2\pi i} \psi^{+\dot A}(t) (z_0+t^2)^{n-1}t \\
\psi^{(1,2)-\dot A}_n &\to \sqrt{2} \int_{t= \pm ia}\frac{\drm t}{2\pi i} \psi^{-\dot A}(t) (z_0+t^2)^{n}\\
\end{align}\end{subequations}
After the twist we have
\begin{subequations}\begin{align}\label{qatwo}
\alpha_{A\dot A, n} &\to \int_{t=\infty}\frac{\drm t}{2\pi}\pd_t X_{A\dot A}(t) (z_0+t^2)^{\frac{n}{2}}\\
\psi^{+\dot A}_n &\to\sqrt{2}\int_{t=\infty}\frac{\drm t}{2\pi i} \psi^{+\dot A}(t) (z_0+t^2)^{\frac{n-2}{2}}t\\
\psi^{-\dot A}_n &\to\sqrt{2}\int_{t=\infty}\frac{\drm t}{2\pi i} \psi^{-\dot A}(t) (z_0+t^2)^{\frac{n}{2}}
\end{align}\end{subequations}

In the $t$ plane, we also find it convenient to define mode operators
that are natural to the $t$ plane,
\begin{subequations}\begin{align}\label{qathree}
\tilde{\alpha}_{A\dot A, n} &\equiv  \int_{t=0}\frac{\drm t}{2\pi} \pd_t X_{A\dot A}(t) t^n\\
\tilde{\psi}^{\alpha\dot A}_r &\equiv \int_{t=0}\frac{\drm t}{2\pi i} \psi^{\alpha\dot A}(t) t^{r-\frac{1}{2}}.
\end{align}\end{subequations}
Note that the bosonic index $n$ is an integer while the fermionic
index $r$ is a half integer.  We have
\begin{subequations}\begin{align}
\tilde{\alpha}_{A \dot A,m}|0\rangle_t=0,\qquad m\ge 0\label{ptthree}\\
\tilde{\psi}^{\alpha\dot A}_r|0\rangle_t=0, \qquad r>0.\label{ptthreep}
\end{align}\end{subequations}
The commutation relations are
\begin{subequations}\begin{align}
\com{\tilde{\alpha}_{A \dot A}}{\tilde{\alpha}_{B \dot B}} &=
      -\epsilon_{A B} \epsilon_{\dot A \dot B} m\delta_{m+n, 0}\label{pttwo}\\
\ac{\tilde{\psi}^{\alpha\dot A}_r}{\tilde{\psi}^{\beta\dot B}_s} &=
      -\epsilon^{\alpha\beta}\epsilon^{AB}\delta_{r+s, 0}
\label{pttwop}
\end{align}\end{subequations}

\subsection{Computing \texorpdfstring{$\gamma^B_{mn}$ and $\gamma^F_{mn}$}{gamma-B and gamma-F }}

In this section we compute the coefficients $\gamma^B_{mn}$ and
$\gamma^F_{mn}$. For this computation we use the relation
(\ref{master}) to relate correlators of operators on the cylinder to
correlators on the $t$ plane. The latter correlators can be computed
very simply, and we thereby find the coefficients $\gamma^B_{mn}$ and
$\gamma^F_{mn}$.

\subsubsection{The Bosonic Coefficients \texorpdfstring{$\gamma^B_{mn}$}{gamma-B}}

Let us compute
\begin{equation}\label{ptone}
\bra{0_{R,-}}\Big (\alpha_{++,_n}\alpha_{--,_m}\Big)
       \sigma^+_2(w_0)\ket{0_R^-}^{(1)}\ket{0_R^-}^{(2)}
  ={}_t\bra{0}\Big (\alpha'_{++,_n}\alpha'_{--,_m}\Big )\ket{0}_t
\end{equation}
where the primes on the operators on the RHS signify the fact that
these operators arise from the unprimed operators by the various maps
leading to the spectral flowed $t$ plane description.  Using the
ansatz (\ref{pfive}) we can write the LHS as
\begin{multline}
\bra{0_{R,-}}\Big(\alpha_{++,_n}\alpha_{--,_m}\Big )
 \exp\left[\sum_{m\ge 1, n\ge 1}\gamma^B_{mn}(- \alpha_{++, -m}\alpha_{--, -n}
                                             + \alpha_{-+, -m}\alpha_{+-, -n})\right]\\
 \times\exp\left[\sum_{m\ge 0,n\ge 1}\gamma^F_{mn}
           \epsilon_{\dot A\dot B}\psi^{+\dot A}_{-m}\psi^{-\dot B}_{-n}\right]\ket{0^-_R}
\end{multline}
Expanding the exponential, and using the commutation relations
(\ref{bcommtwist}), one finds that this LHS equals
\begin{equation}\label{ptsix}
-mn\gamma^B_{mn}\,\braket{0_{R,-}|0^-_R} = -mn\gamma^B_{mn}
\end{equation}
where we have used Equation~\eqref{qstwo}. 

The RHS of \eqref{ptone} can be written by expressing the inserted
operators as contour integrals over circles at large $t$
\begin{equation}\label{ptfive}
{}_t\bra{0}\Big(\int\frac{\drm t_1}{2\pi i}\pd_t X_{++}(t_1) (z_0+t_1^2)^{\frac{n}{2}}\Big)
\Big(\int\frac{\drm t_2}{2\pi i}\pd_t X_{--}(t_2) (z_0+t_2^2)^{\frac{m}{2}}\Big)\ket{0}_t
\end{equation}
with $|t_1|>|t_2|$.  We have
\begin{subequations}\begin{align}
(z_0+t_1^2)^{\frac{n}{2}} &= 
        \sum_{p \ge 0}\choose{\frac{n}{2}}{p}\,z_0^p t_1^{n-2p}\\
(z_0+t_2^2)^{\frac{m}{2}} &= 
        \sum_{q \ge 0}\choose{\frac{m}{2}}{q}\,z_0^q t_2^{m-2q}
\end{align}\end{subequations}
Equating (\ref{ptsix}) and (\ref{ptfive}) gives
\begin{equation}
\gamma_{mn} = -\frac{1}{mn}\sum_{p\ge 0} \sum_{q\ge 0}\choose{\frac{n}{2}}{p}\choose{\frac{m}{2}}{q} 
         z_0^{p+q} {}_t\bra{0}\tilde{\alpha}_{++,n-2p} \tilde{\alpha}_{--,m-2q}\ket{0}_t
\end{equation}
Using the commutation relations (\ref{pttwo}) we get
\begin{equation}
m-2q=-(n-2p) \quad\Rightarrow\quad q=\frac{m+n}{2}-p
\end{equation}
Since $p,q$ are  integral,  either $n,m$ are both even or both odd.
Using Eq. (\ref{ptthree}) we find
\begin{equation}
\gamma_{mn}=\frac{1}{mn}\sum_{p=0}^{\floor{\frac{n}{2}}}
            \choose{\frac{n}{2}}{p}\choose{\frac{m}{2}}{\frac{m+n}{2}-p}z_0^\frac{m+n}{2} (n-2p)
\end{equation}
where the symbol $\floor{~}$ stands for the ``floor'' (i.e. ``integer
part of'').  We can perform this sum using a symbolic manipulation
program like~\textit{Mathematica}\footnote{The sum can be evaluated by
  hand by writing it in terms of the series representation of
  ${}_2F_1(-\frac{m-1}{2},1;\frac{n-1}{2}+2;1)$.}.  With either $n$ or
$m$ even we get
\begin{equation}
\gamma^B_{mn}=0\qquad m\text{ or } n \text{ even}.
\end{equation}
For $n$ and $m$ odd we write
\begin{equation}
m=2m'+1, \qquad n=2n'+1
\end{equation} 
and find
\begin{equation}
\gamma^B_{2m'+1, 2n'+1}
=\frac{2}{(2m'+1)(2n'+1)(1+m'+n')\pi} \frac{z_0^{(1+m'+n')} \Gamma(\frac{3}{2}+m')\Gamma(\frac{3}{2}+n')}
                                           {\Gamma(m'+1)\Gamma(n'+1)}.
\end{equation}
We rewrite and summarize the result as
\begin{equation}\label{gammaB}
\gamma^B_{mn} = \begin{cases}
\frac{4 z_0^{\frac{m+n}{2}}}{mn(m+n)\pi}
    \frac{\Gamma(\frac{m}{2}+1)\Gamma(\frac{n}{2}+1)}
         {\Gamma(\frac{m+1}{2})\Gamma(\frac{n+1}{2})} & m,n\,\text{odd, positive}\\
0 & \text{otherwise}\end{cases}.
\end{equation}
Note that in what follows we frequently abbreviate ``odd and positive''
as $\text{odd}^+$.

\subsubsection{The Fermionic Coefficients \texorpdfstring{$\gamma^F_{mn}$}{gamma-F}}

We follow the same steps to find $\gamma^F_{mn}$.
Let us compute
\begin{equation}\label{ptonep}
\bra{0_{R,-}}\Big(\psi^{++}_n\psi^{--}_m\Big)
          \sigma^+_2(w_0)\ket{0_R^-}^{(1)}\ket{0_R^-}^{(2)}
    ={}_t\bra{0}\Big (\psi'^{++}_n\psi'^{--}_m\Big )\ket{0}_t
\end{equation}
We can write the LHS as
\begin{multline}
\bra{0_{R,-}}\psi^{++}_n\psi^{--}_m
 \exp\left[
-\frac{1}{2}\sum_{ m\ge 1, n\ge 1}\gamma^B_{mn}
                    \epsilon^{AB}\epsilon^{\dot{A}\dot{B}}
                    \alpha_{A\dot{A}, -m}\alpha_{B\dot{B}, -n}
                    \right]\\
 \times\exp\left[\sum_{m\ge0,n\ge 1}\gamma^F_{mn}[\psi^{++}_{-m}\psi^{--}_{-n} - \psi^{+-}_{-m}\psi^{-+}_{-n}]\right]
\ket{0^-_R}
\end{multline}
Expanding the exponential, and using the commutation relations
(\ref{fcommtwist}), one finds that this LHS equals
\begin{equation}
4\gamma^F_{mn}\braket{0_{R,-}|0^-_R} = 4\gamma^F_{mn}
\label{ptsixp}
\end{equation}
The RHS of  (\ref{ptonep}) gives
\begin{equation}\label{ptfivep}
{}_t\bra{0}\Big[\sqrt{2}\int\frac{\drm t_1}{2\pi i}\psi^{++}(t_1) (z_0+t_1^2)^{\frac{n-2}{2}}t_1\Big]
           \Big[\sqrt{2}\int\frac{\drm t_2}{2\pi i}\psi^{--}(t_2) (z_0+t_2^2)^{\frac{m}{2}}\Big]
\ket{0}_t
\end{equation}
with $|t_1|>|t_2|$. We have
\begin{subequations}\begin{align}
(z_0+t_1^2)^\frac{n-2}{2} &= \sum_{p\ge 0}\choose{\frac{n-2}{2}}{p} z_0^p t_1^{n-2-2p}\\
(z_0+t_2^2)^\frac{m}{2} &= \sum_{q\ge 0}\choose{\frac{m}{2}}{q} z_0^q t_2^{m-2q}
\end{align}\end{subequations}
Equating (\ref{ptsixp}) and (\ref{ptfivep}) gives
\begin{equation}
\gamma^{F}_{mn}=\frac{1}{2}\sum_{p\ge 0}\sum_{q\ge 0}
        \choose{\frac{n-2}{2}}{p} \choose{\frac{m}{2}}{q} 
        {}_t\bra{0} \tilde{\psi}^{++}_{n-2p-\frac{1}{2}} \tilde{\psi}^{--}_{m-2q+\frac{1}{2}}
        \ket{0}_t.
\end{equation}
Using the commutation relations~\eqref{pttwop} we get
\begin{equation}
m-2q+\frac{1}{2}=-\big(n-2p-\tfrac{1}{2}\big)
         \quad\Longrightarrow\quad q=\frac{m+n}{2}-p
\end{equation}
From (\ref{ptthreep}) we have that
\begin{equation}
n-2p-\frac{1}{2}>0, ~~~m-2q+\frac{1}{2}<0
\end{equation}
Thus we get
\begin{equation}
\gamma^{F}_{mn}=-\frac{1}{2}
    \sum_{p=0}^{\floor{\frac{n-1}{2}}}
    \choose{\frac{n-2}{2}}{p}\choose{\frac{m}{2}}{\frac{m+n}{2}-p} z_0^{\frac{m+n}{2}}
\end{equation}
With either $n$ or $m$ even we get
\begin{equation}
\gamma^F_{mn}=0\qquad m\text{ or } n \text{ even}.
\end{equation}
For $n$ and $m$ both odd we write
\begin{equation}
m=2m'+1, ~~~n=2n'+1
\end{equation}
and find
\begin{equation}
\gamma^{F}_{2m'+1, 2n'+1}=-\frac{z_0^{(1+m'+n')}\Gamma(\frac{3}{2}+m')\Gamma(\frac{3}{2}+n')}
                                {(2n'+1)(1+m'+n')\pi \Gamma(m'+1)\Gamma(n'+1)}.
\end{equation}
We rewrite and summarize the result as
\begin{equation}\label{gammaF}
\gamma^F_{mn} = \begin{cases}
  -\frac{2z_0^\frac{m+n}{2}}{n(m+n)\pi}
   \frac{\Gamma(\frac{m}{2}+1)\Gamma(\frac{n}{2}+1)}
      {\Gamma(\frac{m+1}{2})\Gamma(\frac{n+1}{2})} & m, n\,\text{odd}^+\\
0 & \text{otherwise}.\end{cases}
\end{equation}
Note that if we compare this result to Equation~\eqref{gammaB}, and
take into account the normalization of $\alpha_m$, the two results
differ by a factor of $\sqrt{\frac{m}{n}}$. The bosons and fermions
behave essentially the same, except for an asymmetry between `+' and
`-', which we attribute to the asymmetry of the twist operator and
Ramond state.

\subsection{The State \texorpdfstring{$|\psi\rangle$}{psi}}

Finally we return to the computation of $|\psi\rangle$
\begin{equation}\label{psip}
|\psi\rangle=-G^-_{A, 0}|\chi\rangle
\end{equation}

We begin by applying the supercharge,
\begin{equation}
G^-_{A, 0}
  = \int_{w=\tau}^{w=\tau+4\pi i}\frac{\drm w}{2\pi i}G^-_{ A}(w) 
  = -\frac{i}{2}\sum_{n=-\infty}^\infty \psi^{-\dot A}_n \alpha_{A\dot A, -n}.
\end{equation}
The positive index modes in the above expression can act on the
exponential in $|\chi\rangle$, generating negative index modes. We
would like to write $|\psi\rangle$ with only negative index modes
acting on $|0^-_R\rangle$; these modes have trivial commutation and
anticommutation relations with each other. Thus we write
\begin{equation}
G^-_{{A}, 0}=
-\frac{i}{2} \sum_{n>0}^\infty \psi^{-\dot A}_{-n} \alpha_{A\dot{A}, n}
-\frac{i}{2} \sum_{n>0}^\infty \psi^{-\dot A}_{n}  \alpha_{A\dot{A}, -n}.
\end{equation}
Recall that $\gamma^B_{mn}, \gamma^F_{mn}$ are nonzero only for odd
indices. Thus we write $n=2n'+1, m=2m'+1$, and find
\begin{subequations}\begin{align}
-\frac{i}{2}\sum_{n>0}^\infty \psi^{-\dot A}_{-n} \alpha_{A\dot A, n}\ket{\chi}
  &=-\frac{i}{2}\sum_{n'\ge 0, m'\ge 0} (2n'+1) \gamma^B_{2m'+1, 2n'+1} 
             \psi^{-\dot A}_{-(2n'+1)} \alpha_{A\dot A, -(2m'+1)}\ket{\chi}\\
-\frac{i}{2} \sum_{n>0}^\infty \psi^{-\dot A}_{n} \alpha_{A\dot A, -n}\ket{\chi}
  &= i\sum_{n'\ge 0, m'\ge 0}\gamma^F_{2m'+1, 2n'+1} 
              \psi^{-\dot A}_{-(2n'+1)}\alpha_{A\dot A, -(2m'+1)}\ket{\chi}
\end{align}\end{subequations}
Thus
\begin{equation}
G^-_{A, 0}|\chi\rangle
=-i\sum_{n'\ge 0, m'\ge 0}\Big((n'+\tfrac{1}{2})\gamma^B_{2m'+1, 2n'+1}-\gamma^F_{2m'+1, 2n'+1}\Big)
               \psi^{-\dot A}_{-(2n'+1)} \alpha_{A\dot A, -(2m'+1)}\ket{\chi}
\end{equation}
Using the values of $\gamma^B, \gamma^F$ from
Equations~\eqref{gammaB} and \eqref{gammaF}, we find
\begin{multline}
(n'+\tfrac{1}{2})\gamma^B_{2m'+1, 2n'+1}-\gamma^F_{2m'+1, 2n'+1}\\
  =z_0^{m'+n'+1}\frac{\Gamma(\frac{3}{2}+m')\Gamma(\frac{3}{2}+n')}
                      {(m'+n'+1)\pi \Gamma(m'+1)\Gamma(n'+1)}
                  \Big(\frac{1}{2m'+1} + \frac{1}{2n'+1}\Big)\\
  =z_0^{m'+n'+1}\frac{2\Gamma(\frac{3}{2}+m')\Gamma(\frac{3}{2}+n')}
                      {\pi (2m'+1)(2n'+1)\Gamma(m'+1)\Gamma(n'+1)}
\end{multline}
The $n'$ and $m'$ sums factorize into the form 
\begin{multline}
G^-_{A, 0}\ket{\chi}
 = -i\Big(
   \sum_{n'\ge 0} \frac{\sqrt{2}z_0^{n'+\frac{1}{2}}\Gamma(\frac{3}{2}+n')}
                       {\sqrt{\pi}(2n'+1)\Gamma(n'+1)}\psi^{-\dot A}_{-(2n'+1)}\Big)\\
   \times\Big(\sum_{m'\ge 0} \frac{\sqrt{2}z_0^{m'+\frac{1}{2}}\Gamma(\frac{3}{2}+m')}
                       {\sqrt{\pi}(2m'+1)\Gamma(m'+1)}\alpha_{A\dot A, -(2m'+1)}
   \Big)\ket{\chi},
\end{multline}
which we can rewrite as
\begin{equation}
G^-_{A, 0}\ket{\chi} = -i\frac{2}{\pi}
\left[\sum_{n\,\text{odd}^+}\frac{z_0^\frac{n}{2}\Gamma(\frac{n}{2}+1)}
                                 {n\Gamma(\frac{n+1}{2})}\psi^{- \dot A}_{-n}\right]
\left[\sum_{m\,\text{odd}^+}\frac{z_0^\frac{m}{2}\Gamma(\frac{m}{2}+1)}
                                 {m\Gamma(\frac{m+1}{2})}\alpha_{A\dot{A},-m}\right]\ket{\chi}
\end{equation}

\subsubsection{The Final State}

Finally, we write down the complete final state. Recall that we break
the state into left and right sectors:
\begin{equation}
\ket{\Psi}_f=\ket{\psi}\ket{\bar{\psi}}.
\end{equation}
The final state on the left sector is give by
\begin{calc}
\ket{\psi}&= -G^-_{A, 0}\ket{\chi}\\
   &= \frac{2i}{\pi}
\left[\sum_{n\,\text{odd}^+}\frac{z_0^\frac{n}{2}\Gamma(\frac{n}{2}+1)}
                                 {n\Gamma(\frac{n+1}{2})}\psi^{-\dot A}_{-n}\right]
\left[\sum_{m\,\text{odd}^+}\frac{z_0^\frac{m}{2}\Gamma(\frac{m}{2}+1)}
                                 {m\Gamma(\frac{m+1}{2})}\alpha_{A\dot{A},-m}\right]\\
  &\times 
\exp\left[-\frac{1}{2}\sum_{p,q\,\text{odd}^+}\gamma^B_{pq}
                      \epsilon^{AB}\epsilon^{\dot{A}\dot{B}}\alpha_{A\dot{A}, -p}\alpha_{B\dot{B}, -q}
          +\sum_{p,q\,\text{odd}^+}\gamma^F_{pq}\epsilon_{\dot A\dot B}\psi^{+\dot A}_{-p}\psi^{-\dot B}_{-q}
                            \right]\ket{0^-_R},
\label{finalstate}
\end{calc}
with a similar expression for
$\ket{\bar{\psi}}=-\bar{G}^-_{{B},0}\ket{\bar{\chi}}$.

\section{Applying the Deformation Operator to Excited States}\label{sec:excited}

We now wish to consider the situation where we have an excitation in
the initial state on the cylinder. These excitations are generated by
operators $\alpha_{A\dot A, n}, \psi^{\alpha\dot A}_n$. Here we must have
$n\le -1$ for bosonic excitations, since there is no momentum and so
the zero mode $\alpha_{A\dot A, 0}$ kills the vacuum. For the
fermions, we have $n\le 0$ for $\psi^{+\dot A}_n$ and $n<0$ for $\psi^{-\dot A}_n$
since we cannot apply the zero mode $\psi^{-\dot A}_0$ to the vacuum
$|0_R^{--}\rangle^{(1)} |0_R^{--}\rangle^{(2)}$ that we have
taken as the starting point before we apply the excitation modes.

Let us note the steps we will have to perform in general:
\begin{enumerate}
\item In Section~\ref{sec:G-contour} we decomposed the action of the
  supercharge into an application of $-(G^{(1)-}_{A,
    0}+G^{(2)-}_{A, 0})$ below the twist and an application of
  $G^-_{A,0}$ above the twist.  Let us first consider the part
  below the twist. We commute these supercharge modes down through any
  excitations $\alpha_{A\dot A, n}, \psi^{\alpha\dot A}_n$ present in the
  initial state. After a supercharge zero mode passes all these modes
  it reaches the state $|0_R^{--}\rangle^{(1)}|0_R^{--}\rangle^{(2)}$
  at the bottom of the cylinder, and we get zero. The commutation
  through the excitations can be computed by using the relations
\begin{subequations}\begin{align}\label{kktwo}
\com{G^{(1)-}_{A,0}}{\alpha^{(1)}_{B\dot B, m}}  
         &= -i \sum_{n=-\infty}^\infty \psi^{(1)-\dot A}_{-n}
\com{\alpha^{(1)}_{A\dot A, n}}{\alpha^{(1)}_{B\dot B, m}} 
   &= -im \epsilon_{AB}\epsilon_{\dot A\dot B}\psi^{(1)-\dot A}_{m}\\
\ac{G^{(1)-}_{A,0}}{\psi^{(1)-\dot B}_{m}} &= 0\\
\ac{G^{(1)-}_{A,0}}{\psi^{(1)+\dot B}_{m}} &= i\epsilon^{\dot A\dot B}\alpha^{(1)}_{A\dot A, m}.
\end{align}\end{subequations}
We have exactly analogous relations for the operators in copy $2$.

\item Having disposed of any supercharge present below the twist, we
  get the twist operator $\sigma_2^+$ acting on a set of bosonic and
  fermionic modes. In the sections below we will take the case where
  we have a single bosonic or fermionic mode below the twist
  $\sigma_2^+$, and will find that this can be expressed as a linear
  combination of single particle modes acting \textit{after} the
  twist.  If we have several modes below the twist, then we can
  perform the same process with each mode, except that there can be in
  addition a ``Wick contraction'' between a pair of bosonic or fermionic
  modes.  This results in a $\co$-number contribution, which we compute as
  well.

\item In this manner we obtain a state where we have $\sigma_2^+$
  acting on the spin down Ramond vacuum, and a set of excitations
  acting after this twist. The action of $\sigma_2^+$ on the Ramond
  vacuum gives the state $|\chi\rangle$ (defined in~\eqref{eq:chiq});
  this state was found in \cite{acm2} and is given in (\ref{pfive}).
  Thus we get a set of modes acting on $|\chi\rangle$. Finally we note
  that we have a part of the amplitude where we must act with the
  supercharge mode applied after all other operators.  This is done
  using the relations
  \begin{subequations}\begin{align}\label{kkthree}
      \com{G^-_{A, 0}}{\alpha_{B\dot B}} &= -\frac{i}{2} \sum_{n=-\infty}^\infty \psi^{-\dot A}_n
      \com{\alpha_{A\dot A, -n}}{\alpha_{B\dot B, m}}
      &=-\frac{1}{2} im \epsilon_{AB}\epsilon_{\dot A\dot B}\psi^{-\dot A}_{m}\\
      \ac{G^{-}_{A,0}}{\psi^{-\dot B}_{m}} &= 0\\
      \ac{G^{-}_{A,0}}{\psi^{+\dot B}_{m}} &=
      i\epsilon^{AB}\alpha^{}_{A\dot A, m}
    \end{align}\end{subequations}
  We can use these relations to commute the supercharge down through
  the modes now present above $\sigma_2^+$, until it reaches the twist
  insertion.  The supercharge applied to the twist, with the Ramond
  vacuum below the twist, results in the state $|\psi\rangle$ given in
  (\ref{finalstate}) \cite{acm2}.  Thus we end up with a set of modes
  on the doubly wound circle, acting on $|\chi\rangle$ and
  $|\psi\rangle$.
\end{enumerate}
In this manner we compute the full state resulting from the action of
the deformation operator on a general initial state containing
excitations.

\subsection{The Action of \texorpdfstring{$\sigma_2^+(w_0)$}{the
    2-twist} on a Single Bosonic Mode}\label{sec:single-boson}

We perform the following computation: we have a single excitation in
the initial state, and the action of $\sigma_2^+$ on this state. The
created state will have the same exponential as in the state
$|\chi\rangle$ created by $\sigma_2^+$ from the vacuum, but in
addition the initial excitation will split into a linear combination
of single particle modes in the final state.
 
Let us find the state
\begin{equation}
|\xi\rangle
 = \sigma_2^+(w_0)\alpha^{(1)}_{A\dot A,n}|0_R^{--}\rangle^{(1)}|0_R^{--}\rangle^{(2)}
\end{equation}
Since the vacuum state $|0_R^{--}\rangle^{(1)}|0_R^{--}\rangle^{(2)}$
is killed by nonnegative modes of the boson, we will take
\begin{equation}\label{lfourt}
n\le -1
\end{equation}

We perform the spectral flows and coordinate maps given in
Section~\ref{sec:app-to-vac}. This brings us to the $t$-plane with all
punctures at $t=\pm ia, t=0$, smoothly closed. Before these steps, the
mode $\alpha^{(1)}_{A\dot A,n}$ is given
in~\eqref{eq:cylinder-modes-before}, and after all the spectral flow
and coordinate maps, it is given in the $t$ plane
by~\eqref{eq:t-modes-before}
\begin{equation}
\alpha^{(1)}_{A\dot A, n}\to  \int_{t=ia}\frac{\drm t}{2\pi} \pd_t X_{A\dot A}(t) (z_0+t^2)^n
\label{lsevenp}
\end{equation}
We will now go through a sequence of steps to find the effect of this
mode in the final state. The essence of these steps in the following.
The initial state modes (\ref{lsevenp}) are defined by a contour
around the point $t=ia$. The final state modes (\ref{qatwo}) are
defined at large $t$. They involve the function
$(z_0+t^2)^{\frac{n}{2}}$ which has branch points in the $t$ plane for
odd $n$, so we cannot directly stretch the contour giving the initial
state modes to get the modes in the final state. Instead, we proceed
from the initial modes to the final modes through a sequence of steps
that expands the initial contour into a linear combination of modes at
large $t$.


The contour in (\ref{lsevenp}) circles the point $t=ia$. As we saw in
Section~\ref{sec:app-to-vac}, there is no singularity at $t=ia$ after
all spectral flows and coordinate maps have been done. Thus if we
define modes
\begin{equation}\label{ften}
\hat\alpha^{(1)}_{A\dot A, n} = \int_{t=ia}\frac{\drm t}{2\pi}\pd_t X_{A\dot A}(t) (t-ia)^n,
\end{equation}
then we will find
\begin{equation}\label{ltenqq}
\hat\alpha_{A\dot A, n}|0\rangle_{ia}=0, ~~~n\ge 0
\end{equation}
where $|0\rangle_{ia}$ is the NS vacuum at the point $t=ia$, and we
have noted that the zero mode vanishes since there is no momentum for
the boson at any stage. The modes (\ref{ften}) have the commutation
relations
\begin{equation}\label{pttwoqq}
\com{\hat\alpha_{A \dot A,m}}{\hat\alpha_{B \dot B,n}}
   =-\epsilon_{A B} \epsilon_{\dot A \dot B} m\delta_{m+n, 0}
\end{equation}

Thus we would like to expand the mode (\ref{lsevenp}) in modes of type (\ref{ften}). Writing
\begin{equation}
z_0+t^2=(t-ia)(t+ia)
\end{equation}
we write the  mode (\ref{lsevenp}) as
\begin{equation}
\alpha^{(1)}_{A\dot A, n}\to \int_{t=ia}\frac{\drm t}{2\pi} \pd_t X_{A\dot A}(t) (t-ia)^n(t+ia)^n
\end{equation}
We have 
\begin{calc}
(t+ia)^n &=\Big(2ia + (t-ia)\Big)^n\\
         &=(2ia)^n \Big ( 1+(2 i a)^{-1} (t-ia)\Big ) ^n\\
         &=\sum_{k\ge 0} \choose{n}{k} (2ia)^{(n-k)} (t-ia)^k
\end{calc}
We thus find
\begin{equation}\label{ktwo}
\alpha^{(1)}_{A\dot A, n}\to 
\sum_{k\ge 0} \choose{n}{k} (2ia)^{(n-k)}\int_{ia} \frac{\drm t}{2\pi}\pd_t X_{A\dot A}(t)(t-ia)^{(n+k)}
   = \sum_{k\ge 0}\choose{n}{k} (2ia)^{(n-k)} \hat{\alpha}^{(1)}_{n+k}.
\end{equation}
Since, there are no other insertions in the $t$-plane, we get a
nonzero contribution only from modes with
\begin{equation}
n+k \le -1~~\Rightarrow~~k \le -n-1.
\end{equation}
(Recall from (\ref{lfourt}) that $n$ is negative.) Thus we have
\begin{equation}\label{lsixt}
\alpha^{(1)}_{A\dot A, n}\to \sum_{k=0}^{-n-1} \choose{n}{k}(2ia)^{(n-k)}\hat{\alpha}^{(1)}_{n+k}.
\end{equation}
Note that this is a finite sum; this is important to have convergence
for sums that we will encounter below.

We can now convert the original $z$-plane modes to $\tilde{\alpha}$
modes defined for large $t$. The contour in the operators
$\hat\alpha_{A\dot A, n}$ circles $t=ia$.  But as we have seen in
Section~\ref{sec:app-to-vac}, there are no singularities at other
points in the $t$ plane, so we can stretch this contour into a contour
at large $t$, which we write as $\int_{t=\infty}$. We get
\begin{equation}
\hat\alpha^{(1)}_{A\dot A, m}\to \int_{t=\infty}\frac{\drm t}{2\pi} \pd_t X_{A\dot A}(t) (t-ia)^m,
\end{equation}
which we can plug into with
\begin{equation}
(t-ia)^m = \sum_{k'\ge 0}\choose{m}{k'}(-ia)^{k'} t^{m-k'}.
\end{equation}
Let us now define modes natural for large $t$ in the $t$ plane,
\begin{equation}\label{qathreeq}
\tilde{\alpha}_{A\dot A, m}\equiv  \int_{t=\infty}\frac{\drm t}{2\pi} \pd_t X_{A\dot A}(t) t^m,
\end{equation}
which satisfy
\begin{equation}\label{ptthree-2}
\com{\tilde{\alpha}_{A \dot A}}{\tilde{\alpha}_{B \dot B}}
  = -\epsilon_{A B} \epsilon_{\dot A \dot B} m\delta_{m+n, 0};\qquad
\tilde{\alpha}_{A \dot A,m}\ket{0}_t=0, \qquad m\ge 0.
\end{equation}
We can expand the modes out to
\begin{equation}
\hat\alpha_{A\dot A, m} = \sum_{k'\ge 0}\choose{m}{k'}(-ia)^{k'}\tilde{\alpha}_{A\dot A, m-k'}.
\end{equation}

Let us now substitute this expansion for $\hat\alpha_{A\dot A, m}$ into (\ref{lsixt}). We get
\begin{equation}
\alpha^{(1)}_{A\dot A, n}\to 
 \sum_{k= 0}^{-n-1} \choose{n}{k} (2ia)^{(n-k)}
 \sum_{k'\ge 0}\choose{n+k}{k'}(-ia)^{k'}\tilde{\alpha}_{A\dot A, n+k-k'}.
\end{equation}
Let us look at the coefficient of $\tilde{\alpha}_{A\dot A, q}$. Thus 
\begin{equation}
q=n+k-k' ~~\Rightarrow ~~k'=n+k-q.
\end{equation}
Since we have $k'\ge 0$, we find
\begin{equation}
n+k-q\ge 0 ~~\Rightarrow~~k\ge q-n
\end{equation}
Thus the $k$ sum runs over the following range
\begin{equation}\begin{aligned}
q-n\le 0:&~~&k&\in [0, -n-1]\\
q-n\ge0:&~~&k&\in [q-n, -n-1]
\end{aligned}\end{equation}
Recall that $n\le -1$. The second of these equations tells us that
there is no summation range at all for $q\ge 0$. Thus we only generate
negative index modes $\tilde\alpha_{A\dot A, q}$ from our expansion:
\begin{equation}\label{qequation}
q\le -1
\end{equation}
We now have
\begin{equation}\begin{aligned}
\alpha^{(1)}_{A\dot A, n}&\to \sum_{q\le -1} 
   \left[\sum_{k= 0}^{-n-1}\choose{n}{k}(2ia)^{(n-k)} \choose{n+k}{n+k-q}(-ia)^{n+k-q}\right]
       \tilde{\alpha}_{A\dot A, q}, \qquad q\le n\\
\alpha^{(1)}_{A\dot A, n}&\to \sum_{q\le -1} 
   \left[\sum_{k= q-n}^{-n-1}\choose{n}{k}(2ia)^{(n-k)}\choose{n+k}{n+k-q}(-ia)^{n+k-q}\right]
      \tilde{\alpha}_{A\dot A, q}, \qquad q\ge n.
\end{aligned}\end{equation}
The sum has the same algebraic expression in both the ranges for $q$, and we find
\begin{equation}\label{ltwthree}
\alpha^{(1)}_{A\dot A, n}\to 
\sum_{q\le -1} \frac{i^{-q}(-1)^n a^{2n-q} \Gamma(-\frac{q}{2})}
                         {2 \Gamma(-n)\Gamma(n+1-\frac{q}{2})}
       \tilde{\alpha}_{A\dot A, q}.
\end{equation}

Now, we wish to convert the $z$-plane modes before the twist into
$z$-plane modes after the twist.  These are the modes on the cylinder
at $\tau\to\infty$, given in (\ref{qatwo})
\begin{equation}\label{ltwone}
\alpha_{A\dot A, p}\to \int_{t=\infty}\frac{\drm t}{2\pi}\pd_t X_{A\dot A}(t) (z_0+t^2)^{\frac{p}{2}} 
\end{equation}
Write
\begin{equation}
t'=\sqrt{t^2+z_0}~~\Rightarrow ~~ t=(t'^2-z_0)^\frac{1}{2}
\end{equation}
with the sign of the square root chosen to give $t'\sim t$ near
infinity. Then we have
\begin{equation}
t^q=(t'^2-z_0)^\frac{q}{2}
  = t'^q \Big(1-z_0 t'^{-2}\Big)^\frac{q}{2}
  = \sum_{k\ge 0}\choose{\frac{q}{2}}{k} (-z_0)^k t'^{q-2k}
  = \sum_{k\ge 0}\choose{\frac{q}{2}}{k} (-z_0)^k (z_0+t^2)^{\frac{q-2k}{2}}.
\end{equation}
Substituting this expansion in (\ref{qathreeq}) we find that
\begin{equation}
\tilde{\alpha}_{A\dot A, q}
  = \int\frac{\drm t}{2\pi}\pd_{t}X_{A\dot A}(t)
    \sum_{k\ge 0}\choose{\frac{q}{2}}{k} (-z_0)^k (z_0+t^2)^\frac{q-2k}{2}.
\end{equation}
Using (\ref{ltwone}) we find 
\begin{equation}\label{ltwtwo}
\tilde{\alpha}_{A\dot A, q}
  = \sum_{k\ge 0}\choose{\frac{q}{2}}{k} (-z_0)^k \alpha_{A\dot A, q-2k}
\end{equation}

We substitute the expansion (\ref{ltwtwo}) into (\ref{ltwthree}), getting
\begin{equation}\label{ltthree}
\alpha^{(1)}_{A\dot A, n}\to \sum_{q\le -1} 
  \frac{i^{-q}(-1)^n  a^{2n-q}  \Gamma(-\frac{q}{2})}
            {2 \Gamma(-n)\Gamma(n+1-\frac{q}{2})}
  \,\sum_{k\ge 0}\choose{\frac{q}{2}}{k} (-z_0)^{k}\alpha_{A\dot A, q-2k}
\end{equation}
Let us look at the coefficient of $\alpha_{A\dot A, p}$ in this sum. This gives
\begin{equation}\label{ltwfive}
p=q-2k
\end{equation}
Note that since $q\le -1$ and $k\ge 0$, we have 
\begin{equation}\label{ltwfour}
p\le -1
\end{equation}
From (\ref{ltwfive}) we see that if $p$ is even only even values of
$q$ contribute to this sum, and if $p$ is odd then only odd values of
$q$ contribute.  For the even $p$ case we write $p=2p', q=2q'$.  From
(\ref{ltwfive}) we set $k=q'-p'$.  Since $k\ge 0$, the range of the
$q'$ sum becomes $p'\le q'\le -1$. We get the sum
\begin{equation}
\sum_{q'=p'}^{-1} \frac{(-1)^{q'+n}a^{2n-2q'}\Gamma(-{q'})}
                       {2 \Gamma(-n)\Gamma(n+1-q')}
                  \choose{q'}{q'-p'} (-z_0)^{q'-p'}
                 = \frac{1}{2}\delta_{n, p'}
\end{equation}
For odd $p$ we write $p=2p'+1, q=2q'+1$. From (\ref{ltwfive}) we again
get $k=q'-p'$.  Since $k\ge 0$, the range of $q'$ is $p'\le q'\le -1$.
We get the sum
\begin{multline}
\sum_{q'=p' }^{-1}\frac{i^{-1}(-1)^{q'+n}  a^{2n-2q'-1}  \Gamma(-{q'}-\frac{1}{2})}
                  {2 \Gamma(-n)\Gamma(n+\frac{1}{2}-q')}\choose{q'+\frac{1}{2}}{q'-p'} (-z_0)^{q'-p'}\\
= \frac{ia^{2(n-p')-1}}{\pi(2n-2p'-1)} 
  \frac{\Gamma(\frac{1}{2}-n)}{\Gamma(-n)}
  \frac{\Gamma(-\frac{1}{2}-p')}{\Gamma(-p')}
\end{multline}

Now that we have converted the initial mode $\alpha_{A\dot A,n}^{(1)}$
to modes at large $t$, we note that we are left with the NS vacuum
$|0\rangle_t$ of the $t$ plane inside these modes. This vacuum just
gave us the state $|\chi\rangle$ on the cylinder~\cite{acm2}
\begin{equation}
\sigma_2^+(w_0) \ket{0_R^{--}}^{(1)}\ket{0_R^{--}}^{(2)} = \ket{\chi}
\end{equation}
Putting the above results together, we see that for $n<0$
\begin{calc}\label{kkone}
&\sigma_2^+(w_0)\alpha^{(1)}_{A\dot A, n} \ket{0_R^{--}}^{(1)} \ket{0_R^{--}}^{(2)}\\
&\qquad = \Bigg(\frac{1}{2}\alpha_{A\dot{A}, 2n} + 
      \sum_{p'\le -1}\Big(\frac{ia^{2(n-p')-1}}{\pi(2n-2p'-1)} 
                 \frac{\Gamma(\frac{1}{2}-n)}{\Gamma(-n)}
                 \frac{\Gamma(-\frac{1}{2}-p')}{\Gamma(-p')}\Big) 
      \,\alpha_{A\dot{A}, 2p'+1}\Bigg)\ket{\chi}\\
&\qquad \equiv \sum_p f^B_{n,p}\alpha_{A\dot A, p}\ket{\chi}
\end{calc}
where we have defined the coefficients $f^B_{n,p}$ for later
convenience.  For $n\ge 0$ we will just have
\begin{equation}
 \sigma_2^+(w_0)\alpha^{(1)}_{A\dot A, n}\ket{0_R^{--}}^{(1)}\ket{0_R^{--}}^{(2)} = 0
\end{equation}
since positive modes annihilate the Ramond vacuum state.

We see that the even modes in the final state get a simple
contribution; this is related to the fact that the twist operator
$\sigma_2$ does not affect such modes when it cuts and joins together
the two copies of the $c=6$ CFT. The odd modes of all levels are
excited.

\subsection{The Action of \texorpdfstring{$\sigma_2^+(w_0)$}{the
    2-twist} on a Single Fermionic Mode}\label{sec:single-fermion}

Let us repeat the above computation for a fermionic mode in the
initial state.  There is a slight difference between the cases of
$\psi^{-\dot A}_n$ and $\psi^{+\dot A}_n$ since the starting vacuum state
$|0_R^{--}\rangle^{(1)} |0_R^{--}\rangle^{(2)}$ breaks the charge
symmetry, and the spectral flows we do also break this symmetry.

Let us start with the first case, $\psi^{(1)-\dot A}_n$.  We start with
Equation~\eqref{eq:t-modes-before}, which we write as
\begin{equation}
\psi^{(1)-\dot A}_n\to \sqrt{2}\int_{ia} \frac{\drm t}{2\pi i} \psi^{-\dot A}_t(t)(t-ia)^n(t+ia)^n
\end{equation}
We perform the spectral flows and coordinate maps in
Section~\ref{sec:app-to-vac} to reach the $t$ plane with all punctures
smoothly closed.  Define natural modes around the point $t=ia$ in the
$t$ plane
\begin{equation}
\hat{\psi}^{-\dot A}_r = \int_{ia} \frac{\drm t}{2\pi i}\psi^{-\dot A}_t(t) (t-ia)^{r-\frac{1}{2}}
\end{equation}
where $r$ is a half-integer. Writing $t+ia=2ia+(t-ia)$, we expand in
powers of $(t-ia)$. Noting that operators $\hat \psi^{-\dot A}_r$ with $r>0$
kill the NS vacuum at $t=ia$, we find
\begin{equation}
\psi^{(1)-\dot A}_n~\to~ \sqrt{2}\sum_{k=0}^{-n-1}\choose{n}{k} (2ia)^{n-k} \hat \psi^{-\dot A}_{n+k+\frac{1}{2}}
\end{equation}

The RHS in the above equation is a finite sum of operators, each given
by a contour integral around $t=ia$. Since there are no singularities
anywhere on the $t$ plane, we can expand each contour to one at large
$t$. We define operators natural for expansion around infinity in the
$t$ plane
\begin{equation}
\tilde \psi^{-\dot A}_r=\int_{t=\infty}\frac{\drm t}{2\pi i}\psi^{-\dot A}_t(t) t^{r-\frac{1}{2}}
\end{equation}
where $r$ is a half-integer. The commutation relations are
\begin{equation}
\ac{\tilde \psi^{\alpha\dot A}_r}{\tilde \psi^{\beta\dot B}_s} =
   -\epsilon^{\alpha\beta}\epsilon^{\dot A\dot B}\delta_{r+s, 0}
\label{pttwop-2}
\end{equation}
We find
\begin{equation}
\hat \psi^{-\dot A}_r=\sum_{k'\ge 0}\choose{r-\frac{1}{2}}{k'}(-ia)^{k'}\,\tilde \psi^{-\dot A}_{r-k'}
\end{equation}

Finally we can expand the operators $\tilde \psi^{-\dot A}_r$ in terms of the
final state modes~\eqref{eq:t-modes-after}, finding
\begin{equation}
\tilde \psi^{-\dot A}_r=\frac{1}{\sqrt{2}} \sum_{k\ge 0}
             \choose{\frac{r-\frac{1}{2}}{2}}{k}(-a^2)^k \psi^{-\dot A}_{r-2k-\frac{1}{2}}.
\end{equation}

We then find
\begin{calc}\label{jtwo}
&\sigma_2^+(w_0) \psi^{(1)-\dot A}_n |0_R^{--}\rangle^{(1)} |0_R^{--}\rangle^{(2)}\\
&\qquad=\Bigg(\frac{1}{2} \psi^{-\dot A}_{2n} + \sum_{p'\le -1}
        \frac{ia^{2(n-p')-1}}{\pi(2n-2p'-1)} 
                 \frac{\Gamma(\frac{1}{2}-n)}{\Gamma(-n)}
                 \frac{\Gamma(-\frac{1}{2}-p')}{\Gamma(-p')}
       \psi^{-\dot A}_{2p'+1}\Bigg)\ket{\chi}\\
&\qquad\equiv\sum_p f^{F-}_{n,p}\psi^{-\dot A}_p\ket{\chi}
\end{calc}

We now turn to the second case, where we have $\psi^{(1)+\dot A}$ before the
twist.  We start with Equation~\eqref{eq:t-modes-before}, which we
write as
\begin{equation}\label{jone}
\psi^{(1)+\dot A}_n\to \sqrt{2}\int_{ia}\frac{\drm t}{2\pi i}\psi^{+\dot A}_t(t)(t-ia)^{n-1}(t+ia)^{n-1}\,t
\end{equation}
We perform the steps in Section~\ref{sec:app-to-vac} as before.  The
natural modes around $t=ia$ are
\begin{equation}
\hat \psi^{+\dot A}_r = \int_{ia}\frac{\drm t}{2\pi i}\psi^{+\dot A}_t(t) (t-ia)^{r-\frac{1}{2}}
\end{equation}
This time we must expand in (\ref{jone}) the factor $t+ia=2ia+(t-ia)$
as well as the factor $t=ia + (t-ia)$. Thus generates two terms
\begin{equation}
\psi^{(1)+\dot A}_n~\to~\frac{1}{\sqrt{2}}\sum_{k=0}^{-n}\choose{n-1}{k}(2ia)^{n-k}
                \hat \psi^{+\dot A}_{n+k-\frac{1}{2}}
               + \sqrt{2}\sum_{k=0}^{-n-1}\choose{n-1}{k} (2ia)^{n-k-1}\hat \psi^{+\dot A}_{n+k+\frac{1}{2}}.
\end{equation}
Define natural modes at large $t$
\begin{equation}
\tilde \psi^{+\dot A}_r=\int_{t=\infty}\frac{\drm t}{2\pi i}\psi^{+\dot A}_t(t)\, t^{r-\frac{1}{2}}
\end{equation}
We find
\begin{equation}
\hat \psi_r^{+\dot A} = \sum_{k\ge 0}\choose{r-\frac{1}{2}}{k} (-ia)^k \tilde \psi^{+\dot A}_{r-k}
\end{equation}
Finally we can expand the modes $\tilde \psi^{+\dot A}$ in terms of the final
state modes in Equation~\eqref{eq:t-modes-after}, finding
\begin{equation}
\tilde \psi_r^{+\dot A}=\frac{1}{\sqrt{2}}\sum_{k\ge 0}\choose{\frac{r-\frac{3}{2}}{2}}{k}\,
                         (-a^2)^k\, \psi^{+\dot A}_{r-2k+\frac{1}{2}}.
\end{equation}
Putting all these expansions together, we find
\begin{calc}\label{eq:sing-ferm-1}
&\sigma_2^+(w_0)\psi^{(1)+\dot A}_n \ket{0_R^{--}}^{(1)}\ket{0_R^{--}}^{(2)}\\
&\qquad = \Bigg(\frac{1}{2}\psi^{+\dot A}_{2n} + 
        \sum_{p'\le -1}\frac{i a^{2(n-p')-1}}{\pi (2n-2p'-1)} 
                            \frac{\Gamma(\frac{1}{2}-n)}{\Gamma(1-n)}
                            \frac{\Gamma(\frac{1}{2}-p')}{\Gamma(-p')}\,\psi^{+\dot A}_{2p'+1}
          \Bigg)\ket{\chi}\\
&\qquad\equiv \sum_p f^{F+}_{n,p} \psi^{+\dot A}_p\ket{\chi}.
\end{calc}

The computations of this section are are very basic to understanding
the effect of the deformation operator: taken by itself, any single
particle mode below the twist insertion $\sigma_2^+(w_0)$ spreads into
a linear combination of single particle modes after the twist, and we
have found the coefficients of this linear combination for both
bosonic and fermionic excitations. In addition the twist creates the
same exponential that arises in the action of the twist of the
vacuum~\eqref{pfive}, so the action of the twist on a single particle
initial state gives rise to states with $1,3,5,\dots$ excitations.

\subsection{Two Bosonic Modes}\label{sec:two-bosons}

Now let us consider the situation where we have two excitations in the
initial twist. Upon the action of the twist $\sigma_2^+$ there will be
two kinds of terms. One, where the modes move separately to the final
state; this contribution can thus be computed by using the expressions
of the last section. The other contribution results from an
interaction between the two modes. Since we are dealing with a theory
of free bosons and free fermions, the only possible interactions
between modes is a ``Wick contraction,'' which produces a $\co$-number
term.

Let us consider the state
\begin{equation}\label{kthree}
\sigma_2^+(w_0)\alpha^{(1)}_{A\dot A, n_1}\alpha^{(1)}_{B\dot B, n_2}  
    \ket{0_R^{--}}^{(1)} \ket{0_R^{--}}^{(2)}
\end{equation}
with $n_1<0, n_2<0$. We wish to move the modes
$\alpha^{(1)}_{A\dot{A},n_i}$ to operator modes acting after the
$\sigma_2^+$ operator. We will get the terms corresponding to each of
these modes moving across separately, but there will also be a term
resulting from the interaction between the two modes.

We follow the sequence of spectral flows and dualities given in
Section~\ref{sec:app-to-vac}. We reach the $t$ plane with all
punctures closed and the operator modes (cf.
Equation~\eqref{eq:t-modes-before})
\begin{equation}
\alpha^{(1)}_{A\dot A, n_1}\alpha^{(1)}_{B\dot B, n_2}\to
\Big(\int_{t=ia}\frac{\drm t_1}{2\pi} \pd_t X_{A\dot A}(t_1) (z_0+t_1^2)^{n_1}\Big)
\Big(\int_{t=ia}\frac{\drm t_2}{2\pi} \pd_t X_{B\dot B}(t_2) (z_0+t_2^2)^{n_2}\Big)
\end{equation}
with the $t_1$ contour outside the $t_2$ contour.

There is no singularity inside the $t_2$ contour, so we can expand the
$t_2$ contour as in Section~\ref{sec:single-boson} to get (cf.
Equation~\eqref{lsixt})
\begin{equation}
\alpha^{(1)}_{B\dot B, n_2}\to 
 \sum_{k_2=0}^{-n_2-1}\choose{n_2}{k_2} (2ia)^{(n_2-k_2)} \hat \alpha^{(1)}_{B\dot{B}, n_2+k_2}
\label{kone}
\end{equation}
For the $t_1$ contour we can get a contribution from both positive and
negative $\hat\alpha$ modes, since the $t_2$ contour gives an operator
inside the $t_1$ contour. Thus we write the general expansion
(\ref{ktwo}) for this contour
\begin{equation}\label{konef}
\alpha^{(1)}_{A\dot A, n_1}\to 
  \sum_{k_1=0}^{\infty} \choose{n_1}{k_1} (2ia)^{(n_1-k_1)}\, \hat \alpha^{(1)}_{A\dot A, n_1+k_1}
\end{equation}
and consider separately two cases:
\begin{enumerate}
\item The range of $k_1$ where $n_1+k_1\le -1$. This gives negative
  index modes just like (\ref{kone}). These modes commute with the
  modes in (\ref{kone}), so we have no interaction between the two
  operators, and we get (cf. (\ref{lsixt}))
  \begin{equation}\label{koneq}
    \alpha^{(1)}_{A\dot A, n_1}\to 
    \sum_{k_1=0}^{-n_1-1}\choose{n_1}{k_1} (2ia)^{(n_1-k_1)}\,\hat \alpha^{(1)}_{A\dot{A}, n_1+k_1}
  \end{equation}

\item The range where $n_1+k_1\ge 0$. Now these modes can annihilate
  the negative modes created by the $t_2$ contour. This results in a
  $\co$-number contribution
  \begin{equation}
    C^B_{A\dot A B\dot B}[n_1,n_2]
    =\smashoperator[l]{\sum_{k_1=-n_1}^{\infty}} 
     \smashoperator[r]{\sum_{k_2=0}^{-n_2-1}}\choose{n_1}{k_1} (2ia)^{n_1-k_1} 
    \choose{n_2}{k_2} (2ia)^{(n_2-k_2)}
    \,\com{\hat \alpha^{(1)}_{A\dot{A}, n_1+k_1}}{\hat \alpha^{(1)}_{B\dot{B}, n_2+k_2}}
  \end{equation}
\end{enumerate}
Using the commutation relation (\ref{pttwoqq}) we get
\begin{calc}
C^B_{A\dot A B\dot B}[n_1,n_2]&= -\epsilon_{A B} \epsilon_{\dot A \dot B}
  \sum_{k_2=0}^{-n_2-1}(-(n_2+k_2))\choose{n_1}{-n_1-n_2-k_2}
                                   \choose{n_2}{k_2}\,(2ia)^{2(n_1+n_2)}\\
&= \epsilon_{A B}\epsilon_{\dot A \dot B}
   \frac{a^{2(n_1+n_2)}\Gamma(-n_1+\frac{1}{2})\Gamma(-n_2+\frac{1}{2})}
             {2\pi (n_1+n_2) \Gamma(-n_1)\Gamma(-n_2)}
\label{wickb}
\end{calc}
Note that this is symmetric in $n_1$ and $n_2$, as it should be since
the modes in (\ref{kthree}) commute and so can be put in either order.

Apart from this $\co$-number term, we still have the contribution where
the two contours in the initial state produce the modes in
Equation~\eqref{kone} and~\eqref{koneq}. We can proceed to expand the
modes $\hat\alpha$ in these sums into modes of type $\tilde\alpha$. We
note from (\ref{qequation}) that each set of modes generates only
negative index modes $\tilde\alpha_q$. Thus we cannot get any
additional $\co$-number contributions from commutators between the
$\tilde\alpha$ modes arising from our two operators.

Next we convert the modes $\tilde\alpha$ to modes of type $\alpha$. From
(\ref{ltwfour}) we see that we again generate only negative index
modes $\alpha_p$ from each of the two sets of modes. Thus there cannot
be any additional $\co$-number contribution from commutation between the
modes $\alpha$ arising from our two operators. In short, the only
$\co$-number contribution we get from ``Wick contraction'' is
(\ref{wickb}), and the remaining part of the state is given by
independently moving the two initial state modes past $\sigma_2^+$ in
the manner given in (\ref{kkone}).

Putting this ``Wick contraction'' term together with the contribution
of the uncontracted terms we get
\begin{multline}\label{eq:two-bosons}
\sigma_2^+(w_0) \alpha^{(1)}_{A\dot A, n_1}\alpha^{(1)}_{B\dot B, n_2}
                                       \ket{0_R^{--}}^{(1)}\ket{0_R^{--}}^{(2)}\\
=\Big[\Big(\sum_{p_1} f^B_{n_1,p_1}\alpha_{A\dot A, p_1}\Big)
 \Big(\sum_{p_2} f^B_{n_2,p_2}\alpha_{B\dot B, p_2}\Big) + C^B_{A\dot A B\dot B}[n_1,n_2]\Big]\,
 \ket{\chi}.
\end{multline}

\subsection{Two Fermionic Modes}\label{sec:two-fermions}

Let us repeat this computation for two fermions in the initial state.
Since the spectral flow treats positive and negative charges
differently, we work with the pair $\psi^{++}$ $\psi^{--}$ and later write
the result for general charges.

Consider 
\begin{equation}\label{kthreeq}
\sigma_2^+(w_0) \psi^{(1),++}_{ n_1}\psi^{(1),--}_{ n_2} \ket{0_R^{--}}^{(1)}\ket{0_R^{--}}^{(2)}
\end{equation}
with $n_1\le0, n_2<0$. We wish to move the modes to those acting after
the $\sigma_2^+$ operator. We will get the terms corresponding to each
of these modes moving across separately, but there will also be a term
resulting from the interaction between the two modes.

Following the sequence of spectral flows and dualities we reach the
$t$ plane with all punctures closed and the operator modes (cf.
Equation~\eqref{eq:t-modes-before})
\begin{equation}
\psi^{(1),++}_{n_1}\psi^{(1),--}_{ n_2} \to
\Big(\sqrt{2}\int_{t=ia}\frac{\drm t_1}{2\pi i} \psi_t^{++}(t_1) (z_0+t_1^2)^{n_1-1}\,t_1\Big)
\Big(\sqrt{2}\int_{t=ia}\frac{\drm t_2}{2\pi i} \psi_t^{--}(t_2) (z_0+t_2^2)^{n_2}\Big )
\end{equation}
with the $t_1$ contour outside the $t_2$ contour.

There is no singularity inside the $t_2$ contour so we get
\begin{equation}
\psi^{(1),--}_{ n_2}\to 
  \sqrt{2}\sum_{k_2=0}^{-n_2-1}\choose{n_2}{k_2}(2ia)^{n_2-k_2}\,{\hat d}^{--}_{n_2+k_2+\frac{1}{2}}
\end{equation}
For $\psi^{(1),++}_{n_1}$ we write the full sum over modes
\begin{equation}
\psi^{(1)+\dot A}_{n_1}~\to~\frac{1}{\sqrt{2}}
   \sum_{k_1=0}^{\infty}\choose{n_1-1}{k_1}(2ia)^{n_1-k_1}\hat \psi^{+\dot A}_{n_1+k_1-\frac{1}{2}}
  + \sqrt{2} \sum_{k_1= 0}^{\infty}\choose{n_1-1}{k_1}(2ia)^{n_1-k_1-1}\hat \psi^{+\dot A}_{n_1+k_1+\frac{1}{2}}
\end{equation}
The anticommutator arising from the first term gives
\begin{equation}
-(2ia)^{2(n_1+n_2)}\sum_{k_2=0}^{-n_2-1}\choose{n_1-1}{-n_1-n_2-k_2}\choose{n_2}{k_2}
\end{equation}
while the anticommutator from the second term gives
\begin{equation}
-2(2ia)^{2(n_1+n_2)}\sum_{k_2=0}^{-n_2-1}\choose{n_1-1}{-n_1-n_2-k_2-1}\choose{n_2}{k_2}.
\end{equation}

The sum of these two contributions can be simplified to give (we now
include the result for the pair $\psi^{+-}, \psi^{-+}$)
\begin{equation}
C^{F,\alpha A  \beta B}[n_1,n_2] = 
  -\epsilon^{\alpha\beta}\epsilon^{A B} 
\frac{a^{2(n_1+n_2)}\Gamma(-n_1+\frac{1}{2})\Gamma(-n_2+\frac{1}{2})}
           {2\pi n_1(n_1+n_2) \Gamma(-n_1)\Gamma(-n_2)}
\end{equation}
\begin{sloppypar}
Note that this is not symmetric in $n_1,n_2$ since the choice of
Ramond vacuum $\ket{0_R^{--}}^{(1)} \ket{0_R^{--}}^{(2)}$ breaks the
symmetry between $+$ and $-$ charge fermions.
\end{sloppypar}

Putting this ``Wick contraction'' term together with the contribution
of the uncontracted terms we get
\begin{multline}\label{eq:two-fermions}
\sigma_2^+(w_0) \psi^{(1),++}_{n_1}\psi^{(1),--}_{n_2}\ket{0_R^{--}}^{(1)}\ket{0_R^{--}}^{(2)}\\
=\Big[\Big(\sum_{p_1} f^{F+}_{n_1,p_1}\psi^{++}_{p_1}\Big)
 \Big(\sum_{p_2} f^{F-}_{n_2,p_2}\psi^{--}_{p_2}\Big) + C^{F,++--}_{n_1,n_2}\Big]\, 
  \ket{\chi}.
\end{multline}

\subsubsection{Summary}

We have computed the $\co$-number ``Wick contraction'' term that
results from the interaction between two initial state modes. Note
that after all the spectral flows we perform, we are dealing with a
theory of free bosons and fermions. Thus even if we had several modes
in the initial state, we can break up the effect of the twist
$\sigma_2^+$ into pairwise ``Wick contractions'' (with value given by
$C^B$ and $C^F$ computed above), and moving any uncontracted modes
past the twist $\sigma_2^+$ using the expressions in
Sections~\ref{sec:single-boson} and~\ref{sec:single-fermion}.  So the
computations of Sections~\ref{sec:single-boson}
and~\ref{sec:single-fermion}, and the present section allow us to find
the effect of $\sigma_2^+$ on any initial state.

\subsection{Complete Action of the Deformation Operator on a Bosonic Mode}\label{sec:complete-two-bosons}

In the last two sections we have focused on the effect of the twist
$\sigma_2^+$. Let us now compute an example where we combine this with
the action of the supercharge described in the beginning of
Section~\ref{sec:excited}.

We start with the state containing one bosonic excitation, and find
the state created by the action of the deformation operator. Thus we
wish to find the state
\begin{equation}
\ket{\psi}_f  = 
\mathcal{T}_{\dot A}\,\alpha_{C\dot C, n}^{(1)}\ket{0_R^{-}}^{(1)}\ket{0_R^{-}}^{(2)}
\end{equation}
with $n\le -1$.

We follow the steps outlined in the beginning of
Section~\ref{sec:excited}.  We have
\begin{equation}
\ket{\psi}_f = \Big(-\sigma_2^+(w_0)\,G^{(1)-}_{A, 0}\,\alpha_{C\dot C, n}^{(1)}
       + G^-_{A,0}\,\sigma_2^+(w_0)\,\alpha_{C\dot C, n}^{(1)}\Big)
   \ket{0_R^{-}}^{(1)}\ket{0_R^{-}}^{(2)}.
\end{equation}
We have
\begin{calc}
\sigma_2^+(w_0)\,G^{(1)-}_{A, 0}\,\alpha_{C\dot C, n}^{(1)}
                                    \ket{0_R^{-}}^{(1)}\ket{0_R^{-}}^{(2)}
  &=\sigma_2^+(w_0)\,\com{G^{(1)-}_{A, 0}}{\alpha_{C\dot C, n}^{(1)}}
                                    \ket{0_R^{-}}^{(1)}\ket{0_R^{-}}^{(2)}\\
  &= (-\epsilon_{AC}\epsilon_{\dot A\dot C})\,i n\,\sigma_2^+(w_0)\,\psi^{(1)-\dot A}_n
                                    \ket{0_R^{-}}^{(1)}\ket{0_R^{-}}^{(2)}
\end{calc}
We can now write down $\sigma_2^+(w_0)\psi^{(1)-\dot A}
_n|0_R^{-}\rangle^{(1)}|0_R^{-}\rangle^{(2)}$ from
(\ref{jtwo}).  For the other term, we first compute
$\sigma_2^+(w_0)\alpha_{C\dot C, m}^{(1)}
|0_R^{-}\rangle^{(1)}|0_R^{-}\rangle^{(2)}$ from (\ref{kkone}), and
apply the operator $G^-_{A, 0}$ using
\begin{calc}
G^-_{A, 0}\alpha_{C\dot C, p}\sigma_2^+\ket{0_R^{-}}^{(1)}\ket{0_R^{-}}^{(2)}
&= \com{G^-_{A, 0}}{\alpha_{C\dot C, p}}\sigma_2^+\ket{0_R^{-}}^{(1)}\ket{0_R^{-}}^{(2)}
   + \alpha_{C\dot C, p}G^-_{A,0}\sigma_2^+\ket{0_R^{-}}^{(1)}\ket{0_R^{-}}^{(2)}\\
&= (-\epsilon_{AC}\epsilon_{\dot A\dot C}) \,\frac{ip}{2}\,\psi^{-\dot A}_p\ket{\chi} +
   \alpha_{C\dot C, p}\ket{\psi}
\end{calc}
where $\ket{\psi}$ is given in (\ref{finalstate}). Putting together
these two contributions, we get
\begin{equation}\label{eq:T-on-boson}
\mathcal{T}_{\dot A}~\alpha_{C\dot C, n}^{(1)}\ket{0_R^{-}}^{(1)}\ket{0_R^{-}}^{(2)}
 = -\epsilon_{AC}\epsilon_{\dot A\dot C}\smashoperator{\sum_{p\,\text{odd}^+}}
   \frac{a^{2n+p}\Gamma(\frac{1}{2}-n)\Gamma(\frac{p}{2})}
             {2\pi\Gamma(-n)\Gamma(\frac{p+1}{2})}\psi^{-\dot A}_{-p}\ket{\chi}
    + \frac{1}{2} \alpha_{C\dot{C}, 2n}\ket{\psi}
\end{equation}
where the states $|\chi\rangle$ and $|\psi\rangle$ are given in
Equations~\eqref{pfive} and~\eqref{finalstate}.

\section{Intertwining Relations}\label{sec:intertwining}

We first present a formal, intuitive derivation of the intertwining
relations, which relate modes of the untwisted sector to modes of the
twisted sector. In Section~\ref{sec:subtleties}, we show how the
derived relations lead to multi-dimensional infinite series whose
value depends on how one takes the limit of the partial sums going to
infinity. We then give a more rigorous derivation of the relations,
which suggests a prescription on how to evaluate the ambiguous series.

\subsection{Basic Derivation}

We are interested in finding a relationship between modes in the
untwisted sector, ``before the twist,'' and modes in the twisted
sector, ``after the twist.'' More precisely we would like to know in
general how to write
\begin{smalleq}
\begin{multline}
\sigma^+_2(z_0) \big(\text{excitations before the twist}\big)\ket{0_R^-}^{(1)}\ket{0_R^-}^{(2)} \\
  = \big(\text{excitations after the twist}\big)\sigma^+_2(z_0)\ket{0_R^-}^{(1)}\ket{0_R^-}^{(2)}\\
  = \big(\text{excitations after the twist}\big)\ket{\chi}.
\end{multline}
\end{smalleq}
In~\cite{acm3}, an algorithm was found for doing just that; however,
the question this section addresses is whether there is a more general
relation between individual modes before the twist operator and after
the twist operator.

Since our argument does not depend on $SU(2)_L$ or
$SU(2)_1\times SU(2)_2$, we suppress the indices on the bosons and
fermions. We begin by noting that before the twist the correct field
expansions are
\begin{equation}\label{eq:bosons-before}
i\pd X^{(1)}(z) = \sum_n \frac{\alpha^{(1)}_n}{z^{n+1}}\qquad
i\pd X^{(2)}(z) = \sum_n \frac{\alpha^{(2)}_n}{z^{n+1}}\qquad
|z|<|z_0|,
\end{equation}
and
\begin{equation}\label{eq:fermions-before}
\psi^{(1)}(z) = \sum_{n\in\ints + \frac{1}{2}} \frac{\psi^{(1)}_n}{z^{n+\frac{1}{2}}}\qquad
\psi^{(2)}(z) = \sum_{n\in\ints + \frac{1}{2}} \frac{\psi^{(2)}_n}{z^{n+\frac{1}{2}}}\qquad
|z|<|z_0|.
\end{equation}
After the twist, the correct expansions are given by
\begin{equation}\label{eq:bosons-after}
i\pd X^{(1)}(z) = \frac{1}{2}\sum_n \frac{\alpha_n}{z^{\frac{n}{2}+1}}\qquad
i\pd X^{(2)}(z) = \frac{1}{2}\sum_n \frac{(-1)^n\alpha_n}{z^{\frac{n}{2}+1}}\qquad
|z|>|z_0|,
\end{equation}
and
\begin{equation}\label{eq:fermions-after}
\psi^{(1)}(z) = \frac{1}{2}\sum_n\frac{\psi_n}{z^{\frac{n}{2}+\frac{1}{2}}}\qquad
\psi^{(2)}(z) = \frac{1}{2}\sum_n\frac{(-1)^n\psi_n}{z^{\frac{n}{2}+\frac{1}{2}}}\qquad
|z|>|z_0|.
\end{equation}
Note that after the twist, there is no unique way of distinguishing
copies $(1)$ and $(2)$, but the above expansions correspond to
\emph{a} way of defining $(1)$ and $(2)$. Following~\cite{acm2,acm3},
we define the modes in the twisted sector with an extra factor of 2 so
that we can work with integers.

Now, recall that the fields $\pd X(z)$ and $\psi(z)$ should be
holomorphic functions except at isolated points where there are other
operator insertions. In particular, there is nothing special that
occurs on the circle $|z|=|z_0|$ (there \emph{is} something special at
the isolated point $z_0$). The curve is the boundary between our
twisted and untwisted mode expansions about the origin $z=0$, but if
one were to do mode expansions about a different point in the complex
plane then the expansions would change across a different curve (that
still passes through $z_0$).  Therefore, \emph{at least away from the
  twist operator at $z_0$}, we expect that the fields $\pd X(z)$ and
$\psi(z)$ should be continuous across $|z|=|z_0|$.

Thus, on the circle $|z|=|z_0|$ ({\em  excluding some neighborhood around
$z=z_0$}) we may identify
\begin{equation}\label{eq:boson-exp-equality}
\sum_n \frac{\alpha^{(1)}_n}{z^{n+1}} = \frac{1}{2}\sum_n \frac{\alpha_n}{z^{\frac{n}{2}+1}}
\qquad
\sum_n \frac{\alpha^{(2)}_n}{z^{n+1}} = \frac{1}{2}\sum_n \frac{(-1)^n\alpha_n}{z^{\frac{n}{2}+1}},
\qquad
|z|=|z_0|
\end{equation}
and
\begin{equation}\label{eq:fermion-exp-equality}
\sum_{n\in\ints + \frac{1}{2}} \frac{\psi^{(1)}_n}{z^{n+\frac{1}{2}}}
 = \frac{1}{2}\sum_n\frac{\psi_n}{z^{\frac{n}{2}+\frac{1}{2}}}
\qquad
\sum_{n\in\ints + \frac{1}{2}} \frac{\psi^{(2)}_n}{z^{n+\frac{1}{2}}}
  = \frac{1}{2}\sum_n\frac{(-1)^n\psi_n}{z^{\frac{n}{2}+\frac{1}{2}}},
\qquad
|z|=|z_0|.
\end{equation}
Multiplying \eqref{eq:boson-exp-equality} by $z^m$ and integrating
along the circle $|z|=|z_0|$, we
get
\begin{calc}\label{eq:boson-coef}
\alpha_m^{(1)} &= \frac{1}{2}\sum_n\alpha_n \int\frac{dz}{2\pi i} z^{m-\frac{n}{2}-1}\\
  &= \frac{1}{2}\sum_n\alpha_n \int_0^{2\pi}\frac{d\theta}{2\pi} (z_0e^{i\theta})^{m-\frac{n}{2}}\\
  &= \frac{1}{2}\alpha_{2m} + \frac{i}{2\pi}
     \sum_{n\,\text{odd}}\frac{z_0^{m-\frac{n}{2}}}{m-\frac{n}{2}}\alpha_n,
\end{calc}
and similarly
\begin{equation}\label{eq:boson-coef-2}
\alpha_m^{(2)} = \frac{1}{2}\alpha_{2m} 
             - \frac{i}{2\pi}\sum_{n\,\text{odd}}\frac{z_0^{m-\frac{n}{2}}}{m-\frac{n}{2}}\alpha_n,
\end{equation}
where the sum over the odds is both positive and negative.  These are
the desired relations between modes before the twist and modes after
the twist. The relations are analogous to the more general Bogolyubov
transformations discussed in condensed matter in~\cite{deGennes}. Note
that the contour in~\eqref{eq:boson-coef} is open, since we must exclude
some infinitesimal neighborhood around $z_0$. It is straightforward to
find the analogous relation for fermions in the R sector:
\begin{equation}\label{eq:fermion-coef}
\psi^{(1,2)}_n = 
\frac{1}{2}\psi_{2n} 
  \pm \frac{i}{2\pi}\sum_{k\,\text{odd}}\frac{z_0^{n-\frac{k}{2}}}{n-\frac{k}{2}}\psi_k.
\end{equation}
Given the delicate nature of the above argument, in particular with
regard to what is happening around $z_0$, one may not be surprised
that there are some hidden subtleties with these relations. Note also
that the derivation would seem to work for any holomorphic field
$\mathcal{O}(z)$, which \emph{cannot be correct} since some modes have
a nontrivial commutator with the twist operator, e.g. $J_0^-$. We should be
careful in what we mean when we write ``$=$'' in the above
expressions.  For example in Equation~\eqref{eq:boson-coef}, we
implicitly mean the operator relation
\begin{equation}
\sigma_2^+(z_0)\alpha_m^{(1)} = \left[\frac{1}{2}\alpha_{2m} + \frac{i}{2\pi}
     \sum_{n\,\text{odd}}\frac{z_0^{m-\frac{n}{2}}}{m-\frac{n}{2}}\alpha_n\right]\sigma_2^+(z_0)
\end{equation}
with the above radial ordering. The usage should be clear from the
context.

Before showing what goes wrong, let us first explore what goes right.
First of all this is the kind of relation we were hoping for: it
relates positive and negative modes before the twist to positive and
negative modes after the twist directly. The method given
in~\cite{acm3} can only relate states to states; it cannot relate an
individual mode before the twist to modes after the twist without
knowing what other excitations one has before the twist. 

An important requirement for Bogolyubov coefficients is that they
respect the commutation relations. We can check that the above
relations are consistent with the commutation relations by computing,
for instance,
\begin{calc}\label{eq:comm-check}
\com{\alpha_m^{(1)}}{\alpha_n^{(1)}} &=
\com{\frac{1}{2}\alpha_{2m} 
   + \frac{i}{2\pi}\sum_{k\,\text{odd}}\frac{z_0^{m-\frac{k}{2}}}{m-\frac{k}{2}}\alpha_k}
    {\frac{1}{2}\alpha_{2n}
   + \frac{i}{2\pi}\sum_{l\,\text{odd}}\frac{z_0^{n-\frac{l}{2}}}{n-\frac{l}{2}}\alpha_l}\\
 &= \frac{m}{2}\delta_{m+n,0} 
   - \frac{z_0^{m+n}}{4\pi^2}\sum_{k\,\text{odd}}\frac{k}{(m-\frac{k}{2})(n+\frac{k}{2})}.
\end{calc}
The sum is divergent; however, if we make the relatively modest
assumption that we should cutoff the sum symmetrically for positive
and negative $k$, then we find\footnote{This mild UV ambiguity might
  be seen as a hint of the other UV issues with the intertwining
  relations; however, this issue does not arise for the fermions, and
  it is of a different character. The UV issues that we discuss at
  length arise with multidimensional series; whereas the above is
  arguably the only reasonable regularization of the series
  in~\eqref{eq:comm-check}.  For example, if one cuts off the positive
  modes at $L$ and the negative modes at $-2L$ then one \emph{does}
  get a different answer, but it is not a consistent truncation of the
  Hilbert space to have a creation operator without its corresponding
  annihilation operator.}
\begin{equation}
\lim_{L\to\infty}\sum_{k\,\text{odd}}^{|k|<L}\frac{k}{(m-\frac{k}{2})(n+\frac{k}{2})}
 = - 2m\pi^2\delta_{m+n,0},
\end{equation}
which gives the correct answer. Similar calculations go through for
the other (anti-)com\-mut\-a\-tions.

Second, it gets the right answer for moving a single mode through the
twist operator. For instance,
\begin{smalleq}
\begin{calc}
\sigma^+_2(z_0)\alpha^{(1)}_{A\dot{A},n}\ket{0_R^-}^{(1)}\ket{0_R^-}^{(2)}
 &= \left(\frac{1}{2}\alpha_{A\dot{A},2n} + \frac{i}{2\pi}
    \sum_{k\,\text{odd}}\frac{z_0^{n-\frac{k}{2}}}{n-\frac{k}{2}}\alpha_{A\dot{A},k}\right)
    \sigma^+_2(z_0)\ket{0_R^-}^{(1)}\ket{0_R^-}^{(2)}\\
 &= \left(\frac{1}{2}\alpha_{A\dot{A}, 2n} + \frac{i}{2\pi}
    \sum_{k\,\text{odd}}\frac{z_0^{n-\frac{k}{2}}}{n-\frac{k}{2}}\alpha_{A\dot{A},k}\right)
    \ket{\chi}\\
 &= \left[\frac{1}{2}\alpha_{A\dot{A}, 2n} + \frac{i}{2\pi}
    \left(\smashoperator[r]{\sum_{k\,\text{odd}^+}}
          \frac{z_0^{n+\frac{k}{2}}}{n+\frac{k}{2}}\alpha_{A\dot{A},-k}
      + \smashoperator{\sum_{k,l\,\text{odd}^+}} \frac{z_0^{n-\frac{k}{2}}}{n-\frac{k}{2}}
                              \gamma^B_{kl}k\alpha_{A\dot{A}, -l}\right)\right]\ket{\chi}.\\
\end{calc}
\end{smalleq}
Making use of the identity in~\eqref{eq:id-1} one sees that
\begin{equation}\label{eq:gamma-id}
\sum_{k\,\text{odd}^+} \frac{z_0^{n-\frac{k}{2}}k\gamma^B_{kl}}{n-\frac{k}{2}}
 = \frac{z_0^{n+\frac{l}{2}}}{n+\frac{l}{2}}
      \left(\frac{\Gamma(\frac{l}{2})\Gamma(-n+\frac{1}{2})}{\Gamma(\frac{l+1}{2})\Gamma(-n)} 
              - 1\right),
\end{equation}
and thus
\begin{equation}\label{eq:single-boson}
\sigma^+_2(z_0)\alpha^{(1)}_{A\dot{A},n}\ket{0_R^-}^{(1)}\ket{0_R^-}^{(2)}
 = \left[\frac{1}{2}\alpha_{A\dot{A},2n} + \frac{i}{2\pi}\sum_{l\,\text{odd}^+}
         \frac{z_0^{n+\frac{l}{2}}}{n+\frac{l}{2}}
    \frac{\Gamma(\frac{l}{2})\Gamma(-n+\frac{1}{2})}{\Gamma(\frac{l+1}{2})\Gamma(-n)}
    \alpha_{A\dot{A},-l}
  \right]\ket{\chi}.
\end{equation}
For $n$ positive the above vanishes as it should, since the positive
even mode in the first term annihilates $\ket{\chi}$ and $\Gamma(-n)$
kills the second term. For $n$ negative this reproduces the result
found in~\cite{acm3} for a single mode in the
initial state.  

Similarly, if one performs the analogous calculation for
fermions with~\eqref{eq:fermion-coef}, then one can use
\begin{equation}
\psi^{\alpha\dot A}_{+k}\ket{\chi} = 2\sum_{p\,\text{odd}^+}
                              \left(\gamma^F_{pk}\delta^\alpha_+\psi^{+\dot A}_{-p}
                                -\gamma^F_{kp}\delta^\alpha_-\psi^{-\dot A}_{-p}\right)
                              \ket{\chi}\qquad k\,\text{odd, positive},
\end{equation}
to find
\begin{subequations}\label{eq:single-fermion}
\begin{align}
\sigma_2^+(z_0)\psi^{(1)+\dot A}_{n}\ket{0_R^-}^{(1)}\ket{0_R^-}^{(2)}
 &= \left[\frac{1}{2}\psi^{+\dot A}_{2n} + \frac{i}{2\pi}
        \sum_{p\,\text{odd}^+}\frac{z_0^{n+\frac{p}{2}}}{n+\frac{p}{2}}
           \frac{\Gamma(\frac{p}{2}+1)\Gamma(-n+\frac{1}{2})}{\Gamma(\frac{p+1}{2})\Gamma(-n+1)}
           \psi^{+\dot A}_{-p}\right]\ket{\chi}\\
\sigma_2^+(z_0)\psi^{(1)-\dot A}_{n}\ket{0_R^-}^{(1)}\ket{0_R^-}^{(2)}
 &= \left[\frac{1}{2}\psi^{-\dot A}_{2n} + \frac{i}{2\pi}
        \sum_{p\,\text{odd}^+}\frac{z_0^{n+\frac{p}{2}}}{n+\frac{p}{2}}
           \frac{\Gamma(\frac{p}{2})\Gamma(-n+\frac{1}{2})}{\Gamma(\frac{p+1}{2})\Gamma(-n)}
           \psi^{-\dot A}_{-p}\right]\ket{\chi}.
\end{align}
\end{subequations}
This agrees with~\cite{acm3}.

\subsection{Problems}\label{sec:problems}

There are two problems with the above derivation. One is that this
formal derivation only makes use of the holomorphicity of the fields,
which means that one could make the same argument for any other
holomorphic field. For instance, consider $J^a(z)$, one would get
\begin{equation}
J^{a(1,2)}_n \overset{?}{=} \frac{1}{2}J^a_{2n}
             \pm \frac{i}{2\pi}\sum_{k\,\text{odd}}\frac{z_0^{n-\frac{k}{2}}}{n-\frac{k}{2}} J^a_k,
\end{equation}
which leads to 
\begin{calc}\label{eq:wrong-J-com}
\com{J^-_0}{\sigma_2^+(z_0)} &= J^-_0\sigma_2^+(z_0) 
              - \sigma_2^+(z_0)\big(J_0^{(1)} + J_0^{(2)}\big)\\
 &\overset{?}{=} 0.
\end{calc}
One should find $\sigma_2^-(z_0)$, not zero. One finds similar
contradictions if one tries to use the same argument for $T(z)$, too.

The second problem, alluded to above, concerns using the intertwining
relations with more than one mode. The simplest instance may be to
consider
\begin{equation}
\sigma_2^+(z_0)\alpha^{(1)}_{++,m}\alpha^{(1)}_{--,-n}\ket{0_R^-}^{(1)}\ket{0_R^-}^{(2)}
 = -m\delta_{m,n}\ket{\chi}\qquad m,n>0.
\end{equation}
If we use Equation~\eqref{eq:boson-coef}, then we get
\begin{calc}\label{eq:two-boson-problem}
- m \delta_{m,n}\ket{\chi}
 &\overset{?}{=}\hspace{-1pt}
\left(\frac{1}{2}\alpha_{++, 2m} 
        + \frac{i}{2\pi}\sum_{k\,\text{odd}}\frac{z_0^{m-\frac{k}{2}}}{m-\frac{k}{2}}
            \alpha_{++, k}\right)\hspace{-5pt}
\left(\frac{1}{2}\alpha_{--, -2n} 
        + \frac{i}{2\pi}\sum_{l\,\text{odd}}\frac{z_0^{-n-\frac{l}{2}}}{-n-\frac{l}{2}}
            \alpha_{--,l}\right)\hspace{-1pt}\ket{\chi}\\
 &= -\frac{m}{2}\delta_{m,n}\ket{\chi}
    -\frac{1}{4\pi^2}\left(\sum_{k\,\text{odd}}\frac{z_0^{m-\frac{k}{2}}}{m-\frac{k}{2}}
                          \alpha_{++, k}\right)
                   \left(\sum_{l\,\text{odd}}\frac{z_0^{-n-\frac{l}{2}}}{-n-\frac{l}{2}}
                         \alpha_{--,l}\right)\ket{\chi}\\
 &= -\frac{m}{2}\delta_{m,n}\ket{\chi}\\
    &\quad-\frac{1}{4\pi^2}\left(\sum_{k\,\text{odd}}\frac{z_0^{m-\frac{k}{2}}}{m-\frac{k}{2}}
                          \alpha_{++, k}\right)\hspace{-5pt}
                   \left[\smashoperator[r]{\sum_{j\,\text{odd}^+}}\alpha_{--,-j}\hspace{-2pt}
                     \left(
                       \frac{z_0^{-n+\frac{j}{2}}}{-n+\frac{j}{2}} + \sum_{l\,\text{odd}^+}
                       \frac{z_0^{-n-\frac{l}{2}}l\gamma_{lj}}{-n-\frac{l}{2}}
                         \right)\hspace{-2pt}\right]\hspace{-1pt}\ket{\chi}\\
\end{calc}
Note that the even--odd cross-terms vanish since they commute and
either the $\alpha_{++,2m}$ or the $\alpha_{++,k}$-sum kills
$\ket{\chi}$ (from~\eqref{eq:single-boson}). Similarly, we need only
look at the commutator of the $\alpha_{++,k}$-sum and the
square-bracketed expression, which gives
\begin{calc}\label{eq:ambig-sum}
&-\left(\sum_{k\,\text{odd}^+}\frac{kz_0^{m-\frac{k}{2}}}{m-\frac{k}{2}}\right)
 \left(\frac{z_0^{-n+\frac{k}{2}}}{-n+\frac{k}{2}} + \sum_{l\,\text{odd}^+}
                       \frac{z_0^{-n-\frac{l}{2}}l\gamma_{lk}}{-n-\frac{l}{2}}
                         \right)\\
&= z_0^{m-n}\sum_{k\,\text{odd}^+}\left(\frac{k}{(m-\frac{k}{2})(n-\frac{k}{2})}
  + \sum_{l\,\text{odd}^+}\frac{z_0^{-\frac{k}{2}-\frac{l}{2}}kl\gamma_{kl}}
                                          {(m-\frac{k}{2})(n+\frac{l}{2})}\right).
\end{calc}
We would like the above expression to evaluate to $2\pi^2 m
\delta_{m,n}$ in order to get the correct answer; however, the above
summations depend sensitively on the order in which one adds the
infinite number of terms. For instance, if we attempt to perform the
$k$-sum \emph{first}, then the first term is divergent and the second
term, using~\eqref{eq:gamma-id}, gives
\begin{equation}
-\sum_{l\,\text{odd}^+}\frac{l}{(m+\frac{l}{2})(n+\frac{l}{2})},
\end{equation}
which is also divergent. 

On the other hand, if we perform the $l$ sum first,
using~\eqref{eq:gamma-id} we find
\begin{calc}
& z_0^{m-n}\sum_{k\,\text{odd}^+}\left(\frac{k}{(m-\frac{k}{2})(n-\frac{k}{2})}
  + \sum_{l\,\text{odd}^+}\frac{z_0^{-\frac{k}{2}-\frac{l}{2}}kl\gamma_{kl}}
                                          {(m-\frac{k}{2})(n+\frac{l}{2})}\right)\\
&= z_0^{m-n}\sum_{k\,\text{odd}^+}\left(\frac{k}{(m-\frac{k}{2})(n-\frac{k}{2})}
  + \frac{k}{(m-\frac{k}{2})(n-\frac{k}{2})}
     \left[\frac{\Gamma(\frac{k}{2})\Gamma(n+\frac{1}{2})}{\Gamma(\frac{k+1}{2})\Gamma(n)} 
            - 1\right]\right)\\
&= z_0^{m-n}\frac{\Gamma(n+\frac{1}{2})}{\Gamma(n)}
  \sum_{k\,\text{odd}^+}\left(\frac{k}{(m-\frac{k}{2})(n-\frac{k}{2})}
  \frac{\Gamma(\frac{k}{2})}{\Gamma(\frac{k+1}{2})}\right),
\end{calc}
which using another identity,
\begin{equation}
\sum_{k\,\text{odd}^+}\frac{k}{(m-\frac{k}{2})(n-\frac{k}{2})}
      \frac{\Gamma(\frac{k}{2})}{\Gamma(\frac{k+1}{2})}
 = 2\pi^2 m \frac{\Gamma(n)}{\Gamma(n+\frac{1}{2})}\delta_{m,n}\qquad m,n>0,
\end{equation}
gives the correct answer.

How should we think of the ambiguity in Equation~\eqref{eq:ambig-sum}?
Any infinite series is implicitly evaluated by determining the limit
of a sequence of \emph{partial sums}. In our case, higher values of
$k$ and $l$ correspond to higher modes, so it is natural to think of
imposing UV cutoffs on the sums, $k<L_1$ and $l<L_2$. We then wish to
take the limit as $L_1, L_2\to\infty$, but there are many different
ways to do that. If we define $b=L_1/L_2$ to parameterize the
different ways of evaluating~\eqref{eq:ambig-sum}, then evaluating the
$k$-sum first corresponds to $b=\infty$, while evaluating the $l$-sum
first corresponds to $b=0$. These are just two of an infinite number
of ways to evaluate the double-sum.

Given that these ambiguous multi-dimensional series are rampant in
this formalism and that frequently the correct method of evaluating
them may be much less obvious,\footnote{For instance, there are cases
  involving triple-sums, where more than one order of evaluating the
  sums give distinct, finite results.} we need a well-motivated
principle that determines the correct way to handle the UV physics.

\subsection{A More Rigorous Derivation}\label{sec:subtleties}
\begin{sloppypar}
  We now present a more rigorous derivation of the intertwining
  relations in Equations~\eqref{eq:boson-coef},
  \eqref{eq:boson-coef-2}, and~\eqref{eq:fermion-coef}. By
  continuously deforming the contour integral for an initial state
  mode outward only where the integrand is holomorphic, we can treat
  the point $z_0$ more carefully. This resolves the two problems
  outlined above.
\end{sloppypar}

\begin{figure}[ht]
\begin{center}
\subfigure[]
  {\includegraphics[height=3cm, clip=true, trim=0 48 0 48]{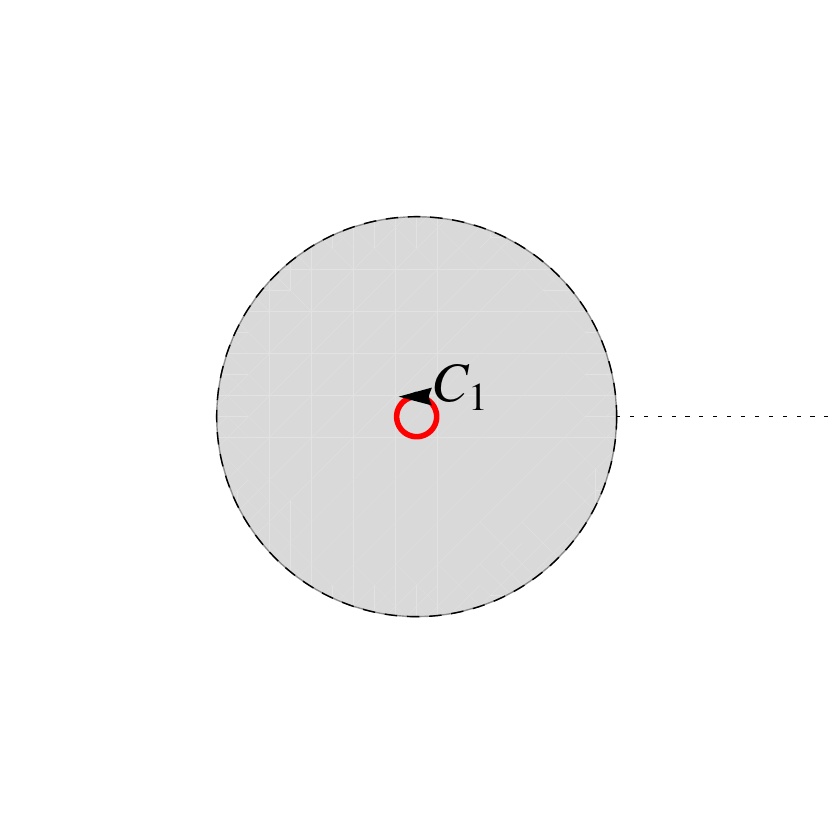}\label{fig:mode1-z}}
\subfigure[]
  {\includegraphics[height=3cm, clip=true, trim=0 48 0 48]{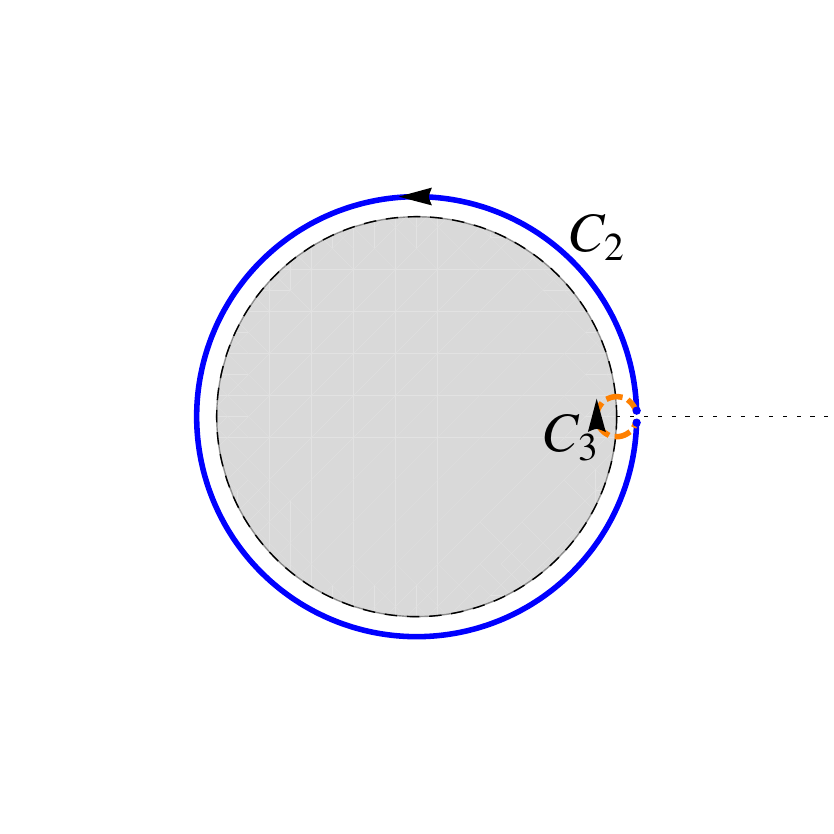}}
  \caption[Deforming the contour in the $z$-plane]{In the $z$-plane,
    showing how the contour $C_1$ (solid, red) in (a) may be deformed
    out and around the branch cut into contours $C_2$ (solid, blue)
    and $C_3$ (dashed, orange) in (b).  The gray circular region is
    the ``before the twist'' region, $|z|<|z_0|$.  The branch cut is
    indicated by the dashed black line extending out from the
    circle.\label{fig:z-plane-contours}}
\end{center}
\end{figure}

\subsubsection{Bosons}

\begin{figure}[tb]
\begin{center}
\includegraphics[height=3cm]{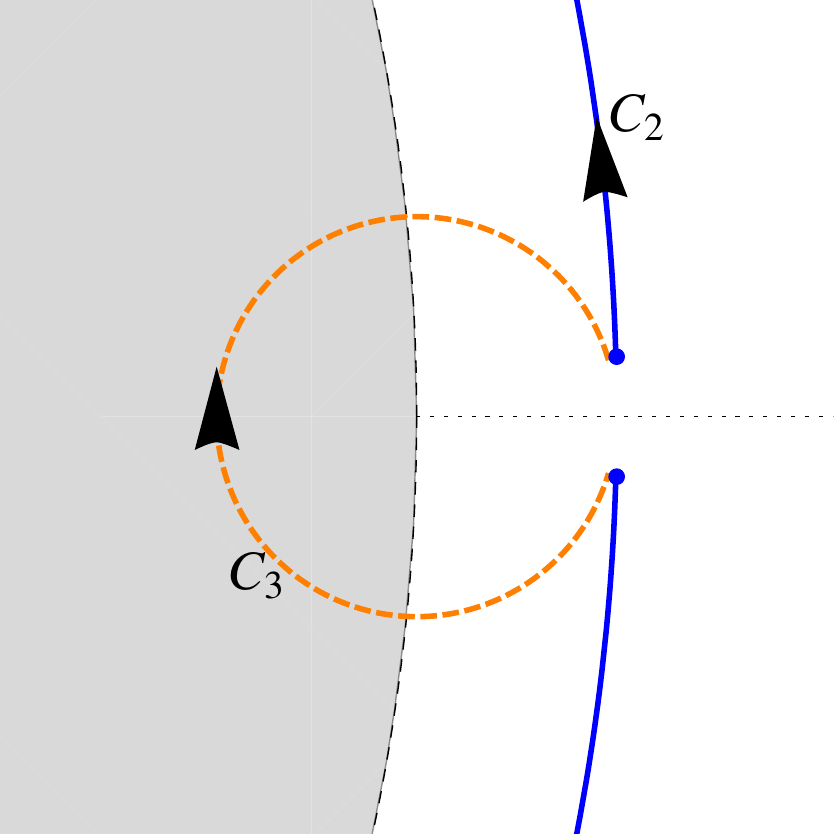}
\caption[A close-up of contours $C_2$ and $C_3$]{A close-up depiction
  of contours $C_2$ and $C_3$ meeting.  Note that we have added an
  artificial gap around the branch cut for illustrative purposes
  only---in fact, both $C_2$ and $C_3$ are full
  circles.\label{fig:close-up}}
\end{center}
\end{figure}

\begin{sloppypar}
Working with the bosons first, let us note that
\begin{equation}
\alpha_n^{(1)} = \oint_{C_1}\frac{\drm z}{2\pi i}i\pd X^{(1)}(z) z^n,
\end{equation}
where $C_1$ is a circular contour with radius less than $|z_0|$ shown
in Figure~\ref{fig:mode1-z}, and $i\pd X^{(1)}(z)$ is a holomorphic
function except at $z=z_0$ (and excluding any other operator
insertions). Thus, we may deform the contour into an open circle $C_2$
of radius larger than $|z_0|$ and a contour, $C_3$, sneaking around
the branch cut starting at $z=z_0$, as shown in
Figures~\ref{fig:z-plane-contours} and~\ref{fig:close-up}.  We take
$C_3$ to be a circle of radius $\veps$, which we eventually take to
zero. The orientations of the contours are shown in the figures.
\end{sloppypar}

We write the above contour integral as
\begin{equation}
\alpha_n^{(1)} = \int_{C_2}\frac{\drm z}{2\pi i}i\pd X^{(1)}(z) z^n 
          + \int_{C_3}\frac{\drm z}{2\pi i}i\pd X^{(1)}(z) z^n.
\end{equation}
The $C_2$ term, is what we have been calculating and is given by (with
$\veps$ corrections)
\begin{calc}
\int_{C_2}\frac{\drm z}{2\pi i} i\pd X(z) z^n &= 
\frac{i}{2}\sum_k\alpha_k \int_{C_2}\frac{\drm z}{2\pi i} z^{n - \frac{k}{2} -1}\\
  &= \frac{1}{4\pi}\sum_k \alpha_k(z_0+\veps)^{n-\frac{k}{2}}
           \int_0^{2\pi}\drm \theta\,e^{i(n-\frac{k}{2})\theta}\\
  &= \frac{1}{2}\alpha_{2n} 
    + \frac{i}{2\pi}\sum_{k\,\text{odd}}\frac{(z_0+\veps)^{n-\frac{k}{2}}}
                                             {n-\frac{k}{2}}\alpha_k.
\end{calc}

\begin{figure}
\begin{center}
\subfigure[]
 {\includegraphics[height=6cm]{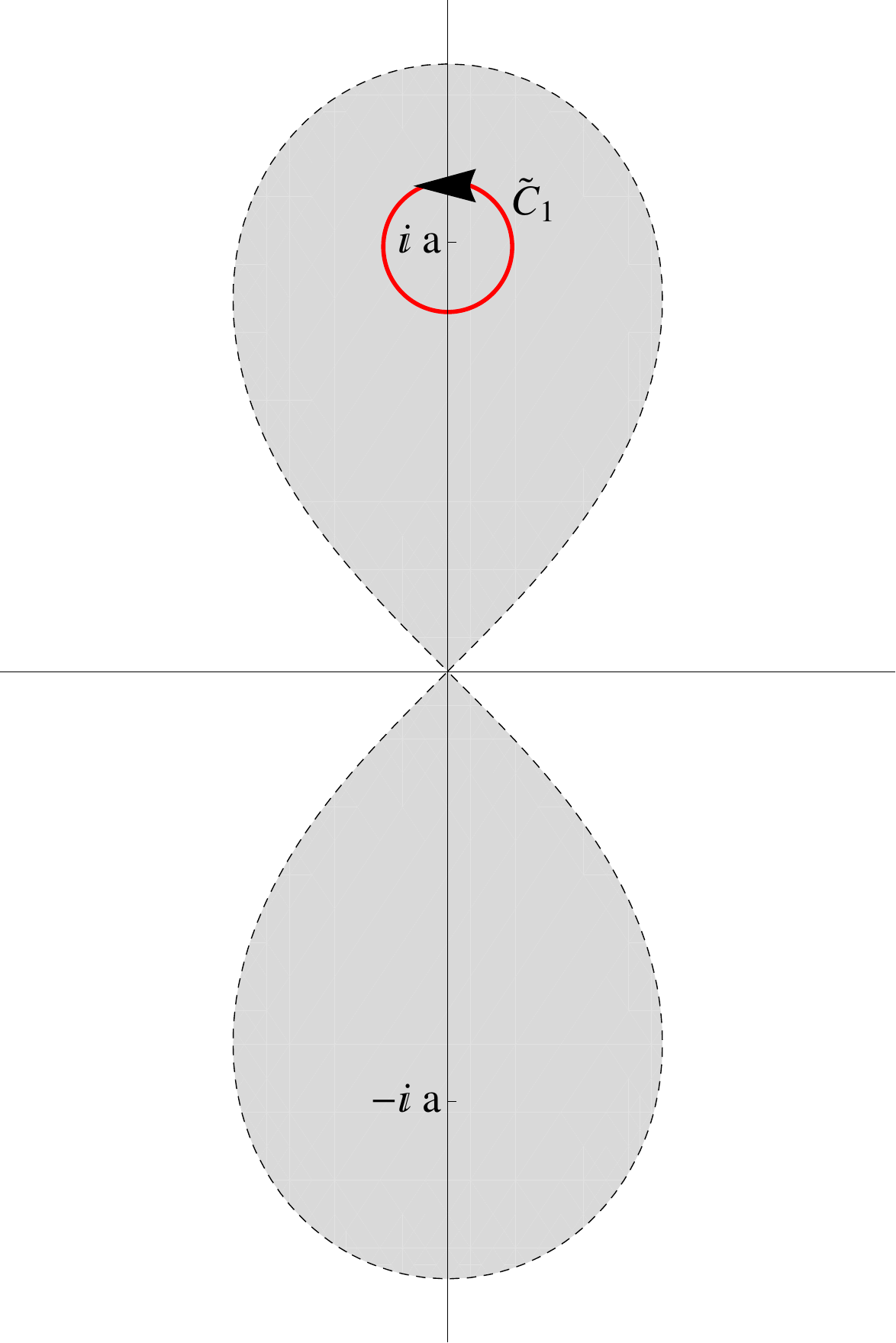}}
\qquad
\subfigure[]
 {\includegraphics[height=6cm]{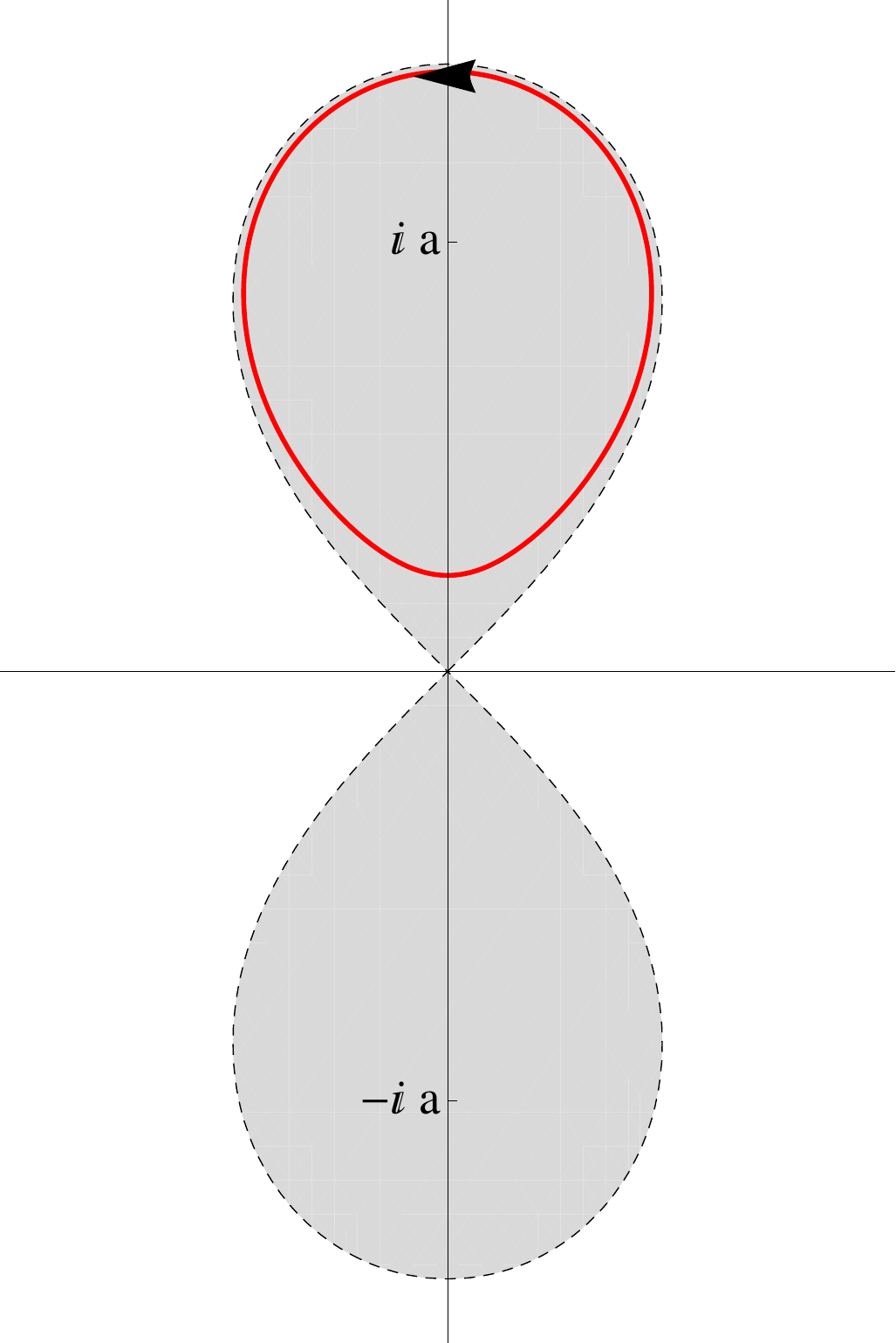}}
\qquad
\subfigure[]
 {\includegraphics[height=6cm]{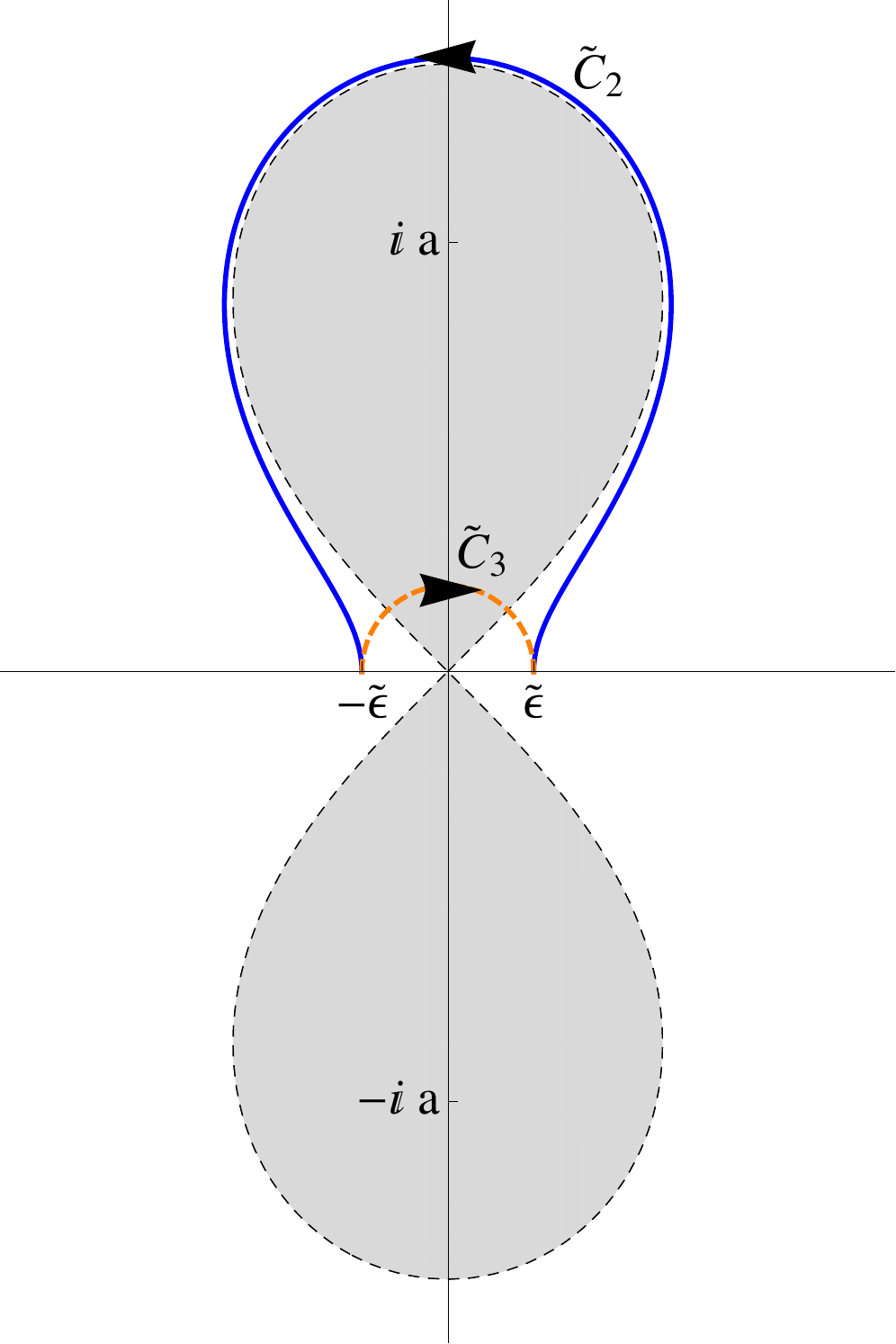}}
 \caption[The contours in the $t$-plane]{We show the $t$-plane where there is no branch cut and
   the fields are single-valued.  We start in~(a) with the $\tilde{C}_1$
   contour (solid, red), which we finally deform out into $\tilde{C}_2$ (solid,
   blue) and $\tilde{C}_3$ (dashed, orange) in~(c).
   \label{fig:t-plane-contours}}
\end{center}
\end{figure}

At this point, it becomes necessary to introduce the covering space
where the fields are well-defined. We map to the covering space
coordinate $t$ via
\begin{equation}
z = z_0 + t^2\qquad a^2 = z_0.
\end{equation}
The points $ia$ and $-ia$ are the two images of the origin $z=0$, one
corresponding to each copy of the fields. Mapping the $z$-plane
contours in Figure~\ref{fig:z-plane-contours} to the $t$-plane results
in Figure~\ref{fig:t-plane-contours}.

We compute the $C_3$ term by going to the $t$-plane
\begin{calc}
\int_{C_3}\frac{\drm z}{2\pi i}\pd X(z)z^n = 
  \int_{\tilde{C}_3}\frac{\drm t}{2\pi i} \pd X(t)(z_0 + t^2)^n
  \xrightarrow[\tilde{\veps}\to 0]{} 0,
\end{calc}
where the radius of the semicircular contour in the $t$-plane,
$\tilde{C}_3$, is given by $\tilde{\veps}$. The contour gives zero
contribution since there is no bosonic insertion at $t=0$ and the
integrand is therefore analytic.  As the length of the contour goes to
zero, therefore, so does the integral. Thus, we see that the story for
the bosons is exactly as stated, and in the limit as $\veps\to 0$ we
reproduce our previous result,
\begin{equation}
  \alpha_n^{(1,2)} = \frac{1}{2}\alpha_{2n} 
  \pm \frac{i}{2\pi}\sum_{k\,\text{odd}}
  \frac{z_0^{n-\frac{k}{2}}}{n-\frac{k}{2}}\alpha_k.
\end{equation}

\subsubsection{Fermions}

For the fermions we have
\begin{equation}
\psi^{(1)}_n = \oint_{C_1}\frac{\drm z}{2\pi i}\psi(z) z^{n-\frac{1}{2}},
\end{equation}
which becomes
\begin{equation}
\psi^{(1)}_n = \int_{C_2}\frac{\drm z}{2\pi i}\psi(z) z^{n-\frac{1}{2}}
              +\int_{C_3}\frac{\drm z}{2\pi i}\psi(z) z^{n-\frac{1}{2}}
\end{equation}
The $C_2$ term is what we have computed for the fermions previously,
and is given by
\begin{calc}
\int_{C_2}\frac{\drm z}{2\pi i}\psi(z) z^{n-\frac{1}{2}} &=
   \frac{1}{2}\sum_k\psi_k\int_{C_2}\frac{\drm z}{2\pi i} z^{n-\frac{k}{2}-1}\\
  &= \frac{1}{2}\psi_{2n} 
  \pm \frac{i}{2\pi}\sum_{k\,\text{odd}}\frac{z_0^{n-\frac{k}{2}}}{n-\frac{k}{2}}\psi_k.
\end{calc}
The $C_3$ contribution also goes to zero for the fermions. When one
goes to the $t$-plane, one finds
\begin{equation}
  \int_{\tilde{C}_3}\frac{\drm t}{2\pi i}\psi(t)\sqrt{2t}(z_0 + t^2)^{n-\frac{1}{2}},
\end{equation}
which acts on the spin field $S(t=0)$. The most singular term in the
OPE between $\psi(t)$ and $S(0)$ is proportional to $1/\sqrt{t}$, and
so one again finds that the semi-circular contour vanishes.

At this point, we see the resolution of the first problem we found
with our formal derivation. If the field has an OPE with the
spin-field in the covering space which is singular enough, then the
$C_3$-contribution is nonvanishing in the limit as $\tilde{\veps}\to
0$. This extra contribution gives exactly the correct answer for
$J_0^-$, for instance, as we demonstrate in
Section~\ref{sec:contour-J}.

\subsubsection{Multiple Contours}

Having resolved the first problem with the formal derivation, we
should now discuss the UV issues that arise with multiple
contours.\footnote{Recall that the definition of the twist operator
  $\sigma_2$ involves a hole in the $z$-plane, whose radius is
  carefully taken to zero~\cite{lm1}. One might suspect that this
  limit has important consequences for these UV issues and that the
  size of the hole plays the role of a UV cutoff. In fact, the hole is
  taken to zero size \emph{before} any of the issues discussed in this
  section, and does not play any role here. The actual issue is the
  interaction between neighboring contours, as discussed.}

Let us write $\veps_i$ for the radius of the $i$th mode's $C_3$
semi-circle. For instance, consider an initial state
\begin{equation}
\alpha^{(1)}_{n_1}\alpha^{(1)}_{n_2}\ket{0_R^-}^{(1)}\ket{0_R^-}^{(2)}.
\end{equation}
Then the $C_3$ part coming from $\alpha^{(1)}_{n_1}$ has radius
$\veps_1$ in the $z$-plane and therefore the semi-circle in the
$t$-plane has radius $\sqrt{\veps_1}$. Similarly, for
$\alpha^{(1)}_{n_2}$. If we require that the $C_2$ parts of
$\alpha_{n_1}$ and $\alpha_{n_2}$ preserve the same ordering, then the
semi-circles in the $t$-plane satisfy
\begin{equation}
\veps_2 < \veps_1.
\end{equation}

We now argue that, in fact, we should have $\veps_2 \ll \veps_1$ in
order to use the intertwining relations as we want to. Consider the
two semi-circular $C_3$ contours at leading order in $\veps_1$ and
$\veps_2$:
\begin{calc}
 \int_{\tilde{C}_3(\veps_1)}\frac{\drm t}{2\pi i} \pd X(t)(z_0 + t^2)^{n_1}
 &\int_{\tilde{C}_3(\veps_2)}\frac{\drm t'}{2\pi i} \pd X(t') (z_0 + {t'}^2)^{n_2}\\
 &\sim \int_{\tilde{C}_3(\veps_1)}\frac{\drm t}{2\pi i} (z_0 + t^2)^{n_1}
\int_{\tilde{C}_3(\veps_2)}\frac{\drm t'}{2\pi i} \frac{ (z_0 + {t'}^2)^{n_2}}{(t-t')^2}\\
 &\sim \int_{\tilde{C}_3(\veps_1)}\frac{\drm t}{2\pi i} z_0^{n_1}
\int_{\tilde{C}_3(\veps_2)}\frac{\drm t'}{2\pi i} \frac{ z_0^{n_2}}{(t-t')^2}\\
 &\sim z_0^{n_1+n_2}\log \frac{1 - \sqrt{\frac{\veps_2}{\veps_1}}}
                              {1+ \sqrt{\frac{\veps_2}{\veps_1}}}\\
 &\sim z_0^{n_1 + n_2} \sqrt{\frac{\veps_2}{\veps_1}} + \cdots.
\end{calc}
This vanishes only if $\veps_2\ll \veps_1$. Note that this argument is
unaffected if the modes are on different copies (the integrals work
out in essentially the same way).

Taking $\veps_2\ll \veps_1$ suggests a particular way to take the
limit for double-sums: if $L_i$ is the cutoff on the
$\alpha_{n_i}^{(1)}$-sum, then we should take
\begin{equation}
L_2 \gg L_1,
\end{equation}
that is, evaluate the $L_2$-sum first and then the $L_1$-sum. This is
exactly the way that we got the correct answer, when the issue was
demonstrated in~\eqref{eq:two-boson-problem}.

\subsubsection{The Prescription}\label{sec:prescription}

We now develop the precise prescription that resolves the UV
ambiguities. Before we state the prescription, we should mention that
there are two kinds of sums over modes. There are the after-the-twist
intertwining sums, which have ordering ambiguities among themselves,
and there can also be before-the-twist sums on modes before the twist
operator.  For example, consider a composite operator like
\begin{equation}\label{eq:Ln}
J_n^{a(1)} = -\frac{1}{4}{(\sigma^{aT})^\alpha}_\beta\sum_{j =-\infty}^\infty
                \psi^{(1)\dg}_{\alpha \dot A, n-j}\psi^{(1)\beta\dot A}_j.
\end{equation}
In fact, these sums also have UV-limit issues when combined with the
intertwining relations as is demonstrated in
Section~\ref{sec:composite}. If we look at the intertwining relation
in~\eqref{eq:boson-coef}, for example, with the implicit cutoff on the
sum,
\begin{equation}\label{eq:boson-coef-cutoff}
\alpha_m^{(2)} = \frac{1}{2}\alpha_{2m} 
                  - \frac{i}{2\pi}\sum_{n\,\text{odd}}^{|n|<L}
                     \frac{z_0^{m-\frac{n}{2}}}{m-\frac{n}{2}}\alpha_n,
\end{equation}
we see that we are approximating a mode $m$ as a linear combination of
modes with UV cutoff $L$. In order for this approximation to become an
exact expansion we must take $L\to\infty$ \emph{with $m$ fixed.} That
is we need to have much higher frequency modes in our sum than the
mode that we are expanding. Therefore, we need to cutoff the
before-the-twist sum in~\eqref{eq:Ln} and ensure that its cutoff is
much less than the after-the-twist cutoff
in~\eqref{eq:boson-coef-cutoff}.

Before proceeding, let us consider more generally what sort of
multi-dimensional series we can get in this formalism. If we have a
bunch of modes in the initial state that we pull across using the
intertwining relations, then there are several different types of
terms that can arise.  There is a term in which all of the positive
modes act on $\ket{\chi}$ separately, and we are left with a product
of one-dimensional sums that result in either~\eqref{eq:single-boson}
or~\eqref{eq:single-fermion}. Then, there are terms where the positive
modes from one sum contract with negative modes of another. This gives
a double sum, like the one in~\eqref{eq:ambig-sum}. If there is a sum
on the before-the-twist modes, then we can get a triple sum if those
two modes contract. This is the most complicated sum possible.

Finally, we are ready to state the prescription that ensures that the
multi-dimensional series converge to the correct answer. The
prescription is
\begin{enumerate}
\item The after-the-twist intertwining sums should be performed from
  innermost contour (right-most mode) to outermost contour (left-most
  mode). This ensures that the $C_3$-terms can be dropped.
\item Any sums on before-the-twist modes should be performed
  \emph{last}. There is no UV-ambiguity among the
  before-the-twist sums.
\end{enumerate}

Note that with this prescription we have a weakened version of our
goal. Because of the UV sensitive series, one cannot directly map
operators to operators since one requires knowledge of what other
modes are around in order to correctly evaluate the series.  While the
modes that we start with before the twist operator may commute, we
need to think about them pulling across the twist operator in a
particular order. It is in this sense, that we have intertwining
relations and not Bogolyubov coefficients.

\section{An Example: Intertwining Relations for \texorpdfstring{$J^a_n$}{Jn}}
\label{sec:example}

There are two equivalent ways of calculating the effect of the twist
operator on composite operators such as $J^a_n$. One way is to use the
contour deformation method described in this section, being careful not
to throw away the small contour $C_3$. The other way is to write $J^a$
as the product of fermion modes and use the intertwining relations for
the fermions, being careful to use the prescriptions described in
Section~\ref{sec:prescription}.

\subsection{The Contour Method}\label{sec:contour-J}

We first describe the contour method mentioned above. If we go through
the argument described in Section~\ref{sec:subtleties} then
\begin{equation}
J^{a(1)}_n = \frac{1}{2}J^a_{2n} + \frac{i}{2\pi}\sum_{k\,\text{odd}}
        \frac{z_0^{n-\frac{k}{2}}}{n-\frac{k}{2}}J^a_k
        + \lim_{\veps\to 0}\int_{C_3}\frac{\drm z}{2\pi i} J^{a(1)}(z)z^n, 
\end{equation}
but this time we find a nonvanishing contribution coming from the
$C_3$ contour. As we pull the $J^{a(1)}_n$ contour out, it acts on the
twist operator (or equivalently, the spin field in the covering space)
and can switch a $\sigma_2^+$ to $\sigma_2^-$. Therefore, let us
consider $\sigma_2^{\alpha}$. 

We can evaluate the $C_3$ contour by going to the $t$-plane, where
$\sigma_2^\alpha(z_0)$ leaves only a spin field $S^\alpha(t=0)$. The
image of $C_3$ in the $t$-plane, $\tilde{C}_3$, is a semi-circle
around the origin as shown in Figure~\ref{fig:t-plane-contours}. Thus,
the $\tilde{C}_3$ contour implicitly acts on the spin field:
\begin{calc}
\int_{C_3}\frac{\drm z}{2\pi i} J^{(1)a}(z)z^n
 &= \left[\int_{\tilde{C}_3}\frac{\drm t}{2\pi i} J^a(t)(z_0 + t^2)^n\right]S^\alpha(0)\\
 &= \frac{1}{2}{(\sigma^{aT})^\alpha}_\beta S^\beta(0)
    \int_{\tilde{C}_3}\frac{\drm t}{2\pi i}\frac{(z_0 + t^2)^n}{t}\\
 &= \frac{1}{2}{(\sigma^{aT})^\alpha}_\beta S^\beta(0)
     \frac{1}{2\pi i}\left(z_0^n \log t \bigg|^{\veps}_{-\veps} + \bigO(\veps)\right)\\
 &=  -\frac{z_0^n}{4}{(\sigma^{aT})^\alpha}_\beta S^\beta(0),
\end{calc}
where we have given the result after taking $\veps \to 0$. When we go
back to the $z$-plane we should write the above result as
\begin{equation}
\sigma_2^\alpha(z_0)J^{a(1)}_n  = \left[\frac{1}{2}J^a_{2n} 
          +\frac{i}{2\pi}\sum_{k\,\text{odd}}\frac{z_0^{n-\frac{k}{2}}}{n-\frac{k}{2}}J^a_k\right]
            \sigma_2^\alpha(z_0)
             - \frac{z_0^n}{4}{(\sigma^{aT})^\alpha}_\beta\sigma_2^\beta(z_0).
\end{equation}
If we had considered $J_n^{a(2)}$, then we obtain a similar result. We
can summarize the two relations and write them in a suggestive form as
\begin{equation}\label{eq:J-intertwining}
\sigma_2^\alpha(z_0)J^{a(1,2)}_n  = \left[\frac{1}{2}J^a_{2n} 
          \pm\frac{i}{2\pi}\sum_{k\,\text{odd}}\frac{z_0^{n-\frac{k}{2}}}{n-\frac{k}{2}}J^a_k\right]
            \sigma_2^\alpha(z_0)
             - \frac{z_0^n}{2}\com{J^a_0}{\sigma_2^\alpha(z_0)}. 
\end{equation}
We see that if we use these intertwining relations, then we get the
correct answer in~\eqref{eq:wrong-J-com}.

While the above intertwining relation is correct, it may not be the
most useful form. Because we have switched the charge on the twist
operator, we now must think about a new state $\ket{\chi}$ created by
the negatively-charged operator acting on the vacuum.

\subsubsection{An Example}

For concreteness, let us consider
\begin{equation}
\sigma_2^+(z_0) J_n^{-(1)}\ket{0^-}^{(1)}\ket{0^-}^{(2)},\qquad n<0,
\end{equation}
then we have
\begin{equation}
\left[\frac{1}{2}J^-_{2n} + \frac{i}{2\pi}
         \sum_{k\,\text{odd}}\frac{z_0^{n-\frac{k}{2}}}{n-\frac{k}{2}}J^-_k
- \frac{z_0^n}{2}J_0^-\right]\ket{\chi}.
\end{equation}
Thus, our first task is to compute the three terms,
\begin{equation}
J_{2n}^-\ket{\chi}\qquad
J_k^-\ket{\chi}\,k\,\text{odd}\qquad 
J_0^-\ket{\chi}.
\end{equation}

There are no real complications in working the terms out. For
instance, for the first term, one starts with\footnote{Note that the
  $J$ after the twist has an extra factor of $1/2$ from before the
  twist, which arises from the fermion--fermion anticommutator having
  an extra factor of $2$ after the twist.}
\begin{equation}
J^-_{2n} = -\frac{1}{4}\sum_j\psi^\dg_{+\dot A, 2n - j}\psi^{-\dot A}_{j},
\end{equation}
then breaks the sum into terms with both modes negative and terms with
odd positive modes that act on $\ket{\chi}$. One can write the result
in the form
\begin{equation}
J^-_{2n}\ket{\chi} = 
-\frac{1}{4}\sum_{2n+1\leq j \leq -1} \psi^\dg_{+\dot A, 2n-j}\psi^{-\dot A}_j\ket{\chi}
 + \sum_{j,p\,\text{odd}^+}\gamma^F_{jp}\psi^\dg_{+\dot A, -p}\psi^{-\dot A}_{2n-j}\ket{\chi}.
\end{equation}
Similarly,  $J^-_0$ becomes
\begin{equation}
J_0^-\ket{\chi} = \sum_{j,p\,\text{odd}^+}\gamma^F_{jp}\psi^\dg_{+\dot A, -j}\psi^{-\dot A}_{-p}\ket{\chi}.
\end{equation}

The $J_k^-$ term is not much work, but there are two distinct cases,
corresponding to whether or not there are terms where both $\psi$s are
raising operators: $k\leq -3$ and $k\geq -1$. One finds
\begin{equation}
J^-_k\ket{\chi} = -\frac{1}{4}\sum_{k+1\leq j \leq -1} \psi^\dg_{+\dot A, k-j}\psi^{-\dot A}_j\ket{\chi}
         + \sum_{j,p\,\text{odd}^+} \gamma^F_{jp}\psi^\dg_{+\dot A, k-j}\psi^{-\dot A}_{-p}\ket{\chi}
         \qquad k\leq -3, \text{odd}
\end{equation}
and
\begin{equation}
J^-_k\ket{\chi} = \sum_{j,p\,\text{odd}^+}\gamma^F_{k+j+1, p}\psi^\dg_{+\dot A, -j-1}\psi^{-\dot A}_{-p}
                 \ket{\chi}
\qquad k\geq -1,\text{odd}.
\end{equation}
The slightly more difficult task is the sum over $k$, which can be written as
\begin{calc}
\sum_{k\,\text{odd}}\frac{z_0^{n-\frac{k}{2}}}{n-\frac{k}{2}}J^-_k\ket{\chi}
 &= \sum_{k\,\text{odd}}^{k\leq -3}\frac{z_0^{n-\frac{k}{2}}}{n-\frac{k}{2}}J^-_k\ket{\chi}
   +\sum_{k\,\text{odd}}^{k\geq -1}\frac{z_0^{n-\frac{k}{2}}}{n-\frac{k}{2}}J^-_k\ket{\chi}\\
 &= 
-\frac{1}{2}\sum_{j,p\,\text{odd}^+}\frac{z_0^{n+\frac{j+p+1}{2}}}{n+\frac{j+p+1}{2}}
       \psi^\dg_{+\dot A, -j-1}\psi^{-\dot A}_{-p}\\
&\qquad+\sum_{j\,\text{odd}}^{j\geq 3}\sum_{p\,\text{odd}^+}
     \left(
       \sum_{k\,\text{odd}}^{-j\leq k\leq -3}\frac{z_0^{n-\frac{k}{2}}}{n-\frac{k}{2}}
            \gamma^F_{k+j+1,p}
     \right)\psi^\dg_{+\dot A, -j-1}\psi^{-\dot A}_{-p}\\
&\qquad+\sum_{j,p\,\text{odd}^+}
       \left(
         \sum_{k\,\text{odd}}^{k\geq -1}\frac{z_0^{n-\frac{k}{2}}}{n-\frac{k}{2}}\gamma^F_{k+j+1,p}
       \right)\psi^\dg_{+\dot A, -j-1}\psi^{-\dot A}_{-p},
\end{calc}
after a few manipulations. Note that all of the terms on the right
implicitly act on $\ket{\chi}$. Finally, using~\eqref{eq:gammaF-sum-1}
one arrives at
\begin{equation}
\sum_{k\,\text{odd}}\frac{z_0^{n-\frac{k}{2}}}{n-\frac{k}{2}}J^-_k\ket{\chi}
 =-\sum_{j,p\,\text{odd}^+}
    \frac{z_0^{n+\frac{j+p+1}{2}}}{2(n+\frac{j+p+1}{2})}
     \frac{\Gamma(\frac{p}{2})\Gamma(-n-\frac{j}{2})}
                {\Gamma(\frac{p+1}{2})\Gamma(-n-\frac{j+1}{2})}
              \psi^\dg_{+\dot A,-j-1}\psi^{-\dot A}_{-p}\ket{\chi}
\end{equation}

Putting it all together, we have
\begin{multline}\label{eq:J-on-vac}
\sigma_2^+(z_0) J_n^{-(1)}\ket{0^-}^{(1)}\ket{0^-}^{(2)} 
  = -\frac{1}{8}\sum_{j=n+1}^{-1}\psi^\dg_{+\dot A, 2n-2j}\psi^{-\dot A}_{2j}\ket{\chi}\\
-\frac{i}{4\pi}
\sum_{j=1}^\infty\sum_{p\,\text{odd}^+}
    \frac{z_0^{n+j+\frac{p}{2}}}{n+j+\frac{p}{2}}
     \frac{\Gamma(\frac{p}{2})\Gamma(-n-j+\frac{1}{2})}
                {\Gamma(\frac{p+1}{2})\Gamma(-n-j)}
              \psi^\dg_{+\dot A,-2j}\psi^{-\dot A}_{-p}\ket{\chi}\\
+\frac{1}{2}\sum_{j,p\,\text{odd}^+}
\left[-\frac{1}{4}\delta_{j+p+2n,0} + \frac{\gamma^F_{2n+p,j}+\gamma^F_{2n+j,p}}{2}
      - z_0^n \frac{\gamma^F_{jp} + \gamma^F_{pj}}{2}\right]\psi^\dg_{+\dot A, -j}\psi^{-\dot A}_{-p}
\ket{\chi},
\end{multline}
where for ease of comparison we have broken the result into
even--even, even--odd, and odd--odd terms. In the above expression, we
define $\gamma^F$ with negative indices to be zero. Furthermore, we
have explicitly symmetrized over $j$ and $p$ in the last term since
\begin{equation}
\psi^\dg_{+\dot A, -j}\psi^{-\dot A}_{-p} = \psi^\dg_{+\dot A, -p}\psi^{-\dot A}_{-j}.
\end{equation}
Below, we compare this result with what one finds when one breaks the
$J^{-(1)}_{n}$ into fermions and uses the fermion intertwining
relations.

\subsection{The Composite Method}\label{sec:composite}

We now show how to reproduce the above result by using the fermion
intertwining relations and our prescription. 

We start by writing
\begin{equation}
J^{a(1)}_n = -\frac{1}{4}{(\sigma^{aT})^\alpha}_\beta\sum_{j =-\infty}^\infty
                \psi^{(1)\dg}_{\alpha\dot A, n-j}\psi^{(1)\beta\dot A}_j,
\end{equation}
and note that the sum on $j$ is a sum over ``before-the-twist'' modes
and therefore should be evaluated last according to the prescription.
We then can use our intertwining relations to write this directly as
\begin{smalleq}\begin{equation}\label{eq:J-sum}
J^{a(1)}_n = -\frac{1}{4}{(\sigma^{aT})^\alpha}_\beta
  \sum_{j =-\infty}^\infty
  \left[\frac{1}{2}\psi^\dg_{\alpha\dot A, 2n-2j} 
           + \frac{i}{2\pi}\sum_{k\,\text{odd}}\frac{z_0^{n-j-\frac{k}{2}}}{n-j-\frac{k}{2}}
                     \psi^\dg_{\alpha\dot A, k}\right]
  \left[\frac{1}{2}\psi^{\beta\dot A}_{2j} + \frac{i}{2\pi}\sum_{l\,\text{odd}}
                             \frac{z_0^{j-\frac{l}{2}}}{j-\frac{l}{2}}\psi_l^{\beta\dot A}\right].
\end{equation}\end{smalleq}
At this point, we can say nothing further until we know what
$J^{a(1)}_n$ acts on. Of course, when faced with an expression like
the above it is rather tempting to evaluate the $j$-sum using
\begin{equation}
\sum_{j=-\infty}^\infty \frac{1}{(n-j-\frac{k}{2})(j-\frac{l}{2})} = 
   -\pi^2\delta_{\frac{l}{2}, n-\frac{k}{2}} \qquad k,l\,\text{odd},
\end{equation}
which immediately leads to the false relation
\begin{equation}
J^{a(1)}_n = \frac{1}{2}J^a_{2n} +
  \frac{i}{2\pi}\sum_{k\,\text{odd}}\frac{z_0^{n-\frac{k}{2}}}{n-\frac{k}{2}}
  J^a_k
\qquad \text{(false relation)}.
\end{equation}
This demonstrates the need for the restriction on before-the-twist
sums.

In order to proceed and compare to~\eqref{eq:J-intertwining}, let us
again consider the state
\begin{equation}\label{eq:J-state}
\sigma_2^+(z_0)J^{-(1)}_n\ket{0_R^-}^{(1)}\ket{0_R^-}^{(2)}.
\end{equation}
Now, one could make the $j$-sum in~\eqref{eq:J-sum} finite by
considering the above; however, our prescription ensures that one gets
the correct answer even if one leaves it as an infinite series.

If we act on the vacuum as in~\eqref{eq:J-state}, then we
get~\eqref{eq:J-sum} acting on $\ket{\chi}$. The $\psi^{\beta\dot A}$ is
the rightmost mode and so the $l$-sum should be evaluated first,
followed by the $k$-sum, and then finally the $j$-sum. We can
use~\eqref{eq:single-fermion} to quickly read off the result of the
$l$-sum. There are now two distinct terms to consider for the $k$-sum.
There is the possibility of the $\psi^\dg_{\alpha\dot A}$ contracting with
the result of the $l$-sum, and the $\psi^\dg$ can pass through and act
on $\ket{\chi}$ (and we can again use~\eqref{eq:single-fermion}). The
contraction gives zero. For other composite operators, however, the
contraction term can be nonzero.

Following the above procedure, we get
\begin{equation}\begin{split}
-\frac{1}{2}\sum_{j=-\infty}^\infty
  &\left[\frac{1}{2}\psi^\dg_{+\dot A, 2(n-j)} + \frac{i}{2\pi}\sum_{p\,\text{odd}^+}
    \frac{z_0^{n-j+\frac{p}{2}}}{n-j+\frac{p}{2}}
    \frac{\Gamma(\frac{p}{2})\Gamma(-n+j+\frac{1}{2})}{\Gamma(\frac{p+1}{2})\Gamma(-n+j)}
    \psi^\dg_{+\dot A, -p}\right]\\
  &\left[\frac{1}{2}\psi^{-\dot A}_{2j} + \frac{i}{2\pi}\sum_{q\,\text{odd}^+}
    \frac{z_0^{j+\frac{q}{2}}}{j+\frac{q}{2}}
    \frac{\Gamma(\frac{q}{2})\Gamma(-j+\frac{1}{2})}{\Gamma(\frac{q+1}{2})\Gamma(-j)}
    \psi^{-\dot A}_{-q}\right].
\end{split}\end{equation}
There are four terms from the above multiplication. 

The even--even term is
\begin{equation}
(\text{even--even})=
-\frac{1}{8}\sum_{j=n+1}^{-1}\psi^\dg_{+\dot A, 2n-2j}\psi^{-\dot A}_{2j}.
\end{equation}
One can show that the two even--odd cross-terms are identical, and sum
to
\begin{equation}
(\text{even--odd}) =
-\frac{i}{4\pi}\sum_{j=-\infty}^{-1}
\sum_{p\,\text{odd}^+}\frac{z_0^{n-j+\frac{p}{2}}}{n-j+\frac{p}{2}}
    \frac{\Gamma(\frac{p}{2})\Gamma(-n+j+\frac{1}{2})}{\Gamma(\frac{p+1}{2})\Gamma(-n+j)}
    \psi^\dg_{+\dot A, -p}\psi^{-\dot A}_{2j}
\end{equation}
One may write the odd--odd term as
\begin{equation}
(\text{odd--odd}) = \frac{1}{8\pi^2}\sum_{p,q\,\text{odd}^+}\psi^\dg_{+\dot A, -p}\psi^{-\dot A}_{-q}
        z_0^{n+\frac{p+q}{2}}
        \frac{\Gamma(\frac{p}{2})\Gamma(\frac{q}{2})}{\Gamma(\frac{p+1}{2})\Gamma(\frac{q+1}{2})}
        S(n,p,q),
\end{equation}
where
\begin{equation}
S(n,p,q) = \sum_{j=n+1}^{-1}\frac{1}{(n-j+\frac{p}{2})(j+\frac{q}{2})}
                       \frac{\Gamma(-n+j+\frac{1}{2})\Gamma(-j+\frac{1}{2})}
               {\Gamma(-n+j)\Gamma(-j)}.
\end{equation}

Comparing the above to the three terms in Equation~\eqref{eq:J-on-vac},
one finds agreement provided
\begin{equation}
\frac{z_0^{n+\frac{j+p}{2}}}{8\pi^2}\frac{\Gamma(\frac{j}{2})\Gamma(\frac{p}{2})}
                                          {\Gamma(\frac{j+1}{2})\Gamma(\frac{p+1}{2})}
                                          S(n,j,p)
 = \frac{1}{2}\left[-\frac{1}{4}\delta_{j+p+2n,0} + \frac{\gamma^F_{2n+p,j}+\gamma^F_{2n+j,p}}{2}
      - z_0^n \frac{\gamma^F_{jp} + \gamma^F_{pj}}{2}\right].
\end{equation}
This equation follows from the identity
\begin{multline}\label{eq:hard-id}
\sum_{k=0}^\mu\frac{1}{(\mu-k-\frac{\alpha}{2})(k-\frac{\beta}{2})}
   \frac{\Gamma(\mu -k + \frac{3}{2})\Gamma(k+\frac{3}{2})}
        {\Gamma(\mu - k + 1)\Gamma(k+1)}\\
=\pi - \frac{\pi}{\frac{\alpha + \beta}{2} -\mu}
\left(\frac{\Gamma(\frac{\alpha+3}{2})\Gamma(\frac{\alpha}{2}-\mu)}
           {\Gamma(\frac{\alpha}{2}+1)\Gamma(\frac{\alpha-1}{2}-\mu)}
+\frac{\Gamma(\frac{\beta+3}{2})\Gamma(\frac{\beta}{2}-\mu)}
           {\Gamma(\frac{\beta}{2}+1)\Gamma(\frac{\beta-1}{2}-\mu)}
\right),
\end{multline}
where the left-hand side is $S(n,p,q)$ with $\mu = -n-2$, $\alpha= p
-2$, and $\beta = q-2$. If we call the above sum $F(\mu, \alpha,
\beta)$, then one can prove the identity by showing that both sides of
Equation~\eqref{eq:hard-id} obey the four-term recursion relation
\begin{multline}
  f_0(\mu,\alpha, \beta) F(\mu, \alpha, \beta) 
+ f_1(\mu,\alpha, \beta) F(\mu+1, \alpha, \beta)\\
+ f_2(\mu,\alpha, \beta) F(\mu+2, \alpha, \beta) 
+ f_3(\mu,\alpha, \beta) F(\mu+3, \alpha, \beta) 
 = 0,
\end{multline}
with
\begin{subequations}
\begin{smalleq}\begin{align}
f_0(\mu, \alpha,\beta) &= (-3+\alpha -2 \mu ) (\alpha +\beta -2 \mu ) (3-\beta +2 \mu ) (\alpha +\beta -6 (3+\mu ))\\
f_1(\mu, \alpha,\beta) &= 3 (\alpha +\beta -2 (1+\mu )) \big(-2 \beta ^2 (2+\mu )+\alpha ^2 (\beta -2 (2+\mu ))+\beta  (79+8 \mu  (9+2 \mu ))\notag\\
&+\alpha  \left(79+\beta ^2+8 \mu  (9+2 \mu )-2 \beta  (12+5 \mu )\right)-2 (128+\mu  (175+4 \mu  (20+3 \mu )))\big)\\
f_2(\mu, \alpha,\beta) &= 3 (\alpha +\beta -2 (2+\mu )) \big(\beta ^2 (5+2 \mu )+\alpha ^2 (5-\beta +2 \mu )+4 (2+\mu ) \left(43+32 \mu +6 \mu ^2\right)\notag\\
&\quad-2 \beta  (47+\mu  (39+8 \mu ))-\alpha  \left(94+\beta ^2-2 \beta  (12+5 \mu )+2 \mu  (39+8 \mu )\right)\big)\\
f_3(\mu, \alpha,\beta) &= (\alpha +\beta -6 (2+\mu )) (\alpha -2 (3+\mu )) (\beta -2 (3+\mu )) (\alpha +\beta -2 (3+\mu )).
\end{align}\end{smalleq}
\end{subequations}
Thus the identity holds by induction once one confirms that it holds
for $\mu=0,1,2$. It may be helpful to use an algebraic manipulation
program such as \textit{Mathematica} to show the above.

\section{Discussion}\label{sec:orbifold-future}

We developed technology for computing the effect of the blow-up mode
deformation operator on untwisted states of the D1D5 CFT.  We would be
remiss, if we did not note that the methods presented here should
generalize to higher order twist operators.

In Section~\ref{sec:app-to-vac}, we showed that one could manipulate
the supercharge contour so that it acts before and after the twist. We
then showed that the twist operator acts on the vacuum to create a
squeezed state in terms of the twisted sector modes. The coefficients
of the bosonic and fermionic excitations appear in
Equations~\eqref{gammaB} and~\eqref{gammaF} and the full state appears
in Equation~\eqref{finalstate}. We can analyze the large $m$ and $n$
limit of the coefficients, by noting the asymptotic behavior
\begin{equation}
\frac{\Gamma(x+1)}{\Gamma(x+\frac{1}{2})} 
           \sim \sqrt{x}\,e^{-\frac{1}{8x} + \bigO\left(\frac{1}{x^2}\right)},
\end{equation}
which can be derived from Stirling's approximation.  The large $m$ and
$n$ behavior is
\begin{subequations}\begin{align}
\gamma^B_{mn} &\sim \frac{2z_0^{\frac{m+n}{2}}}{(m+n)\pi}\frac{1}{\sqrt{mn}}\\
\gamma^F_{mn} &\sim -\frac{z_0^{\frac{m+n}{2}}}{(m+n)\pi}\sqrt{\frac{m}{n}}.
\end{align}\end{subequations}
Let us note that the behavior of $\gamma^B_{mn}$ is more properly
multiplied by $\sqrt{mn}$ if we want unit-normalized raising operators
$a^\dg_m = \alpha_{-m}/\sqrt{m}$. Thus, when $m\sim n$ the fermions
and bosons behave essentially the same. The asymmetry between $m$ and
$n$ results from the asymmetry between $+$ and $-$ $J_0^3$-charge in
our setup.

Let us note that the state appearing in Equation~\eqref{finalstate} is
before the integral over the $z$ and $\bar{z}$ coordinates that give
rise to an energy-conserving delta-function. No terms can contribute
in Equation~\eqref{finalstate}, and thus the Ramond vacuum is
invariant under the deformation to first order in perturbation theory.
This is expected from supersymmetry. It is convenient to have an
off-shell expression for later calculations, however.

In Section~\ref{sec:excited}, we computed the effect of the
deformation operator on some simple excited states. First, a single
bosonic and fermionic excitation, as given in Equations~\eqref{kkone}
and~\eqref{eq:sing-ferm-1}. In both, expressions there is a
``diagonal'' component in which an initial $\alpha_m^{(1)}$ mode
becomes $\alpha_{2m}$ in the final state. Recalling our convention for
the twisted sector, these both have the same energy and moding. Then,
there is the off-diagonal contribution which consists of only odd
modes. Because the exponential gives pairs of odd modes, the only part
that conserves $L_0$ is the diagonal contribution. Again, it is still
useful to have the off-shell expression. We then computed the effect
of the twist operator on two bosonic and two fermionic modes in
Sections~\ref{sec:two-bosons} and~\ref{sec:two-fermions}. The
calculation culminates in Equations~\eqref{eq:two-bosons}
and~\eqref{eq:two-fermions}, before the application of the
supercharge. In Section~\ref{sec:complete-two-bosons}, we included the
effect of the supercharge in Equation~\eqref{eq:T-on-boson}. The
supercharge changes some bosonic excitations to fermionic excitations,
but it does not change our conclusion about going on-shell for a
single bosonic excitation.

While the (untwisted) Ramond vacuum and single bosonic or fermionic
excitations do not have very exciting physics when we go on-shell, let
us consider two bosonic excitations. Looking at the expression
in~\eqref{eq:two-bosons}, we see that there is no even--odd problem
like for a single mode and so there is an off-diagonal contribution
on-shell. Applying the supercharge does not alter the weights of the
modes, and does not effect our argument. For concreteness consider the
case in which we have two initial bosonic modes with combined weight
$10$ and for which there is no $\co$-number contribution .  We first
have the two $f^B$ terms. We could take them both to be
$\alpha_{-1}$-modes, for instance. This contributes weight 1 out of
the original 10. The remaining 9, can be made up out of the
exponential in $\ket{\chi}$ in many different ways.  For instance, one
could have 9 pairs of additional $\alpha_{-1}$ modes. Therefore, even
on-shell there is a nonzero probability for the two original
excitations to fragment into as many as 20 bosonic lower weight modes.
This is all only first-order in perturbation theory!

In Section~\ref{sec:intertwining}, we developed a different method to
compute the effect of the twist operator on excited states. The
technique makes the Bogolyubov-nature of the deformation operator more
transparent. There were some UV ambiguous sums that were tamed with
the prescription in Section~\ref{sec:prescription}. With the
prescription the results are self-consistent and agree with those of
the previous sections.

Using the technology and understanding developed here, we hope to
address some important outstanding questions concerning the D1D5 CFT
and black hole physics. In particular, we hope to elucidate the nature
of the proposed ``non-renormalization'' theorem~\cite{dmw02}; how
states of the CFT fragment, thereby ``scrambling''
information~\cite{susskind, preskill}; black hole formation; and the
in-falling observer. All of these are important and deep issues that
will probably require lots of work to resolve; however, the work shown
here should serve as a step toward these goals. 

All of the calculations here and in~\cite{acm2, acm3, ac2} have been
off-shell. A more modest next step would be to analyze the physics
that arise on-shell. Another task is to systematically analyze the
combinatorial factors that arise at different orders in perturbation
theory. We defer these steps for future calculations.
\chapter{Conclusions and Outlook}
\label{ch:conclude}

The D1D5 system has been and remains indispensable in the theoretical
study of black holes in string theory. In Chapter~\ref{ch:d1d5}, we
gave an overview of the D1D5 system and its CFT description.
Describing the same process using both supergravity and CFT
descriptions is one of the best ways to gain understanding of black
holes. We have focused on the CFT description here, emphasizing its
utility in understanding some of the mysteries of the gravitational
description.

In Chapter~\ref{ch:coupling}, we outlined how to perturbatively relax
the decoupling limit allowing the supergravity fields in flat space to
interact with the AdS/CFT. The results of this chapter apply equally
well to other incarnations of the AdS--CFT correspondence so we do not
restrict ourselves to $\AdS_3$--$\text{CFT}_2$.  The general
interaction term is given in Equation~\eqref{eq:general-S-int}. This
can be used to compute absorption, emission, or even more general
scattering processes. While some similar calculations were performed
in~\cite{dmw99}, we have made a much more general and systematic
treatment. In particular we have taken the effect of the neck into
account. 

In Chapter~\ref{ch:emission}, we applied the technique of
Chapter~\ref{ch:coupling} to reproduce the ergoregion emission from
JMaRT geometries from a CFT amplitude.  Since ergoregion emission from
the JMaRT geometries and Hawking radiation from a black hole use the
same vertex operator, we argued that the ergoregion emission \emph{is}
the Hawking radiation from the particular microstate. This point of
view was first advanced in~\cite{cm1, cm2, cm3}, but there were
several important missing links. First, the coupling between the
asymptotic flat space and the CFT was fixed by demanding agreement
with Hawking radiation, which may be a bit unsatisfying. Second, only
partial spectrums had been reproduced from a subset of the JMaRT
solitons.  In Chapter~\ref{ch:coupling}, we fix the coupling by
demanding agreement between gravity and CFT for a two-point function
\emph{within} AdS; this step does not require us to relax the
decoupling limit. This addresses the first issue. In
Chapter~\ref{ch:emission}, we reproduce the \emph{full} spectrum from
\emph{all} of the JMaRT geometries.

In our computation of the full spectrum, there are several points of
interest. One is that we demonstrate how spectral flow and Hermitian
conjugation can be used to relate a single CFT calculation to many
different gravitational processes. In particular, for $\kappa = 1$, we
first computed the emission of a single quantum out of the geometry.
We then related this amplitude to the JMaRT ergoregion emission. A
second point is that in our calculation of emission for $\kappa>1$, we
had to make a conjecture of the form of supergravity excitations in
$\kappa$-orbifolded geometries. This conjecture seems natural from the
perspective that all of the modes for $\kappa=1$ get ``divided by''
$\kappa$. It would be interesting to understand this last issue better.

Finally, in Chapter~\ref{ch:orbifold} we describe technology that
allows one to compute the effect of deforming the CFT off its orbifold
point. More specifically we studied a single application of the
marginal 2-twist deformation operator. A surprisingly rich structure
emerges from the twisting. The effect of the twist operator is to very
similar to a Bogolyubov transformation. Much of this material should
go through with only minor alteration to higher-twist operators. So
this is relevant to anyone interested in describing the physics of
$S_{N_1N_5}$-twist operators using a Hamiltonian description. A lot of
the results reported in this chapter require more calculation and
analysis to elucidate the black hole physics of interest, which we put
off for future papers. Note, however, that the basic physics of
interest we can already see, as discussed in
Section~\ref{sec:orbifold-future}. At first order in perturbation
theory, we find that a few high-energy excitations can fragment into
many low-energy excitations.

There are many puzzles left for black holes in string theory. We
expect that the D1D5 CFT will be useful in resolving them. Many
outstanding questions about the fuzzball proposal and black holes
revolve around dynamical questions: How do black holes thermalize and
what is the role of quasi-normal modes? How does one get into a
fuzzball state? What does an infalling observer experience? What
happens to a infalling spherical shell of matter? The marginal
deformation that we add to the CFT can be treated in time-dependent
perturbation theory. It is hoped, then, that the D1D5 CFT and the
results in Chapter~\ref{ch:orbifold} will address some of those
questions.

\clearpage
\backmatter

\appendix

\chapter{Notation and Conventions for the Orbifold CFT} 
\label{ap:CFT-notation}

Here, we carefully define the notation and conventions used throughout
this paper, and to be used in future work. The system we describe
lives in $M_{4,1}\times T^4\times S^1$. We restrict our attention to
the case where the compact space is a torus, although one may also
consider K3.

The base space of the CFT is the $S^1$ and time ($\re$), with fields
living in the orbifolded target space:
${(T^4)^{N_1N_5}}/{S_{N_1N_5}}$. The expressions in this appendix are
exclusively given for the complex plane.

\section{Symmetries and Indices}

The symmetries of our theory are $SU(2)_L\times SU(2)_R$ and the
$SO(4)_I\simeq SU(2)_1\times SU(2)_2$ rotations of the torus. Indices
correspond to the following representations
\begin{gather*}
\alpha, \beta \qquad \text{doublet of $SU(2)_L$} \hspace{60pt}
\dot{\alpha}, \dot{\beta} \qquad \text{doublet of $SU(2)_R$}\\
A, B \qquad \text{doublet of $SU(2)_1$}\hspace{60pt}
\dot{A}, \dot{B} \qquad \text{doublet of $SU(2)_2$}\\
i, j \qquad \text{vector of $SO(4)$}.
\end{gather*}
One can project vectors of $SO(4)$ into the doublets of $SU(2)_1\times
SU(2)_2$, using the usual Pauli spin matrices and the identity matrix
\[
(\sigma^i)^{\dot{A}A},\qquad \sigma^4 = i\id_2.
\]
The indices are such that, for instance, $(\sigma^2)^{\dot{2}1}= i$. 

We use indices $a,b,c=1,2,3$ for the triplet of any $SU(2)$. Their
occurrence is rare enough that which $SU(2)$ is being referred to
should be unambiguous. Note that the $SU(2)$ generators
${(\sigma^a)_\alpha}^\beta$ naturally come with one index raised and
one index lowered. On the other hand the Clebsch-Gordan coefficients
to project a vector of $SO(4)$ into two $SU(2)$'s naturally come with
both indices raised (or lowered), as above.

We raise and lower all $SU(2)$ doublet indices in the same way so that
\begin{equation}
\epsilon_{\alpha\beta}v^\beta = v_\alpha\qquad v^\alpha = \epsilon^{\alpha\beta}v_\beta,
\end{equation}
where 
\begin{equation}
\epsilon_{12} = -\epsilon_{21} = \epsilon^{21} = -\epsilon^{12} = 1,
\end{equation}
and therefore
\begin{equation}
\epsilon_{\alpha\beta}\epsilon^{\beta\gamma} = \delta_\alpha^\gamma.
\end{equation}

\section{Field Content}

The bosonic field content of each copy of the CFT consists of a vector
of $SO(4)_I$, $X^i(z,\bar{z})$, giving the position of that effective
string in the torus. The fermions on the left sector have indices in
$SU(2)_L\times SU(2)_2$, while the fermions in the right sector have
indices in $SU(2)_R\times SU(2)_2$:
\begin{equation}
\psi^{\alpha \dot{A}}(z)\qquad \bar{\psi}^{\dot{\alpha}\dot{A}}(\bar{z}).
\end{equation}
These fermions are complex, so there are two complex fermions and
their Hermitian conjugates in the left sector and two complex fermions
and their Hermitian conjugates in the right sector. 

Note that we use the abbreviated notation
\begin{equation}
[X]^{\dot{A}A} = \frac{1}{\sqrt{2}}X^i(\sigma^i)^{\dot{A}A}.
\end{equation}

\section{Currents}

The holomorphic currents of our theory that form a closed OPE algebra
are an $SU(2)_L$ current, $J^a(z)$; the supersymmetry currents,
$G^{\alpha A}(z)$; and the stress--energy $T(z)$. The right sector has
the corresponding anti-holomorphic currents. Obviously, in this case,
the index $a$ on $J$ transforms in $SU(2)_L$. 

For each copy of the CFT, the currents are realized in terms of the
fields as
\begin{subequations}\label{eq:currents-def}
\begin{align}
J^a(z) &= \frac{1}{4}\epsilon_{\dot{A}\dot{B}}
	\psi^{\alpha \dot{A}}\epsilon_{\alpha\beta}{(\sigma^{*a})^\beta}_\gamma	\psi^{\gamma\dot{B}}\\
G^{\alpha A}(z) &= \psi^{\alpha \dot{A}}[\pd X]^{\dot{B}A}\epsilon_{\dot{A}\dot{B}}\\
T(z) &= \frac{1}{2}\epsilon_{\dot{A}\dot{B}}\epsilon_{AB}[\pd X]^{\dot{A}A}[\pd X]^{\dot{B}B}
	+ \frac{1}{2}\epsilon_{\alpha\beta}\epsilon_{\dot{A}\dot{B}}
			\psi^{\alpha\dot{A}}\pd\psi^{\beta\dot{B}}.
\end{align}
\end{subequations}
Note that the $SO(4)_I$ of the torus is an outer automorphism and so
while we can make a generator that acts appropriately on the fermions
we cannot make one that also acts appropriately on the bosons. 

%
 
\section{Hermitian Conjugation}

Because we work in a Euclidean time formalism, one must address
Hermitian conjugation carefully so that it is consistent with the
physical, real-time formalism.

A quasi-primary field of weight $(\Delta, \bar{\Delta})$ is Hermitian
conjugated as~\cite{difrancesco}
\begin{equation}
[\mathcal{O}(z,\bar{z})]^\dg = \bar{z}^{-2\Delta}z^{-2\bar{\Delta}}
	\mathcal{O}^\dg\big(\tfrac{1}{\bar{z}}, \tfrac{1}{z}\big),
\end{equation}
where $\mathcal{O}^\dg(z, \bar{z})$ has the opposite charges under
$SU(2)_L\times SU(2)_R\times SO(4)_I$.

The fermions Hermitian conjugate as
\begin{equation}
\big(\psi^{\alpha\dot{A}}\big)^\dg(z)
	=
	-\epsilon_{\alpha\beta}\epsilon_{\dot{A}\dot{B}}\psi^{\beta\dot{B}}(z)
	= - \psi_{\alpha\dot{A}}(z).
\end{equation}
This reality condition ensures that there are only four real degrees
of freedom in both the left and right sectors. The specific sign can
be determined from the basic fermion correlator and demanding
a positive-definite norm.

The bosons conjugate as
\begin{equation}
\big(X^i\big)^\dg(z, \bar{z}) = X^i(z, \bar{z})\qquad
\big([X]^{\dot{A}A}\big)^\dg(z,\bar{z}) 
	= -\epsilon_{\dot{A}\dot{B}}\epsilon_{AB}[X]^{\dot{B}B}(z,\bar{z}).
\end{equation}

The stress energy tensor and the $SU(2)_L$ current are both Hermitian
and so conjugate trivially; whereas, the supercurrents conjugate as
\begin{equation}
\big(G^{\alpha A}\big)^\dg(z) = -\epsilon_{\alpha\beta}\epsilon_{AB} G^{\beta B}(z).
\end{equation}
Again, the specific sign is determined by requiring the norm to be
positive-definite.

The Ramond vacua conjugate as
\begin{equation}
\big(\vac_R^\alpha\big)^\dg = {}^R_\alpha\hspace{-4pt}\bvac
\qquad {}_\alpha^R\hspace{-4pt}\braket{\varnothing|\varnothing}_R^\beta = \delta_\alpha^\beta.
\end{equation}

\section{OPE}

We normalize the fields so that the basic correlators are
\begin{subequations}
\begin{align}
\vev{X^i(z)X^j(w)} &= -2\delta^{ij}\log |z-w|\label{eq:bos-norm1}\\
\vev{\psi^{\alpha \dot{A}}(z)\psi^{\beta \dot{B}}(w)} &= 
	-\frac{\epsilon^{\alpha\beta}\epsilon^{\dot{A}\dot{B}}}{z-w},
\end{align}
\end{subequations}
where it is also useful to note that Equation~\eqref{eq:bos-norm1}
implies
\begin{equation}
\vev{[X]^{\dot{A}A}(z)[X]^{\dot{B}B}(w)} =2 \epsilon^{\dot{A}\dot{B}}\epsilon^{AB}\log |z-w|.
\end{equation}
From which, the commonly used
\begin{equation}
[\pd X(z)]^{\dot{A}A}[\pd X(w)]^{\dot{B}B}\sim 
	\frac{\epsilon^{\dot{A}\dot{B}}\epsilon^{AB}}{(z-w)^2},
\end{equation}
immediately follows.

The OPE current algebra for a single copy of the $\Nsc =4$ CFT is
\begin{subequations}\label{eq:currents-OPE}
\begin{align}
J^a(z)J^b(w) &\sim \frac{c}{12}\frac{\delta^{ab}}{(z-w)^2} 
			+ i{\epsilon^{ab}}_c\frac{J^c(w)}{z-w}\\
J^a(z)G^{\alpha A}(w) &\sim \tfrac{1}{2}{(\sigma^{*a})^\alpha}_\beta\frac{G^{\beta A}(w)}{z-w}\\
G^{\alpha A}(z)G^{\beta B}(w) &\sim
         -\frac{c}{3}\frac{ \epsilon^{AB}\epsilon^{\alpha\beta}}{(z-w)^3}
	+ \epsilon^{AB}\epsilon^{\beta\gamma}{(\sigma^{*a})^\alpha}_\gamma
		\left[\frac{2J^a(w)}{(z-w)^2} + \frac{\pd J^a(w)}{z-w}\right]
		-\epsilon^{AB}\epsilon^{\alpha\beta}\frac{T(w)}{z-w} \\
T(z)J^a(w) &\sim \frac{J^a(w)}{(z-w)^2} + \frac{\pd J^a(w)}{z-w}\\
T(z)G^{\alpha A}(w) &\sim \frac{\tfrac{3}{2}G^{\alpha A}(w)}{(z-w)^2}
			+\frac{\pd G^{\alpha A}(w)}{z-w} \\
T(z)T(w) &\sim \frac{c}{2}\frac{1}{(z-w)^4} + 2\frac{T(w)}{(z-w)^2} + \frac{\pd T(w)}{z-w},
\end{align}
\end{subequations}
which agrees with the above correlators for $c=6$.

For convenient reference, we include the OPEs of the currents with the
basic primary fields, $\pd X$ and $\psi$:
\begin{subequations}\begin{align}
J^a(z)\psi^{\alpha\dot{A}}(w) &\sim \frac{1}{2}{(\sigma^{*a})^\alpha}_\beta 
 	\frac{\psi^{\beta \dot{A}}(w)}{z-w}\\
G^{\alpha A}(z)[\pd X(w)]^{\dot{B}B} &\sim \epsilon^{AB}
	\left(\frac{\psi^{\alpha \dot{B}}(w)}{(z-w)^2} 
		+ \frac{\pd \psi^{\alpha \dot{B}}(w)}{z-w}\right)\\
G^{\alpha A}(z)\psi^{\beta \dot{A}}(w) &\sim 
	\epsilon^{\alpha\beta}\frac{[\pd X(w)]^{\dot{A}A}}{z-w}\\
T(z)[\pd X(w)]^{\dot{A}A} &\sim \frac{[\pd X(w)]^{\dot{A}A}}{(z-w)^2} 
	+ \frac{[\pd^2 X(w)]^{\dot{A}A}}{z-w}\\
T(z)\psi^{\alpha \dot{A}}(w) &\sim \frac{\tfrac{1}{2}\psi^{\alpha \dot{A}}(w)}{(z-w)^2}
	+ \frac{\pd\psi^{\alpha \dot{A}}(w)}{z-w}.
\end{align}\end{subequations}

\section{Mode Algebra}

We define the modes corresponding to the above currents according to
their weight, $\Delta$, by
\begin{equation}\begin{split}
\mathcal{O}_m  &= \oint\frac{\drm z}{2\pi i}\mathcal{O}(z) z^{\Delta + m-1}\\
\mathcal{O}(z) &= \sum_m \mathcal{O}_m\, z^{-(\Delta + m)}.
\end{split}\end{equation}
The weight may be read off from the OPE of the current with the
stress--energy tensor. Fermionic currents have half-integer weight. In
the NS sector, fermions are periodic in the plane and therefore we
need integer powers of $z$. This means the fermionic currents have
modes labeled by half-integer $m$.

Using the OPE current algebra above, one finds that the modes form an
algebra:
\begin{subequations}\label{eq:mode-algebra}
\begin{align}
\com{J^a_m}{J^b_n} 
  &= \tfrac{c}{12}m\delta^{ab}\delta_{m+n, 0} +i{\epsilon^{ab}}_cJ^c_{m+n}\\
\com{J^a_m}{G^{\alpha A}_n}
  &= \frac{1}{2}{(\sigma^{*a})^\alpha}_\beta G^{\beta A}_{m+n}\\
\ac{G^{\alpha A}_m}{G^{\beta B}_n} 
  &=-\tfrac{c}{6}(m^2-\tfrac{1}{4})\epsilon^{AB}\epsilon^{\alpha\beta}\delta_{m+n,0}
 +(m-n)\epsilon^{AB}\epsilon^{\beta\gamma}{(\sigma^{*a})^\alpha}_\gamma J^a_{m+n}
 -\epsilon^{AB}\epsilon^{\alpha\beta}L_{m+n}\\
\com{L_m}{J^a_n} &= -n J^a_{m+n}\\
\com{L_m}{G^{\alpha A}_n} &= (\tfrac{m}{2}-n)G^{\alpha A}_{m+n}\\
\com{L_m}{L_n} &= c\tfrac{m^3-m}{12}\delta_{m+n, 0} + (m-n)L_{m+n}.
\end{align}
\end{subequations}

The infinite-dimensional algebra has a finite, anomaly-free subalgebra
which is of primal importance for the AdS--CFT correspondence. The
anomaly-free subalgebra has a basis of $\{J_0^a,\,G^{\alpha
  A}_{\pm\frac{1}{2}},\,L_0,\, L_{\pm 1}\}$. The smaller subalgebra
spanned by $\{J^3_0,\, L_0\}$ is the Cartan subalgebra, which means we
may label states and operators by their charge $m$ and their weight
$h$.

For reference, we provide the mode algebra of the two canonical
primary fields. The $\pd X$'s modes are $\alpha_n$.
\begin{subequations}\begin{align}
\com{\alpha_m^{\dot{A}A}}{\alpha_n^{\dot{B}B}} &= m \epsilon^{\dot{A}\dot{B}}
	\epsilon^{A B}\delta_{n+m,0}\\
\ac{\psi^{\alpha\dot{A}}_m}{\psi^{\beta\dot{B}}_n} &= 
	-\epsilon^{\alpha\beta}\epsilon^{\dot{A}\dot{B}}\delta_{m+n,0}\\
\com{J^a_m}{\psi^{\alpha\dot{A}}_n} &= \tfrac{1}{2}{(\sigma^{*a})^\alpha}_\beta
		\psi^{\beta\dot{A}}_{m+n}\\
\com{G^{\alpha A}_m}{\alpha^{\dot{B}B}_n} &= -n\epsilon^{AB}\psi^{\alpha\dot{B}}_{m+n}\\ 
\ac{G^{\alpha A}_m}{\psi^{\beta\dot{A}}_n} &= \epsilon^{\alpha\beta}\alpha_{m+n}^{\dot{A}A}\\
\com{L_m}{\alpha^{\dot{A}A}_n} &= -n\alpha_{m+n}^{\dot{A}A}\\
\com{L_m}{\psi^{\alpha\dot{A}}_n} 
	&= -(\tfrac{m}{2}+n)\psi^{\alpha\dot{A}}_{m+n}.
\end{align}\end{subequations}

\section{Useful Identities}

These identities are useful for relating vectors of $SO(4)_I$ to tensors
in $SU(2)_1\times SU(2)_2$:
\begin{subequations}\begin{align}
\epsilon_{\dot{A}\dot{B}}\epsilon_{AB}(\sigma^i)^{\dot{A}A}(\sigma^j)^{\dot{B}B} 
	&= -2\delta^{ij}\\
(\sigma^i)^{\dot{A}A}(\sigma^i)^{\dot{B}B} &= -2\epsilon^{\dot{A}\dot{B}}\epsilon^{AB}.
\end{align}\end{subequations}

It is useful to know how to relate the $(+,-,3)$ basis for the triplet
of $SU(2)$ to the $(1,2,3)$ basis:
\begin{subequations}
\begin{align}
\delta^{++} &= \delta^{--} = \delta_{++} = \delta_{--} = 0\\
\delta^{+-} &= \delta^{-+} = 2\qquad \delta_{+-}= \delta_{-+} = \tfrac{1}{2}\\
\epsilon^{+-3} &= -2i\\
\sigma^+ &= \begin{pmatrix}0 & 2 \\ 0 & 0 \end{pmatrix}\qquad
\sigma^- = \begin{pmatrix}0 & 0 \\ 2 & 0 \end{pmatrix}.
\end{align}
\end{subequations}
One can raise and lower the `$3$' index with impunity.

\section{\texorpdfstring{$n$}{n}-twisted Sector Mode Algebra}

In the $n$-twisted sector, by which we mean modes whose contour orbits
a twist operator, we can only define the modes by summing over all
$n$-copies of the field. This allows us to define fractional
modes. The modes are defined by~\cite{lm2}
\begin{equation}
\mathcal{O}_\frac{m}{n} = \oint_0\frac{\drm z}{2\pi i}\sum_{k=1}^n\mathcal{O}^{(k)}(z)
	e^{2\pi i\frac{m}{n}(k-1)}z^{\Delta + \frac{m}{n} -1}.
\end{equation}
One can confirm that the integrand is $2\pi$-periodic and therefore
well-defined. If one lifts to a covering space using a map that
locally behaves as
\begin{equation}
z = b t^n,
\end{equation}
then the mode in the base $z$-space can be related to a mode in the
$t$-space:
\begin{equation}\label{eq:frac-mode-cover}
\mathcal{O}_{\frac{m}{n}}^{(z)} = b^\frac{m}{n} n^{1-\Delta}\mathcal{O}_{m}^{(t)},
\end{equation}
where $\Delta$ is the weight of the field.

To compute a correlator in the twisted sector, one may either work in
the base space with the summed-over-copies modes or one may work in
the covering space with the opened-up mode. If one works in the base
space, then one should use the algebra with the \emph{total} central
charge
\begin{equation}
c_\text{tot.} = nc;
\end{equation}
the algebra is otherwise unchanged. If one works in the covering space
then one uses the central charge of a single copy of the CFT, but must
remember to write all of the factors that come in lifting to the
cover. These two methods give identical answers.

\section{Spectral Flow}\label{sec:spectral-flow}

The $\Nsc = 4$ algebra is a vector space at every point $z$ in the
complex plane, spanned by the local operators $\{J^a(z), G^{\alpha
A}(z), T(z)\}$. This vector space closes under the OPE. It is possible
to make a $z$-dependent change of basis and preserve the algebra.

Making an $SU(2)_L$ transformation in the `3' direction to the local
operators by an angle, $\eta(z)=i\alpha\log z$, at every point $z$ is
called ``spectral flow'' by $\alpha$ units. While this may look like a
nontrivial transformation, the new algebra is isomorphic to the old
algebra~\cite{spectral}.

It is important to remember that $\log z$ has a branch cut, which we
put on the real axis for the following discussion.  Let us suppose we
start in the NS sector, where the local operators are periodic in the
complex plane. Let us spectral flow the local operators by $\alpha$
units. Suppose we start on the (positive imaginary side of the)
positive real axis, where $\eta = 0$ and the new operators are the
same as the old operators. As we make a counter-clockwise circle in
the complex plane, the angle between the old operators and the new
operators increases. Across the branch cut on the real axis, where
before the operators were continuous, now there is a large, finite
$SU(2)_L$ transformation.

More illustratively, consider how the fermions behave under spectral
flow, as described above:
\begin{equation}
\psi^{\pm \dot{A}}(z)\mapsto {\psi^{\pm \dot{A}}}'(z) 
	= e^{\pm\frac{i}{2}\eta(z)}\psi^{\pm\dot{A}}(z)
	= z^{\mp\frac{\alpha}{2}}\psi^{\pm\dot{A}}(z).
\end{equation}
We see that except for even $\alpha$, there is a branch cut. Moreover,
if we spectral flow by an odd number of units, then the new operators
$\psi'(z)$ have the opposite periodicity from $\psi(z)$. In general,
one expects that an operator with charge $m$ under $SU(2)_L$
transforms as
\begin{equation}\label{eq:nice-spec-flow}
\mathcal{O}(z) \mapsto z^{-\alpha m}\mathcal{O}(z);
\end{equation}
however, the superconformal algebra and its $SU(2)_L$ subalgebra, in
particular, is anomalous which leads to nontrivial transformations of
some operators.

Since the spectral flowed algebra and the original algebra are
isomorphic, there is a bijective mapping from states living in the
representations of one algebra to states living in the representation
of its spectral flow. Since the NS sector and the R sector are related
by spectral flow, we can map problems in one sector into problems in
the other.

The operator which maps states into their spectral flow images, we
call $\mathcal{U}_\alpha$,
\begin{equation}
\ket{\psi'} = \mathcal{U}_\alpha\ket{\psi}.
\end{equation} 
Formally, then, we may write the action of spectral flow on operators
as
\begin{equation}
\mathcal{O}'(z) = \mathcal{U}_\alpha\mathcal{O}(z)\mathcal{U}^{-1}_\alpha,
\end{equation}
so that amplitudes are invariant under spectral flow.  The spectral
flow operator, $\mathcal{U}_\alpha$ may be roughly defined as an
``improper gauge transformation''~\cite{spectral, spectral-yu,
  vafa-warner}.

The spectral flow operator is most naturally defined in the context of
bosonized fermions. We can bosonize the fermions as (conventions
chosen to be consistent with~\cite{lm2})
\begin{equation}
\psi^{+\dot{1}} = e^{-i\phi_6}\qquad
\psi^{+\dot{2}} = e^{i\phi_5} \qquad
\psi^{-\dot{1}} = e^{-i\phi_5}\qquad
\psi^{-\dot{2}} = -e^{i\phi_6},
\end{equation}
which gives the $SU(2)_L$ current in the form\footnote{There are
  implicit cocycles on the exponentials, which make unrelated fermions
  anticommute. Thus, the order of exponentials in expressions
  matters.}
\begin{equation}
J^3(z) = \frac{i}{2}\big(\pd\phi_5(z) - \pd\phi_6(z)\big)\qquad
J^+(z) = e^{-i\phi_6}e^{i\phi_5}(z)\qquad
J^-(z) = e^{-i\phi_5}e^{i\phi_6}(z).
\end{equation}
The fields $\phi_5$ and $\phi_6$ are the (holomorphic half of) real
bosons normalized such that
\begin{equation}
\vev{\phi_i(z)\phi_j(w)} = -\delta_{ij}\log(z-w).
\end{equation} 
They may be expanded as
\begin{equation}
\phi_i = q_i - \frac{i}{2}p_i\,\log z + (\text{modes}),
\end{equation}
where $q_i$ and $p_i$ are the zero-mode position and momentum which
satisfy
\begin{equation}
\com{q_i}{p_j} = i\delta_{ij}.
\end{equation}

With this bosonization, the spectral flow operator can be written as
~\cite{vafa-warner}
\begin{equation}
\mathcal{U}_\alpha = e^{i\alpha(q_5 - q_6)}.
\end{equation}
We see that spectral flow corresponds to increasing and decreasing the
zero mode momentum of the fields $\phi_5$ and $\phi_6$ used to
bosonize the fermions. The Baker--Campbell--Hausdorff formula implies
\begin{equation}
e^{i\alpha q_i}e^{i\beta p_j} = e^{-i\alpha\beta\delta_{ij}}
       e^{i\beta p_j}e^{i\alpha q_i},
\end{equation}
which one can use to confirm that this operator has the correct action
on fermions.

From this perspective, one can see that any operator that is ``pure
exponential'' in $\phi_5$ and $\phi_6$ transforms as in
Equation~\eqref{eq:nice-spec-flow}. If one considers any of the chiral
primaries of the $\Nsc=4$ orbifold theory, one finds that all of the
chiral primaries are ``pure exponential'' and therefore transform
using Equation~\eqref{eq:nice-spec-flow}.

We define the periodicity of the fermions in the complex plane via the
parameter $\beta_\pm$:
\begin{equation}
\psi^{\pm\dot{A}}(ze^{2\pi i}) = e^{i\pi\beta_\pm}\psi^{\pm\dot{A}}(z)\qquad
\bar{\psi}^{\pm\dot{A}}(\bar{z}e^{2\pi i}) = e^{i\pi\bar{\beta}_\pm}\bar{\psi}^{\pm\dot{A}}(z).
\end{equation}
Obviously $\beta_\pm$ is only defined modulo 2 under addition.  We use
spectral flow to alter the fermion content of the CFT states and to go
from the NS sector, $\beta_\pm =0$, to the R sector $\beta_\pm =
1$. The effect of spectral flow by $\alpha$ units on the left and
$\bar{\alpha}$ units on the right is
\begin{equation}
\beta_\pm \mapsto \beta'_\pm = \beta_\pm \pm \alpha\qquad
\bar{\beta}_\pm \mapsto \bar{\beta}'_\pm = \bar{\beta}_\pm \pm \bar{\alpha}.
\end{equation}
 
One finds that the the currents transform under spectral flow as
follows
\begin{equation}\begin{split}
J^3(z) &\mapsto J^3(z) - \frac{c\alpha}{12z}\\
J^\pm(z) &\mapsto z^{\mp\alpha}J^\pm(z)\\
G^{\pm A}(z) &\mapsto z^{\mp\frac{\alpha}{2}}G^{\pm A}(z)\\
T(z) &\mapsto T(z) -\frac{\alpha}{z}J^3(z)+\frac{c\alpha^2}{24z^2},
\end{split}\end{equation}
which gives rise to the transformation of the modes,
\begin{equation}\begin{split}
J^3_m &\mapsto J^3_m - \frac{c\alpha}{12}\delta_{m,0}\\
J^\pm_m &\mapsto J^\pm_{m\mp\alpha}\\
G^{\pm A}_m &\mapsto G^{\pm A}_{m\mp\frac{\alpha}{2}}\\
L_m &\mapsto L_m -\alpha J^3_m+\frac{c\alpha^2}{24}\delta_{m,0}.
\end{split}\end{equation}

Spectral flow also acts on states, changing the weight and charge by
\begin{subequations}\label{eq:spectral-flow-hj}
\begin{align}
h\mapsto h' &= h + \alpha m + \frac{c\alpha^2}{24}\\
m\mapsto m' &= m + \frac{c\alpha}{12},
\end{align}
\end{subequations}
which can be read off from $L_0$ and $J^3_0$.  Frequently, one can
deduce the spectral-flowed state from the weight and charge.

Note that spectral flow by $\alpha_1$ units followed by spectral flow
by $\alpha_2$ units is equivalent to spectral flow by $\alpha_1 +
\alpha_2$ units. Therefore, spectral flow forms an abelian group, and
\begin{equation}
\mathcal{U}_\alpha^{-1} = \mathcal{U}_{-\alpha}.
\end{equation}

\section{Ramond Sector}

The CFT in the complex plane naturally has periodic fermions, which
corresponds to the NS sector. One can, however, spectral flow by an
odd number of units to the Ramond sector. If one starts with the NS
vacuum and then spectral flows with $\alpha = -1$, then the state in
the R sector has
\begin{equation}
h = \frac{1}{4} \qquad m = \frac{1}{2}.
\end{equation}
From the weight we see that this must be the R ground state. Let us
call this state
\begin{equation}
\vac_R^+.
\end{equation}
Since, we are now in the Ramond ground state we have fermion zero
modes, and therefore may act with $J^-_0$, which gives us the state
\begin{equation}\label{eq:Rminus-expanded}
\vac_R^- = J^-_0\vac_R^+ 
	 = -\tfrac{1}{2}\epsilon_{\dot{A}\dot{B}}\psi^{-\dot{A}}_0\psi^{-\dot{B}}_0\vac_R^+
	 = \psi_0^{-\dot{2}}\psi_0^{-\dot{1}}\vac_R^+.
\end{equation}
The normalization is fixed by the commutation relations of
$J_0^a$. Since $J_0^-$ has zero weight, one can be sure that this
state is also a member of the R vacuum. Acting twice with $J^-_0$
annihilates the state, from which one concludes that these states from
a doublet of $SU(2)_L$,
\begin{equation}
\vac_R^\alpha,
\end{equation}
and one also can determine that
\begin{equation}\label{eq:Rplus-expanded}
\vac_R^+ = J_0^+\vac_R^-
	= \tfrac{1}{2}\epsilon_{\dot{A}\dot{B}}\psi^{+\dot{A}}_0\psi^{+\dot{B}}_0\vac_R^-
	= \psi^{+\dot{1}}_0\psi^{+\dot{2}}_0\vac_R^-.
\end{equation}

What happens if we act on these states not with a pair of fermion zero
modes in the current $J$, but with a single fermion zero mode
directly? Since one cannot raise the charge of the state $\vac_R^+$ or
lower the charge of the state $\vac_R^-$, one must have
\begin{equation}
\psi^{+\dot{A}}_0\vac_R^+ = 0 \qquad \psi^{-\dot{A}}_0\vac_R^- = 0.
\end{equation}
This can also be seen from Equations~\eqref{eq:Rminus-expanded} and
\eqref{eq:Rplus-expanded}. However, one ought to be able to contract the
fermion zero mode index with the R vacuum doublet index to form
\begin{equation}
\vac_R^{\dot{A}} = \frac{1}{\sqrt{2}}\epsilon_{\alpha\beta}\psi_0^{\alpha \dot{A}}\vac_R^\beta,
\end{equation}
where the normalization is determined from the fermion mode
anticommutation relations. Since there are four fermion zero modes (in
the left sector), we expect four Ramond vacuum. We see that those
vacua form a doublet of $SU(2)_L$ and a doublet of $SU(2)_2$.

Of course, the same story holds on the right sector of the theory as
well, which gives a total of 16 Ramond vacua:
\begin{equation}
\vac_R^{\alpha\dot{\alpha}} \qquad \vac_R^{\dot{A}\dot{\alpha}} \qquad
\vac_R^{\alpha\dot{A}} \qquad \vac_R^{\dot{A}\dot{B}}.
\end{equation}
Note that we must be very careful to always write the index
corresponding to the left zero modes first and the index corresponding
to the right zero modes second.

What are the images of the Ramond vacua in the NS sector? From the
action of spectral flow on the charge and weight of a state, one can
conclude that the Ramond vacua are one unit of spectral flow from
chiral primary ($h=m$) states in the NS sector; or equivalently,
negative one units of spectral flow from anti-chiral primary ($h=-m$)
states. Therefore, there is a one-to-one correspondence between the R
vacua and the NS chiral primary states. There are four left chiral
primary states in the NS sector,
\begin{equation}
\vac_{NS} \qquad
\psi^{+ \dot{A}}_{-\frac{1}{2}}\vac_{NS}\qquad 
\epsilon_{\dot{A}\dot{B}}\psi^{+\dot{A}}_{-\frac{1}{2}}\psi^{+\dot{B}}_{-\frac{1}{2}}\vac_{NS},
\end{equation}
and there are also four states in the right sector. Thus a total of 16
chiral primary states in the NS sector that get mapped onto the 16
Ramond vacua via spectral flow.  These are all of the chiral primary
states for a single strand of the CFT. In the twisted sector, there
are more chiral primary states which correspond to Ramond ``vacua'' in
the twisted sector.

\chapter{Cartesian to Spherical Clebsch--Gordan Coefficients}
\label{ap:spherical}

In this section, we outline our conventions for relating irreducible
spherical tensors to ordinary Cartesian tensors in flat space. This
fixes the factors in going from Equation~\eqref{eq:general-S-int} to
Equation~\eqref{eq:general-decay-rate} for the D1D5 case, and explains
how we define the correctly normalized differential operator, so that
Equation~\eqref{eq:diff-op-norm} is satisfied.

Our goal is to show how to construct the coefficients, $Y_{l,m_\psi,
m_\phi}^{j_1\dots j_l}$, that define the differential operator in
Equation~\eqref{eq:diff-op} such that it satisfies
Equation~\eqref{eq:diff-op-norm},
\[
Y^{k_1k_2\cdots k_l}_{l,m_\psi,m_\phi}\pd_{k_1}\pd_{k_2}\cdots\pd_{k_l}
		\big[r^{l'} Y_{l', m_\psi', m_\phi'}(\Omega_3)\big]
	= \delta_{ll'}\delta_{m_\psi m'_\psi}\delta_{m_\phi m'_\phi}.
\]
We take the spherical harmonics as a starting point. The Cartesian
coordinates for the noncompact space are related to the angular
coordinates via
\begin{equation}\begin{split}
x^1 &= r \cos\theta\cos\psi\\
x^2 &= r \cos\theta\sin\psi\\
x^3 &= r \sin\theta\cos\phi\\
x^4 &= r \sin\theta\sin\phi,
\end{split}\end{equation}
where the $(\theta, \psi, \phi)$ are restricted to
\begin{equation}
\theta\in \big[0,\tfrac{\pi}{2}\big)\qquad
\psi,\phi \in [0,2\pi).
\end{equation}

Spherical harmonics can be written as a homogeneous polynomial of the
Cartesian unit vector components of the form
\begin{equation}
Y_{l,m_\psi,m_\phi}(\Omega_3) = \frac{1}{r^l}
     \mathcal{Y}^{l,m_\psi, m_\phi}_{j_1\cdots j_l}
     x^{j_1}\cdots x^{j_l} 
  = \mathcal{Y}^{l,m_\psi, m_\phi}_{j_1\cdots j_l}n^{j_1}\cdots n^{j_l}.
\end{equation}
The $\mathcal{Y}$ must be pairwise symmetric and traceless. One can
compute these tensors by the usual methods of breaking up a tensor
into irreducible components, or by inspection of the explicit form of
the spherical harmonics in angular components.

Given the spherical harmonic normalization
\begin{equation}\label{eq:sphere-harm-norm}
\int Y_{l,m_\psi, m_\phi}^* Y_{l',m_\psi', m_\phi'} \drm \Omega 
   = \delta_{ll'}\delta_{m_\psi m_\psi'}\delta_{m_\phi m_\phi'},
\end{equation}
one can determine an orthogonality relation for the $\mathcal{Y}$'s:
\begin{equation}\label{eq:ortho-1}
\big(\mathcal{Y}^{l,m_\psi, m_\phi}_{j_1\cdots j_l}\big)^*
\mathcal{Y}^{l',m'_\psi,m'_\phi}_{k_1\cdots k_{l'}}
\int (n^{j_1}\cdots n^{j_l})(n^{k_1}\cdots n^{k_{l'}})\drm \Omega
= \delta_{ll'}\delta_{m_\psi m_\psi'}\delta_{m_\phi m_\phi'}.
\end{equation}
The integral over $l+l'$ unit vectors defines a natural inner product
on the Clebsch--Gordan coefficients $\mathcal{Y}$.

We label the integral
\begin{equation}
I^{j_1\cdots j_lk_1\cdots k_{l'}},
\end{equation}
and note that $I$ must be symmetric in all of its indices.
Furthermore, from the symmetry of the integral one must conclude that
$I$ vanishes unless every index appears an even number of times. For
instance,
\begin{equation}
I^{j} = \int n^j \drm\Omega = 0.
\end{equation}
As a corollary, $I$ vanishes unless it has an even number of indices.
Having picked off the easiest properties, let's without further
comment give the general form. Let $a_i$ be the total number of times
the index $i$ appears in the collection of indices on $I$, then
\begin{smalleq}\begin{calc}
I^{[a_1,a_2,a_3,a_4]} &=\int (n^1)^{a_1}(n^2)^{a_2}(n^3)^{a_3}(n_4)^{a_4}
                           \drm\Omega_3\\
	&= 
\left[\int_0^{\frac{\pi}{2}}\cos^{a_1 + a_2+1}\theta\sin^{a_3+a_4+1}\theta\drm\theta\right]
	\left[\int_0^{2\pi}\cos^{a_1}\psi\sin^{a_2}\psi\drm\psi\right]
	\left[\int_0^{2\pi}\cos^{a_3}\phi\sin^{a_4}\phi\drm\phi\right].\\
\end{calc}\end{smalleq}
We recognize the above definite integrals as different representations
of the beta function:
\begin{subequations}
\begin{align}
\int_0^\frac{\pi}{2}\cos^\alpha\theta\sin^\beta\theta\,\drm\theta
	&= \frac{1}{2}B\big(\tfrac{\alpha+1}{2},\tfrac{\beta+1}{2}\big)\\
\int_0^{2\pi}\cos^\alpha\theta\sin^\beta\theta\,\drm\theta
	&= \frac{1}{2}[1 + (-1)^\alpha + (-1)^{\alpha + \beta} + (-1)^{\beta}]
	 B\big(\tfrac{\alpha+1}{2},\tfrac{\beta+1}{2}\big),
\end{align}
\end{subequations}
where the second equation follows from the first. Therefore, one sees
that \emph{provided all the $a_i$ are even}
\begin{calc}
I^{[a_1, a_2, a_3, a_4]} &=2  B\big(\tfrac{a_1+a_2+2}{2},\tfrac{a_3+a_4+2}{2}\big)
	 B\big(\tfrac{a_1+1}{2},\tfrac{a_2+1}{2}\big)
	 B\big(\tfrac{a_3+1}{2},\tfrac{a_4+1}{2}\big)\\
&= 2\pi^2\left(\frac{1}{\pi^2}
\frac{\Gamma\big(\tfrac{a_1+1}{2}\big)\Gamma\big(\tfrac{a_2+1}{2}\big)
      \Gamma\big(\tfrac{a_3+1}{2}\big)\Gamma\big(\tfrac{a_4+1}{2}\big)}
	{\Gamma\big(\tfrac{a_1+a_2+a_3+a_4+4}{2}\big)}\right)\\
&= \frac{2\pi^2(a_1-1)!(a_2-1)!(a_3-1)!(a_4-1)!}
         {2^{a_1+a_2+a_3+a_4-4}(\frac{a_1}{2}-1)!(\frac{a_2}{2}-1)!
           (\frac{a_3}{2}-1)!(\frac{a_4}{2}-1)!(\frac{a_1+a_2+a_3+a_4}{2}+1)!}.
\end{calc}
Note that in the last line one must use the limit
\begin{equation}
\lim_{x\to 0} \frac{(x-1)!}{(\frac{x}{2}-1)!} = \frac{1}{2},
\end{equation}
in the event that some of the $a_i$ are zero.

The orthogonality condition from Equation~\eqref{eq:ortho-1} is
\begin{equation}
I^{j_1\cdots j_lk_1\cdots k_{l'}}
 \big(\mathcal{Y}^{l, m_\psi, m_\phi}_{j_1\cdots j_l}\big)^*
 \mathcal{Y}^{l', m_\psi', m_\phi'}_{k_1\cdots k_{l'}}
 =  \delta_{ll'}\delta_{m_\psi m_\psi'}\delta_{m_\phi m_\phi'},
\end{equation}
which motivates the choice of
\begin{equation}
Y_{l, m_\psi, m_\phi}^{j_1\cdots j_l} 
  \propto I^{j_1\cdots j_lk_1\cdots k_l}
    \big(\mathcal{Y}^{l, m_\psi, m_\phi}_{k_1\cdots k_l}\big)^*.
\end{equation}
We can think of the $2l$-index $I$ as defining an inner product on the
space of symmetric traceless $l$-index tensors, spanned by the
$\mathcal{Y}^{l,m_\psi, m_\phi}_{j_1\dots j_l}$. Then, we can think of
$Y_{l,m_\psi, m_\phi}$ as (proportional to) the dual of
$\mathcal{Y}^{l,m_\psi,m_\phi}$.

Since the $l$ derivatives are symmetrized and the spherical harmonic's
Cartesian form is also symmetrized, we get a factor of $l!$. One finds
that
\begin{equation}
I^{j_1\cdots j_lk_1\cdots k_l}
    \big(\mathcal{Y}^{l, m_\psi, m_\phi}_{k_1\cdots k_l}\big)^*
	\pd_{j_1}\cdots \pd_{j_l} 
    r^l Y_{l, m_\psi, m\phi}(\theta, \psi, \phi)
	\bigg|_{r\to 0}
 = l!
\end{equation}
and therefore we define
\begin{equation}
Y_{l, m_\psi, m_\phi}^{j_1\cdots j_l} 
  = \tfrac{1}{l!} I^{j_1\cdots j_lk_1\cdots k_l}
    \big(\mathcal{Y}^{l, m_\psi, m_\phi}_{k_1\cdots k_l}\big)^*.
\end{equation}

The first few normalized spherical harmonics are given by
\begin{subequations}\label{eq:spherical-harm}
\begin{align}
\sqrt{2}\pi Y_{0, 0, 0} &= 1\\
\sqrt{2}\pi Y_{1, 1, 0} &= \sqrt{2}\frac{x^1+ix^2}{r} = \sqrt{2}\cos\theta\, e^{i\psi}\\
\sqrt{2}\pi Y_{1, 0, 1} &= \sqrt{2}\frac{x^3+ix^4}{r} = \sqrt{2}\sin\theta\,e^{i\phi}\\
\sqrt{2}\pi Y_{2, 2, 0} &= \sqrt{3}\left(\frac{x^1+ix^2}{r}\right)^2 
	= \sqrt{3}\cos^2\theta\, e^{i\psi}\\
\sqrt{2}\pi Y_{2, 0, 2} &= \sqrt{3}\left(\frac{x^3 + ix^4}{r}\right)^2 
	= \sqrt{3}\sin^2\theta\, e^{2i\phi}\\
\sqrt{2}\pi Y_{2, 1, 1} &= \sqrt{6}\frac{(x^1 +ix^2)(x^3+ix^4)}{r^2}
	= \sqrt{6}\sin\theta\cos\theta e^{i\psi + i\phi}\\
\sqrt{2}\pi Y_{2, 1,-1} &= \sqrt{6}\frac{(x^1+ix^2)(x^3-ix^4)}{r^2} 
	= \sqrt{6}\sin\theta\cos\theta\, e^{i\psi -i\phi}\\
\sqrt{2}\pi Y_{2, 0, 0} &= \sqrt{3}\frac{(x^1+ix^2)(x^1-ix^2) - (x^3+ix^4)(x^3-ix^4)}{r^2}
	= \sqrt{3}(\cos^2\theta - \sin^2\theta)
\end{align}
\end{subequations}
Since
\begin{equation}
Y_{l,m_\psi, m_\phi}^* = Y_{l,-m_\psi, -m_\phi},
\end{equation}
one can find the rest of the $l=0,1,2$ spherical harmonics by complex
conjugation. The first few Clebsch--Gordan coefficients are given by
\begin{subequations}
\begin{align}
\sqrt{2}\pi \mathcal{Y}^{0,0,0} &= 1\\
\sqrt{2}\pi \mathcal{Y}^{1,1,0}_j &= \sqrt{2}(\delta^1_j + i \delta^2_j)\\
\sqrt{2}\pi \mathcal{Y}^{1,0,1}_j &= \sqrt{2}(\delta^3_j + i \delta^4_j)\\
\sqrt{2}\pi \mathcal{Y}^{2,2,0}_{ij} 
	&= \sqrt{3}[\delta^1_i\delta^1_j -\delta^2_i\delta^2_j + 2 i\delta^1_i\delta^2_j]\\
\sqrt{2}\pi\mathcal{Y}^{2,0,2}_{ij}
	&= \sqrt{3}[\delta^3_i\delta^3_j -\delta^4_i\delta^4_j + 2 i\delta^3_i\delta^4_j]\\
\sqrt{2}\pi\mathcal{Y}^{2,1,1}_{ij}
	&= \sqrt{6}[\delta^1_i\delta^3_j - \delta^2_i\delta^4_j + i\delta^1_i\delta^4_j
			+ i\delta^2_i\delta^3_j]\\
\sqrt{2}\pi\mathcal{Y}^{2,1,-1}_{ij}
	&= \sqrt{6}[\delta^1_i\delta^3_j + \delta^2_i\delta^4_j - i\delta^1_i\delta^4_j
			+ i\delta^2_i\delta^3_j]\\
\sqrt{2}\pi\mathcal{Y}^{2,0,0}_{ij}
	&= \sqrt{3}[\delta^1_i\delta^1_j + \delta^2_i\delta^2_j 
	- \delta^3_i\delta^3_j-\delta^4_i\delta^4_j].
\end{align}
\end{subequations}
For compactness, we have neglected to symmetrize the indices above.
Finally, we should list the first few $I$'s. The zero-index $I$ is
\begin{subequations}
\begin{equation}
I_{(0)} = 2\pi^2;
\end{equation}
the 2-index $I$ can be written as a matrix,
\begin{equation}
I_{(2)}^{ij} = \frac{\pi^2}{2}
\begin{pmatrix}
1 & & & \\
 & 1 & & \\
 & & 1 & \\
 & & & 1
\end{pmatrix}^{ij}.
\end{equation}
The following two components of the 4-index $I$ suffice to deduce the
rest of the components from symmetry:
\begin{equation}
I_{(4)}^{1111} = \frac{\pi^2}{4}\qquad
I_{(4)}^{1122} = \frac{\pi^2}{12}.
\end{equation}
\end{subequations}
Finally, we write down the $l=0$ and $l=1$ coefficients for the
derivative operators:
\begin{subequations}
\begin{align}
Y_{0,0,0} &= \sqrt{2}\pi\\
Y_{1,1,0}^j &= \frac{\pi}{2}(\delta^1_j + i\delta^2_j)\\
Y_{1,0,1}^j &= \frac{\pi}{2}(\delta^3_j + i\delta^4_j),
\end{align}
and the $l=2$ coefficients
\begin{align}
\begin{split}
Y_{2,2,0}^{ij} &= \frac{1}{2!}\frac{\sqrt{3}}{\sqrt{2}\pi}(I^{ij11} - I^{ij22} - 2i I^{ij12})\\
	&= \frac{1}{\sqrt{6}}\frac{\pi}{4}
\begin{pmatrix}
1 & -i & & \\
-i & -1 & & \\
 & & & \\
 & & & 
\end{pmatrix}^{ij}
\end{split}\\
Y_{2,0,2}^{ij} &= \frac{1}{\sqrt{6}}\frac{\pi}{4}
\begin{pmatrix}
 & & & \\
 & & & \\
& &  1 & -i  \\
& & -i & -1 \\
\end{pmatrix}^{ij}\\
Y_{2,1,1}^{ij} &= \frac{\pi}{8\sqrt{3}}
\begin{pmatrix}
 & & 1 & -i\\
 & & -i & -1\\
1 & -i & & \\
-i & -1 & &
\end{pmatrix}^{ij}\\
Y_{2,1,-1}^{ij} &= \frac{\pi}{8\sqrt{3}}
\begin{pmatrix}
 & & 1 & -i\\
 & & i & 1\\
1 & i & & \\
-i & 1 & &
\end{pmatrix}^{ij}\\
Y_{2,0,0}^{ij} &= \frac{\pi}{8}\sqrt{\frac{3}{2}}
\begin{pmatrix}
1 & & & \\
 & 1 & & \\
 & & -1 & \\
 & &  & -1
\end{pmatrix}^{ij}
\end{align}
\end{subequations}
Again, the remaining coefficients can be found by complex
conjugation. Clearly, it is not too difficult to find higher $l$
coefficients as needed.
\chapter{Normalizing the CFT state and the Vertex Operator}
\label{ap:norm}

Here we present the normalization calculation for the $\kappa=1$
initial state of Chapter~\ref{ch:emission}, along with that of the
vertex operator for emission of $\phi_{ij}$.

\section{Normalizing the Initial State}

To find the normalization constant $\mathcal{C}_L$, we take the
Hermitian conjugate to find
\begin{equation}
\mathop{\vphantom{a}}_{A\dot{A}}^L\hspace{-3pt}
\bra{\phi_{N+1}^{\frac{l}{2},\frac{l}{2}-k}}
  = -\mathcal{C}_L^*\, 
  \mathop{\vphantom{a}}_{NS}\hspace{-3pt}\bvac\tilde{\sigma}_{l+1}\epsilon_{\dot{A}\dot{B}}
		\psi_{\frac{1}{2}}^{-\dot{B}}\epsilon_{AB}G^{+B}_\frac{1}{2}(J_0^+)^k L_1^N,
\end{equation}
and then compute the norm,
\begin{smalleq}\begin{calc}
\mathop{\vphantom{a}}_{A\dot{A}}^L \hspace{15pt}&\hspace{-18pt}
\braket{\phi_{N+1}^{\frac{l}{2},\frac{l}{2}-k}|\phi_{N+1}^{\frac{l}{2},\frac{l}{2}-k}}_L^{B\dot{B}}
	\\
  &= -|\mathcal{C}_L|^2 \epsilon_{\dot{A}\dot{C}}\epsilon_{AC}
\,\mathop{\vphantom{a}}_{NS}\hspace{-3pt}\bvac\tilde{\sigma}^0_{l+1}
	\psi_{\frac{1}{2}}^{-\dot{C}} G^{+C}_\frac{1}{2}(J_0^+)^k
			L_1^N L_{-1}^N
	(J_0^-)^k G^{-B}_{-\frac{1}{2}}
	\psi_{-\frac{1}{2}}^{+\dot{B}}\sigma^0_{l+1}\vac_{NS}\\
  &= -|\mathcal{C}_L|^2 \epsilon_{\dot{A}\dot{C}}\epsilon_{AC}
\,\mathop{\vphantom{a}}_{NS}\hspace{-3pt}\bvac\tilde{\sigma}^0_{l+1}
	\psi_{\frac{1}{2}}^{-\dot{C}} G^{+C}_\frac{1}{2}(J_0^+)^k
		\left(\prod_{j=1}^N j(2L_0 + j-1)\right)
	(J_0^-)^k G^{-B}_{-\frac{1}{2}}
	\psi_{-\frac{1}{2}}^{+\dot{B}}\sigma^0_{l+1}\vac_{NS}\\
  &= -\frac{N!(N+l+1)!}{(l+1)!}|\mathcal{C}_L|^2 \epsilon_{\dot{A}\dot{C}}\epsilon_{AC}
\,\mathop{\vphantom{a}}_{NS}\hspace{-3pt}\bvac\tilde{\sigma}^0_{l+1}
	\psi_{\frac{1}{2}}^{-\dot{C}} G^{+C}_\frac{1}{2}(J_0^+)^k
	(J_0^-)^k G^{-B}_{-\frac{1}{2}}
	\psi_{-\frac{1}{2}}^{+\dot{B}}\sigma^0_{l+1}\vac_{NS}\\
  &= -\frac{k!\,l!}{(l-k)!}\frac{N!(N+l+1)!}{(l+1)!}|\mathcal{C}_L|^2 
		\epsilon_{\dot{A}\dot{C}}\epsilon_{AC}
	\,\mathop{\vphantom{a}}_{NS}\hspace{-3pt}\bvac\tilde{\sigma}^0_{l+1}
	\psi_{\frac{1}{2}}^{-\dot{C}} 
		G^{+C}_\frac{1}{2}G^{-B}_{-\frac{1}{2}}
	\psi_{-\frac{1}{2}}^{+\dot{B}}\sigma^0_{l+1}\vac_{NS}\\
  &= -\delta_A^B\frac{k!\,l!}{(l-k)!}\frac{N!(N+l+1)!}{(l+1)!}|\mathcal{C}_L|^2 
	\epsilon_{\dot{A}\dot{C}}
	\,\mathop{\vphantom{a}}_{NS}\hspace{-3pt}\bvac\tilde{\sigma}^0_{l+1}
	\psi_{\frac{1}{2}}^{-\dot{C}} 
		(L_0 + J_0^3)
	\psi_{-\frac{1}{2}}^{+\dot{B}}\sigma^0_{l+1}\vac_{NS}.
\end{calc}\end{smalleq}
Proceeding with the calculation, one finds
\begin{smalleq}\begin{calc}
\mathop{\vphantom{a}}_{A\dot{A}}^L\hspace{-3pt}
\braket{\phi_{N+1}^{\frac{l}{2},\frac{l}{2}-k}|\phi_{N+1}^{\frac{l}{2},\frac{l}{2}-k}}_L^{B\dot{B}}
  &= -(l+1)\delta_A^B\epsilon_{\dot{A}\dot{C}}
        \frac{k!\,l!}{(l-k)!}\frac{N!(N+l+1)!}{(l+1)!}|\mathcal{C}_L|^2
	\hspace{-3.07pt}\mathop{\vphantom{a}}_{NS}\hspace{-3pt}\bvac\tilde{\sigma}^0_{l+1}
	\psi_{\frac{1}{2}}^{-\dot{C}} 
	\psi_{-\frac{1}{2}}^{+\dot{B}}\sigma^0_{l+1}\vac_{NS}\\
  &= -\delta_A^B\frac{N!(N+l+1)!k!}{(l-k)!}|\mathcal{C}_L|^2 
	\epsilon_{\dot{A}\dot{C}}
	\,\mathop{\vphantom{a}}_{NS}\hspace{-3pt}\bvac\tilde{\sigma}^0_{l+1}
	\psi_{\frac{1}{2}}^{-\dot{C}} 
	\psi_{-\frac{1}{2}}^{+\dot{B}}\sigma^0_{l+1}\vac_{NS}\\
   &= \delta_A^B\,\delta_{\dot{A}}^{\dot{B}}
      \frac{N!(N+l+1)!k!}{(l-k)!}(l+1)|\mathcal{C}_L|^2,
\end{calc}\end{smalleq}
where we have used the fact that the chiral primary twist operators
are correctly normalized. The factor of $l+1$ comes from the fermion
anticommutator, since in the twisted sector there are $l+1$ copies of
the fermion field that go into what we call $\psi$.  One can
understand this factor most easily by using
Equation~\eqref{eq:frac-mode-cover}. By demanding that
\begin{equation}
\mathop{\vphantom{a}}_{A\dot{A}}^L\hspace{-3pt}
\braket{\phi_{N+1}^{\frac{l}{2},\frac{l}{2}-k}|\phi_{N+1}^{\frac{l}{2},\frac{l}{2}-k}}_L^{B\dot{B}} = \delta_A^B\delta_{\dot{A}}^{\dot{B}},
\end{equation}
we conclude that the normalized state (with the left \emph{and} right parts)
is
\begin{equation}\begin{split}
\ket{\phi}^{A\dot{A}B\dot{B}} 
      &= \sqrt{\frac{(l-k)!(l-\bar{k})!}{N!\bar{N}!(N+l+1)!(\bar{N}+l+1)!k!\bar{k}!(l+1)^2}}\\
	&\qquad\times L_{-1}^N(J_0^-)^k  G^{-A}_{-\frac{1}{2}}
	\psi_{-\frac{1}{2}}^{+\dot{A}}\,\,
	\bar{L}_{-1}^{\bar{N}}(\bar{J}_0^-)^{\bar{k}} \bar{G}^{\dot{-}B}_{-\frac{1}{2}}
	\bar{\psi}_{-\frac{1}{2}}^{\dot{+}\dot{B}}\sigma^0_{l+1}\vac_{NS}.
\end{split}\end{equation}
In this computation we use the identity
\begin{equation}
\prod_{j=1}^k \big(j l -j(j-1)\big)
	=  \frac{k!\, l!}{(l-k)!}.
\end{equation}

\section{Normalizing the Vertex Operator}

The left part of the vertex operator is given by
\begin{equation}
\mathcal{V}_{L; l,k}^{A\dot{A}}(z)
	= N_L\left(\big(J_0^+\big)^kG_{-\frac{1}{2}}^{+A}\psi_{-\frac{1}{2}}^{-\dot{A}}
		\tilde{\sigma}^0_{l+1}(z)\right)_z.
\end{equation}
We need to normalize the vertex operator. To that end, we begin by
writing its Hermitian conjugate:
\begin{equation}
{\mathcal{V}^{A\dot{A}}_{L;l,k}}^\dg(z) = -(-1)^{k+1}\epsilon_{AB}\epsilon_{\dot{A}\dot{B}}\,N_L^*\,
	\left(\big(J_0^-\big)^k G_{-\frac{1}{2}}^{-B}\psi_{-\frac{1}{2}}^{+\dot{B}}
			\sigma^0_{l+1}\left(z\right)\right)_{z}
                      = \epsilon_{AB}\epsilon_{\dot{A}\dot{B}}\mathcal{V}_{l,2l-k}^{B\dot{B}}(z),
\end{equation}
where the second equality is the condition needed to ensure the total
interaction action is Hermitian.

The factor of $(-1)^{k+1}$ comes from the $G$ and the $J_0$'s. We
illustrate below with $J_0^+$:
\begin{calc}
\left[\big(J_0^+\big)_z\right]^\dg &= \left[\oint_z\frac{\drm z'}{2\pi i}J^+(z')\right]^\dg\\
  &= -\oint_{\bar{z}}\frac{\drm {\bar{z}}'}{2\pi i}J^-\left(\tfrac{1}{{\bar{z}}'}\right)
          \frac{1}{{\bar{z}}'^2}\\
  &= -\oint_{\frac{1}{\bar{z}}}\frac{\drm \xi}{2\pi i}J^-(\xi)\\
  &= -\big(J_0^-\big)_{\frac{1}{\bar{z}}},
\end{calc}
where when making the change of variables $\xi = 1/\bar{z}'$ there are
two minus signs. One comes from the Jacobian $\drm\bar{z}' = - 1/\xi^2
\drm \xi$, and the other comes from making the contour
counter-clockwise. The $G_{-\frac{1}{2}}$ behaves in the same way;
however, the $\psi_{-\frac{1}{2}}$ is different:
\begin{calc}
\left[\big(\psi^{-\dot{A}}_{-\frac{1}{2}}\big)_z\right]^\dg 
  &= \left[\oint_z\frac{\drm z'}{2\pi i}\frac{\psi^{-\dot{A}}(z')}{z'-z}\right]^\dg\\
  &= -(-\epsilon_{-+}\epsilon_{\dot{A}\dot{B}})
       \oint_{\bar{z}}\frac{\drm \bar{z}'}{2\pi i}\,\psi^{+\dot{B}}\big(\tfrac{1}{\bar{z}'}\big)\,
       \frac{1}{\bar{z}'(\bar{z}'-\bar{z})}\\
  &= -\epsilon_{\dot{A}\dot{B}}\oint_{\frac{1}{\bar{z}}}\frac{\drm \xi}{2\pi i}\psi^{+\dot{B}}(\xi)
        \frac{1}{\xi\left(\frac{1}{\xi} - \bar{z}\right)}\\
  &= \frac{\epsilon_{\dot{A}\dot{B}}}{\bar{z}}
      \oint_\frac{1}{\bar{z}}\frac{\drm \xi}{2\pi i}
         \frac{\psi^{+\dot{B}}(\xi)}{\xi - \frac{1}{\bar{z}}}\\
  &= \frac{\epsilon_{\dot{A}\dot{B}}}{\bar{z}}
     \big(\psi^{+\dot{B}}_{-\frac{1}{2}}\big)_{\frac{1}{\bar{z}}};
\end{calc}
it receives an extra minus sign from the integrand.

We use the notation
\begin{equation}
\vev{\cdot} = {}_{NS}\bvac\cdot\vac_{NS}
\end{equation}
for the NS-vacuum expectation value.  Proceeding with the
normalization, the 2-point function is given by
\begin{smalleq}\begin{calc}
\Vev{{\mathcal{V}_{L;l,k}^{A\dot{A}}}^\dg(z)\mathcal{V}_{L;l,k}^{B\dot{B}}(0)}
	&= \hspace{-2pt}(-1)^{k+2}|N_L|^2\epsilon_{AC}\epsilon_{\dot{A}\dot{C}}\hspace{-2pt}
		\Vev{\hspace{-1pt}
                    \left(\hspace{-2pt}
                    \big(J_0^-\big)^k\hspace{-1pt} 
                     G_{-\frac{1}{2}}^{-C}\psi_{-\frac{1}{2}}^{+\dot{C}}
			\sigma^0_{l+1}\left(z\right)\right)_{z}\hspace{-4pt}
		\left(\hspace{-2pt}
                      \big(J_0^+\big)^k\hspace{-1pt}
                      G_{-\frac{1}{2}}^{+B}\psi_{-\frac{1}{2}}^{-\dot{B}}
		\tilde{\sigma}^0_{l+1}(0)\right)_0\hspace{-1pt}}\\
	&= -|N_L|^2\epsilon_{AC}\epsilon_{\dot{A}\dot{C}}
		\Vev{\left(\psi_{-\frac{1}{2}}^{+\dot{C}}\sigma^0_{l+1}\left(z\right)\right)_{z}
		\left(G_{-\frac{1}{2}}^{-C}\big(J_0^-\big)^k\big(J_0^+\big)^kG_{-\frac{1}{2}}^{+B}
		\psi_{-\frac{1}{2}}^{-\dot{B}}\tilde{\sigma}^0_{l+1}(0)\right)_0} \\
	&= -|N_L|^2\epsilon_{AC}\epsilon_{\dot{A}\dot{C}}\frac{k!\,l!}{(l-k)!}
		\Vev{\left(\psi_{-\frac{1}{2}}^{+\dot{C}}\sigma^0_{l+1}\left(z\right)\right)_{z}
		\left(G_{-\frac{1}{2}}^{-C}G_{-\frac{1}{2}}^{+B}
		\psi_{-\frac{1}{2}}^{-\dot{B}}\tilde{\sigma}^0_{l+1}(0)\right)_0} \\
	&= |N_L|^2\delta_A^B\epsilon_{\dot{A}\dot{C}}\frac{k!\,l!}{(l-k)!}
		\Vev{\left(\psi_{-\frac{1}{2}}^{+\dot{C}}\sigma^0_{l+1}\left(z\right)\right)_{z}
		\left(L_{-1}\psi_{-\frac{1}{2}}^{-\dot{B}}\tilde{\sigma}^0_{l+1}(0)\right)_0} \\
	&= |N_L|^2\delta_A^B\epsilon_{\dot{A}\dot{C}}\frac{k!\,l!}{(l-k)!}
		\lim_{v\to 0}\pd_v
		\Vev{\left(\psi_{-\frac{1}{2}}^{+\dot{C}}\sigma^0_{l+1}\left(z\right)\right)_{z}
		\left(\psi_{-\frac{1}{2}}^{-\dot{B}}\tilde{\sigma}^0_{l+1}(v)\right)_v} \\
	&= |N_L|^2\delta_A^B\epsilon_{\dot{A}\dot{C}}\epsilon^{\dot{C}\dot{B}}
		\frac{k!\,l!}{(l-k)!}
		\lim_{v\to 0}\pd_v
		\frac{l+1}{(z-v)^{l+1}}\\
	&= |N_L|^2\delta_A^B\delta_{\dot{A}}^{\dot{B}}
		\frac{k!\,(l+1)!}{(l-k)!}
		(l+1)\frac{1}{z^{l+2}}.
\end{calc}\end{smalleq}
The factor of $l+1$ has the same origin as in the normalization of the
initial state.  Using the above, one finds
\begin{equation}
N_L = \sqrt{\frac{(l-k)!}{k!\,(l+1)!(l+1)}},
\end{equation}
and thus the left part of the vertex operator is
\begin{equation}
\mathcal{V}_{L; l,k}^{A\dot{A}}(z)
	= \sqrt{\frac{(l-k)!}{k!\,(l+1)!(l+1)}}
  \left(\big(J_0^+\big)^kG_{-\frac{1}{2}}^{+A}\psi_{-\frac{1}{2}}^{-\dot{A}}
		\tilde{\sigma}^0_{l+1}(z)\right)_z.
\end{equation}
The normalization is chosen such that the vertex operator in the
complex plane satisfies
\begin{equation}\begin{split}
\vev{\mathcal{V}^{A\dot{A}B\dot{B}}_{l,-m_\psi, -m_\phi}(z)
\mathcal{V}^{C\dot{C}D\dot{D}}_{l,m_\psi, m_\phi}(0)}
 &= \frac{\epsilon^{AC}\epsilon^{\dot{A}\dot{C}}\epsilon^{BD}\epsilon^{\dot{B}\dot{D}}}
	{|z|^{l+2}}\\
\vev{\mathcal{V}^{ij}_{l,-m_\psi, -m_\phi}(z)
	\mathcal{V}^{kl}_{l,m_\psi, m_\phi}(0)} 
	&= \frac{\delta^{ik}\delta^{jl}}{|z|^{l+2}}.
\end{split}\end{equation}

\chapter{Computing Correlation Functions of
  \texorpdfstring{$S_{N_1N_5}$}{SN1N5}-twist Operators}
\label{ap:twist-corr}

We use the methods developed in~\cite{lm1, lm2} to compute correlation
functions of $S_{N_1N_5}$-twist operators. We call the physical space
where the problem is posed the ``base space'' which has the
coordinates complex $z$ and $\bar{z}$. In the base space, the basic
fields are multi-valued.  The basic method for computing the
correlators of twist operators consists of using a meromorphic mapping
to a covering space where there is one set of single-valued fields.
Having single-valued fields comes at the expense of introducing
curvature in the covering space.  Fortunately, we can conformally map
the curved covering space to a manifold with metric in a fiducial
(flat) form. Because of the conformal curvature anomaly, the path
integral is not invariant under conformal mappings; however, the path
integral changes in a calculable way. Note that we do not compute the
combinatorial factors that result from properly symmetrizing over the
copies of the orbifold theory.

\section{Basic Method}

Suppose we compute the path integral of our conformal field theory on
the manifold with fiducial metric $\widehat{ds}^2$, which we call
$\hat{Z}$. The path integral of the same conformal field theory on a
manifold with metric $ds^2 = e^{\phi} \widehat{ds}^2$ which we call
$Z$, is related to $\hat{Z}$ by~\cite{liouville}
\begin{equation}
Z = e^{S_L}\hat{Z},
\end{equation}
where the Liouville action, $S_L$, is given by
\begin{equation}
S_L = \frac{c}{96\pi}\int\drm^2 t\sqrt{-\hat{g}}\bigg[
	\hat{g}^{\mu\nu}\pd_\mu\phi\pd_\nu\phi + 2 \hat{R}\phi\bigg].
\end{equation}
$\hat{g}$ is the fiducial metric, $\hat{R}$ is the Ricci scalar of the
\emph{fiducial} metric, and $c$ is the central charge of a
\emph{single} copy of the fields ($c=6$ in our case).

We now outline the precise steps needed to compute correlators of
twist operators, as given in~\cite{lm1}. The problem is posed in terms
of some twist operators inserted at various points $z_i$ in the base
space. To make the fields well-defined, we cut a small hole of radius
$\veps_i$ around each $z_i$ and demand that the fields have the
correct twisted boundary conditions around that hole. Furthermore, we
regulate the path integral by putting the correlator on a disc of
radius $1/\delta$, which encloses all of the $z_i$ except twist
operators that are inserted at infinity. To define the boundary
conditions on the edge of the disc, we glue a second flat disc onto
the edge of the first, giving the base space the topology of a sphere.
We insert any twist operators located at infinity at the center of
this second disc.  We work on the base space with metric
\begin{equation}\label{eq:base-metric}
ds^2 = \begin{cases}
dzd\bar{z}, & |z| < \frac{1}{\delta}\\
d\tilde{z}d\bar{\tilde{z}}, & |\tilde{z}| < \frac{1}{\delta}
\end{cases},\hspace{30pt}
\tilde{z} = \frac{1}{\delta^2}\frac{1}{z}.
\end{equation}
The path integral with the various regularizations on the above metric
we write as
\begin{equation}
Z^{(s)}_{\veps, \delta}[\sigma_{n_1}(z_1)\cdots\sigma_{n_N}(z_N)].
\end{equation}

We can map to a covering space with metric $ds'^2$ with coordinates $t$
and $\bar{t}$ via a meromorphic function $z = z(t)$. Then, the path
integral is given by
\begin{equation}\label{eq:corr-to-covering-space}
Z^{(s)}_{\veps,\delta}[\sigma_{n_1}(z_1)\cdots\sigma_{n_N}(z_N)] 
   = Z^{(s')}_{\veps,\delta},
\end{equation}
where
\begin{equation}
ds^2 = dzd\bar{z} \quad \Longrightarrow \quad 
	{ds'}^2 = \der{z}{t}\der{\bar{z}}{\bar{t}}\,dtd\bar{t}.
\end{equation}
When we write $Z^{(s')}$ with no square brackets we mean that there
are no insertions in the path integral. Throughout this discussion, we
assume that we are computing the correlator of bare twists which leave
no insertions in the covering space. If this is not the case as in
Chapter~\ref{ch:emission}, then the right hand side of
Equation~\eqref{eq:corr-to-covering-space} is multiplied by the
separately computed correlator of those insertions in the covering
space~\cite{lm1, lm2}.

The sizes of the various holes in the $t$ coordinates are determined
by the mapping. The boundary conditions at the edges of the holes are
defined by pasting in flat pieces of manifold in the covering
space.\footnote{If the twist operators are not bare twists, then there
  would also be some operator insertions in the covering space;
  however, here we just discuss bare twists.}

The induced metric on the covering space, ${ds'}^2$, is conformally
related to the fiducial metric we choose to work with
\begin{equation}\label{eq:covering-metric}
\widehat{ds}^2 = \begin{cases}
dtd\bar{t}, & |t| <\frac{1}{\delta'}\\
d\tilde{t}d\bar{\tilde{t}}, & |\tilde{t}| < \frac{1}{\delta'}
\end{cases},\hspace{30pt}
\tilde{t} = \frac{1}{\delta'^2}\frac{1}{t}.
\end{equation}
Thus, we can write
\begin{equation}
Z_{\veps,\delta}^{(s)}[\sigma_{n_1}(z_1)\cdots\sigma_{n_N}(z_N)] 
    = e^{S_L}Z^{(\hat{s})}_{\veps,\delta, \delta'}.
\end{equation}
The correlator of the regularized twist operators is defined by
\begin{equation}\label{eq:def-corr}
\vev{\sigma_{n_1}(z_1)\cdots\sigma_{n_N}(z_N)}_{\veps, \delta} 
   = \frac{Z^{(s)}_{\veps,\delta}[\sigma_{n_1}(z_1)\cdots\sigma_{n_N}(z_N)]}
	{\big(Z_{\delta}\big)^{s}}
   = e^{S_L} \frac{Z_{\veps, \delta, \delta'}^{(\hat{s})}}{\big(Z_\delta\big)^s},
\end{equation}
where the $s$ in the denominator is the number of Riemann sheets or
the number of copies involved in the correlator (not to be confused
with the metric). The path integral in the numerator, then, is also
only over the twisted copies of the CFT, while the path integral with
no insertions in the denominator, $Z_\delta$, is the path integral of
a single copy of the CFT with the metric in
Equation~\eqref{eq:base-metric}.

The path integrals on the right-hand side of
Equation~\eqref{eq:def-corr} have no insertions and are on a metric of
identical form. For the case where the covering space manifold has
spherical topology, the path integrals cancel out up to the $\delta$
and $\delta'$ dependence. Ultimately, we want to normalize the twist
operators (and correlator) and take the limit of $\veps, \delta,
\delta'\to 0$. We normalize the $\veps$-dependence by demanding that
the two-point function of two twist operators is fixed. The
$\delta'$-dependence cancels out as it should, and any
$\delta$-dependence that remains corresponds to twist operators at
infinity. This $\delta$-dependence can be cancelled out by appropriate
normalization of the two-point function with one twist at infinity.

We should mention that there is an important final step to compute the
correlation function. The twist operators $\sigma_n$ are implicitly
labeled by a certain $n$-cycle element of $S_{N_1N_5}$; however, we
are modding out by $S_{N_1N_5}$, so we should really label the twists
with \emph{conjugacy classes}~\cite{lm1}. This fact is taken into
account combinatorial in the body of the text. One can understand the
structure and relation of these combinatorial factors to the genus of
the covering space via diagrammatic methods developed
in~\cite{rastelli2}.

\section{Spherical Correlation Functions of Twist Operators}

Our goal in this appendix is to simplify and algorithmize correlation
functions of bare twist operators using the Liouville action method.
We restrict our consideration to correlation functions that have
spherical genus, because they have a couple of simplifications and
give the leading contribution in a large-$N_1N_5$
expansion~\cite{lm1}.

Consider a correlator of the form
\begin{equation}\label{eq:gen-corr}
  Z = \vev{\sigma_{p_1}(z_1)\sigma_{p_2}(z_2)\cdots\sigma_{p_M}(z_M)
    \sigma_{q_0}(\infty)\sigma_{q_1}(\infty)\sigma_{q_2}(\infty)
    \cdots\sigma_{q_{N-1}}(\infty)},
\end{equation}
where these are the normalized twist operators that have unit
correlator with themselves at unit separation in the $z$-plane. We
come to the normalization later. The $p_i$ and $q_j$ are the lengths
of the $S_N$-cycles for the corresponding twist operators. We have $M$
twists in the finite $z$-plane and $N$ twists at $z=\infty$. This is
the most general correlator of bare twists one could consider. Note
that one could alternatively consider only twists in the finite
$z$-plane and then take the limit as some of the $z_i$ go to infinity.

The genus of the covering space is determined by the Riemann--Hurwitz
formula
\begin{equation}\label{eq:RW-formula}
g = \frac{1}{2}\sum_{i=1}^{M}(p_i-1) 
   + \frac{1}{2}\sum_{j=0}^{N-1}(q_j-1) - s +1,
\end{equation}
where $s$ is the total number of sheets, or copies, involved in the
correlator. We restrict our attention to the sphere, $g=0$.  In which
case, we do not have to compute the path integral, as we see when we
normalize the twist operators.

\section{Properties of the Map}

One needs to find the map to the covering space. This is the difficult
part of the problem. One can use a differential equation approach, as
demonstrated in~\cite{lm1} or other methods. We do not treat this
aspect of the problem here.

The map must have certain properties.  It should be meromorphic with
three types of points where the map's local properties become
important: images of the twists, infinite images of infinity, and
finite images of infinity. We parameterize the behavior as follows
\begin{equation}\begin{aligned}\label{eq:gen-map}
z - z_i &\approx a_i (t-t_i)^{p_i}\hspace{50pt} 
                    & z &\approx z_i, t \approx t_i\\
z &\approx b_0 t^{q_0} 
                    & z &\to \infty, t\to \infty\\
z &\approx \frac{b_j}{(t-t_\infty^j)^{q_j}} 
                    & z &\to \infty, t \approx t_\infty^j,
\end{aligned}\end{equation}
where we only care about the leading term displayed above. In the end,
the correlator depends only on the $a_i$, $b_j$, $p_i$, and $q_j$.

We should specify exactly how many different images of infinity the
map has. We claim that we can arrange the map so that $q_0$ through
$q_{N-1}$ are as above and all additional finite images of infinity
are of order $1$, and furthermore
\begin{equation}
q_0 + \sum_{j=1} q_j = \sum_{j=0}^{N-1} q_j + \sum_{j=N}^{F}q_j = s,
\end{equation}
where $q_N$ through $q_F$ are all unity. This means that the number of
distinct images of infinity is
\begin{equation}
F= s + N - \sum_{j=0}^{N-1} q_j.
\end{equation}

\section{Computing the Liouville Action}

The correct way to define the correlator in
Equation~\eqref{eq:gen-corr} is to cut holes in the $z$ plane around
the $z_i$ with radius $\veps$ and then a large cut at
$|z|=1/\delta$~\cite{lm1}.  The small holes around the $z_i$ have the
appropriate boundary conditions for the twist operators. Upon the cut
at $|z|=1/\delta$, we glue a second flat disc with complex coordinate
$\tilde{z}$. The center of this second disc has a hole of size
$\tilde{\veps}$, which has the boundary conditions appropriate for the
$N$ twists at infinity.  The metric on this manifold is
\begin{equation} 
\begin{split}
ds^2 &= \left\{ 
\begin{array}{rl} 
 dzd\bar{z} & \qquad |z| <\frac{1}{\delta}\\ 
 \\
 d\tilde{z}d\bar{\tilde{z}} & \qquad |\tilde{z}| < \frac{1}{\delta}
\end{array} \right. \\
z &= \frac{1}{\delta^2\tilde{z}}.
\end{split} \label{Eqn:BaseMetric}
\end{equation} 

\begin{figure}[ht]
\begin{center}
\includegraphics[width=6cm]{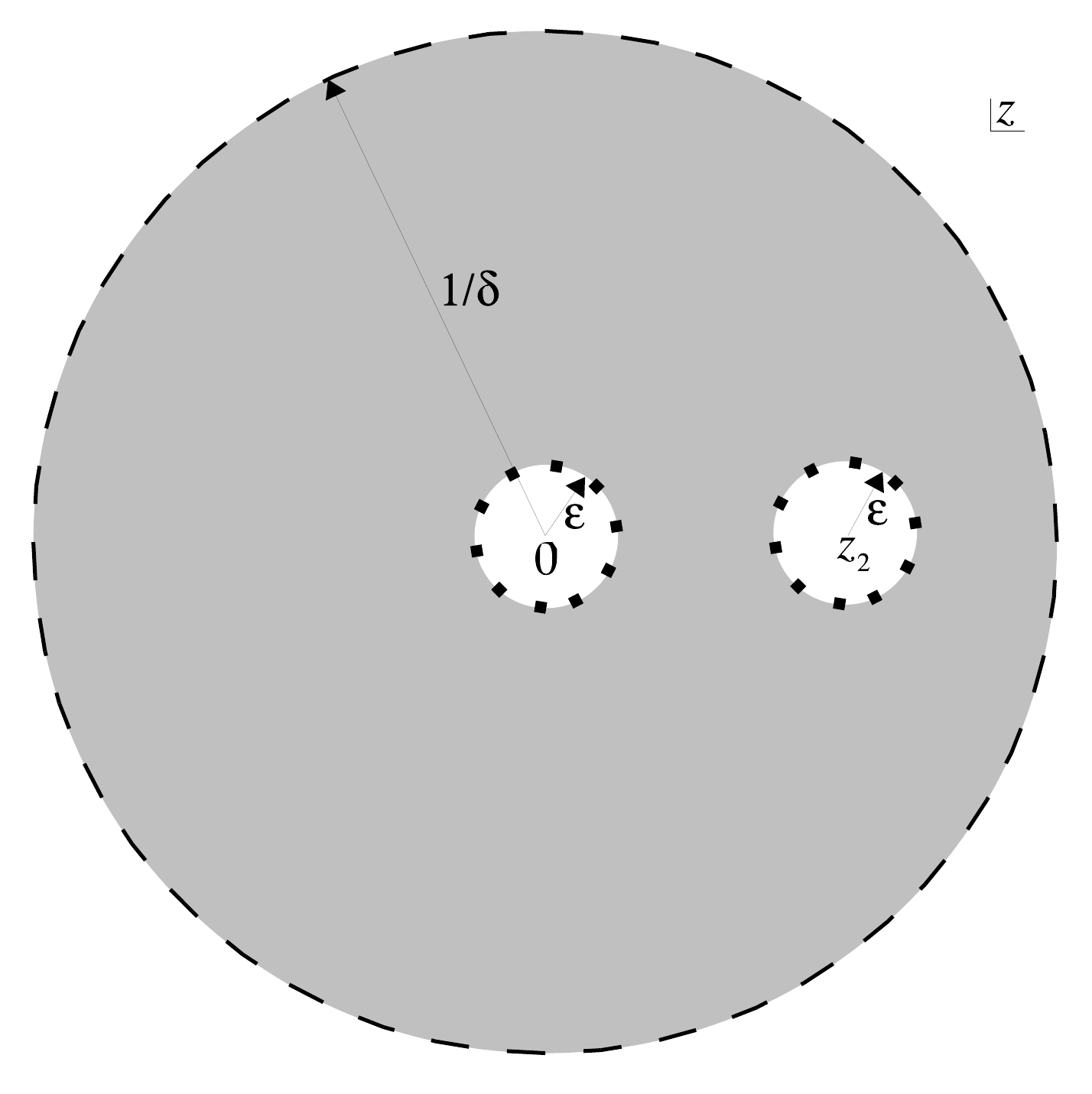}
\includegraphics[width=6cm]{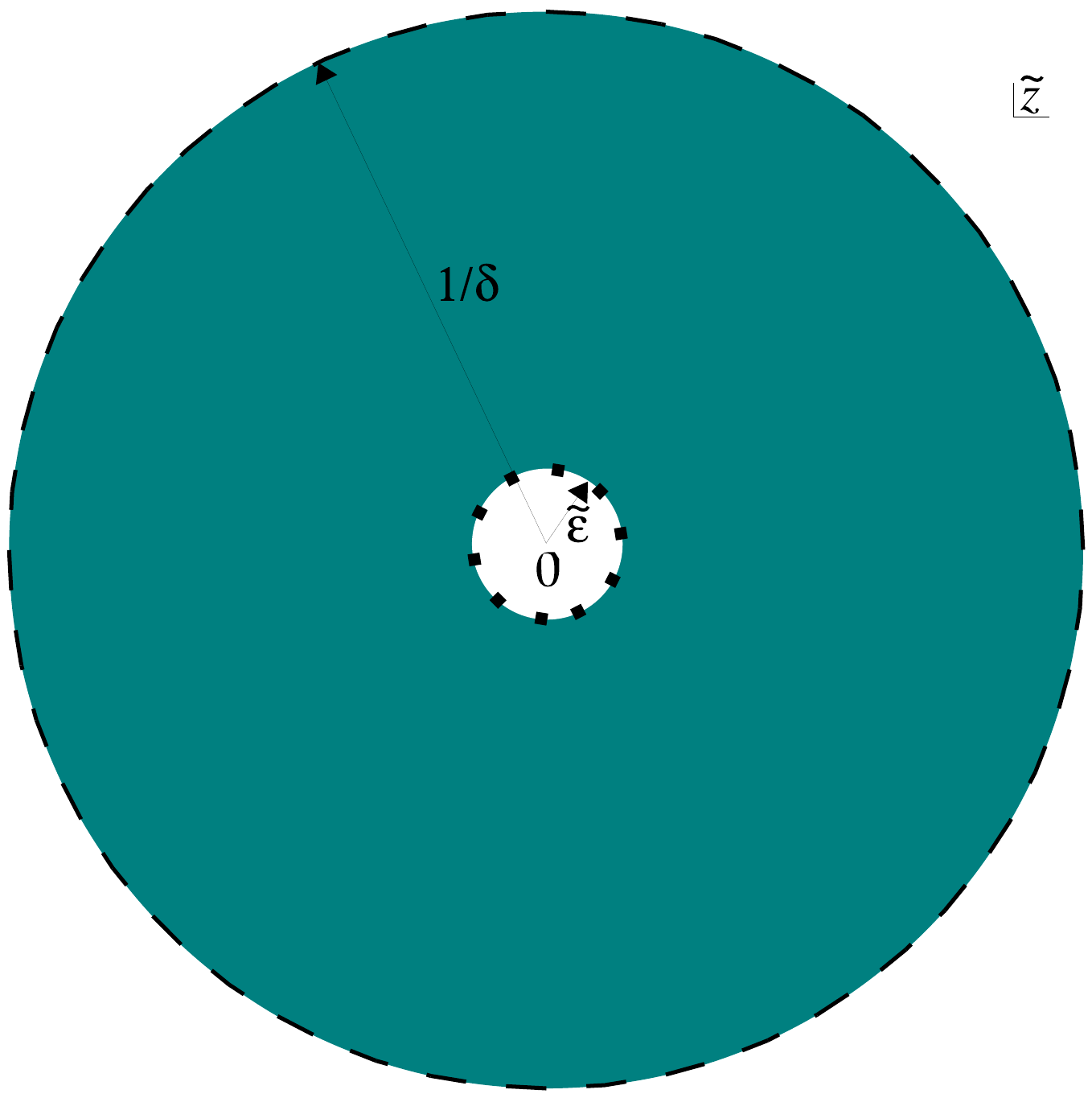}\\
\includegraphics[width=6cm]{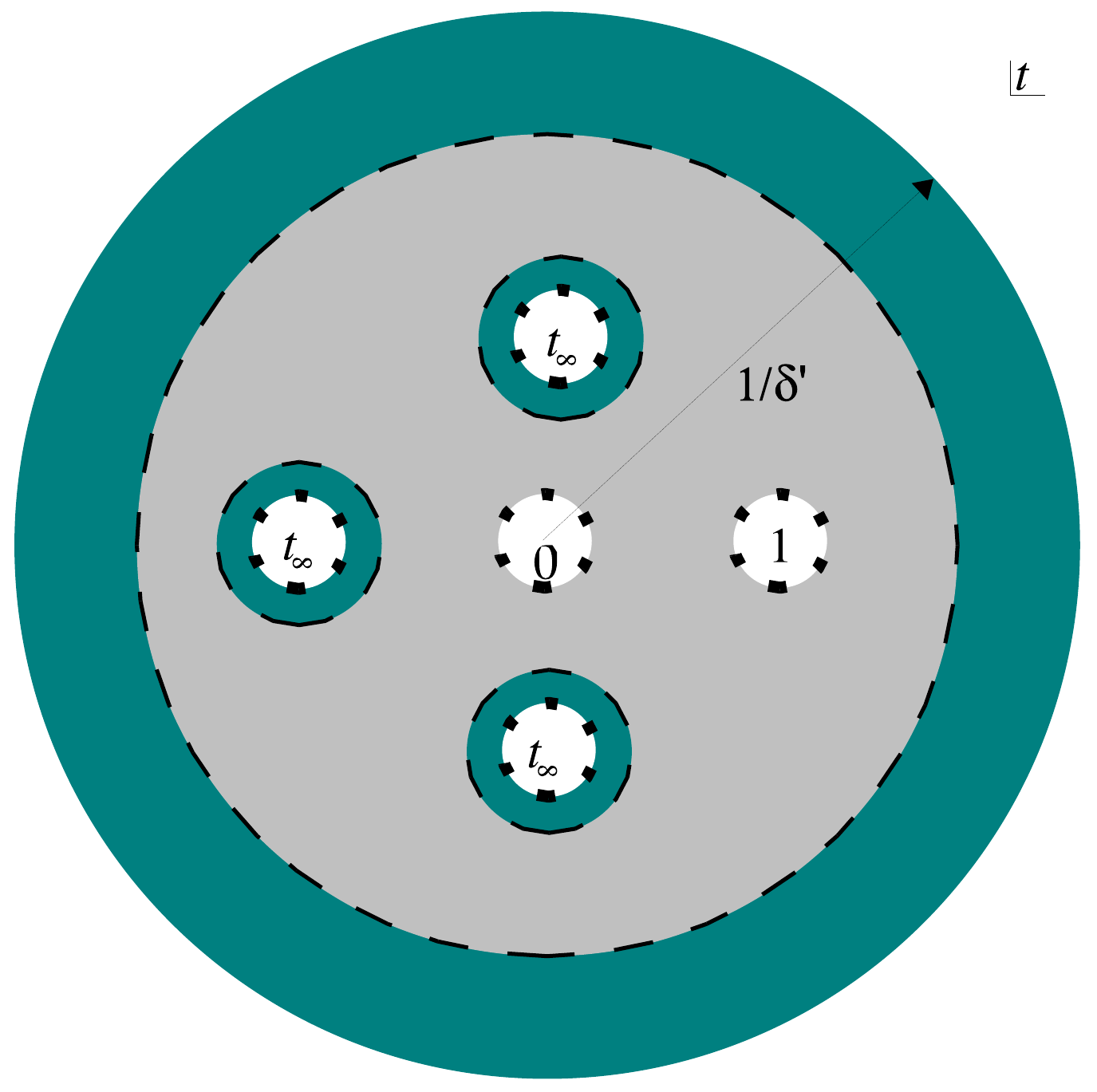}
\includegraphics[width=6cm]{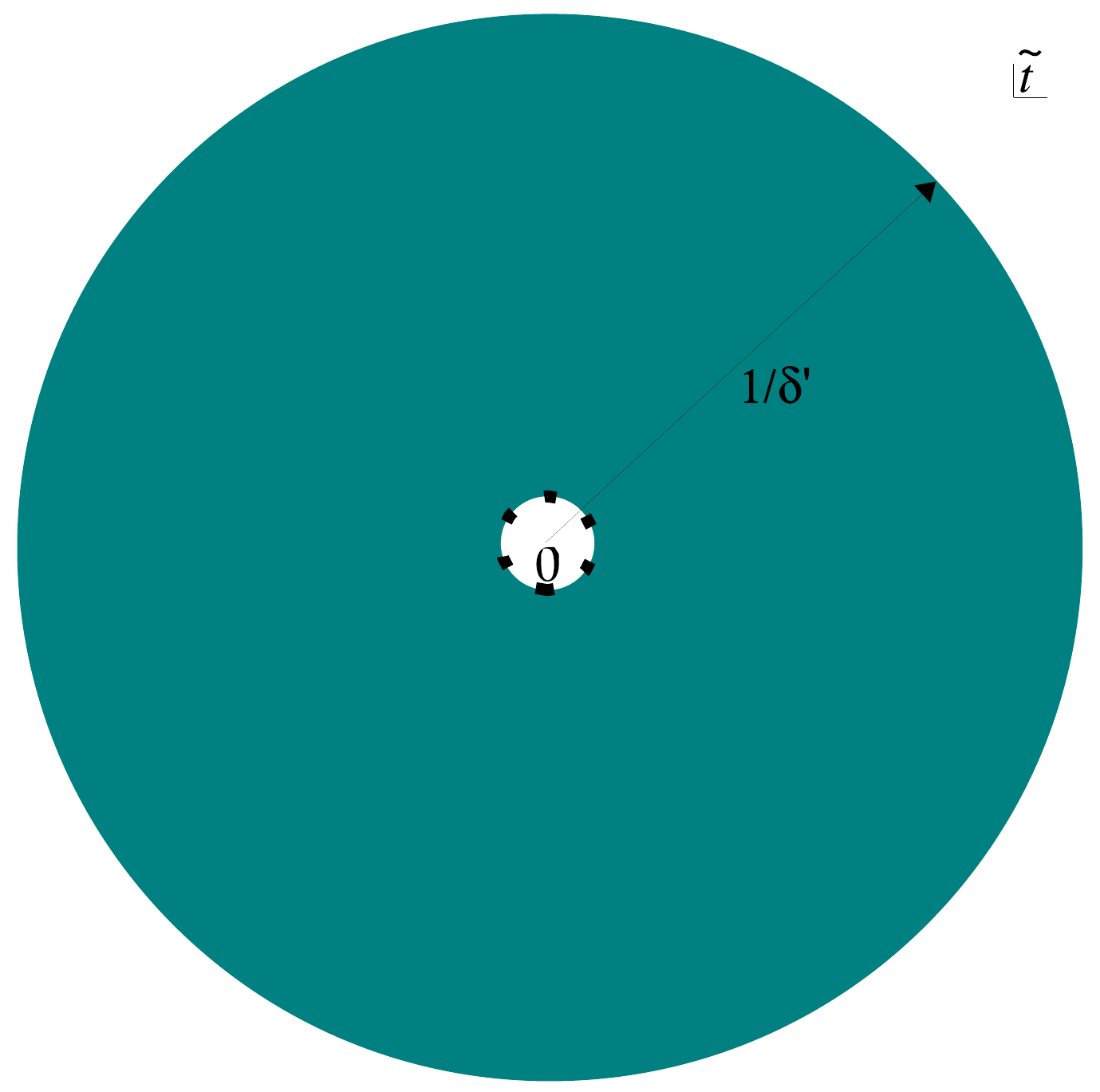}
\end{center}
\caption[The map to the covering space]{The figure depicts the effect
  of the map, and the various regions which contribute to the
  Liouville action for a correlator with two twist operators in the
  finite $z$-plane and some twist operators at infinity. This can be
  thought of as a special case of the correlator needed to compute the
  twist Jacobian factor T in Chapter~\ref{ch:emission}. The relative
  sizes are not accurate. The top two circles show the $z$-sphere,
  while the bottom two show the covering space, or $t$-sphere.}
\label{fig:map-regions}
\end{figure}

Note that there is a ring of curvature at $|z|=1/\delta$. The base
space with holes for the correlator in Chapter~\ref{ch:emission} is
pictured in Figure~\ref{fig:map-regions}.

We map to a covering space with coordinates $t$ and $\bar{t}$ using
the map given in Equation~\eqref{eq:gen-map}. The manifold still has
holes from the images of the twist operators. We close all of the
holes in the manifold by pasting in flat patches. Then, the covering
space manifold, which we call $\Sigma$, becomes compact with the
topology of a sphere. The Riemann--Hurwitz
formula~\eqref{eq:RW-formula} determines the genus of the covering
space. We restrict our attention to $g=0$, but in other cases one can
get higher genus covering spaces, as discussed in~\cite{lm1}. The $N$
twist operators which corresponded to a single hole in the base space
at $\tilde{z}=0$ map to different holes in the covering space, all but
one in the finite-$t$ plane and one at infinity. These holes we also
close with flat patches. The covering space for the case needed in
Section~\ref{sec:T} is pictured in Figure~\ref{fig:map-regions}.

The metric induced on the manifold $\Sigma$ from the base metric
\eqref{Eqn:BaseMetric} is conformally related to the fiducial metric
for the $t$-sphere,
\begin{equation} 
\begin{split}
\widehat{ds}^2 &= \left\{ 
\begin{array}{rl} 
dtd\bar{t} & \qquad |t| <\frac{1}{\delta'}\\
\\
d\tilde{t}d\bar{\tilde{t}} & \qquad |\tilde{t}| <\frac{1}{\delta'}\\
\end{array} \right. \\
t &= \frac{1}{{\delta'}^2\tilde{t}}.
\end{split}  \label{Eqn:FiducialMetric}
\end{equation}
We choose $\delta'$ such that the outermost image of
$|z|=\frac{1}{\delta}$ is contained in the first half of the
$t$-sphere.

The Liouville field $\phi$ is defined by
\begin{equation}
ds^2 = e^\phi\,\widehat{ds}^2.
\end{equation}
We break the manifold $\Sigma$ into a ``regular region,'' which is the
image of the ``first half'' of the $z$-sphere with the holes cut out;
the ``annuli,'' which consist of the images of the second half of the
$z$-sphere with the hole around $\tilde{z}=0$ cut out; the ``second
half'' of the $t$-sphere with the hole cut out; and finally, the flat
patches which we pasted in to make the manifold compact. Using this
categorization, we can write $\phi$ as
\begin{equation}\label{eq:def-phi}
\phi = \begin{cases}
\log \left|\der{z}{t}\right|^2;  & \text{regular region}\\
\log\left|\der{\tilde{z}}{t}\right|^2; & \text{annuli}\\
\log\left|\der{\tilde{z}}{\tilde{t}}\right|^2; & \text{second half of the $t$-sphere}\\
\text{(constant)}; \hspace{30pt}& \text{flat patches}
\end{cases}.
\end{equation}
Note that $\phi$ is a continuous function over $\Sigma$, but its first
derivative is discontinuous across the boundaries between the above
regions.

The goal of this section is to compute the Liouville action,
\begin{equation}
S_L = \frac{c}{96\pi}\int_{\hat{\Sigma}}\drm^2 t\sqrt{\hat{g}}
\left[\pd_\mu\phi\pd_\nu\phi\hat{g}^{\mu\nu} + 2 \hat{R}\phi\right],
\end{equation}
in the limit of small $\veps$, $\tilde{\veps}$, $\delta$, and
$\delta'$. Any regularization dependence drops out once we normalize
the twist operators to have unit 2-point function with themselves at
unit separation. It is important to note that the curvature and metric
in the above equation are on the fiducial manifold $\hat{\Sigma}$, and
not the more complicated curved manifold $\Sigma$. This means, for
instance, the only curvature contribution comes from the ring of
concentrated curvature at $|t|=1/\delta'$ where the two discs of the
fiducial metric are glued together.

There are the following nonzero contributions to the Liouville action:
\begin{enumerate}
\item a contribution $S_L^{(1)}$ from the regular region $R$ 
	(light gray in the figure).
\item a contribution $S_L^{(2)}$ from the annulus $A$, between 
	the outer image of $|z|=\frac{1}{\delta}$ and $|t|=1/\delta'$.
\item a contribution $S_L^{(3)}$ from the ring of curvature 
	concentrated at $|t|=1/\delta'$.
\item a contribution $S_L^{(4)}$ from the second half of the $t$-sphere,
	 but outside of the patched hole around $\tilde{t}=0$,
\item a contribution $S_L^{(5)}$ from the annuli around the finite 
	images of infinity between the image of $|\tilde{z}|=\tilde{\veps}$ 
	and the image of $|z|=|\tilde{z}|=1/\delta$.
\end{enumerate}
The flat patches (and the various boundaries between the regions) do
not contribute. In the flat regions, there is no curvature of the
fiducial metric and $\phi$ is constant; therefore, there can be no
contribution. On their boundaries, $\pd\phi$ is nonzero, but bounded
and is integrated over a set of measure zero. Therefore, the edges of
the filled holes do not contribute either. There can be no curvature
contribution beyond $S_L^{(3)}$, but one may worry about possible
kinetic term contributions from various boundaries; however, these
cannot contribute as long as $\pd\phi$ is bounded, which means that as
long as $\phi$ is continuous, there is no contribution from the
kinetic term on boundaries. Note that the fourth and fifth
contributions are zero for the calculations performed in~\cite{lm1}.

\subsection{Kinetic Contributions to the Liouville Action}

We break the $t$-sphere into the following regions where we need to
compute the contributions to the Liouville action kinetic term:
\begin{enumerate}
\item the regular region, $R$. $R$ is defined as the image of the
  first half of the $z$-sphere under the map.
\item the outer annulus, $A_0$. $A_0$ is bounded by $|z|> 1/\delta$
  and $|t|< 1/\delta'$.
\item the second half of the $t$-sphere, $S$. $S$ is bounded by
  $|\tilde{z}|>\tilde{\veps}$ and $|\tilde{t}|<1/\delta'$.
\item the annuli surround the finite images of infinity, $A_j$. These
  annuli are bounded by the images of $|\tilde{z}|< 1/\delta$ and
  $|\tilde{z}|>\tilde{\veps}$.
\end{enumerate}

We can integrate by parts, by noting
\begin{equation}
\pd_\mu\phi\pd^\mu\phi = \pd_\mu\big(\phi\pd^\mu\phi\big) - \phi\pd_\mu\pd^\mu\phi.
\end{equation}
The second term vanishes for any holomorphic transformation.
Therefore, we are left with
\begin{equation}
S_L = \frac{c}{96\pi}\int_\Sigma\pd_\mu\big(\phi\pd^\mu\phi\big)\drm A
    = \frac{c}{96\pi}\int_{\pd\Sigma} g^{\mu\nu}n_\mu \phi\pd_\nu\phi\, \drm\ell.
\end{equation}
In our case, the metric is given by
\begin{equation}
ds^2 = dzd\bar{z} \qquad\Longrightarrow\qquad
 g_{z\bar{z}} = \frac{1}{2}\qquad g^{z\bar{z}} = 2.
\end{equation}
Thus, we may write
\begin{equation}
S_L = 2\frac{c}{96\pi}\int_{\pd\Sigma} n_{\bar{z}}\phi\pd_z\phi\,\drm\ell
	+ 2\frac{c}{96\pi}\int_{\pd\Sigma} n_z\phi\pd_{\bar{z}}\phi\,\drm\ell
    = I + \bar{I},
\end{equation}
where we recognize that the second term is just the complex conjugate
of the first. Let us suppose that $\pd\Sigma$ has two types of
contours: internal and external. We need to treat these separately,
\begin{equation}
I = \sum I_\mathrm{int} + I_\mathrm{ext}.
\end{equation}
Note that there will only be one external contour. It is convenient at
this point to introduce real coordinates:
\begin{equation}
\begin{split}
z &= x+ iy\qquad z = x -iy\\
ds^2 &= dx^2 + dy^2\\
n_{\bar{z}} &= \frac{n_x + in_y}{2}\qquad n_z = \frac{n_x-in_y}{2}.
\end{split}
\end{equation}
Consider one of the internal boundaries of $\Sigma$. Let's
parameterize the contour (counter-clockwise, of course) as
\begin{equation}\begin{split}
z &= F(s)\qquad \bar{z} = \bar{F}(s)\qquad s\in [0,1)\\
F(s) &= f(s) + ig(s)\qquad F(0) = F(1)\\
\drm \ell &= |F'(s)|\,\drm s,
\end{split}\end{equation}
where $F$ is a holomorphic function and $f$ and $g$ are real-valued
functions.

The normal should point \emph{out of $\Sigma$} and therefore
\emph{into the hole}. We claim, that this means that (since 
the contour is counterclockwise)
\begin{equation}
n_x = -\frac{g'(s)}{\sqrt{f'(s)^2+g'(s)^2}}= -\frac{g'(s)}{|F'(s)|}\qquad
n_y = \frac{f'(s)}{\sqrt{f'(S)^2+g'(s)^2}}= \frac{f'(s)}{|F'(s)|},
\end{equation}
which leads one to conclude
\begin{equation}
n_{\bar{z}} = \frac{1}{2|F'(s)|}\big(-g'(s) + if'(s)\big)
	    = \frac{i}{2}\frac{F'(s)}{|F'(s)|}.
\end{equation}
Plugging in,
\begin{calc}
I_\mathrm{int} &= 2\frac{c}{96\pi}\int_{\mathrm{hole}}\left(\frac{iF'(s)}{2|F'(s)|}\right)
	\,\phi\pd\phi\, |F'(s)|\drm s\\
	&= i\frac{c}{96\pi}\int_\mathrm{hole}\phi\pd\phi\,F'(s)\drm s\\
	&= i\frac{c}{96\pi}\int_\mathrm{hole}\phi\pd\phi\, \drm z,
\end{calc}
where again the contour is counter clockwise.

What changes for the external boundary? For the external boundary, the
normal points \emph{outward and away from the enclosed area}, which
means that $n_x$ and $n_y$ flip signs. Therefore, we pick up an extra
minus sign. We can absorb this extra minus sign by remembering to use
a convention: internal boundaries counter clockwise, external
boundaries clockwise. Then, we always can use the formula
\begin{equation}
  S^\text{kinetic}_L = \frac{c}{96\pi} \left[i\int_{\pd\Sigma} \phi\pd\phi\,\drm z + \mathrm{c.c.}\right],
\end{equation}
where $\mathrm{c.c.}$ indicates the complex conjugate.

\subsubsection{The Regular Region}

Here we compute the contributions to the regular region. We integrate
by parts in which case we need only compute the contributions from the
various curves making up the boundary of $R$.

We begin with the boundaries from the twists, defined by
\begin{equation}
|z-z_j| = \veps \Longrightarrow |t-t_j| 
      = \left(\frac{\veps_j}{|a_i|}\right)^\frac{1}{p_i}.
\end{equation}
The Liouville scalar $\phi$ in the regular region is given by
\begin{equation}
\phi = \log \der{z}{t} + \log\der{\bar{z}}{\bar{t}} 
     = 2\log \left|\der{z}{t}\right|.
\end{equation}
Near the twists, then, we have
\begin{equation}
\phi = 2\log \left(p_i|a_i||t-t_i|^{p_i - 1}\right)\qquad
\pd\phi = \frac{p_i-1}{t-t_i}.
\end{equation}
The contribution from the twist boundaries, then, is given by
\begin{calc}
S_L^i &= -\frac{c}{12}(p_i - 1)\left[\log |a_i| + \log p_i 
       + (p_i -1)\log\left(\frac{\veps_i}{|a_i|}\right)^\frac{1}{p_i}\right]\\
     &= -\frac{c}{12}(p_i - 1)\left[\frac{1}{p_i}\log |a_i| + \log p_i 
       + \frac{p_i - 1}{p_i}\log \veps_i\right].
\end{calc}

Next, we have the outer boundary, defined by
\begin{equation}
|z| = \frac{1}{\delta} \Longrightarrow 
    |t| = \frac{1}{(|b_0|\delta)^\frac{1}{q_0}}.
\end{equation}
The field $\phi$ is given by
\begin{equation}
\phi = 2\log\left( q_0|b_0| |t|^{q_0-1}\right)\qquad
\pd\phi = \frac{q_0 -1}{t}
\end{equation}
The contribution (remembering the extra minus sign for an outer
boundary) is
\begin{equation}
S_L^{\infty, 0} = \frac{c}{12}(q_0 - 1)
   \left[\frac{1}{q_0}\log |b_0| + \log q_0
     - \frac{ q_0 - 1}{q_0} \log \delta\right].
\end{equation}

Finally, we have the boundaries from the finite images of infinity,
defined by
\begin{equation}
|z| = \frac{1}{\delta} \Longrightarrow 
   |t-t_\infty^j| = (|b_j|\delta)^\frac{1}{q_j}
\end{equation}
where
\begin{equation}
\phi = 2 \log\left[\frac{|b_j|q_j}{|t-t_\infty^j|^{q_j + 1}}\right]\qquad
\pd\phi = -\frac{q_j +1}{t-t_\infty^j}.
\end{equation}
Thus, the contribution from these boundaries is given by
\begin{equation}
S_L^{\infty, j} = -\frac{c}{12}(q_j +1)\left[\frac{1}{q_j}\log |b_j|
  - \log q_j + \frac{q_j + 1}{q_j}\log\delta\right].
\end{equation}

\subsubsection{The ``Outer'' Annulus \texorpdfstring{$A_0$}{A0}}

For $A_0$, the field is defined by
\begin{equation}
\phi = \log \der{\tilde{z}}{t} + \log \der{\bar{\tilde{z}}}{\bar{t}}
     = 2\log \left|\der{\tilde{z}}{t}\right|.
\end{equation}
In this region $\phi$, always has the same form:
\begin{equation}
\phi = 2 \log \left(\frac{q_0}{|b_0|\delta^2}|t|^{-(q_0+1)}\right)\qquad
\pd\phi = - \frac{q_0 + 1}{t}.
\end{equation}
The contribution to the Liouville action from the outer annulus, then,
is given by
\begin{equation}
S_L = \frac{c}{12}(q_0 + 1)^2\log |t|\biggr|^\text{outer}_\text{inner},
\end{equation}
where the outer boundary is given by
\begin{equation}
|t| = \frac{1}{\delta'},
\end{equation}
and the inner boundary is given by
\begin{equation}
|z| = \frac{1}{\delta} \Longrightarrow |t| = (\delta|b_0|)^{-\frac{1}{q_0}}.
\end{equation}
Therefore, the contribution is given by
\begin{equation}
S_L = \frac{c}{12}(q_0 + 1)^2\left[\frac{1}{q_0}\log | b_0| 
      + \frac{1}{q_0}\log\delta - \log\delta'\right].
\end{equation}

\subsubsection{The ``Inner'' Annuli, \texorpdfstring{$A_j$}{Aj}}

For the inner annuli, the field is given by
\begin{equation}
\phi = \log\der{\tilde{z}}{t} + \log\der{\bar{\tilde{z}}}{\bar{t}}
     = 2 \log\left|\der{\tilde{z}}{t}\right|.
\end{equation}
From the local behavior of the map, one finds that
\begin{equation}
\phi = 2\log\frac{q_j}{|b_j|\delta^2}|t-t_\infty^j|^{q_j -1}\qquad
\pd\phi = \frac{q_j -1 }{t}.
\end{equation}
As above, the form of the field does not change over the annuli, hence
\begin{equation}
S_L^j = \frac{c}{12}(q_j - 1)^2\log |t-t_\infty^j|\biggr|^{\text{outer}}_\text{inner},
\end{equation}
where the outer boundary is 
\begin{equation}
|\tilde{z}| = \frac{1}{\delta}\Longrightarrow |t-t_\infty^j|
    = (|b_j|\delta)^\frac{1}{q_j}
\end{equation}
and the inner boundary is given by
\begin{equation}
|\tilde{z}| = \tilde{\veps} \Longrightarrow
 |t-t_\infty^j| = (|b_j|\tilde{\veps}\delta^2)^\frac{1}{q_j}.
\end{equation}
Thus, 
\begin{equation}
S_L^j = -\frac{c}{12}\frac{(q_j -1)^2}{q_j}\log(\tilde{\veps}\delta).
\end{equation}

\subsubsection{Second Half of the \texorpdfstring{$t$}{t}-sphere, \texorpdfstring{$S$}{S}}

For the second half of the $t$-sphere, the field $\phi$ is defined via
\begin{equation}
\phi = \log \der{\tilde{z}}{\tilde{t}} 
      + \log \der{\bar{\tilde{z}}}{\bar{\tilde{t}}}
      = 2 \log \left|\der{\tilde{z}}{\tilde{t}}\right|.
\end{equation}
Again, the field has the same form over this region:
\begin{equation}
\phi = 2\log \frac{{\delta'}^{2q_0} q_0}{\delta^2|b_0|}|\tilde{t}|^{q_0 - 1}\qquad
\pd\phi = \frac{q_0 - 1}{\tilde{t}}.
\end{equation}
The contribution to the Liouville action from $S$, is given by
\begin{equation}
S_L = \frac{c}{12} (q_0 - 1)^2\log |\tilde{t}|\biggr|^\text{outer}_\text{inner}.
\end{equation}
The outer boundary is given by 
\begin{equation}
|\tilde{t}|_\text{outer} = 1/\delta',
\end{equation}
while the inner boundary is given by
\begin{equation}
|\tilde{z}| = \tilde{\veps} \Longrightarrow
|\tilde{t}|_\text{inner} = \frac{1}{{\delta'}^2}
   \big(\delta^2\tilde{\veps}|b_0|\big)^\frac{1}{q_0}.
\end{equation}
Thus, the contribution to the Liouville action from the second half of
the $t$-sphere is given by
\begin{equation}
S_L^{S} = -\frac{c}{12}\frac{(q_0-1)^2}{q_0}\left[\log|b_0|
  + \log\tilde{\veps} + 2\log\delta - q_0\log\delta'\right]
\end{equation}

\subsubsection{The Total Kinetic Contribution}

Adding up the sundry contributions to the kinetic part of the
Liouville action, one finds that the total contribution to the
Liouville action from the kinetic term is
\begin{equation}\begin{split}
S_L^\text{kinetic} &= -\frac{c}{12}\Biggr\{
\sum_i \frac{p_i - 1}{p_i}\log |a_i|
+\sum_j \frac{q_j +1}{q_j}\log |b_j|
-\frac{5 q_0 -1}{q_0} \log |b_0|\\
&\qquad+\sum_i(p_i - 1)\log p_i
- \sum_j (q_j +1)\log q_j
-(q_0 -1)\log q_0\\
&\qquad + \sum_i \frac{(p_i -1)^2}{p_i}\log \veps_i
  + \left[\sum_j \frac{(q_j -1)^2}{q_j} + \frac{(q_0-1)^2}{q_0}\right]
             \log\tilde{\veps}\\
&\qquad + 2\left[\sum_j \frac{q_j^2 + 1}{q_j} 
         + \frac{ q_0^2 -4 q_0 + 1}{q_0}\right]\log\delta\\
&\qquad + 4 q_0 \log\delta'\Biggr\}
\end{split}\end{equation}

\subsection{The Curvature Contribution}

All of the curvature in the fiducial metric is concentrated on a ring
at $|t|=1/\delta'$. Since we have restricted our consideration to
correlation functions that are on a sphere, we may write
\begin{equation}
S_L^\text{curvature} = \frac{c}{48\pi}\int\drm^2 t\, \hat{R}\phi
  = \frac{c}{6}\phi\bigg|_{|t|=\frac{1}{\delta'}}.
\end{equation}
The value the field takes is
\begin{equation}
\phi = 2 \log \left(\frac{q_0{\delta'}^{q_0+1}}{|b_0|\delta^2}\right),
\end{equation}
and thus
\begin{calc}
S_L^\text{curvature} &= \frac{c}{3}
   \log \left(\frac{q_0{\delta'}^{q_0+1}}{|b_0|\delta^2}\right)\\
   &= -\frac{c}{12}\bigg[4\log |b_0| - 4 \log q_0 + 8\log \delta
     - 4(q_0 +1)\log\delta'\bigg]
\end{calc}

Adding this to the kinetic contribution, one finds
\begin{equation}\begin{split}
S_L^\text{total} &= -\frac{c}{12}\Biggr\{
\sum_i \frac{p_i - 1}{p_i}\log |a_i|
+\sum_j \frac{q_j +1}{q_j}\log |b_j|
-\frac{ q_0 -1}{q_0} \log |b_0|\\
&\qquad+\sum_i(p_i - 1)\log p_i
- \sum_j (q_j +1)\log q_j
-(q_0 +3)\log q_0\\
&\qquad + \sum_i \frac{(p_i -1)^2}{p_i}\log \veps_i
  + \left[\sum_j \frac{(q_j -1)^2}{q_j} + \frac{(q_0-1)^2}{q_0}\right]
             \log\tilde{\veps}\\
&\qquad + 2\left[\sum_j \frac{q_j^2 + 1}{q_j} 
         + \frac{ q_0^2 + 1}{q_0}\right]\log\delta\\
&\qquad -4\log\delta'\Biggr\}
\end{split}\end{equation}

\section{The Unnormalized Correlator}

The correlator of twists is defined by
\begin{equation}
Z = \vev{\cdot} = \frac{\vev{\cdot}_{\delta,\veps}}
     {\braket{\varnothing|\varnothing}_{\delta,\veps}}
  = e^{S_L}\frac{Z'_{\delta'}}{(Z_\delta)^s},
\end{equation}
where $s$ is the number of sheets involved in the correlator and
$\cdot$ is the twist operators in the correlator.  $Z_\delta$ is the
partition function on the $z$-sphere, and $Z'_{\delta'}$ is the
partition function on the $t$-sphere with the fiducial metric. One can
extract the dependence on the cutoff from the partition function and
write
\begin{equation}
Z_\delta = Q \delta^{-\frac{c}{3}} \qquad
Z'_{\delta'} = Q' {\delta'}^{-\frac{c}{3}}.
\end{equation}
The cutoff-independent factors $Q$ and $Q'$ depend on the precise
shape that the $z$-sphere and the covering space have, respectively.
We have chosen the metric on the base space and the covering space in
such a way that $Q=Q'$; however, it will be instructive to leave
$Q'$'s in for now.

Thus, the unnormalized correlator is given by
\begin{equation}
Z = e^{S_L}\frac{Q'{\delta'}^{-\frac{c}{3}}}{Q^s \delta^{-s\frac{c}{3}}}.
\end{equation}
One can immediately see that the $\delta'$-dependence cancels out, as
it should.

\section{Normalizing the Twist Operators}

We normalize the twist operators by demanding that they have unit
correlator with themselves at unit separation in the $z$-plane.  For
the two-point function
\begin{equation}
\vev{\sigma_n(0)\sigma_n(z_2)},
\end{equation}
the map is of the form
\begin{equation}
z = z_2\frac{t^n}{t^n - (t-1)^n}
\end{equation}
which behaves as
\begin{equation}\begin{aligned}
z &\approx (-1)^{n+1} z_2 t^n &  z&\approx 0, t\approx 0\\
z-z_2 &\approx z_2(t-1)^n & z&\approx z_2, t\approx 1\\
z &\approx \frac{z_2}{n}t & z&\to\infty, t\to\infty\\
z &\approx -\frac{z_2}{n}\frac{\alpha_j}{(1-\alpha_j)^2}\frac{1}{t-t_\infty^j}
  & z&\to\infty, t\to t_\infty^j.
\end{aligned}\end{equation}
The finite images of infinite satisfy
\begin{equation}
t_\infty^j = \frac{1}{1-\alpha_j}\qquad
(\alpha_j)^n = 1,
\end{equation}
where the $\alpha_j$ are the $n$ roots of unity, and $\alpha_0=1$.
Therefore, $j$ runs from $0$ to $n-1$.

For the two-point function, we identify the important parameters
\begin{equation}\begin{aligned}
a_1 &= (-1)^{n+1}z_2 & p_1&=n\\
a_2 &= z_2 & p_2&=n\\
b_0 &= \frac{z_2}{n} & q_0 &= 1\\
b_j &= \frac{z_2}{n}\frac{\alpha_j}{(1-\alpha_j)^2} & q_j &= 1.
\end{aligned}\end{equation}

The Liouville action is given by
\begin{smalleq}\begin{equation}
S_L^\text{total} = -\frac{c}{12}\Biggr\{2\frac{n-1}{n}\log|z_2|
                  + 2\sum_{j=1}^{n-1}\log |b_j| + 2(n-1)\log n
                  + \frac{(n-1)^2}{n}\log\veps_1\veps_2
                  + 4n\log\delta
                  - 4\log\delta'\Biggr\}.
\end{equation}\end{smalleq}
We compute
\begin{calc}
\sum_{j=1}^{n-1}\log |b_j| &= (n-1)\log\frac{z_2}{n} 
       + \log\left|\prod_{j=1}^{n-1}\frac{\alpha_j}{(1-\alpha_j)^2}\right|\\
       &= (n-1)\log\frac{z_2}{n} - \log n^2\\
       &= (n-1)\log z_2 - (n+1)\log n,
\end{calc}
and therefore
\begin{equation}
S_L^\text{total} = -\frac{c}{12}\Biggr\{
 2(n-1)\frac{n+1}{n}\log |z_2| - 4\log n
 + \frac{(n-1)^2}{n}\log \veps_1\veps_2
 + 4n\log\delta - 4\log\delta'\Biggr\}.
\end{equation}

The unnormalized two point correlator, then, is given by
\begin{equation}
\vev{\sigma_n^{\veps_1}(0)\sigma_n^{\veps_2}}_\delta
 = \frac{Q'}{Q^n}|z_2|^{-\frac{c}{6}\left(n - \frac{1}{n}\right)}
   n ^{\frac{c}{3}}(\veps_1\veps_2)^{-\frac{c}{12}\frac{(n-1)^2}{n}}.
\end{equation}
Note that since $s=n$ for the correlator, the $\delta$-dependence
cancelled out as it should. Therefore, the normalized twist operator
is defined by
\begin{equation}
\sigma_n = \sqrt{\frac{Q^n}{Q'}}n^{-\frac{c}{6}}
  \veps^{\frac{c}{12}\frac{(n-1)^2}{n}}\sigma_n^{\veps}
\end{equation}

\section{The General Correlator}

We are now ready to put all of the pieces together to compute the
normalized correlator of twists in Equation~\eqref{eq:gen-corr} for
spherical genus.

The correlator of the normalized twists is given by
\begin{calc}
Z^\text{norm.} &= \frac{Q^{\frac{1}{2}\sum_{i=1}^M p_i 
    + \frac{1}{2}\sum_{j=0}^{N-1} q_j}}{{Q'}^{\frac{M+N}{2}}}
   \left(\prod_{i=1}^M p_i\prod_{j=0}^{N-1}q_j\right)^{-\frac{c}{6}}
   \prod_{i=1}^M \veps_i^{\frac{c}{12} \frac{(p_i -1)^2}{p_i}}
   \tilde{\veps}^{\frac{c}{12}\sum_{j=0}^{N-1} \frac{(q_j-1)^2}{q_j}}
   \,Z\\
   &=\frac{Q^{\frac{1}{2}\sum_{i=1}^M p_i 
    + \frac{1}{2}\sum_{j=0}^{N-1} q_j - s}}{{Q'}^{\frac{M+N}{2}-1}}\hspace{-1pt}
   \left(\prod_{i=1}^M p_i\prod_{j=0}^{N-1}q_j\right)^{-\frac{c}{6}}
   \prod_{i=1}^M \veps_i^{\frac{c}{12} \frac{(p_i -1)^2}{p_i}}
   \tilde{\veps}^{\frac{c}{12}\sum_{j=0}^{N-1} \frac{(q_j-1)^2}{q_j}}
   \,\left(\frac{\delta^s}{\delta'}\right)^\frac{c}{3}e^{S_L}\\
   &= \frac{Q^{\frac{M+N}{2}-1+g}}{{Q'}^{\frac{M+N}{2} -1}}
   \left(\prod_{i=1}^M p_i\prod_{j=0}^{N-1}q_j\right)^{-\frac{c}{6}}
   \prod_{i=1}^M \veps_i^{\frac{c}{12} \frac{(p_i -1)^2}{p_i}}
   \tilde{\veps}^{\frac{c}{12}\sum_{j=0}^{N-1} \frac{(q_j-1)^2}{q_j}}
   \,\left(\frac{\delta^s}{\delta'}\right)^\frac{c}{3}e^{S_L}\\
   &= Q^g\left(\frac{Q}{Q'}\right)^{\frac{M+N}{2}-1}
   \left(\prod_{i=1}^M p_i\prod_{j=0}^{N-1}q_j\right)^{-\frac{c}{6}}
   \prod_{i=1}^M \veps_i^{\frac{c}{12} \frac{(p_i -1)^2}{p_i}}
   \tilde{\veps}^{\frac{c}{12}\sum_{j=0}^{N-1} \frac{(q_j-1)^2}{q_j}}
   \,\left(\frac{\delta^s}{\delta'}\right)^\frac{c}{3}e^{S_L}
\end{calc}
We see that as long as we restrict our attention to correlators whose
covering space is a sphere, and we carefully ensure that $Q=Q'$, then
we never have to compute the path integral. This is exactly what we
are doing, and thus
\begin{equation}
Z^\text{norm.} = 
   \left(\prod_{i=1}^M p_i\prod_{j=0}^{N-1}q_j\right)^{-\frac{c}{6}}
   \prod_{i=1}^M \veps_i^{\frac{c}{12} \frac{(p_i -1)^2}{p_i}}
   \tilde{\veps}^{\frac{c}{12}\sum_{j=0}^{N-1} \frac{(q_j-1)^2}{q_j}}
   \,\left(\frac{\delta^s}{\delta'}\right)^\frac{c}{3}e^{S_L}.
\end{equation}
The $\delta'$, $\veps_j$, and $\tilde{\veps}$ dependence all cancels
out. The power of $\delta$ is
\begin{calc}
\delta: \qquad &\frac{c}{3}s - \frac{c}{6}\sum_{j=0}^{F}\frac{q_j^2+1}{q_j}\\
  &= \frac{c}{3}s
    - \frac{c}{6}s - \frac{c}{6}\sum_{j=0}^F\frac{1}{q_j}\\
  &= \frac{c}{6}\left( s - \sum_{j=1}^F \frac{1}{q_j}\right)\\
  &= \frac{c}{6}\sum_{j=0}^F\left(q_j - \frac{1}{q_j}\right)\\
  &= 4\sum_{j=0}^{N-1}\Delta_{q_j}
\end{calc}
If there are no twists at infinity then the power vanishes,
appropriately. Typically, one divides this dependence out when one
inserts twists at infinity, but we leave them explicitly here, for
now. Let us call the total weight of all the operators at infinity
\begin{equation}
\Delta_\infty = \sum_{j=0}^{N-1}\Delta_{q_j}
\end{equation}

We can rewrite the correlator in the slightly nicer form:
\begin{smalleq}\begin{calc}\label{eq:gen-corr-answer}
Z^\text{norm.} &= \delta^{4\Delta_\infty}
  \left(\prod_{i=1}^M p_i\prod_{j=0}^{N-1}q_j\right)^{-\frac{c}{6}}
  \,\left(\prod_{i=1}^M |a_i|^\frac{1}{p_i}p_i\right)^{-\frac{c}{12}(p_i -1)}
  \prod_{j=1}^F\left(\frac{|b_j|^{\frac{1}{q_j}}}{q_j}\right)^{-\frac{c}{12}(q_j +1)}\hspace{-2pt}
  |b_0|^{\frac{c}{12}\frac{q_0-1}{q_0}}q_0^{\frac{c}{12}(q_0+3)}\\
  &= \delta^{4\Delta_\infty}\hspace{-4pt}
  \left(\prod_{i=1}^M p_i^{-\frac{c}{12}(p_i + 1)}\right)\hspace{-5pt}
  \left(\smashoperator{\prod_{j=1}^{N-1}} q_j^{\frac{c}{12}(q_j -1)}\right)\hspace{-4pt}
  q_0^{\frac{c}{12}(q_0 + 1)}\hspace{-4pt}
  \left(\prod_{i=1}^M |a_i|^{-\frac{c}{12}\frac{p_i -1}{p_i}}\right)\hspace{-5pt}
  \left(\prod_{j=1}^{F} |b_j|^{-\frac{c}{12}\frac{q_j+1}{q_j}}\right)\hspace{-4pt}
       |b_0|^{\frac{c}{12}\frac{q_0 -1}{q_0}}\\
  &=\delta^{4\Delta_\infty}
  \left(\prod_{i=1}^M p_i^{-\frac{c}{12}(p_i + 1)}\right)
  \left(\prod_{j=0}^{N-1} q_j^{\frac{c}{12}(q_j -1)}\right)
  \left(\prod_{i=1}^M |a_i|^{-\frac{c}{12}\frac{p_i -1}{p_i}}\right)
  \left(\prod_{j=0}^{F} |b_j|^{-\frac{c}{12}\frac{q_j+1}{q_j}}\right)
       |b_0|^\frac{c}{6}q_0^\frac{c}{6}\\
\end{calc}\end{smalleq}
It would be nice if we could say something about the $a_i$ and $b_j$.
For instance, we should think about what happens if we multiply the
whole map by a constant scale factor. If one wants to consider a case
when there are no twists at infinity, then one should set the $q_j =
1$ and $F=s$ to obtain the simpler formula
\begin{equation}
  Z^\text{norm.} =   \left(\prod_{i=1}^M p_i^{-\frac{c}{12}(p_i + 1)}\right)
  \left(\prod_{i=1}^M |a_i|^{-\frac{c}{12}\frac{p_i -1}{p_i}}\right)
  \left(\prod_{j=0}^{s} |b_j|^{-\frac{c}{6}}\right)
  |b_0|^\frac{c}{6}.
\end{equation}
Note that the $b_0$ dependence cancels out.

\subsection{Checking the Correlator}

What happens if we multiply the map by a complex prefactor $\alpha$? 
This corresponds to scaling and rotating the positions of all of the
twist operators, but preserving the origin. If the map is of the form
\begin{equation}
z = f(t),
\end{equation}
for some holomorphic function $f$, then we consider the transformation
\begin{equation}
z = f(t) \qquad \mapsto \qquad z' = \alpha f(t).
\end{equation}
The only effect is rescaling all of the $a_i$ and $b_j$ by
$\alpha$. Thus, we observe that
\begin{equation}
Z^\text{norm.} \mapsto Z' = |\alpha|^D Z^\text{norm.},
\end{equation}
where
\begin{calc}
D &=  -\frac{c}{12}\sum_{i=1}^M \frac{p_i-1}{p_i} -\frac{c}{12}\sum_{j=0}^F \frac{q_j + 1}{q_j}
	+ \frac{c}{6}\\
  &= \frac{c}{12}\sum_{i=1}^M \frac{1}{p_i} - \frac{c}{12}\sum_{j=0}^F \frac{1}{q_j} 
	- \frac{c}{12}(M+F-2)\\
  &= 2\Delta_\infty - 2\sum_{i=1}^M \Delta_{p_i} - \frac{c}{12}(s + F-2) 
	+ \frac{c}{12}\sum_{i=1}^M (p_i-1)\\
  &= 2\Delta_\infty - 2\sum_{i=1}^M \Delta_{p_i}.
\end{calc}
Therefore, we see that the effect of scaling by $\alpha$ is
\begin{equation}
Z' = \frac{1}{|\alpha|^{2\sum\Delta_{p_i}-2\Delta_\infty}}Z,
\end{equation}
as it should be. This is a good check that the formula is correct.

\chapter{The Exponential Ansatz}\label{ap:exp-ansatz}

In this appendix we check that the exponential ansatz in
Equation~\eqref{pfive} for the bosonic excitations is indeed correct;
the fermionic case works in a similar way.

The bosonic fields completely decouple from the fermionic fields.
Further, the set $\{\alpha_{++, m}, \alpha_{--, n}\}$ decouples from
the set $\{ \alpha_{+-,m},\alpha_{-+, n}\}$. Thus in what follows we
write only the modes $\{\alpha_{++, m}, \alpha_{--, n}\}$.

We have from the basic relation (\ref{master})
\begin{equation}
\bra{0_{R,-}}
\Big(\alpha_{++, n_1}\alpha_{--, n_2}\dots \Big)
  \sigma^+_2(w_0)\ket{0^-_R}^{(1)}\ket{0^-_R}^{(2)}
 ={}_t\bra{0}\Big (\alpha'_{++, n_1}\alpha'_{--, n_2}\dots \Big )|0\rangle_t
\label{apptwo}
\end{equation}
where the operators $\alpha'$ arise from following the various
coordinate changes and spectral flows that bring us from the original
operators $\alpha$ on the cylinder to operators on the $t$ plane (with
the state $|0\rangle_t$ at $t=0$).  If we can understand all
amplitudes of this type, then we will have a complete understanding of
the state $\sigma_2^+(w_0)|0^-_R\rangle^{(1)}|0^-_R\rangle^{(2)}$. Our
ansatz for this state is 
\begin{equation}
\sigma_2^+(w_0)|0^-_R\rangle^{(1)}|0^-_R\rangle^{(2)}
     =e^{-\sum_{m>0,n>0}\gamma^B_{mn}\alpha_{++,m}\alpha_{--,n}}|0^-_R\rangle
\label{appfive}
\end{equation}

The initial operators $\alpha$ are given by eq.(\ref{qaone}), and
their map to the final form in the $t$ plane is given by
eq.(\ref{qatwo}). We can expand the latter form in terms of natural
modes on the $t$ plane (\ref{qathree}) 
\begin{calc}
\alpha_{A\dot A, n}&\to \int_{t=\infty} \frac{\drm t}{2\pi i}
                         \pd_t X_{A\dot A}(t) (z_0+t^2)^{\frac{n}{2}}\\
&= \int_{t=\infty}\frac{\drm t}{2\pi i} \pd_t X_{A\dot A}(t)
        \sum_{k\ge 0} \choose{\frac{n}{2}}{k} z_0^k t^{n-2k} \\
&= \sum_{k\ge 0} \choose{\frac{n}{2}}{k} z_0^k \tilde{\alpha}_{A\dot A, n-2k}
\label{appsix} 
\end{calc} 
All we need to know is that this is a linear relation
\begin{equation}\label{appthree}
\alpha_{A\dot A, n}=\sum_{p=-\infty}^\infty B_{n,p}\tilde{\alpha}_{A\dot A,p}
\end{equation}
with some constant coefficients $B_{np}$. Since the relation
(\ref{appthree}) is linear in the field operators, we will have as
many operators inserted between the parenthesis $( )$ on the RHS of
(\ref{apptwo}) as on the LHS. But on the RHS we just have these mode
operators sandwiched between the $t$ plane vacuum state. Thus the
amplitude will be evaluated by Wick contractions between these
operators. From this fact we can immediately note two things: we must
have an even number of insertions, and there must be an equal number
of $\alpha_{++}$ and $\alpha_{--}$ modes.

Let us start with the simplest case: two operator insertions, which is
computation we encountered in finding $\gamma^B_{mn}$. We have (with
$n_1>0, n_2>0$) 
\begin{equation} \label{appone}
\langle 0_{R,-}|\Big (\alpha_{++, n_1}\alpha_{--, n_2}\Big)
     \sigma^+_2(w_0)|0^-_R\rangle^{(1)}|0^-_R\rangle^{(2)}
   ={}_t\langle 0 |\Big (\alpha'_{++, n_1}\alpha'_{--, n_2}\Big )
                  |0\rangle_t
\end{equation}
With the ansatz (\ref{appfive}), the LHS gives
\begin{multline}
\langle 0_{R,-}|\Big(\alpha_{++, n_1}\alpha_{--, n_2}\Big)
\sigma^+_2(w_0)|0^-_R\rangle^{(1)}\otimes |0^-_R\rangle^{(2)}\\
=\langle 0_{R,-}|\Big(\alpha_{++, n_1}\alpha_{--, n_2}\Big)
e^{-\sum_{m_1>0, m_2>0}\gamma^B_{m_1m_2}\alpha_{++,-m_1}\alpha_{--,-m_2}}
   |0^-_R\rangle
\end{multline}
The contribution to this amplitude comes from expanding the
exponential to first order giving 
\begin{equation} 
\langle 0_{R,-}|\Big(\alpha_{++, n_1}\alpha_{--, n_2}\Big)(-1)\hspace{-10pt}
\sum_{m_1>0,  m_2>0}\hspace{-10pt}
   \gamma^B_{m_1m_2}\alpha_{++,-m_1}\alpha_{--,-m_2}|0^-_R\rangle
=(-1) n_1n_2 \gamma^B_{n_2n_1} 
\end{equation}
The RHS of~\eqref{appone} gives
\begin{equation}
{}_t\langle 0 |\sum_{p_1, p_2} B_{n_1,p_1}B_{n_2,p_2} 
\tilde{\alpha}_{++,p_1}\tilde{\alpha}_{--,p_2}|0\rangle_t=\sum_{p_1>0} p_1
B_{n_1,p_1}B_{n_2, -p_1} 
\end{equation}
Thus we get the relation 
\begin{equation} \label{appseven}
(-1)n_1n_2\gamma^B_{n_2n_1}=\sum_{p_1>0} p_1 B_{n_1,p_1}B_{n_2, -p_1}
\end{equation}
which gives the $\gamma^B_{mn}$ in~\eqref{gammaB} when we
use~\eqref{appsix}.

Let us now consider the next simplest case: four operator insertions.
The LHS of (\ref{apptwo}) gives ($n_1, n_2, n_3, n_4>0$) 
\begin{equation}
\langle 0_{R,-}|
\Big(\alpha_{++, n_1}\alpha_{--, n_2}\alpha_{++, n_3}\alpha_{--, n_4}\Big)
   \sigma^+_2(w_0)|0^-_R\rangle^{(1)}|0^-_R\rangle^{(2)} 
\end{equation} 
We must now expand the exponential in the ansatz (\ref{appfive}) to
second order, getting for the LHS of (\ref{apptwo})
\begin{multline}
{}^{(2)}\langle 0_{R,-}|
\Big(\alpha_{++, n_1}\alpha_{--, n_2}\alpha_{++, n_3}\alpha_{--, n_4}\Big) 
\frac{1}{2!}(-1)^2\\
\times
\Big(\sum_{m_1>0, m_2>0}\gamma^B_{m_1m_2}\alpha_{++,-m_1}\alpha_{--,-m_2}\Big)
\Big(\sum_{m_3>0, m_4>0}\gamma^B_{m_3m_4}\alpha_{++,-m_3}\alpha_{--,-m_4}\Big)
    |0^-_R\rangle
\end{multline}
This gives
\begin{equation}
\frac{1}{2!}(-1)^2(2!)n_1n_2n_3n_4
\Big (\gamma^B_{n_2n_1}\gamma^B_{n_4n_3} 
      + \gamma^B_{n_4n_1}\gamma^B_{n_2n_3}\Big)
\label{appeight}
\end{equation}
where the factor $(2!)$ comes from the fact that the set $\alpha_{++,
  n_1}\alpha_{--,n_2}$ can contract with the operators from either of
the two $\gamma^B$ factors.

The RHS of (\ref{apptwo}) gives
\begin{multline}
{}_t\langle 0 |\sum_{p_1, p_2,p_3,p_4} 
               B_{n_1,p_1}B_{n_2,p_2}B_{n_3,p_3}B_{n_4,p_4}
 \tilde{\alpha}_{++,p_1}\tilde{\alpha}_{--,p_2}
 \tilde{\alpha}_{++,p_3}\tilde{\alpha}_{--,p_4}|0\rangle_t\\
= \Big(\sum_{p_1>0} p_1 B_{n_1,p_1}B_{n_2, -p_1}\Big)
  \Big(\sum_{p_3>0} p_3 B_{n_3,p_3}B_{n_4, -p_3}\Big)\\
 +\Big(\sum_{p_1>0} p_1 B_{n_1,p_1}B_{n_4, -p_1}\Big)
  \Big(\sum_{p_3>0} p_3 B_{n_3,p_3}B_{n_2, -p_3}\Big)
\end{multline}
On using (\ref{appseven}) this gives
\begin{equation}
n_1n_2n_3n_4\Big(\gamma^B_{n_2n_1}\gamma^B_{n_4n_3} 
   + \gamma^B_{n_4n_1}\gamma^B_{n_2n_3} \Big)
\end{equation}
which agrees with (\ref{appeight}).

Thus we have verified the ansatz to order four in the bosonic field
operators. Proceeding in this way we can verify the complete
exponential ansatz.

The fermionic case is similar. Modes on the cylinder map linearly to
the modes for the case where we are on the $t$ plane and we have the
NS vacuum at $t=0$. On this $t$ plane the modes must appear in pairs
to allow the amplitude to be nonvanishing, thus we must have an even
number of modes in each term in the ansatz. Our ansatz allows all
modes that are nonvanishing on the chosen vacuum state; thus the
situation is similar to the bosonic case where we allowed all negative
index bosonic operators in the ansatz. Thus the fermionic part of the
ansatz can be verified in the same way as the bosonic part.

\chapter{Some Useful Series}\label{ap:series}

In this appendix, we collect some series we find useful in the main
text. Some of the identities can be proved using hypergeometric
series, while others we know of no conclusive proof; however, we are
confident they are correct after numerical study.  These series arise
when considering a single boson and fermion.
\begin{align}\label{eq:id-1}
\sum_{k\,\text{odd}^+}\frac{1}{(n-\frac{k}{2})(k+l)}\frac{\Gamma(\frac{k}{2}+1)}
                                                         {\Gamma(\frac{k+1}{2})} 
 &= \frac{\pi l\Gamma(\frac{l+1}{2})}{4\Gamma(\frac{l}{2}+1)}\frac{1}{n+\frac{l}{2}}
      \left(\frac{\Gamma(\frac{l}{2})\Gamma(-n+\frac{1}{2})}{\Gamma(\frac{l+1}{2})\Gamma(-n)} 
              - 1\right)\\
\sum_{k\,\text{odd}^+} \frac{1}{k(p+k)(n-\frac{k}{2})}
         \frac{\Gamma(\frac{k}{2}+1)}{\Gamma(\frac{k+1}{2})}
 &= -\frac{\pi\Gamma(\frac{p+1}{2})}{4\Gamma(\frac{p}{2}+1)}\frac{1}{n+\frac{p}{2}}
    \left(
   \frac{\Gamma(\frac{p}{2}+1)\Gamma(-n+\frac{1}{2})}{\Gamma(\frac{p+1}{2})\Gamma(1-n)} -1\right)
\end{align}

These series arise when considering a single fermion:
\begin{subequations}
\begin{align}
\sum_{k\,\text{odd}^+}\frac{z_0^{n-\frac{k}{2}}\gamma^F_{pk}}{n-\frac{k}{2}}
 &=  \frac{z_0^{n+\frac{p}{2}}}{2(n+\frac{p}{2})}
  \left(\frac{\Gamma(\frac{p}{2}+1)\Gamma(-n+\frac{1}{2})}{\Gamma(\frac{p+1}{2})\Gamma(-n+1)} 
                   - 1\right)\\
\sum_{k\,\text{odd}^+}\frac{z_0^{n-\frac{k}{2}}\gamma^F_{kp}}{n-\frac{k}{2}}
 &=-\frac{z_0^{n+\frac{p}{2}}}{2(n+\frac{p}{2})}
  \left(\frac{\Gamma(\frac{p}{2})\Gamma(-n+\frac{1}{2})}{\Gamma(\frac{p+1}{2})\Gamma(-n)} 
                   - 1\right).\label{eq:gammaF-sum-1}
\end{align}
\end{subequations}

Some other useful series are
\begin{subequations}\begin{align}
\sum_{k\,\text{odd}^+}\frac{\Gamma(\frac{k}{2})}{\Gamma(\frac{k+1}{2})(m+\frac{k}{2})}
 &= \pi\frac{\Gamma(m+\frac{1}{2})}{\Gamma(m+1)}\\
\sum_{l\,\text{odd}^+}\frac{l}{(m-\frac{l}{2})(n-\frac{l}{2})}
      \frac{\Gamma(\frac{l}{2})}{\Gamma(\frac{l+1}{2})}
 &= 2\pi^2 m \frac{\Gamma(n)}{\Gamma(n+\frac{1}{2})}\delta_{m,n}\qquad m,n>0
\end{align}\end{subequations}

\begin{singlespace}
\phantomsection\pdfbookmark[-1]{\bibname}\null
\addcontentsline{toc}{chapter}{\bibname}
\bibliography{dissertation}
\end{singlespace}

\end{document}